\documentclass[final]{jfp-epi}
\usepackage{jfp-cup2epi}
\usepackage{mathpartir}
\usepackage{caption}
\usepackage{subcaption}
\usepackage{listings}
\usepackage{wrapfig}
\usepackage{hhline}
\usepackage{etoc}
\usepackage{microtype}

\usepackage{xcolor}
\usepackage{bbold}
\usepackage{xspace}
\usepackage{xparse}
\usepackage{fancyvrb}
\usepackage{tcolorbox}
\usepackage{stmaryrd}
\usepackage{cleveref}
\usepackage{amsmath}
\usepackage{amsthm}
\usepackage{thm-restate}
\usepackage{booktabs, array}

\newtheorem{definition}{Definition}
\newtheorem{corollary}{Corollary}

\newtheorem{lemma}{Lemma}
\newtheorem{theorem}{Theorem}
\newtheorem{proposition}{Proposition}
\usepackage{smaller}

\DeclareMathAlphabet{\mathttbf}{\encodingdefault}{\ttdefault}{bx}{n}

\usepackage{enumitem}
\usepackage{tikz}

\usepgflibrary{arrows}

\newcommand*{\commentpound}{\protect\scalebox{0.9}{\protect\raisebox{0.04em}{\#}}\kern0.2pt
}

\newcommand{\crefrangeconjunction}{--}
\crefname{ln}{line}{lines}

\crefname{id}{\commentpound}{\commentpound}
\crefname{figure}{Figure}{Figures}
\crefname{table}{Table}{Tables}

\newsavebox{\listingbox}

\lstdefinelanguage{Pat}{morekeywords={def, interface, guard, receive, from, free, fail,
  interface, spawn, let, in, if, else},sensitive=t, comment=[l]{\#\ },escapeinside={(*}{*)},morestring=[d]{"},showstringspaces=false
}

\definecolor{Gainsboro}{RGB}{220,220,220}
\definecolor{Gray}{RGB}{127,127,127}

\usepackage{stackengine}
\stackMath

\newcounter{sqindex}

\newcommand{\res}[1]{(\nu #1)}
\newcommand{\resconf}[2]{\res{#1}(#2)}
\newcommand{\tya}{A}
\newcommand{\tyb}{B}

\newcommand{\basety}{C}

\newcommand{\pata}{E}
\newcommand{\patb}{F}
\newcommand{\patc}{G}

\newcommand{\config}[1]{\mathcal{#1}}
\newcommand{\ctx}[1]{\mathcal{#1}}

\newcommand{\tyunit}{\one}
\newcommand{\nil}{\calcwd{nil}}
\newcommand{\cons}[2]{#1 \; \texttt{{::}} \; #2}
\newcommand{\tysum}[2]{#1 + #2}
\newcommand{\tysump}[2]{(#1 + #2)}
\newcommand{\typair}[2]{#1 \times #2}
\newcommand{\typairp}[2]{(#1 \times #2)}
\newcommand{\tylist}[1]{\mkwd{List}(#1)}
\newcommand{\midspace}{\,\mid\,}

\newcommand{\var}[1]{\textit{#1}}
\newcommand{\mkwd}[1]{\textsf{#1}}
\newcommand{\calcwd}[1]{\textbf{\textsf{#1}}}
\newcommand{\caseof}[2]{\calcwd{case} \; #1 \; \calcwd{of} \; \{ #2 \}}
\newcommand{\caselof}[3]{\calcwd{caseL} \; (#1 : #2) \; \calcwd{of} \; \{ #3 \}}
\newcommand{\caselofdecl}[2]{\calcwd{caseL} \; #1 \; \calcwd{of} \; \{ #2 \}}
\newcommand{\caseltwo}[3]{\calcwd{caseL} \; (#1 : #2) \; \calcwd{of} \; \{ }
\newcommand{\caselnotype}[2]{\calcwd{caseL} \; #1 \; \calcwd{of} \; \{ #2 \}}

\newcommand{\inl}[1]{\calcwd{inl} \; #1}
\newcommand{\inr}[1]{\calcwd{inr} \; #1}

\newcommand{\caseofcore}[5]
  {\caseof{#1}{\inl{#2} \mapsto #3; \inr{#4} \mapsto #5}}
\newcommand{\caselofcore}[6]
  {\caselof{#1}{#2}{\nil \mapsto #3; \cons{#4}{#5} \mapsto #6}}
\newcommand{\caselofcoredecl}[5]
  {\caselofdecl{#1}{\nil \mapsto #2; \cons{#3}{#4} \mapsto #5}}
\newcommand{\one}{\mathbf{1}}
\newcommand{\hastype}{\,{:}\,}
\newcommand{\seq}[1]{\overrightarrow{#1}}
\newcommand{\tseq}[4][]{#2 \vdash_{#1} #3 \hastype #4}

\newcommand{\annarrow}{\overset{\mathsf{ann}}{\hookrightarrow}}
\newcommand{\anntmseq}[5][]
    {#2 \vdash_{#1} #3 \hastype #4 \annarrow #5}
\newcommand{\anngseq}[6][]{\jarg{#2} #3 \vdash_{#1} #4 \hastype #5 \annarrow #6}

\newcommand{\annprogseq}[2]{\vdash #1 \annarrow #2}
\newcommand{\anndefseq}[3][]{\vdash_{#1} #2 \annarrow #3}

\newcommand{\elabarrow}{\overset{\mathsf{elab}}{\rightsquigarrow}}

\newcommand{\elabsynth}[5][]{#2 \vdash_{#1} #3 \,\synth\, #4 \elabarrow #5}
\newcommand{\elabgdsynth}[6][]{\jarg{#2} #3 \vdash_{#1} #4 \,\synth\, #5 \elabarrow #6}
\newcommand{\elabchk}[5][]{#2 \vdash_{#1} #3 \,\chk\, #4 \elabarrow #5}

\newcommand{\vseq}[3]{#1 \vdash #2 \hastype #3}
\newcommand{\gseq}[5][]{\jarg{#2} #3 \vdash_{#1} #4 \hastype #5}
\newcommand{\cseq}[3][]{#2 \vdash_{#1} #3}

\newcommand{\msgtag}[1]{{\ensuremath{\color{purple} \mathttbf{#1}}}}
\newcommand{\msg}[2]{\msgtag{#1}\texttt{(}#2\texttt{)}}
\newcommand{\msgg}[2]{#1\texttt{(}#2\texttt{)}}

\newcommand{\kwspawn}{\calcwd{spawn}}
\newcommand{\kwreceive}{\calcwd{receive}}
\newcommand{\send}[3]{#1 \,{!}\, \msg{#2}{#3}}
\newcommand{\receive}[4]{\kwreceive \; \msg{#1}{#2} \; \calcwd{from} \; #3 \mapsto #4}
\newcommand{\receivethree}[3]{\kwreceive \; \msg{#1}{#2} \; \calcwd{from} \; #3 \mapsto}

\newcommand{\spawn}[1]{\kwspawn\;#1}

\newcommand{\kwfree}{\calcwd{free}}
\newcommand{\kwfail}{\calcwd{fail}}
\newcommand{\sugarfree}[1]{\kwfree\;#1}
\newcommand{\free}[1]{\kwfree \mapsto #1}
\newcommand{\sugarfail}[1]{\kwfail\;#1}
\newcommand{\sugarfailiface}[2][\iface]{\kwfail^{#1}\;#2}
\newcommand{\mbnew}{\calcwd{new}\xspace}
\newcommand{\guard}[2]{\calcwd{guard} \; #1 \; \{ #2 \}}
\newcommand{\guardone}[1]{\calcwd{guard} \; #1 \; \{}

\newcommand{\guardann}[3]{\calcwd{guard} \, #1 \, {:} \, #2 \, \{ #3 \}}
\newcommand{\guardifaceann}[4][\iface]{\calcwd{guard}^{#1} \, #2 \, {:} \, #3 \, \{ #4 \}}
\newcommand{\guardanntwo}[2]{\calcwd{guard} \, #1 \, {:} \, #2 \, \{}
\newcommand{\sentmsg}[3]{{#1}\leftarrow\msg{#2}{#3}}
\newcommand{\sentmsgg}[3]{{#1}\leftarrow\msgg{#2}{#3}}

\newcommand{\tyenv}{\Gamma}
\newcommand{\plainenv}{\Delta}
\newcommand{\tyenvb}{\Pi} 

\newcommand{\un}[1]{\mkwd{un}(#1)}

\newcommand{\mbone}{\mathbb{1}}
\newcommand{\mbzero}{\mathbb{0}}

\newcommand{\tysend}[1]{\texttt{!}#1}
\newcommand{\tyrecv}[1]{\texttt{?}#1}
\newcommand{\tysendp}[1]{\texttt{!}(#1)}
\newcommand{\tyrecvp}[1]{\texttt{?}(#1)}
\newcommand{\tysendm}{\tysend{\msgtag{m}}}

\newtcolorbox{fixmebox}{colback=black!5!white,
colframe=black!75!black,fonttitle=\small\bfseries,
size=small,sharp corners=all,title=FIXME}

\newtcolorbox{todobox}{colback=red!5!white,
colframe=red!75!black,fonttitle=\small\bfseries,
size=small,sharp corners=all,title=TODO}

\newtcolorbox{notebox}{colback=blue!5!white,
colframe=blue!75!black,fonttitle=\small\bfseries,
size=small,sharp corners=all,title=NOTE}

\newcommand{\pnf}[1]{\vDash #1}
\newcommand{\pnftwo}[2]{#1 \vDash #2}
\newcommand{\pnflittwo}[2]{#1 \vDash_{\mkwd{lit}} #2}
\newcommand{\defeq}{\triangleq}

\newcommand{\totheleft}[1]{\begin{flushleft}#1\end{flushleft}}

\newcommand{\header}[1]{
  \begin{flushleft}
    \textbf{#1}
  \end{flushleft}
}
\newcommand{\headersig}[2]{
  \begin{flushleft}
    \textbf{#1} \hfill \framebox{#2}
  \end{flushleft}
}

\newcommand{\headertwo}[2]{
  \begin{flushleft}
    \textbf{#1} \hfill #2
  \end{flushleft}
}

\newcommand{\ceval}{\longrightarrow}
\newcommand{\cevalp}{\,\longrightarrow_{\mathcal{P}}\,}
\newcommand{\cevalpnopad}{\longrightarrow_{\mathcal{P}}}
\newcommand{\cevalnopad}{\longrightarrow}
\newcommand{\prog}{\mathcal{P}\xspace}
\newcommand{\pprog}{\withann{\mathcal{P}}}
\newcommand{\progdef}[3]{(#1, #2, #3)}

\newcommand{\progseq}[1]{\vdash #1}
\newcommand{\defseq}[2][]{\vdash_{#1} #2}

\newcommand{\dom}[1]{\mkwd{dom}(#1)}

\newcommand{\many}[1]{{#1^{\star}}}

\newcommand{\bl}{\begin{array}{l}}
\newcommand{\blt}{\begin{array}[t]{l}}
\newcommand{\el}{\end{array}}

\newcommand{\mbrecvone}{\tyrecv{\mbone}}
\newcommand{\mbsendone}{\tysend{\mbone}}
\newcommand{\mbrecvzero}{\tyrecv{\mbzero}}
\newcommand{\mbsendzero}{\tysend{\mbzero}}
\newcommand{\deriv}[1]{\mathbf{#1}}
\newcommand{\derivd}{\deriv{D}}

\newcommand{\sem}[1]{\llbracket#1\rrbracket}
\newcommand{\multiset}[1]{\langle #1 \rangle}
\newcommand{\mseta}{\textsf{A}}
\newcommand{\msetb}{\textsf{B}}

\newcommand{\incll}[1][]{\sqsubseteq_{#1}}

\newcommand{\subpat}[2]{#1 \incll[] #2}
\newcommand{\subpatone}{\sqsubseteq}

\newcommand{\subpatconstr}[2]{#1 \mathop{\mathtt{<:}} #2}

\newcommand{\subusage}{\leq}
\newcommand{\subtype}{\leq}
\newcommand{\subtypetwo}[2]{#1 \leq #2}

\newcommand{\tyequiv}{\simeq}

\newcommand{\tyequivtwo}[2]{ #1 \tyequiv #2}
\newcommand{\envwithout}{\mathop{-}}
\newcommand{\envwithouttwo}[2]{#1 \envwithout #2}
\newcommand{\without}{\mathop{/}}
\newcommand{\defmsg}{\algomsg{m}}

\newcommand{\nma}{u}
\newcommand{\nmb}{v}
\newcommand{\nmc}{w}

\newcommand{\fn}[1]{\mkwd{fn}(#1)}
\newcommand{\fv}[1]{\mkwd{fv}(#1)}

\NewDocumentCommand{\letinone}{o m}{\calcwd{let}\: #2 {\IfValueT{#1}{{: #1}}} =
}
\NewDocumentCommand{\letintwo}{o m m}{\calcwd{let}\: #2 {\IfValueT{#1}{{: #1}}} = #3 \: \calcwd{in}}
\NewDocumentCommand{\letin}{o m m m}{\calcwd{let}\: #2 {\IfValueT{#1}{{: #1}}} = #3 \: \calcwd{in} \: #4
}

\newcommand{\patplus}{\oplus}
\newcommand{\patplustwo}[2]{#1 \patplus #2}
\newcommand{\patconcat}{\odot}
\newcommand{\patconcattwo}[2]{#1 \patconcat #2}

\newcommand{\set}[1]{\{ #1 \}}

\newcommand{\pcomb}{\bowtie}
\newcommand{\pcombtwo}[2]{#1 \pcomb #2}
\newcommand{\pluscombtwo}[2]{#1 + #2}

\newcommand{\hastypetwo}[2]{#1 \mathop{:} #2}

\newcommand{\scomb}{\triangleright}
\newcommand{\scombtwo}[2]{#1 \scomb #2}

\newcommand{\tyint}{\mkwd{Int}}
\newcommand{\tystring}{\mkwd{String}}
\newcommand{\tyfun}[3][]{#2 \xrightarrow{#1} #3}
\newcommand{\wcirc}{\circ}
\newcommand{\bcirc}{\bullet}
\newcommand{\linear}{\square}
\newcommand{\unrestricted}{\blacksquare}
\newcommand{\tyannfun}[2]{\tyfun[\linann]{#1}{#2}}
\newcommand{\tylinfun}[2]{\tyfun[\linear]{#1}{#2}}
\newcommand{\tyunrfun}[2]{\tyfun[\unrestricted]{#1}{#2}}

\newcommand{\linann}{\diamond}
\newcommand{\patvara}{\alpha}
\newcommand{\patvarb}{\beta}

\newcommand{\envjoin}{\fatsemi}
\newcommand{\envmerge}{\sqcap}

\newcommand{\bidirsep}{\mathop{\mbox{\smaller$\blacktriangleright$}}}

\newcommand{\joinseq}[4]{#1 \envjoin #2 \,\bidirsep\, #3; #4}

\newcommand{\mergeseq}[4]{#1 \envmerge #2 \,\bidirsep\, #3; #4}
\newcommand{\manymergeseq}[4]{#1 \envmerge \cdots \envmerge #2 \,\bidirsep\, #3; #4}

\newcommand{\combineseq}[4]{#1 + #2 \,\bidirsep\, #3; #4}
\newcommand{\combineseqmany}[3]{#1 \,\bidirsep\, #2; #3}

\newcommand{\constrs}{\Phi}
\newcommand{\constr}{\phi}

\newcommand{\checkenvseq}[4]{\mkwd{check}(#1, #2, #3) = #4}

\newcommand{\flaglambda}[4][\linann]{\lambda^{\!#1} (#2){:}\, #3 \,.\,#4}
\newcommand{\unlambda}[3]{\flaglambda[\unrestricted]{#1}{#2}{#3}}
\newcommand{\linlambda}[3]{\flaglambda[\linear]{#1}{#2}{#3}}

\newcommand{\pmbty}{\varsigma}

\newcommand{\ppretya}{\pi}
\newcommand{\ppretyb}{\rho}
\newcommand{\ptya}{\tau}
\newcommand{\ptyb}{\sigma}
\newcommand{\ppata}{\gamma}
\newcommand{\ppatb}{\delta}
\newcommand{\penv}{\Theta}

\newcommand{\erase}[1]{\mkwd{erase}(#1)}

\newcommand{\mergesol}[3]{#1 \mathop{\approx} { #2 \mathop{\hat{\envmerge}} #3}}
\newcommand{\apppatsubst}[2][\patsubst]{#1(#2)}
\newcommand{\patsubst}{\Xi}

\newenvironment{proofcase}[1]
  {\totheleft{\textit{Case \textsc{#1}}}}
  {}
\newenvironment{proofcase*}[1]
  {\totheleft{\textit{Case} #1}}
  {}
\newenvironment{subcase}[1]
  {\totheleft{\textit{Subcase #1}}}
  {}

\newcommand{\sigs}{\mathcal{S}}
\newcommand{\psigs}{\withann{\mathcal{S}}}
\newcommand{\proginterfaces}{\mathcal{I}}

\newcommand{\siglookup}[2][\prog]{#1(\algomsg{#2})}

\newcommand{\proglookup}[2][\prog]{#1(#2)}
\newcommand{\pproglookup}{\proglookup[\pprog]}

\newcommand{\jarg}[1]{\{ #1 \} \;}
\newcommand{\synthseq}[5][]{#2 \;\synth_{#1}\; #3 \,\bidirsep\, #4;\, #5}
\newcommand{\chkseq}[5][]{#2 \;\chk_{#1} \; #3 \,\bidirsep\, #4;\, #5}
\newcommand{\chkgseq}[7][]{\jarg{#2} #3 \,\chk_{#1}\, #4 \,\bidirsep\, #5;\, #6;\, #7}
\newcommand{\chkgseqcon}[8][]{\jarg{#2; #3} #4 \,\chk_{#1}\, #5 \,\bidirsep\, #6;\, #7;\, #8}
\definecolor{dRed}{rgb}{0.85, 0.0, 0.0}

\newcommand{\algodefseq}[3][\pprog]{\vdash_{#1} #2 \,\bidirsep\, #3}
\newcommand{\algoprogseq}[2]{\vdash #1 \,\bidirsep\, #2}

\definecolor{dBlue}{rgb}{0.0, 0.0, 0.65}
\definecolor{dGreen}{rgb}{0.0, 0.55, 0.65}
\newcommand{\chkcol}{dBlue}
\newcommand{\synthcol}{dRed}

\newcommand{\synth}{{\color{\synthcol} \Rightarrow}}
\newcommand{\chk}{{\color{\chkcol} \Leftarrow}}

\newcommand{\subtyseq}[3]{\subtypetwo{#1}{#2} \, \bidirsep \, #3}
\newcommand{\unrseq}[2]{\mkwd{unr}(#1) \,\bidirsep\, #2}
\newcommand{\equivsymb}{\sim}
\newcommand
    {\equivseq}[3]
    {#1 \equivsymb #2 \, \bidirsep \, #3}

\newcommand{\algomsg}{\msgtag}

\newcommand{\pv}[1]{\mkwd{pv}(#1)}
\newcommand{\aps}{\apppatsubst}

\newcommand{\tmc}{L}
\newcommand{\tma}{M}
\newcommand{\tmb}{N}

\newcommand{\withann}{\widehat}
\newcommand{\ptma}{\withann{\tma}}
\newcommand{\ptmb}{\withann{\tmb}}
\newcommand{\pvala}{\withann{\vala}}
\newcommand{\pvalb}{\withann{\valb}}

\definecolor{shade}{RGB}{195,195,195}
\definecolor{shadered}{RGB}{220,50,50}
\definecolor{shadeblue}{RGB}{100,100,220}
\newcommand{\shade}[1]{{\setlength{\fboxsep}{0pt}\colorbox{shade}{\ensuremath{#1}}}}

\newcommand{\gd}{G}
\newcommand{\pgd}{\withann{\gd}}

\newcommand{\pretya}{T}
\newcommand{\pretyb}{U}
\newcommand{\usage}{\eta}
\newcommand{\usable}{\wcirc}
\newcommand{\returnable}{{\bcirc}}
\newcommand{\usageann}[2]{#2^{#1}}

\newcommand{\usagecombtwo}[2]{{#1} \scomb {#2}}
\newcommand{\mbcomb}{\mathop{\mbox{\smaller$\boxplus$}}}
\newcommand{\mbcombtwo}[2]{#1 \mbcomb #2}
\newcommand{\mbtya}{J}
\newcommand{\mbtyb}{K}

\newcommand{\retty}[1]{\usageann{\returnable}{#1}}
\newcommand{\usablety}[1]{\usageann{\usable}{#1}}

\newcommand{\usagety}[1]{\usageann{\usage}{#1}}

\newcommand{\framestack}{\Sigma}
\newcommand{\emptystack}{\epsilon}
\newcommand{\mvframe}{\sigma}
\newcommand{\fframe}[2]{\langle #1, #2 \rangle}

\newcommand{\frameconcat}[2]{#1 \cdot #2}

\newcommand{\thread}[2]{\llparenthesis\, #1, #2 \,\rrparenthesis}
\newcommand{\returnablepred}[1]{\mkwd{returnable}(#1)}

\newcommand{\makereturnable}[1]{\lfloor #1 \rfloor}
\newcommand{\makeusable}[1]{\lceil #1 \rceil}

\newcommand{\vala}{V}
\newcommand{\valb}{W}

\newcommand{\fndef}[4]{\calcwd{def} \: #1(#2){:}\: #3 \: \{ #4 \}}
\newcommand{\fndefthree}[3]{\calcwd{def} \: #1(#2){:}\: #3 \: \{}
\newcommand{\fnapp}[2]{#1(#2)}
\newcommand{\pdef}{\withann{D}}

\newcommand{\stackseq}[3]{#1 \vdash #2 \,\bidirsep\, #3}

\newcommand{\strip}[1]{|{#1}|}

\newcommand{\minusage}[2]{\mkwd{min}(#1, #2)}
\newcommand{\langname}{\ensuremath{\mkwd{Pat}}\xspace}
\newcommand{\secref}[1]{\S\ref{#1}}
\newcommand{\secrefp}[1]{(\secref{#1})}
\newcommand{\irrelevantpred}[1]{\mkwd{irrelevant}(#1)}
\newcommand{\strictsubty}{\preccurlyeq}
\newcommand{\strictsubtytwo}[2]{#1 \preccurlyeq #2}

\newcommand{\cruftpred}[1]{\mkwd{cruft}(#1)}

\newcommand{\then}{\, . \,}
\newcommand{\mbbase}[1]{\mkwd{base}(#1)}
\newcommand{\qqquad}{\quad \qquad}

\newcommand{\tradsynth}[3]{#1 \vdash #2 \,\synth\, #3}
\newcommand{\tradchk}[3]{#1 \vdash #2 \,\chk\, #3}

\newcommand{\waiting}[3]{\mkwd{waiting}(#1, #2, #3)}

\newcommand{\msgset}{\mathcal{M}}

\newcommand{\iface}{I}
\newcommand{\ifacemv}{\iota}
\newcommand{\contya}{R}
\newcommand{\contyb}{S}
\newcommand{\conmb}[1][\iface]{\mkwd{Mailbox}(#1)}
\newcommand{\mbnewiface}[1][\iface]{\calcwd{new}\!\texttt{[}{#1}\texttt{]}\xspace}

\newcommand{\varann}[2]{{#1}^{#2}}

\newcommand{\interfaces}[1]{\mkwd{interfaces}(#1)}
\newcommand{\sendann}[4]{{#1} \,{!}^{#2}\, \msg{#3}{#4}}

\newcommand{\contyenv}{\Omega}
\newcommand{\tysendiface}[2][\iface]{\texttt{!}^{#1}#2}
\newcommand{\tyrecviface}[2][\iface]{\texttt{?}^{#1}#2}

\newcommand{\sig}{\mathcal{S}}

\newcommand*{\Checkmark}{\ensuremath{\checkmark\kern-0.6em\checkmark}}

\newcommand*{\yes}{\ensuremath{\bullet}}

\newcommand*{\no}{\ensuremath{\wcirc}}

\newcommand*{\eg}{\textit{e.g.}}

\newcommand*{\ie}{\textit{i.e.}}

\newcommand{\promote}[1]{{\uparrow{(#1)}}}
\newcommand{\demote}[1]{{\downarrow{(#1)}}}

\newcommand{\subst}[3]{#1 \{ #2 / #3 \}}

\newif\ifhighlight
\highlightfalse

\newcommand{\ADDED}[1]{\ifhighlight
    {\color{blue}#1}\else
    {#1}\fi
}

\setcopyright{none}
\etocdepthtag.toc{mtchapter}
\etocsettagdepth{mtchapter}{subsection}
\etocsettagdepth{mtappendix}{none}
\AtBeginDocument{\providecommand\BibTeX{{\normalfont B\kern-0.5em{\scshape i\kern-0.25em b}\kern-0.8em\TeX}}}

\begin{document}
\lefttitle{Special Delivery: Programming with Mailbox Types (Extended version)}

\title[Special Delivery: Programming with Mailbox Types]{Special Delivery}
\subtitle{Programming with Mailbox Types (Extended version)}
\author{Simon Fowler}
\orcid{0000-0001-5143-5475}
\affiliation{\institution{University of Glasgow}
\country{UK}
\authoremail{Simon.Fowler@glasgow.ac.uk}
}

\author{Duncan Paul Attard}
\orcid{0000-0002-2448-5394}
\affiliation{\institution{University of Malta}
\country{Malta}
\authoremail{Duncan.Attard@um.edu.mt}
}

\author{Danielle Marshall}
\orcid{0000-0002-4284-3757}
\affiliation{\institution{Royal Holloway, University of London}
\country{UK}
\authoremail{Danielle.Marshall@rhul.ac.uk}
}

\author{Simon J. Gay}
\orcid{0000-0003-3033-9091}
\affiliation{\institution{University of Glasgow}
\country{UK}
\authoremail{Simon.Gay@glasgow.ac.uk}
}

\author{Phil Trinder}
\orcid{0000-0003-0190-7010}
\affiliation{\institution{University of Glasgow}
\country{UK}
\authoremail{Phil.Trinder@glasgow.ac.uk}
}

\begin{abstract}
The asynchronous and unidirectional communication model supported by
\emph{mailboxes} is a key reason for the success of actor languages like Erlang
and Elixir for implementing reliable and scalable distributed systems.
Although actors eliminate many of the issues stemming from shared memory
concurrency, they remain vulnerable to communication errors such as protocol
violations and deadlocks. Behavioural types make it possible to detect
communication errors early in the development process, but most work has
addressed channel-based languages rather than actor languages.

\emph{Mailbox types} are a recent behavioural type system for actors
first introduced for a process calculus by de'Liguoro and Padovani in 2018,
and capture the contents of a mailbox as a commutative regular expression.
Due to aliasing and nested
evaluation contexts, moving from a process calculus to a programming language is
challenging.
This paper presents \langname, the first programming language design
incorporating mailbox types, and describes its algorithmic type system.
\langname\ is a higher-order functional language with sums, products, and lists,
along with interfaces that allow fine-grained reasoning about mailbox contents.
Typechecking in \langname\ detects four classes of behavioural
error: protocol violation, unexpected message, forgotten reply, and
self-deadlock, as well as the usual data type errors.

\langname's type system makes essential use of quasi-linear typing to tame
the challenges that aliasing poses for static typing. We introduce
a \emph{co-contextual} algorithmic type system, based around a novel use
of backwards bidirectional typing, and prove the algorithmic type system
sound and complete with respect to the declarative type system.
We implement a mailbox type checker, and use it to demonstrate the
expressiveness of \langname\ on a factory automation case study
and a series of 11 examples from the Savina actor benchmark suite.
These examples show that \langname\ matches the expressiveness of previous work on
dynamically checked session types for actors, and our experimental results show
that typechecking times are below 100ms.
Our work establishes a foundation for applying mailbox typing to practical actor
languages such as Erlang, Elixir and Scala/Akka.
\end{abstract}

\maketitle

\section{Introduction}\label{sec:introduction}

Software is increasingly concurrent and distributed, but coordinating
concurrent computations introduces a host of additional correctness
issues like communication mismatches and deadlocks. Communication-centric
languages such as Go, Erlang, and Elixir make it possible to avoid
many of the issues stemming from shared memory concurrency by
structuring applications as lightweight processes that
communicate through explicit message passing. There are two main
classes of communication-centric language. In \emph{channel-based}
languages like Go, processes communicate over channels, where a
\lstinline+send+ in one process is paired with a \lstinline[language=]+receive+
in the recipient
process. In \emph{actor} languages like Erlang or Elixir, a message is
sent to the \emph{mailbox} of the recipient process.
A mailbox is an incoming message queue; in certain actor languages, the
recipient can select which message from the mailbox to handle next.

Although communication-centric languages eliminate many concurrency 
issues, some remain. For example, a process may still receive a message
that it is not equipped to handle, or wait for a message that it will
never receive.  Such communication errors often occur sporadically and
unpredictably after deployment, making them difficult to locate and
fix.

\emph{Behavioural type systems} \citep{DBLP:journals/csur/HuttelLVCCDMPRT16}
encode correct communication behaviour to support \emph{correct-by-construction}
concurrency. Behavioural type systems, in particular \emph{session types}
\citep{DBLP:conf/concur/Honda93, DBLP:conf/parle/TakeuchiHK94, HondaVK98, GayV25},
have been extensively applied to specify communication protocols in
channel-based languages \citep{DBLP:journals/ftpl/AnconaBB0CDGGGH16}.
There has, however, been far less application of behavioural typing to actor
languages. Existing work either imposes restrictions on the actor model to
retrofit session types \citep{MostrousV11, TaboneF21, TaboneF22, HFDG21,
FowlerH26} or
relies on dynamic typing \citep{NeykovaY16}.
\ADDED{In general, session types are built around the idea of ordered
communication channels. Channels are a significantly different communication model
to the many-to-one unordered communication model supported by mailboxes, and
therefore using session types in an actor language typically requires users to
rewrite their applications.}
We discuss these systems further in \secref{sec:related}.

Our approach is based on
\emph{mailbox types}, a behavioural type system for actor mailboxes first
introduced in the context of a process calculus~\citep{dP18:mbc}. We present the first
programming language design incorporating mailbox types and we detail an
algorithmic type system, an implementation, and a range of benchmarks and a
factory case study. Due to aliasing and nested evaluation contexts, the move
from a process calculus to a programming language is challenging. We make
essential and novel use of quasi-linear typing~\citep{Kobayashi99:quasilinear,
EnnalsSM04} to tame some of the complexity introduced by aliasing, and our
algorithmic type system is co-contextual~\citep{ErdwegBKKM15,
KuciEBBM17:cocontextual}, achieved through a novel use of backwards
bidirectional typing~\citep{Zeilberger18:bbt}.

\subsection{Actor languages: easy to distribute, difficult to type}
\begin{figure}
    \centering
    \begin{subfigure}{0.27\textwidth}
        \centering
        \includegraphics[width=0.8\textwidth]{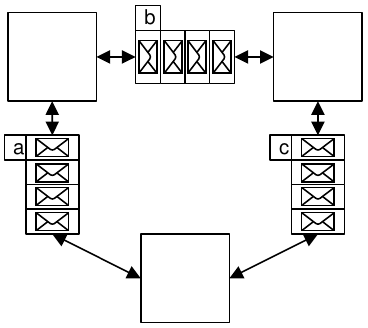}
        \caption{Channels}
        \label{fig:chans-actors:channels}
    \end{subfigure}
    \qquad\quad
    \begin{subfigure}{0.27\textwidth}
        \centering
        \includegraphics[width=0.8\textwidth]{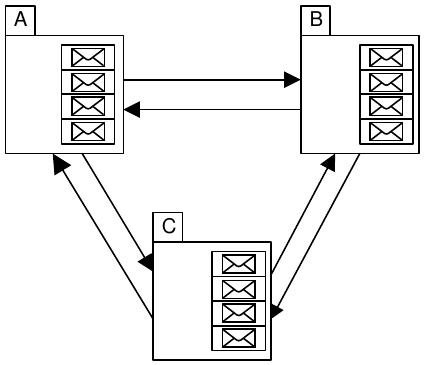}
        \caption{Actors}
        \label{fig:chans-actors:actors}
    \end{subfigure}
\caption{Channel- and actor-based communication~\citep{FowlerLW17:acca}}
    \label{fig:chans-actors}
\end{figure}

Channel-based languages support anonymous
processes that communicate over named channels, whereas actor
languages provide named processes each equipped with a mailbox.
Figure~\ref{fig:chans-actors} contrasts the approaches, and is taken from
a detailed comparison~\citep{FowlerLW17:acca}.

Actor languages have proven to be effective for implementing
reliable and scalable distributed systems~\citep{trinder2017scaling}.
Communication in actor languages is asynchronous and
unidirectional: many actors may \emph{send} messages to an actor $A$, whereas
only $A$ may \emph{receive} from its mailbox.
Mailboxes provide \emph{data locality} as each message
is stored with the process that will handle it.
In channel-based languages, since it is possible to send channel names over
other channels, such languages must either sacrifice locality and reduce
performance, or rely on complex distributed algorithms~\citep{HuYH08:sj,
Chaudhuri09:dist-cml}.

Although it is straightforward to add a type system to a channel-based language,
adding a type system to an actor language is more difficult, as process
names (process IDs or PIDs) must be parameterised by a type that supports all
messages that can be received. The type is therefore less precise, requiring
subtyping~\citep{HeWT14:takka} or
synchronisation~\citep{TasharofiDJ13:actors-plus, BoerCJ07} to avoid a total loss
of modularity~\citep{FowlerLW17:acca}.

The situation becomes even more pronounced when considering behavioural type
systems: communication errors might be prevented in channel-based languages by
giving one end of a channel the session type
${!}\mkwd{Int}.{!}\mkwd{Int}.{?}\mkwd{Bool}.\mkwd{End}$ (send two integers, and
receive a Boolean), and the other end the \emph{dual} type
${?}\mkwd{Int}.{?}\mkwd{Int}.{!}\mkwd{Bool}.\mkwd{End}$. Behavioural type
systems for actor languages are much less straightforward due to the asymmetric
communication model provided by mailboxes.
In practice, designers of session type systems for actor languages either
emulate session-typed channels~\citep{MostrousV11}, or use \emph{multiparty
session types} to govern the communication actions performed by a
process~\citep{FowlerH26, NeykovaY16, HFDG21}, requiring a developer to rewrite
their application to structure its communication using session types.

\subsection{Mailbox types}\label{sec:mailbox_types}
\citet{dP18:mbc} observe that session types require a \emph{strict ordering}
of messages, whereas most actor systems use \emph{selective receive} to
process messages out-of-order.
Concentrating on \emph{unordered} interactions enables behavioural typing for
mailboxes with many senders.

\paragraph*{Mailbox typing by example: a future variable.}
Rather than reasoning about the \emph{behaviour of a process}, mailbox types
reason about the \emph{contents of a mailbox}. Consider a \emph{future
variable}, which is a placeholder for the result of a concurrent computation. A
future can receive many \lstinline+get+ messages that are only fulfilled
\emph{after} a \lstinline+put+ message initialises the future with a value.
After the future is initialised, it fulfils all \lstinline+get+ messages by
sending its value; a second \lstinline+put+ message is explicitly disallowed.
We can implement a future straightforwardly in Erlang:

\noindent
\begin{minipage}{0.43\textwidth}
\begin{lstlisting}[language=erlang, morekeywords={reply}]
-module future.
empty_future() ->
  receive
    { put, X } -> full_future(X)
  end.
full_future(X) ->
  receive
    { get, Pid } ->
      Pid ! { reply, X },
      full_future(X);
    { put, _ } ->
      erlang:error("Multiple writes")
  end.
\end{lstlisting}
\end{minipage}
\hfill
\begin{minipage}{0.5\textwidth}
\begin{lstlisting}[language=erlang,firstnumber=14,morekeywords={reply}]
client() ->
  Future = spawn(future, empty_future, []),
  Future ! { put, 5 },
  Future ! { get, self() },
  receive
    { reply, Result } ->
      io:fwrite("~w~n", [Result])
  end.
\end{lstlisting}
\end{minipage}

The \lstinline{empty_future} function awaits a \lstinline{put} message to set
the value of the future (lines 3--5), and transitions to the \lstinline{full_future} state.
A \lstinline{full_future} receives \lstinline{get} messages (lines 7--13) containing a process
ID (PID) used to reply with the future's value.
The \lstinline{client} function spawns a future (line 15), sends a
\lstinline{put} message followed by a \lstinline{get} message (lines 16--17),
and awaits the result (lines 18--21). The program prints the number 5.

Several communication errors can arise in this example:

\begin{itemize}
    \item \textbf{Protocol violation.}
        Receiving two \lstinline{put} messages will result in a \emph{runtime}
        error.
\item \textbf{Unexpected message.}
        Sending a message other than \lstinline{get} or \lstinline{put} to the
        future will silently succeed, but the message will never be retrieved, resulting in a memory leak.
\item \textbf{Forgotten reply.}
        If the future fails to send a \lstinline{reply} message after
        receiving a \lstinline{get} message, the client will be left
        waiting forever.
\item \textbf{Self-deadlock.}
        If the client attempts to receive a \lstinline{reply} message before
        sending a \lstinline{get} message, it will be left waiting forever.
\end{itemize}

\noindent
All of the above issues can be solved by mailbox typing. We can write the
following types:
{\small
\[
    \begin{array}{rcl}
        \mkwd{EmptyFuture} & \defeq &
            \tyrecvp{\msg{Put}{\mkwd{Int}} \patconcat
                \many{\msg{Get}{\mkwd{ClientSend}}}}\\
\mkwd{FullFuture} & \defeq & \tyrecv{\many{\msg{Get}{\mkwd{ClientSend}}}} \\
\mkwd{ClientSend} & \defeq & \tysend{\msg{Reply}{\mkwd{Int}}} \\
        \mkwd{ClientRecv} & \defeq & \tyrecv{\msg{Reply}{\mkwd{Int}}} \\
        \end{array}
\]
}

A mailbox type combines a \emph{capability} (either $\texttt{!}$ for an output
capability, analogous to a PID in Erlang; or $\texttt{?}$ for
an input capability) with a \emph{pattern}. A pattern is a \emph{commutative
regular expression}: in the context of a send mailbox type, the pattern will
describe the messages that must be sent; in the context of a receive mailbox
type, it describes the messages that the mailbox may contain.

A mailbox name (\eg, \lstinline{Future}) may have
different types at different points in the program.
$\mkwd{EmptyFuture}$ types an input capability of an empty future mailbox,
and denotes that the mailbox
may contain a single $\msgtag{Put}$ message with an $\mkwd{Int}$ payload, and
potentially many ($\many{}$) $\msgtag{Get}$ messages each with a $\mkwd{ClientSend}$ payload.
$\mkwd{FullFuture}$ types an input capability of the future after a
$\msgtag{Put}$ message has been received, and requires that the mailbox
\emph{only} contains $\msgtag{Get}$ messages.
$\mkwd{ClientSend}$ is an output mailbox type which requires that a
$\msgtag{Reply}$ message must be sent; $\mkwd{ClientRecv}$ is an input
capability for receiving the $\msgtag{Reply}$.
For each mailbox name, sends and receives must ``balance out'': if a message is
sent, it must eventually be received.

\citet{dP18:mbc} introduce a small extension of the asynchronous
$\pi$-calculus~\citep{AmadioCS98}, which they call the \emph{mailbox calculus}, and
endow it with mailbox types. They express a future in
the mailbox calculus as follows, where the mailbox is denoted $\var{self}$.

{\small
\[
\bl
{
\begin{array}{rcl}
    \mkwd{emptyFuture}(\var{self}) & \defeq &
    \var{self} \mathop{\tt ?} \msgtag{Put}(\var{x}) \then \mkwd{fullFuture}(\var{self}, \var{x}) \\
    \mkwd{fullFuture}(\var{self}, \var{x}) & \defeq & \calcwd{free}\; \var{self} \then \calcwd{done} \\
                                        & +      &
                                        \var{self} \mathop{\tt ?} \msgtag{Get}(\var{sender}) \then
                                        (\send{\var{sender}}{Reply}{\var{x}} \parallel \mkwd{fullFuture}(\var{self}, \var{x}))
                                        \\
                                        & +      & \var{self} \mathop{\tt ?} \msgtag{Put}(\var{x}) \then 
                                        \calcwd{fail} \; \var{self}
\end{array}
} \vspace{0.5em} \\
{\begin{array}{l@{\hspace {0em}}l}
        \resconf{\var{future}}{& \mkwd{emptyFuture}(\var{future}) \parallel
        \send{\var{future}}{Put}{5} \parallel \\
       & \resconf{\var{self}}{\send{\var{future}}{Get}{\var{self}} \parallel
    (\var{self}\mathop{\tt ?}\msgtag{Reply}(\var{x}) \then \calcwd{free}\; \var{self} \then
    \mkwd{print}(\mkwd{intToString}(\var{x})}}))
\end{array}
}
\el
\]
}

A process calculus is useful for expressing the essence of concurrent
computation, but there is a large gap between a process calculus and a
programming language design. The biggest issue is the separation of static
and dynamic terms: a programming language specifies the \emph{program} that a
user writes, whereas a process calculus provides a snapshot of the system at a
given time.
A particular difference comes with name generation: in a process calculus, we
can write name restrictions directly; in a programming language, we
instead have a language construct (like $\mbnew$) that is evaluated
to create a fresh name at runtime.
Further complexities come with nested evaluation contexts, sequential
evaluation, and aliasing. We explore these challenges in greater detail
in~\secref{sec:whatstheissue}.

We propose \langname\footnote{\url{https://en.wikipedia.org/wiki/Postman_Pat}},
a functional programming language that supports mailbox types, in which we
express the \emph{future} example as follows ($\var{self}$ is again the
future's mailbox):

{\small
\begin{minipage}{0.5\textwidth}
\[
\bl
    \fndefthree{\mkwd{emptyFuture}}{\var{self}: \mkwd{EmptyFuture}}{\one} \\
        \quad \guardanntwo{\var{self}}{\msgtag{Put} \patconcat \many{\msgtag{Get}}} \\
        \qquad
            \receivethree{Put}{x}{\var{self}} \\
            \qqquad {\mkwd{fullFuture}(\var{self}, \var{x})} \\
        \quad \} \\
    \}\vspace{0.5em}
    \\
    \fndefthree{\mkwd{fullFuture}}{\var{self}: \mkwd{FullFuture}, \var{value}: \mkwd{Int}}{\one} \\
        \quad \guardanntwo{\var{self}}{\many{\msgtag{Get}}} \\
        \qquad \free{()} \\
        \qquad
            \receivethree{Get}{\var{user}}{\var{self}} \\
        \qquad \quad \send{\var{user}}{Reply}{\var{value}}; \\
        \qquad \quad \mkwd{fullFuture}(\var{self}, \var{value}) \\
        \quad \} \\
    \} \\
    \el
\]
\vspace{0.5em}
\end{minipage}
\hfill
\begin{minipage}{0.45\textwidth}
\[
    \bl
    \fndefthree{\mkwd{client}}{}{\one} \\
        \quad \letintwo{\var{future}}{\mbnew} \\
        \quad \spawn{\mkwd{emptyFuture}(\var{future})}; \\
        \quad \letintwo{\var{self}}{\mbnew} \\
        \quad \send{\var{future}}{Put}{5}; \\
        \quad \send{\var{future}}{Get}{\var{self}}; \\
        \quad \guardanntwo{\var{self}}{\msgtag{Reply}} \\
        \qquad \receivethree{Reply}{\var{result}}{\var{self}} \\
        \qquad \quad \sugarfree{\var{self}};\\
        \qquad \quad \mkwd{print}(\mkwd{intToString}(\var{result})) \\
        \quad \} \\
    \}
    \el
\]
\end{minipage}
}

The {\langname} program has a similar structure to the Erlang example with
$\mkwd{client}$, \mkwd{emptyFuture} and \mkwd{fullFuture} functions, and the
mailbox types are similar to those in the mailbox calculus specification. There
are, however, some differences compared with the Erlang future. The main
difference is that in {\langname} mailboxes are first-class: we create a new
mailbox with $\mbnew$, and receive from it using the $\calcwd{guard}$
expression. A \calcwd{guard} expression acts on a mailbox
and may contain two kinds of clause: $\free{\tma}$ frees the mailbox if it
contains no messages and there are no other references to it, and evaluates
$\tma$; and $\receive{m}{\seq{x}}{\var{y}}{\tma}$ retrieves a message with tag
$\msgtag{m}$ from the mailbox, binding its payloads to $\seq{x}$ and re-binding
the mailbox variable (with an updated type) to $\var{y}$ in continuation $\tma$.
In the above code, $\sugarfree{\var{self}}$ is syntactic sugar
(see~\secref{sec:declarative}).

{\langname} has all of the key characteristics of a programming
language, unlike the mailbox calculus. Static and dynamic terms are
distinguished, \ie, we \emph{do not} need to write name restrictions with dynamic
names known \emph{a priori}. \langname is an expression-based language that
supports $\calcwd{let}$-bindings, which enable sequential composition and
nested evaluation contexts, along with data types and higher-order
functions.

\paragraph*{Scope of guarantees.}
Crucially \emph{all of the classes of concurrency error described earlier
(protocol violations, unexpected messages, forgotten replies, and
self-deadlocks) result in type errors}.
As with most behavioural type systems for higher-order functional languages,
ruling out inter-process deadlocks requires additional machinery such as
priorities~\cite{Kobayashi06} or restricting communication topologies to make
cycles impossible to construct~\cite{Wadler14, LindleyM15}.
The mailbox calculus~\cite{dP18:mbc} achieves
deadlock-freedom through dependency graphs, which cannot be applied in a
language with aliasing and nested evaluation~\secrefp{sec:dgs}, so we leave a
full study of inter-process deadlock detection for \langname to future work.

\subsection{Contributions}
Despite being a convincing proposal for behavioural typing for actor languages,
mailbox typing has received little attention since its introduction in 2018.
The overarching contribution of this paper, therefore, is the first design and
implementation of a concurrent programming language with support for mailbox
types.  Concretely, we make four main contributions:

\begin{enumerate}
    \item We introduce a declarative type system for
        \langname~\secrefp{sec:declarative}, a functional programming language
        supporting mailbox types, making essential and novel use of quasi-linear
        types.
We show type preservation, mailbox conformance, and a progress result.
Mailbox conformance ensures that a process will not receive a message
        from a mailbox other than those described by the mailbox's type, and
        rules out protocol violations.
Our progress result shows the absence of self-deadlocks, and also
        implies the absence of unexpected messages and of forgotten replies.
\item We introduce a co-contextual algorithmic type system for
        \langname~\secrefp{sec:algorithmic},
        making use of backwards bidirectional typing. We prove that the
        algorithmic type system is sound and complete with respect to the
        declarative type system.
\item We extend $\langname$ with sum, product, and list types; 
        higher-order functions; and mailbox interfaces~\secrefp{sec:extensions}.
\item We outline our implementation~\secrefp{sec:implementation},
      and demonstrate the expressiveness of \langname by encoding all
      of the examples from~\citet{dP18:mbc}, and all 11 of the
      Savina benchmarks~\citep{ImamS14a:savina} used
      by~\citet{NeykovaY16} in their evaluation of multiparty session types for
      actor languages~\secrefp{sec:implementation:examples}. We also detail
      a larger factory case study.
\end{enumerate}

This paper is a significantly extended and revised version of a paper of the
same name that was published at
ICFP'23~\citep{DBLP:journals/pacmpl/FowlerASGT23}.  
This version includes thoroughly revised formal core languages that remove
unnecessary annotations in the declarative type system and enable a simplified
metatheory for the algorithmic type system.
We also incorporate all definitions
omitted from the original conference paper, add many more examples
throughout, and give more proof details for the main results.
Other highlights include:
\begin{itemize}
    \item Full technical details of the sum and product type extensions~\secrefp{sec:extensions}.
    \item A new extension to allow \langname to express and type
        lists~\secrefp{sec:extensions:lists}, and
        a revised evaluation showing how to use lists for the relevant Savina
        examples~\secrefp{sec:implementation:examples}.
    \item Full declarative typing rules for extensions of \langname with
        first-class functions~\secrefp{sec:extensions:lambdas} and mailbox
        interfaces~\secrefp{sec:extensions:interfaces}, and full details of how
        to use contextual type information to typecheck these
        features~\secrefp{sec:extensions:typechecking-contextual}.
    \item A more complete description of the implementation of the \langname\
        typechecker~\secrefp{sec:implementation_overview}.
    \item A new extensive worked example showing how to express Dijkstra's
            classic \emph{Sleeping Barber} concurrency problem in \langname,
            along with a
            more complete discussion of the implementation of the factory
            case study~\secrefp{sec:implementation:examples}.
\end{itemize}

A yet more comprehensive version~\citep{extended} contains full proofs.
The \langname\ typechecker is available on GitHub
(\url{https://www.github.com/mailbox-types/pat-lang}).

\section{Mailbox types in a programming language: what are the issues?}\label{sec:whatstheissue}
Session typing was originally studied in the context of process calculi
(\eg,~\citep{HondaVK98, Vasconcelos12}), but later work~\citep{GayV10,
Wadler14,FowlerKDLM21} introduced session types for languages based
on the linear $\lambda$-calculus.
In contrast, the more relaxed view of linearity in the mailbox
calculus~\cite{dP18:mbc} makes
language integration far more challenging: a mailbox name may be used
several times to \emph{send} messages, but only once to
\emph{receive} a message. The intuition is that while \emph{sends} simply add
messages to a mailbox, \emph{receives} determine the later behaviour of the
actor. To illustrate, consider the following code that shows part of the future
example from~\secref{sec:introduction}:

    {\small
\[
    \begin{array}{p{1em} l}
            1. & \fndefthree{\mkwd{client}}{}{\one} \\
            2. & \quad \letintwo{\var{future}}{\mbnew} \\
            3. &  \quad \spawn{\mkwd{emptyFuture}(\var{future})}; \\
            4. &  \quad \letintwo{\var{self}}{\mbnew} \\
            5. &  \quad \send{\var{future}}{Put}{5}; \\
            6. &  \quad \send{\var{future}}{Get}{\var{self}}; \\
            7. &  \quad \guardanntwo{\var{self}}{\msgtag{Reply}} \\
            8. & \qquad \receivethree{Reply}{\var{result}}{\var{self}} \\
            9. &  \qquad \quad \sugarfree{\var{self}};\\
            10. & \qquad \quad \mkwd{print}(\mkwd{intToString}(\var{result})) \\
        11. &  \quad \} \\
    12. & \}
            \el
\]
    }
The $\mkwd{client}$ definition uses the $\var{future}$ mailbox twice
to send a message (lines 5 and 6), and similarly uses the $\var{self}$ mailbox
twice: once as a message payload (line 6), and once to receive a message
(line 7).
In the mailbox calculus, a name remains constant and cannot be aliased; this is
at odds with idiomatic programming where expressions are aliased with
$\calcwd{let}$ bindings or function application. Moreover functional languages
provide nested evaluation contexts and sequential evaluation.

\subsection{Challenge: mailbox name aliasing}\label{sec:by-example:aliasing}

Ensuring appropriate mailbox use is challenging in the presence of aliasing.
\ADDED{For example, in the $\mkwd{fullFuture}$ function described earlier,
there is a strong expectation of how resources are used:
the $\var{self}$ mailbox name is consumed by the $\calcwd{guard}$
expression and only re-bound as $\var{self}$ in the $\calcwd{receive}$ clause;
notably the \var{self} variable is \emph{not} available in the $\calcwd{free}$
clause as we would otherwise be able to send to a mailbox after it was deallocated.}

We \ADDED{could try to} write a function that attempts to use a mailbox after it
has been freed:

{\small
\[
    \bl
        \fndefthree{\mkwd{unsafeUse1}}{x: \tyrecv{\many{\msg{Msg}{}}}}{\one} \\
\quad \guardanntwo{x}{\many{\msgtag{Msg}}} \\
            \qquad \receivethree{Msg}{}{z} \\
            \qquad \quad \send{x}{Msg}{}; \\
            \qquad \quad \mkwd{unsafeUse1}(z) \\
            \qquad \kwfree \mapsto \send{x}{Msg}{} \\
            \quad \} \\
        \}
    \el
\]
}

\ADDED{Such} errors cannot arise in a fully linear type system, since it is not
possible to use a resource after it has been consumed.
We could require that a name cannot be used after it has been guarded upon by
insisting that the subject and body of a $\calcwd{guard}$ expression are typable
under disjoint type environments. Indeed, such an approach correctly rules out
the error in the previous example.
Alas, the check can easily be circumvented by introducing an alias $a$ for $x$:

{\small
\[
    \bl
        \fndefthree{\mkwd{unsafeUse2}}{x: \tyrecv{\many{\msg{Msg}{}}}}{\one} \\
            \quad \letintwo{a}{x} \\
            \quad \guardanntwo{a}{\many{\msgtag{Msg}}} \\
            \qquad \receivethree{Msg}{}{z} \\
            \qquad \quad \send{x}{Msg}{}; \\
            \qquad \quad \mkwd{unsafeUse2}(z) \\
            \qquad \kwfree \mapsto \send{x}{Msg}{} \\
            \quad \} \\
        \}
    \el
\]
}

Here, the new name prevents the typechecker from realising that it has been used
in the body of the guard.
Worse, we must also handle nested evaluation contexts,
meaning that the next use of a mailbox variable is not necessarily contained
within a subexpression of the guard:

{\small
\[
    \bl
       \fndefthree{\mkwd{unsafeUse3}}{x: \tyrecv{\many{\msg{Msg}{}}}}{\one} \\
\quad \calcwd{let} \; \_ = \\
            \qquad \guardanntwo{x}{\many{\msgtag{Msg}}} \\
            \quad \qquad \receivethree{Msg}{}{z} \\
            \qquad \qquad \send{x}{Msg}{}; \\
            \qquad \qquad \mkwd{unsafeUse3}(z) \\
            \qquad \quad \kwfree \mapsto \send{x}{Msg}{} \\
            \quad \}
            \; \calcwd{in}
            \; \send{x}{Msg}{} \\
        \}
    \el
\]
}

Much of the intricacy arises from using a mailbox name many times as
an output capability. In each process, we can avoid the name hygiene problems
above using three principles:

\begin{enumerate}
    \item No two distinct variables should represent the same underlying mailbox
        name.
    \item Once let-bound to a different name, a mailbox variable is considered
        out of scope.
    \item A mailbox name cannot be used after it has been used in a
        $\calcwd{guard}$ expression.
\end{enumerate}

These principles ensure syntactic hygiene: the first and second handle the disconnect
between static names and their dynamic counterparts, allowing us to reason that
two syntactically distinct variables indeed refer to different mailboxes.
The third ensures that a mailbox name is correctly `consumed' by a
$\calcwd{guard}$ expression, allowing us to correctly update its type.

\paragraph*{Aliasing through communication.}
We further need to consider the possibility that aliasing is introduced as a
consequence of communication.
Consider the following example, where mailbox $a$ receives the message
$\msg{m}{b}$, and $b$ is already free in the continuation of the
$\calcwd{receive}$ clause:

{\small
\begin{minipage}{0.6\textwidth}
\[
    \begin{array}{rcl}
        \sentmsg{a}{m}{b}
        &
        \parallel
        &
        {
            \bl
                \guardanntwo{a}{\msgtag{m}} \\
                    \quad \receivethree{m}{x}{\var{y}} \\
                    \qquad \send{b}{n}{x}; \\
                    \qquad \sugarfree{y} \\
                \}
            \el
        }
    \end{array}
\]
\end{minipage}
{\Large
$\longrightarrow$
}
\begin{minipage}{0.2\textwidth}
    \[
        \bl
                \qquad \send{b}{n}{b}; \\
                \qquad \sugarfree{a} \\
        \el
    \]
\end{minipage}
}

\ADDED{
In the above example,
$\sentmsg{a}{m}{b}$ refers to a message $\msg{m}{b}$ that has been sent to
mailbox $a$.
}
Here, although the code suggests that $x$ and $b$ are distinct, aliasing is
introduced through communication (violating principle 1).

\subsection{Mailbox calculus solution: dependency graphs}\label{sec:dgs}
The mailbox calculus~\cite{dP18:mbc} uses a \emph{dependency graph} (DG) both to avoid issues
with aliasing and to eliminate cyclic dependencies and hence deadlocks.
\ADDED{In a dependency graph, the vertices are mailbox names, and an edge arises
between two vertices if one name depends on another. Specifically, a dependency
arises between two names $a$ and $b$ if $b$ appears as a payload in a message
addressed to $a$ (meaning that a process must receive from $a$ before being able
to use $b$) or if $b$ appears in the continuation of a process that must first
receive from $a$.
}
As an example, the mailbox calculus process
$(\nu a)(\nu b)(\send{a}{m}{b} \parallel \calcwd{free} \: \var{b} \parallel
\var{a}?\msg{m}{x} \then \calcwd{free} \: \var{a})$ would have DG $(\nu
a)(\nu b)(\set{a, b})$ due to the dependency arising from sending $b$ over $a$.

Alas, a language implementation \emph{cannot} use
this approach as it relies on knowing runtime names directly. To see why,
consider the following \langname program, which evaluates
to an analogous configuration:

{\small
\[
    \bl
    \letintwo{a}{\mbnew} \\
    \letintwo{b}{\mbnew} \\
    \send{a}{m}{b}; \\
    \spawn{(\sugarfree{b})}; \\
    \guardanntwo{a}{\msgtag{m}} \\
    \quad \receive{m}{x}{\var{a}}{\sugarfree{a}} \\
    \}
    \el
\]
}

The first issue is how to create a scoped DG from $\mbnew$: one option is
to introduce a scoped construct
$\calcwd{let}\:\calcwd{mailbox}\:x\:\calcwd{in}\:\tma$, but this approach
fails as soon as we rename $x$ using a $\calcwd{let}$-binder.
A more robust approach could be to follow~\citet{AhmedFM07}
and~\citet{Padovani19} and endow mailbox types with a type-level identity by
giving $\calcwd{new}$ an existential type and introducing a scoped
$\calcwd{unpack}$ construct.
However there still remain two issues: first, it is unclear how to extend DGs to
capture the more complex scoping and sequencing induced by nested contexts.
Second, each mailbox type would require a static identity (\eg,
${!}^{\iota}\msgtag{Msg}$) which
becomes too restrictive, since we would need to include identities in message
payload types when communicating names. As an example, each client of the
future from~\secref{sec:introduction} would require a separate message type.

\subsection{The \langname solution: quasi-linear typing}\label{sec:quasi-linear-typing}
The many-sender, single-receiver pattern is closely linked to \emph{quasi-linear
typing}~\citep{Kobayashi99:quasilinear}; our
formulation is closer to that of~\citet{EnnalsSM04}.  Quasi-linear types were
originally designed to overcome some limitations of full linear types in the
contexts of both memory management and programming convenience, and allow a
value to be used once as a \emph{first-class} (returnable) value, but several
times as a \emph{second-class} value~\citep{OsvaldEWAR16}. A second-class value
can be \emph{consumed} within an expression, for example as the subject of a
send operation, but cannot escape the scope in which it is defined.

This distinction maps directly onto the many-sender, single-receiver communication model used
by the mailbox calculus. We augment mailbox types with a \emph{usage}: either
$\returnable$, a \emph{returnable} reference that allows a
type to appear in the return type of an expression; or $\usable$, a
`second-class' reference. The subject of a $\calcwd{guard}$ must be
returnable.  With usage information we can ensure that:

\begin{enumerate}
    \item there is \emph{only one} returnable reference for each mailbox name in a
        process;
    \item only returnable references can be renamed, avoiding problems with
        aliasing; and
    \item the returnable reference is the final lexical use of a mailbox name
        in a process.
\end{enumerate}

Quasi-linear types rule out all three of the previous examples.
In \mkwd{unsafeUse1}, $x$ is consumed by the $\calcwd{guard}$ expression and
cannot be used thereafter.  In \mkwd{unsafeUse2}, since $x$ is the subject of
a \calcwd{let} binding, it must be returnable and therefore cannot be used in
the body of the binding. In \mkwd{unsafeUse3}, since $x$ is used as the
subject of a $\calcwd{guard}$ expression, that use must be returnable and
therefore the last lexical occurrence of $x$, ruling out the use of $x$ in the
outer evaluation context.
Quasi-linear typing cannot account for inter-process deadlocks, but can still
rule out self-deadlocks.

\paragraph*{Ruling out aliasing through communication.}
Quasi-linear types alone do not safeguard against introducing aliasing through
communication, and we cannot use DGs for the reasons stated above.  However,
treating all received names as second-class, coupled with some simple syntactic
restrictions (\eg, by ensuring that either all message payloads or all variables
free in the body of the $\calcwd{receive}$ clause have base types) eliminates
unsafe aliasing.

\paragraph*{Summary.} Quasi-linear types and the lightweight syntactic
checks outlined above ensure that mailboxes are used safely in a
concurrent language that allows aliasing, and obviate the need for
the static global dependency graph used in the mailbox calculus. We
show that the checks are not excessively restrictive by expressing all
of the examples shown by~\citet{dP18:mbc}, and all of the 11 Savina
benchmarks~\citep{ImamS14a:savina} used by~\citet{NeykovaY16} to
demonstrate the expressiveness of behavioural type systems for actor
languages~\secrefp{sec:implementation:examples}.
 \section{\langname: a core language with mailbox types}\label{sec:declarative}
This section introduces $\langname$, a core functional programming language
with mailbox types, along with a declarative type system and an operational
semantics. We begin by defining \langname as a first-order language, and later
show how it extends to a higher-order language
in~\secref{sec:extensions:lambdas}.

\subsection{Syntax}

\begin{figure}[t]
      {\small
    \header{Syntax of types}
      \[
      \begin{array}{lrcl}
          \text{Mailbox types} & \mbtya, \mbtyb & ::= & \tysend{\pata} \midspace \tyrecv{\pata} \\
          \text{Mailbox patterns} & \pata, \patb & ::= & \mbzero \midspace \mbone \midspace
          \msgtag{m} \midspace \patplustwo{\pata}{\patb} \midspace \patconcattwo{\pata}{\patb} \midspace \many{\pata}
                                  \\
          \text{Base types} & \basety & ::= & \one \midspace \mkwd{Int} \midspace \mkwd{String} \midspace \cdots \\
          \text{Types} & \pretya, \pretyb & ::= & \basety \midspace \mbtya \\
          \text{Usage annotations} & \usage & ::= & \usable \midspace \returnable \\
          \text{Usage-annotated types} & \tya, \tyb & ::= & \basety \midspace \usagety{\mbtya} \\
          \text{Type environments} & \tyenv, \tyenvb & ::= & \cdot \midspace \tyenv, x : \tya
      \end{array}
      \]
      \header{Syntax of terms}
  \[
      \begin{array}{lrcl}
          \text{Variables} & x, y, z \\
          \text{Definition names} & f \\
          \text{Definitions} & D & ::= & \fndef{f}{\seq{x: \tya}}{\tyb}{\tma} \\
          \text{Values} & \vala, \valb & ::= & x \midspace c \\
          \text{Computations} & \tmc, \tma, \tmb & ::=  & \vala \midspace
          \letin{x}{\tma}{\tmb} \midspace \fnapp{f}{\seq{V}} \\
                       &         & \mid & \spawn{\tma} \midspace \mbnew
                       \midspace \send{\vala}{m}{\seq{\valb}} \midspace
                       \guard{\vala}{\seq{G}} \midspace \sugarfail{\vala} \\
\text{Guards} & G & ::= & \free{\tma} \midspace \receive{m}{\seq{x}}{y}{\tma}
    \end{array}
\]
}

  \caption{The syntax of  \langname, a core language with mailbox types}
  \label{fig:declarative:syntax}
\end{figure}

Figure~\ref{fig:declarative:syntax} shows the syntax for \langname. We defer
discussion of types to~\secref{sec:declarative-typing-rules}.

\paragraph*{Programs and definitions.}
A program $\prog = \progdef{\sigs}{\seq{D}}{\tma}$ consists of
a \emph{signature} $\sigs$ that maps message tags (\eg, \msgtag{Put} and
\msgtag{Get} from the future example) to payload types;
a set of \emph{definitions} $\seq{D}$; and an
\emph{initial term} $\tma$.
Each definition $\fndef{f}{\seq{x : \tya}}{\tyb}{\tma}$ is
a function with name $f$, annotated arguments $\seq{x : \tya}$, return type
$\tyb$, and body $\tma$. We write $\prog(f)$ to retrieve the
definition for function $f$, and $\prog(\msgtag{m})$ to
retrieve the payload types for message $\msgtag{m}$.

\paragraph*{Values.}
It is convenient to introduce a syntactic distinction between values and
computations, inspired by \emph{fine-grain call-by-value}~\citep{LPT03}, in
order to simplify our typing rules.
Values $\vala, \valb$ include variables $x$ and constants $c$; we assume that
the set of constants includes at least the unit value $()$ of type $\one$.

\paragraph*{Terms.}
The functional fragment of the language is largely standard.
Every value is a trivial computation. The only evaluation context is
$\letin{x}{\tma}{\tmb}$, which evaluates term $\tma$ and binds its result to $x$
in $\tmb$.
Function application $\fnapp{f}{\seq{V}}$ applies function $f$ to arguments $\seq{V}$.
As usual, we use $\tma; \tmb$ as sugar for
$\letin{x}{\tma}{\tmb}$, where $x$ does not occur free in $\tmb$.

In the concurrent fragment of the language, $\spawn{M}$ spawns term $M$ as a
separate process, and $\mbnew$ creates a fresh mailbox name. Term
$\send{\vala}{m}{\seq{\valb}}$ sends message $\msgtag{m}$ with payloads
$\seq{W}$ to mailbox $\vala$.

The $\guard{\vala}{\seq{\gd}}$ construct inspects mailbox $\vala$ and potentially
invokes a guard in $\seq{\gd}$.
Note that although our examples have included mailbox patterns in
$\calcwd{guard}$ expressions for clarity, and some annotations are required by
our algorithmic type system~\secrefp{sec:algorithmic}, annotations are not
required in the core language used by the declarative type system.

In an untyped system, the $\sugarfail{\vala}$ construct represents an
irrecoverable failure and would be invoked if an unexpected message arrives
(similar to raising an error in the Erlang future example
in~\secref{sec:introduction}).
In the typed setting, the $\sugarfail{\vala}$ construct eliminates a mailbox
$\vala$ of type $\tyrecv{\mbzero}$; we will see in~\secref{sec:metatheory} that
the type system ensures that $\sugarfail{\vala}$ can never be evaluated.

As an example, consider the following modification of the $\mkwd{fullFuture}$
example from~\secref{sec:introduction} that includes an additional $\calcwd{receive}$ guard
for a $\msgtag{Put}$ message. This additional guard is not supported by the
pattern $\many{\msgtag{Get}}$ or the $\mkwd{FullFuture}$ type. 

{\small
\[
    \bl
\fndefthree{\mkwd{fullFuture}}{\var{self}: \mkwd{FullFuture}, \var{value}: \mkwd{Int}}{\one} \\
        \quad \guardanntwo{\var{self}}{\many{\msgtag{Get}}} \\
        \qquad \free{()} \\
        \qquad
            \receivethree{Get}{\var{user}}{\var{self}} \\
        \qquad \quad \send{\var{user}}{Reply}{\var{value}}; \\
        \qquad \quad \mkwd{fullFuture}(\var{self}, \var{value}) \\
        \qquad
            \receivethree{Put}{\var{x}}{\var{self}} \\
            \qqquad \sugarfail{\var{self}} \\
        \quad \} \\
    \} \\
    \el
\]
}

As we will see shortly, given the $\many{\msgtag{Get}}$ annotation on the guard
and the fact that the $\var{self}$ parameter has annotation $\mkwd{FullFuture}$, 
in order to type this program, the type system would require the occurrence of
$\var{self}$ in the subject of the $\calcwd{guard}$ to have type
$\tyrecvp{\mbone \patplus (\msgtag{Get} \patconcat \many{\msgtag{Get}})
\patplus (\msgtag{Put} \patconcat \mbzero)}$, which is equivalent to
$\tyrecv{\many{\msgtag{Get}}}$. Therefore, the binding occurrence of
$\var{self}$ in the $\msgtag{Put}$ case would have type $\mbrecvzero$ and
so can be eliminated using $\kwfail$. \langname's \emph{mailbox conformance}
property ensures that a $\calcwd{fail}$ construct will never be evaluated.

\paragraph*{Guards.}
The $\free{M}$ guard is triggered when a mailbox is empty and there are no
more references to it in the system;
and $\receive{m}{\seq{x}}{y}{\tma}$ is triggered when the mailbox contains
a message with tag $\msgtag{m}$, binding its payloads to $\seq{x}$ and
the continuation mailbox name with updated mailbox type to $y$ in continuation
term $\tma$.

We write $\sugarfree{V}$ as syntactic sugar for $\guard{V}{\free{()}}$.
It only makes sense to free the \emph{input} reference to a mailbox;
output references are consumed upon sending a message.

\subsection{Type system}\label{sec:declarative-typing-rules}

This section describes a declarative type system for $\langname$. We begin
by discussing mailbox types in more depth, in particular showing how to define
subtyping and equivalence.

\subsubsection{Types}
A mailbox type consists of a \emph{capability}, either \emph{output} $\texttt{!}$
or \emph{input} $\texttt{?}$, and a \emph{pattern}. A system can contain multiple
references to a mailbox as an output capability, but only one as an input
capability.
A \emph{pattern} is a \emph{commutative} regular expression, \ie, a regular expression
where composition is unordered. The $\mbone$ pattern is the unit of pattern
composition $\patconcat$, denoting the empty mailbox.
The $\mbzero$ pattern denotes the \emph{unreliable} mailbox, which has received
an unexpected message. It is not possible to send to, or receive from, an
unreliable mailbox, but we will show that reduction does not cause a mailbox to
become unreliable.
The pattern $\msgtag{m}$ denotes a mailbox containing a single message
$\msgtag{m}$.
Note that unlike in~\secref{sec:introduction}, mailbox types in our formalism do
not pair a message tag with its payload; instead, tags are associated with
payload types via the program signature. This design choice allows us to more
easily compare the declarative system with the algorithmic system
in~\secref{sec:algorithmic}, and unlike~\citet{dP18:mbc} means we need not
define types and subtyping coinductively.
Pattern choice $\pata \patplus \patb$ denotes that the mailbox
contains either messages conforming to pattern $\pata$ or \ADDED{messages conforming to pattern} $\patb$. Pattern
composition $\pata \patconcat \patb$ denotes that the mailbox contains messages
conforming to $\pata$ and \ADDED{messages conforming to} $\patb$ (in either order).
Finally, $\many{\pata}$ denotes replication
of $\pata$, so for example $\many{\msgtag{m}}$ denotes that the mailbox can
contain zero or more instances of message $\msgtag{m}$.
Mailbox patterns obey the usual laws of commutative regular expressions:
$\mbone$ is the unit for $\patconcat$, while $\mbzero$ is the unit for
$\patplus$ and is cancelling for $\patconcat$. Composition $\patconcat$ is
associative, commutative, and distributes over $\patplus$; and $\patplus$ is
associative and commutative. We treat both $\patplus$ and $\patconcat$ as
right-associative, so a pattern
$\msgtag{m} \patplus \msgtag{n} \patplus \msgtag{p}$ should be read as
$\msgtag{m} \patplus (\msgtag{n} \patplus \msgtag{p})$.

\paragraph*{Pattern semantics.}
It follows that different syntactic representations of patterns may have the
same meaning, \eg, patterns $\mbone \patplus \mbzero
\patplus (\msgtag{m} \patconcat \msgtag{n})$ and $\mbone \patplus (\msgtag{n}
\patconcat \msgtag{m})$.
Following~\citet{dP18:mbc}, we define a set-of-multisets semantics for mailbox
patterns: the intuition is that each multiset defines a configuration
of messages that could be present in the mailbox. For example the semantic
representation of both of the patterns above is $\set{\multiset{},
\multiset{\msgtag{m}, \msgtag{n}}}$. We let $\mseta, \msetb$ range over
multisets and formally define pattern semantics as follows:

{
\begin{mathpar}
    \sem{\mbzero} = \emptyset

    \sem{\mbone} = \{ \multiset{} \}

    \sem{E \patplus F} = \sem{E} \cup \sem{F}

    \sem{E \patconcat F} = \{ \mseta \uplus \msetb \mid \mseta \in \sem{E}, \msetb
    \in \sem{F} \}

    \sem{\msgtag{m}} = \{ \multiset{\msgtag{m}} \}

    \sem{\many{E}} = \sem{\mbone} \cup \sem{E} \cup \sem{E \patconcat E} \cup
    \cdots
\end{mathpar}
}

The pattern $\mbzero$ is interpreted as an empty set; $\mbone$ as a singleton
set containing an empty multiset; $\patplus$ as set union; $\patconcat$ as
pointwise multiset union; $\msgtag{m}$ as a singleton set containing the
singleton multiset; and $\many{\pata}$ as the possibly infinite set containing
any number of compositions of interpretations of $\pata$.

\paragraph*{Usage annotations.}
A type $\pretya$ can be a \emph{base type} $\basety$, or a mailbox type
$\mbtya$. As discussed in~\secref{sec:whatstheissue},
\emph{quasi-linearity} is used to avoid aliasing issues. \emph{Usage-annotated}
types $\tya, \tyb$ annotate mailbox types with a usage: either second-class
($\usable$), or returnable ($\returnable$). There are no restrictions on the
use of a base type, so all base types are treated as returnable.
Only values with a returnable type can be returned from an evaluation frame.

\subsubsection{Operations and predicates on types}
We say that a type is \emph{returnable}, written $\returnablepred{\tya}$, if
$\tya$ is a base type $\basety$ or a returnable mailbox type $\retty{\mbtya}$.
The $\makereturnable{-}$ operator produces a returnable type from a
non-annotated type, or changes an arbitrary mailbox type to be returnable.
Similarly, the $\makeusable{-}$ operator produces a type with a second-class
usage annotation:

{
\begin{mathpar}
    \makereturnable{\basety} = \basety

    \makereturnable{\pretya} = \retty{\pretya}

    \makereturnable{\usagety{\mbtya}} = \retty{\mbtya}

    \makeusable{\basety} = \basety

    \makeusable{\pretya} = \usablety{\pretya}

    \makeusable{\usagety{\mbtya}} = \usablety{\mbtya}
\end{mathpar}
}

We also extend the operators to type environments in the usual way.
\ADDED{The $\mbbase{\tya}$ predicate holds if $\tya$ is some base type
$\basety$. We extend the predicate to environments:
$\mbbase{\tyenv}$ holds if for all $x : \tya \in \tyenv$, it follows that
$\mbbase{\tya}$.
} 

\paragraph*{Subtyping.} Subtyping is crucial for mailbox typing. Rather than
being an additional feature to increase the expressiveness of the system,
subtyping is the core technical mechanism that allows us to determine whether an
inferred pattern is contained within a specification, for example checking that
$\msgtag{Get} \patconcat \msgtag{Get}$ is contained within a specification
$\many{\msgtag{Get}}$.
Subtyping relies on \emph{pattern inclusion}.
A pattern $\pata$ is \emph{included} in a pattern $\patb$, written $\pata \incll
\patb$, if every multiset in the semantics of $\pata$ also occurs in the
semantics of pattern $\patb$, \ie, $\pata \incll \patb \defeq \sem{\pata}
\subseteq \sem{\patb}$.

\begin{definition}[Subtyping]
    The \emph{subtyping} relation is defined by the following rules:

    {
    \begin{mathpar}
        \inferrule
        { }
        { \basety \subtype \basety }

        \inferrule
        { \pata \subpatone \patb \\ \usage_1 \subusage \usage_2 }
        { \usageann{\usage_1}{\tyrecv{\pata}} \subtype \usageann{\usage_2}{\tyrecv{\patb}}  }

        \inferrule
        { \patb \subpatone \pata \\ \usage_1 \subusage \usage_2 }
        { \usageann{\usage_1}{\tysend{\pata}} \subtype \usageann{\usage_2}{\tysend{\patb}}  }
    \end{mathpar}
}

    \emph{Usage subtyping} is defined as the smallest reflexive
      relation defined by axioms $\usage \subusage \usage$ and ${\returnable}
      \subusage {\usable}$.
We write $\tya \tyequiv \tyb$, read ``$\tya$ is equivalent to $\tyb$'',
if both $\tya \subtype \tyb$ and $\tyb \subtype
\tya$, \ie, either $\tya, \tyb$ are the same base type, or are mailbox types
with the same capability and pattern semantics.
\end{definition}

Base types are subtypes of themselves. As with previous accounts
of subtyping in actor languages~\citep{HeWT14:takka}, subtyping is
\emph{covariant} for
mailbox types with a receive capability: an input reference to
a mailbox can safely be replaced with another that can receive more messages.
Likewise subtyping is \emph{contravariant} for mailboxes with a
send capability: an output reference to a mailbox can safely be replaced with
another that can send a smaller set of messages.
Intuitively, as returnable usages are more powerful than second-class usages,
returnable types can be used when only a second-class type is required.

Following~\citet{dP18:mbc} we introduce names for particular classes of mailbox
types:

\begin{definition}[Relevant, reliable, usable]
        A mailbox type $\mbtya$ is \emph{relevant} if $\mbtya \not\subtype \mbsendone$, and
                \emph{irrelevant} otherwise;
\emph{reliable} if $\mbtya \not\subtype \mbrecvzero$ and
            \emph{unreliable} otherwise; and
\emph{usable} if \ADDED{$\mbsendzero \not\subtype \mbtya$} and
            \emph{unusable} otherwise.
A \emph{type environment} $\tyenv$ is reliable if all input mailbox
            types $\tyrecv{\pata}$ in $\tyenv$ are reliable.
\end{definition}

Relevant mailbox names \emph{must} be used, whereas irrelevant names need
not be. Likewise reliable and usable names \emph{can} be used, whereas
unreliable and unusable names cannot.
In the remainder of the paper we assume that all output mailbox types are
usable: we therefore forbid types such as $\tysend{\mbzero}$
or $\tysendp{\msgtag{m} \patconcat \mbzero}$.

\begin{definition}[Unrestricted and linear types]
    We say that a type $\tya$ is \emph{unrestricted}, written $\un{\tya}$, if
$\tya$ is a base type
    or $\tya = \usablety{\mbsendone}$.
    Otherwise, we say that $\tya$ is \emph{linear}.
\end{definition}

Our type system ensures that variables with a linear type must be used, whereas
variables with an unrestricted type can be discarded.
We extend subtyping to type environments.

\begin{definition}[Environment subtyping]
    Environment subtyping $\tyenv_1 \subtype \tyenv_2$ is
    the preorder relation on environments defined as follows:
{
    \begin{mathpar}
        \inferrule
        { }
        { \tyenv \subtype \tyenv }

        \inferrule
        { \tyenv_1 \subtype \tyenv_2 \\
          \tyenv_2 \subtype \tyenv_3 }
        { \tyenv_1 \subtype \tyenv_3}

        \inferrule
        { \un{\tya} }
        { \tyenv, x : \tya \subtype \tyenv }

        \inferrule
        { \tya \subtype \tyb }
        { \tyenv, x : \tya \subtype \tyenv, x : \tyb }
    \end{mathpar}
    }
\end{definition}

The subtyping relation includes a notion of weakening, allowing an environment
$\tyenv$ to be a subtype environment of $\tyenv'$ if it contains additional
entries of unrestricted type.

\paragraph*{Type combination.}
Mailbox types ensure that sends and receives ``balance out'', meaning that every
send is matched with a receive. For example, using a mailbox at type
$\tysend{\msgtag{Put}}$ and $\tyrecvp{\msgtag{Put} \patconcat
\many{\msgtag{Get}}}$ results in a mailbox type $\tyrecvp{\many{\msgtag{Get}}}$.
The key technical device used to achieve this goal is \emph{type
combination}: combining a mailbox type $\tysend{\pata}$ and a
mailbox type $\tysend{\patb}$ results in an output mailbox type that must send
\emph{both} $\pata$ \emph{and} $\patb$; combining an input and an output
capability results in an input capability that no longer needs to receive the
output pattern. We can also combine identical base types.
Note that it is \emph{not} possible to combine two input capabilities as this
would permit simultaneous reads of the same mailbox, which would be unsafe.

\begin{definition}[Type combination]
  \emph{Type combination} $\mbcombtwo{\pretya}{\pretyb}$ is the commutative
  partial binary operator defined by the following axioms:

  {
\begin{mathpar}
    \mbcombtwo{\basety}{\basety} = \basety

    \mbcombtwo{\tysend{\pata}}{\tysend{\patb}} = \tysendp{\pata \patconcat \patb}

    \mbcombtwo{\tysend{\pata}}{\tyrecvp{\pata \patconcat \patb}} = \tyrecv{\patb}

    \mbcombtwo{\tyrecvp{\pata \patconcat \patb}}{\tysend{\pata}} = \tyrecv{\patb}
\end{mathpar}
}
\end{definition}

Following~\citet{CrafaP17}, for convenience we identify types up to
commutativity and associativity, \eg, we do not distinguish between
$\retty{\tyrecvp{\msgtag{A} \patconcat \msgtag{B}}}$ and
$\retty{\tyrecvp{\msgtag{B} \patconcat \msgtag{A}}}$, though since these
patterns are semantically equivalent we can always rewrite one into another
using subtyping.
We may however need to use subtyping to rewrite a type into a form that allows two
mailbox types to be combined (\eg, to combine $\tysend{\msgtag{A}}$ and
$\tyrecv{\many{\msgtag{A}}}$, we would need to use subtyping to rewrite
the latter type to $\tyrecvp{\msgtag{A} \patconcat \many{\msgtag{A}}}$).

The following \emph{usage combination} operator is used to enforce
quasi-linearity. Importantly, it is not commutative because a $\usable$
variable use must occur \emph{before} a $\returnable$ use (ensuring that the
returnable use is the variable's last lexical occurrence). Furthermore, note
that $\scombtwo{\returnable}{\returnable}$ is undefined (ensuring that there is
only one returnable instance of a variable per thread).

\begin{definition}[Usage combination]
    The \emph{usage combination operator} is the partial binary
    operator defined by the axioms
    $\usagecombtwo{\usable}{\usable} = \usable$ and
    $\usagecombtwo{\usable}{\returnable} = \returnable$.
\end{definition}

We can now define usage-annotated type and environment combination, which is
used extensively throughout the declarative type system in order to enforce
quasi-linearity and mailbox typing within a single process.
Since usage combination is not commutative, typing multiple concurrent processes
at runtime uses an alternative environment combination operator, described
in~\secref{sec:metatheory}.

\begin{definition}[Usage-annotated type combination]
  The \emph{usage-annotated type combination operator} $\scombtwo{\tya}{B}$ is
  the partial binary operator defined by the axioms
  $\scombtwo{\basety}{\basety} = {\basety}$ and
  $
      \scombtwo
        {\usageann{\usage_1}{\mbtya}}
        {\usageann{\usage_2}{\mbtyb}}
        =
        {\usageann{\usagecombtwo{\usage_1}{\usage_2}}{(\mbcombtwo{\mbtya}{\mbtyb})}}$.
\end{definition}

\begin{definition}[Environment combination ($\tyenv$)]
    The \emph{usage-annotated environment combination operator} $\scombtwo{\tyenv_1}{\tyenv_2}$ is the
  smallest partial operator on type environments closed under the
  following rules:

  {\small
  \begin{mathpar}
      \inferrule
      { }
      { \scombtwo{\cdot}{\cdot} = \cdot}

      \inferrule
      { x \not\in \dom{\tyenv_2} \\
        \scombtwo{\tyenv_1}{\tyenv_2} = \tyenv
      }
      {
        \scombtwo{(\tyenv_1, x : \tya)}{\tyenv_2} = \tyenv,
        x : \tya
      }

      \inferrule
      { x \not\in \dom{\tyenv_1} \\
        \scombtwo{\tyenv_1}{\tyenv_2} = \tyenv
      }
      {
        \scombtwo{\tyenv_1}{(\tyenv_2, x : \tya)} =
        \tyenv, x : \tya
      }

      \inferrule
      {
          \scombtwo{\tyenv_1}{\tyenv_2} = \tyenv
      }
      {
        \scombtwo{(\tyenv_1, x : \tya)}{(\tyenv_2, x : \tyb)} =
            \tyenv, x : (\scombtwo{\tya}{\tyb})
      }
  \end{mathpar}
}
\end{definition}

We use usage-annotated type combination when combining the types of a
variable used in two successive
evaluation frames (\ie, in the subject and body of a
$\calcwd{let}$ expression).
We also require \emph{disjoint} combination, where two environments are only
able to share variables of base type:

\begin{definition}[Disjoint environment combination]
  Disjoint environment combination $\pluscombtwo{\tyenv_1}{\tyenv_2}$ is the
  smallest partial operator on type environments closed under the
  following rules:

{\small
\begin{mathpar}
    \inferrule
    { }
    { \pluscombtwo{\cdot}{\cdot} = \cdot }

    \inferrule
    { x \not\in \dom{\tyenv_2} \\
      \pluscombtwo{\tyenv_1}{\tyenv_2} = \tyenv
    }
    { \pluscombtwo{(\tyenv_1, x : \tya)}{\tyenv_2} = \tyenv, x : \tya }

    \inferrule
    { x \not\in \dom{\tyenv_1} \\
      \pluscombtwo{\tyenv_1}{\tyenv_2} = \tyenv
    }
    { \pluscombtwo{\tyenv_1}{(\tyenv_2, x : \tya)} = \tyenv, x : \tya }

    \inferrule
    { \pluscombtwo{\tyenv_1}{\tyenv_2} = \tyenv }
    { \pluscombtwo{(\tyenv_1, x : \basety)}{(\tyenv_2, x : \basety)} = \tyenv, x : \basety }
\end{mathpar}
}
\end{definition}

\subsubsection{Pattern residual and pattern normal form}

\begin{figure}[t]
    {\small
\headersig{Pattern residual}{$\pata \without \msgtag{m}$}
\begin{mathpar}
  \inferrule
  {}
  { \mbzero \without \defmsg \defeq \mbzero }

  \inferrule
  {}
  { \mbone \without \defmsg \defeq \mbzero }

  \inferrule
  {}
  { \msgtag{m} \without \msgtag{m} \defeq \mbone }

  \inferrule
  { \msgtag{m} \ne \msgtag{n} }
  { \algomsg{m} \without \algomsg{n} \defeq \mbzero }

  \inferrule
  {}
  { (\patplustwo{\pata}{\patb}) \without \defmsg
    \defeq (\pata \without \defmsg) \patplus (\patb \without \defmsg) }

  \inferrule
  {}
  { (\many{\pata}) \without \msgtag{m} \defeq (\pata \without \msgtag{m}) \patconcat \many{\pata} }

  \inferrule
  {}
  { (\pata \patconcat \patb) \without \defmsg \defeq
    ((\pata \without \defmsg) \patconcat \patb) \patplus
    (\pata \patconcat (\patb \without \defmsg))
  }
\end{mathpar}
\headertwo
    {Pattern normal form (PNF)}
    {\framebox{\ADDED{$\pnf{\pata}$}}~\framebox{$\pnflittwo{\pata}{\patb}$}~\framebox{$\vphantom{\vDash_\mkwd{lit}}\pnftwo{\pata}{\patb}$}}
\begin{mathpar}
  \inferrule
  { \pnftwo{\pata}{\pata} }
  { \pnf{\pata} }

  \inferrule
  {}
  { \pnflittwo{\pata}{\mbone} }

  \inferrule
  { \patb \tyequiv \pata \without \defmsg }
  { \pnflittwo{\pata}{\defmsg \patconcat \patb} }

  \inferrule
  { \pnflittwo{\pata}{\patb_1} \\ \pnftwo{\pata}{\patb_2} }
  { \pnftwo{\pata}{\patb_1 \patplus \patb_2} }

  \inferrule
  { \pnflittwo{\pata}{\patb} }
  { \pnftwo{\pata}{\patb} }
\end{mathpar}
}

\caption{Pattern residual and pattern normal form}
  \label{fig:pnf}
\end{figure}

Before looking at the typing rules, it is useful to discuss the concepts of a
\emph{pattern residual} and \emph{pattern normal form}, whose formal
descriptions are given in Figure~\ref{fig:pnf}.

\paragraph*{Pattern residual.}
Whenever we receive a message from a mailbox, the type of the receive reference
to that mailbox should be updated to reflect that the message is no longer
present. For instance, in our future example, the \mkwd{EmptyFuture} type is
$\tyrecvp{\msgtag{Put} \patconcat \many{\msgtag{Get}}}$, but the
\mkwd{FullFuture} type (used after receiving the $\msgtag{Put}$ message)
is $\tyrecvp{\many{\msgtag{Get}}}$ (that is, the $\mkwd{EmptyFuture}$ type
\emph{without} the $\msgtag{Put}$ message).

The \emph{pattern residual} $\pata \without \msgtag{m}$ calculates the pattern
$\pata$ after $\msgtag{m}$ is consumed, and corresponds to the Brzozowski
derivative~\citep{B64} over a commutative regular expression.
The residual of $\mbzero$, $\mbone$, or $\msgtag{n}$ (where $\msgtag{n} \ne
\msgtag{m}$) with respect to a message tag $\msgtag{m}$ is the unreliable
pattern $\mbzero$.
The residual of $\msgtag{m}$ with respect to $\msgtag{m}$ is $\mbone$. The
residual operator distributes over $\patplus$, and the residual of
composition is the disjunction of the residual of each subpattern.
In our example, $(\msgtag{Put} \patconcat \many{\msgtag{Get}}) \without
\msgtag{Put} \tyequiv \many{\msgtag{Get}}$.

\paragraph*{Pattern normal form.}
When typing $\calcwd{guard}$ expressions, we require mailbox
types to be in \emph{pattern normal form} (PNF). A pattern $\pata$ is in PNF if
it is in the form $\pata_1 \patplus \cdots \patplus \pata_n$ where each $\pata_i$ is
$\mbone$ (used for typing a
$\calcwd{free}$ guard) or of the form $\msgtag{m} \patconcat \patb$ (used for
typing a $\calcwd{receive}$ guard, where $\patb$ is equivalent to $\pata
\without \msgtag{m}$).

We express PNF using three judgements:
Judgement $\pnftwo{\pata}{\patb}$ can be read ``pattern $\patb$ is in pattern
normal form with respect to pattern $\pata$''.
Judgement
$\pnflittwo{\pata}{\patb}$ is similar but requires $\patb$ to be free of
pattern choice constructors ($\patplus$). Judgement $\pnf{\pata}$ can be read
``pattern $\pata$ is in pattern normal form'', and holds if
$\pnftwo{\pata}{\pata}$.

As an example, pattern
$\mbone \patplus (\msgtag{m} \patconcat \msgtag{n})$ is in PNF:
we could guard on a mailbox with this pattern by including a $\calcwd{free}$
guard and a $\calcwd{receive}$ guard that receives a message with tag
$\msgtag{m}$.
However, the equivalent pattern
$(\mbone \patconcat \mbone) \patplus \mbzero \patplus (\msgtag{m} \patconcat \msgtag{n})$ is \emph{not} in PNF.

The definition of PNF relies on the definition of the pattern residual, and PNF
is later used in the typing rule for $\calcwd{guard}$ expressions.

\subsubsection{Typing rules}

\begin{figure}[t]
{\small
    \headertwo
        {Typing rules for programs and definitions}
        {\framebox{$\progseq{\prog}$}~\framebox{$\defseq{D}$}}
\begin{mathpar}
    \inferrule
    {
        \prog = \progdef{\sigs}{\seq{D}}{\tma} \\
        (\defseq[\prog]{D_i})_{i \in 1..n} \\
          \tseq[\prog]{\cdot}{\tma}{\tyunit}
    }
    { \progseq{\prog} }

    \inferrule
    {
        \tseq[\prog]{\seq{x : \tya}}{\tma}{\tyb}
    }
    { \defseq[\prog]{\fndef{f}{\seq{x : \tya}}{\tyb}{\tma}} }
\end{mathpar}

\headersig{Typing rules for values and computations}{$\tseq[\prog]{\tyenv}{\tma}{\tya}$}
\begin{mathpar}
    \inferrule
    [T-Var]
    { }
    { \vseq{\hastypetwo{x}{\tya}}{x}{\tya}  }

    \inferrule
    [T-Const]
    { c \text{ has base type } \basety }
    { \vseq{\cdot}{c}{\basety} }

    \inferrule
    [T-App]
    {
\prog(f) = \fndef{f}{\seq{x : \tya}}{\tyb}{\tma} \\
        (\vseq{\tyenv_i}{\vala_i}{\tya_i})_{i \in 1..n}
    }
    { \tseq{\tyenv_1 + \cdots + \tyenv_n}{\fnapp{f}{\vala_1, \ldots, \vala_n}}{\tyb} }

    \inferrule
    [T-Let]
    {
        \tseq{\tyenv_1}{\tma}{\tya} \\
        \returnablepred{\tya} \\
        \tseq{\tyenv_2, x : \tya}{\tmb}{\tyb}
    }
    { \tseq{\scombtwo{\tyenv_1}{\tyenv_2}}{\letin{x}{\tma}{\tmb}}{\tyb} }

    \inferrule
    [T-Spawn]
    {
      \tseq{\tyenv}{\tma}{\one}
    }
    {
    \tseq{\makeusable{\tyenv}}{\spawn{\tma}}{\one} }

    \inferrule
    [T-New]
    { }
    { \tseq{\cdot}{\mbnew}{\retty{\mbrecvone}} }

    \inferrule
    [T-Send]
    {
      \proglookup{\algomsg{m}} = \seq{\pretya} \\
      \tseq{\tyenv}{\vala}{\usablety{\tysend{\algomsg{m}}}} \\
      (\tseq{\tyenv'_i}{\valb_i}{\makeusable{\pretya_i}})_{i \in 1..n}
    }
    { \tseq
        {\tyenv + \tyenv'_1 + \cdots +  \tyenv'_n }
        {\send{\vala}{m}{\seq{\valb}}}
        {\one}
    }

    \inferrule
    [T-Guard]
    {
      \tseq{\tyenv_1}{\vala}{\retty{\tyrecv{\pata}}} \\
      \gseq{\pata}{\tyenv_2}{\seq{G}}{\tya} \\
      \pnf{\pata}
    }
    { \tseq{{\tyenv_1} + {\tyenv_2}}{\guard{\vala}{\seq{G}}}{\tya} }

    \inferrule
    [T-Fail]
    { \tseq{\tyenv}{\vala}{\retty{\tyrecv{\mbzero}}} }
    { \tseq{\tyenv}{\sugarfail{\vala}}{\tya} }

    \inferrule
    [T-Sub]
    { \subtypetwo{\tyenv}{\tyenv'} \\
      \subtypetwo{\tya}{\tyb} \\
      \tseq{\tyenv'}{\tma}{\tya} }
    { \tseq{\tyenv}{\tma}{\tyb}  }
\end{mathpar}

\headertwo
    {Typing rules for guards}
    {
        \framebox{$\gseq[\prog]{\pata}{\tyenv}{\seq{\gd}}{\tya}$}
        ~
        \framebox{$\gseq[\prog]{\pata}{\tyenv}{\gd\vphantom{\seq{\gd}}}{\tya}$}
    }
\begin{mathpar}
    \inferrule
    [TG-GuardSeq]
    {
        (\gseq{\pata_i}{\tyenv}{\gd_i}{\tya})_{i \in 1..n}
    }
    {
        \gseq
            {\pata_1 \patplus \cdots \patplus \pata_n}
            {\tyenv}
            {\seq{G}}
            {\tya}
    }\\
\inferrule
    [TG-Free]
    {
        \tseq{\tyenv}{\tma}{\tya}
    }
    { \gseq{\mbone}{\tyenv}{\free{\tma}}{\tya} }

    \inferrule
    [TG-Recv]
    {
      \prog(\msgtag{m}) = \seq{\pretya} \\
      \mbbase{\seq{\pretya}} \vee \mbbase{\tyenv} \\
      \tseq
        {\tyenv,
            y : \usageann{\returnable}{\tyrecv{E}},
            \seq{x} : \seq{\makeusable{\pretya}}}
        {\tma}
        {\tyb}
    }
    {
      \gseq{\algomsg{m} \patconcat
      E}{\tyenv}{\receive{m}{\seq{x}\!}{\!y}{\tma}}{\tyb}
    }
  \end{mathpar}
}
\caption{Declarative typing rules for \langname}
\label{fig:declarative-typing}
\end{figure}

Figure~\ref{fig:declarative-typing} shows declarative typing rules for
\langname. As the system is declarative it helps to read the rules top-down.

\paragraph*{Programs and definitions.}
A program is typable if all of its definitions are typable, and its body has the
unit type $\one$.
A definition $\fndef{f}{\seq{x : \tya}}{\tyb}{\tma}$ is typable if
$\tma$ has type $\tyb$ under environment $\seq{x : \tya}$.

\paragraph*{Terms.}
Term typing has the judgement $\tseq[\prog]{\tyenv}{\tma}{\tya}$, which states
that when defined in the context of program $\prog$, under environment $\tyenv$,
term $\tma$ has type $\tya$. We omit the $\prog$ parameter in the rules for readability.
Rule \textsc{T-Var} types a variable in a singleton environment; we
account for weakening using environment subtyping in \textsc{T-Sub}.
Rule \textsc{T-Const} types a constant under an empty environment; we assume an
implicit schema mapping constants to types, and assume the existence of at least
the unit value $()$ of type $\one$.
Rule \textsc{T-App} types application of a function $f$ according to the
definition of $f$ in
$\prog$. Each argument must be typable under a disjoint type
environment to avoid aliasing mailbox names in the body of the
function.

Rule \textsc{T-Let} types sequential composition. The subject of the
$\calcwd{let}$ expression must be returnable: since
$\scombtwo{\tyenv_1}{\tyenv_2}$ is defined, we know that if the subject
(typable using $\tyenv_1$) contains
a returnable variable, then it cannot appear in $\tyenv_2$.
This avoids aliasing and unsafe usage errors.

Rule \textsc{T-Spawn} types the $\spawn{\tma}$ construct, which spawns a
unit-typed term $M$ as a new process.  The type environment used to type $M$ can
contain any number of returnable types, but the conclusion of the rule  ``masks'' any
returnable types as second-class.
Intuitively, this is because there is no need to impose
an ordering on how a variable is used in a separate process: while
within a single process a \calcwd{guard} on some name $x$ should not precede a
send on $x$, there is no such restriction if the two expressions are executing
in concurrent processes.
Rule \textsc{T-New} creates a fresh mailbox with type $\retty{\mbrecvone}$,
capturing the requirement that subsequent sends and receives must ``balance
out'' to an empty mailbox.

Rule \textsc{T-Send} types a send expression $\send{V}{m}{\seq{W}}$, where a
message $\msgtag{m}$ with payloads $\seq{W}$ is sent to a mailbox $V$.
Value $\vala$ must be a reference
with type $\usablety{\tysend{\msgtag{m}}}$, meaning that it can be used to send
message $\msgtag{m}$. The mailbox only needs to be \emph{second-class}, but
subtyping means that we can also send to a returnable name. All payloads
\ADDED{$\seq{\valb}$} must be subtypes of the types defined by the signature for message
$\msgtag{m}$, and payloads must be typable under separate environments
to avoid aliasing when receiving a message.
Unlike in session-typed functional programming languages, sending is a
side-effecting operation of type $\one$; behavioural typing is instead
enforced using environment combination.

Rule \textsc{T-Guard} types the expression $\guard{\vala}{\seq{\gd}}$,
which retrieves from mailbox $\vala$ using guards $\seq{\gd}$.
The first premise ensures that under a type environment $\tyenv_1$, mailbox $V$
has type $\retty{\tyrecv{\pata}}$: the mailbox should have a receive capability
with pattern $\pata$, and must be returnable. Demanding that the mailbox is
returnable rules out unsafe usage errors since we cannot use the mailbox name
in the continuation.
The second premise states that under type environment $\tyenv_2$, guards $\seq{G}$ all return a value of type
$\tya$ and correspond to pattern $\pata$.  The final premise, $\pnf{\pata}$, ensures
that $\pata$ is in pattern normal form.

Rule \textsc{T-Fail} types the $\sugarfail{\vala}$ construct, which is the
elimination form for an unreliable mailbox with type $\mbrecvzero$. The type
system will ensure that this construct is never evaluated, and so it may have an
arbitrary return type.

Finally, rule \textsc{T-Sub} allows the use of
subtyping. Subtyping on type environments is crucial when constructing
derivations, \eg, two patterns may have the same semantics but differ
syntactically. Applying \textsc{T-Sub} makes it possible to rewrite
mailbox types so that they can be combined by the type combination
operators. We also allow the usual use of subsumption on expression types.

\paragraph*{Guards.}

Given a pattern of the form $\pata_1 \patplus \cdots \patplus \pata_n$,
rule \textsc{TG-GuardSeq} types a sequence of guards, ensuring that each guard
$\gd_i$ is typable using pattern literal $\pata_i$ under the same type
environment and with the same return type.
Rule \textsc{TG-Free} types a
guard of the form $\free{M}$, where $M$ has type $\tya$.

Finally, rule \textsc{TG-Recv} types a guard of the form
$\receive{m}{\seq{x}}{y}{\tma}$, which retrieves a message with tag $\msgtag{m}$
from the mailbox, binding its payloads (whose types are retrieved from the
signature for message $\msgtag{m}$) to $\seq{x}$, and re-binding the mailbox 
(with an updated type) to $y$ in continuation $\tma$. The payloads are made
\emph{second-class} rather than returnable, as otherwise the payloads could
interfere with the names in the enclosing context and potentially violate the
constraints required by quasi-linearity.
\ADDED{The rule also introduces a conservative check to avoid aliasing by
communication~\secrefp{sec:whatstheissue}: either all received payloads must
have base types, or all free variables in the environment must have base types.
We discuss a more liberal version of this condition
in~\secref{sec:extensions:interfaces}.}

\paragraph*{Example.}

We end this section by showing the derivation for the $\mkwd{client}$ definition
from the example in~\secref{sec:introduction}, which creates a future mailbox
and a self mailbox, initialises the future with a number, and then requests and
prints the result.  In the following, we abbreviate $\var{future}$ to $\var{f}$,
$\var{self}$ to $s$, and $\var{result}$ to $r$. We assume that the program
includes a signature $\sigs = [\msgtag{Put} \mapsto \tyint, \msgtag{Get} \mapsto
\tysend{\msgtag{Reply}}, \msgtag{Reply} \mapsto \tyint]$, the
$\mkwd{emptyFuture}$ and $\mkwd{fullFuture}$ definitions
from~\secref{sec:introduction}, as well as the primitives
$\mkwd{print} : \tystring \to \one$ and
$\mkwd{intToString} : \tyint \to \tystring$.
We split the derivation into three subderivations. Since it is easier to read
derivations top-down, we start by typing the $\calcwd{guard}$ expression. In the
following, we refer to the $\calcwd{receive}$ guard as $G$, and name the first derivation $\derivd_1$:

{\scriptsize
    \begin{mathpar}
        \inferrule
        {
            \inferrule
            { }
            {
                \tseq
                { \var{s} : \retty{\tyrecvp{\msgtag{Reply} \patconcat \mbone}}}
                { \var{s} }
                { \retty{\tyrecvp{\msgtag{Reply} \patconcat \mbone}} }
            }
            \\
            \inferrule
            {
                \inferrule
                {
                    \inferrule
                    { \inferrule*
                      { }
                      { \tseq{\var{s} :
                      \retty{\tyrecv{\mbone}}}{\var{s}}{\retty{\tyrecv{\mbone}}} }
                    }
                    { \tseq
                        {\var{s} : \retty{\tyrecv{\mbone}}}
                        {\sugarfree{\var{s}}}
                        {\one}
                    }
                    \\
                    \tseq
                        {\var{r} : \tyint}
                        {\mkwd{print}(\mkwd{intToString}(\var{r}))}
                        {\one}
                }
                {
                    \tseq
                    { \var{s} : \retty{\mbrecvone}, \var{r} : \tyint}
                    { \sugarfree{\var{s}}; \mkwd{print}(\mkwd{intToString}(\var{r})) }
                    { \one}
                }
            }
            { \gseq
                {\msgtag{Reply} \patconcat \mbone}
                {
                    \cdot
                }
                {
                    {
                        \bl
                    \receivethree{Reply}{\var{r}}{\var{s}} \\
                        \quad \sugarfree{\var{s}};
                        \mkwd{print}(\mkwd{intToString}(\var{r}))
                        \el
                    }
                }
                {\one}
            }
            \\
            \pnf{\msgtag{Reply} \patconcat \mbone}
        }
        {
            \tseq
                {\var{s} : \retty{\tyrecvp{\msgtag{Reply} \patconcat \mbone}} }
                {
                    \guard{\var{s}}{G}
                }
                {\one}
        }
    \end{mathpar}
}

The type of the $\var{s}$ mailbox in the subject of the $\calcwd{guard}$
expression is $\retty{\tyrecvp{\msgtag{Reply} \patconcat \mbone}}$
denoting that the mailbox can contain a single $\msgtag{Reply}$
message. The $\calcwd{receive}$ guard binds $s$
at type $\retty{\mbrecvone}$ and $r$ at $\tyint$, freeing $s$ and using $r$ in
the $\mkwd{print}$ expression.
The above derivation is used within derivation $\deriv{D}_2$:

{\scriptsize
\begin{mathpar}
    \inferrule
    {
        \inferrule
        {
            \inferrule*
            { }
            { \tseq
                {\var{f} : \usablety{\tysend{\msgtag{Put}}}}
                {\var{f}}
                {\usablety{\tysend{\msgtag{Put}}}}
            }
            \\
            \inferrule*
            { }
            { \tseq{\cdot}{5}{\tyint} }
        }
        { \tseq
            {\var{f} : \usablety{\tysend{\msgtag{Put}}}}
            {\send{\var{f}}{Put}{5}}
            {\one}
        }
        \\
        \inferrule*
        {
            \inferrule
            {
                \inferrule*
                {
                    \inferrule*
                    { }
                    {
                        \tseq
                        {\var{f} : \usablety{\tysend{\msgtag{Get}}}}
                        {\var{f}}
                        {\usablety{\tysend{\msgtag{Get}}}}
                    }
                }
                { \tseq
                    {\var{f} : \retty{\tysend{\msgtag{Get}}}}
                    {\var{f}}
                    {\usablety{\tysend{\msgtag{Get}}}}
                }
                \\
                \inferrule*
                { }
                { \tseq
                    {\var{s} : \usablety{\tysend{\msgtag{Reply}}}}
                    {\var{s}}
                    {\usablety{\tysend{\msgtag{Reply}}}}
                }
            }
            { \tseq
                {\var{f} : \retty{\tysend{\msgtag{Get}}}, \var{s} : \usablety{\tysend{\msgtag{Reply}}}}
                {\send{\var{f}}{Get}{\var{s}}}
                {\one}
            }
            \\
            \deriv{D}_1
        }
        {
            \tseq
            { \var{f} : \retty{\tysend{\msgtag{Get}}},
              \var{s} : \retty{\mbrecvone}
            }
                {
                    {
                        \bl
                            \send{\var{f}}{Get}{\var{s}}; \\
                            \guard{\var{s}}{G}
                        \el
                    }
                }
                { \one }
        }
    }
    { \tseq
        {\var{f} : \retty{\tysendp{\msgtag{Put} \patconcat \msgtag{Get}}},
         \var{s} : \retty{\mbrecvone}
        }
        {
            {
                \bl
                    \send{\var{f}}{Put}{5}; \;
                    \send{\var{f}}{Get}{\var{s}}; \\
                    \guard{\var{s}}{\cdots}
                \el
            }
        }
        {\one}
    }
\end{mathpar}
}

Here $\var{f}$ is used to send a \msgtag{Put} and then a \msgtag{Get} with  $\var{s}$ of type
$\usablety{\tysend{\msgtag{Reply}}}$
as a payload.
As the two sends to the $\var{f}$ mailbox are sequentially composed, the
type of $\var{f}$ at the root of the subderivation is
 $\retty{\tysendp{\patconcattwo{\msgtag{Put}}{\msgtag{Get}}}}$. Since \var{s} is used at type $\retty{\tyrecvp{\msgtag{Reply} \patconcat
\mbone}}$ in $\derivd_1$, the send and receive patterns balance out to the empty
mailbox type $\retty{\tyrecv{\mbone}}$.
Finally, we can construct the derivation for the entire term:

{\scriptsize
\begin{mathpar}
    \inferrule
    {
        \inferrule*
        { }
        { \tseq{\cdot}{\mbnew}{\retty{\mbrecvone}} }
        \quad
        \inferrule*
        {
            \inferrule*
                {
                    \inferrule*
                    {
                        \inferrule*
                        {}
                        {
                            \tseq
                            {\var{f} : \retty{\tyrecvp{\msgtag{Put} \patconcat \many{\msgtag{Get}}}}}
                            { \mkwd{emptyFuture}(\var{f}) }
                            { \one }
                        }
                    }
                    { \tseq
                        {\var{f} : \usablety{\tyrecvp{\msgtag{Put} \patconcat \many{\msgtag{Get}}}}}
                        {\spawn{\mkwd{emptyFuture}(\var{f})}}
                        {\one}
                    }
                }
                {
                    \tseq
                    {\var{f} : \usablety{\tyrecvp{(\msgtag{Put} \patconcat
                    \msgtag{Get}) \patconcat \mbone}}}
                        {\spawn{\mkwd{emptyFuture}(\var{f})}}
                        {\one}
                }
            \\
            \inferrule*
            {
                \inferrule*
                { }
                { \tseq{\cdot}{\mbnew}{\retty{\mbrecvone}}}
                \\
                \deriv{D}_2
            }
            {
                \tseq
                { \var{f} : \retty{\tysendp{\msgtag{Put} \patconcat
                \msgtag{Get}} } }
                {
                    \bl
                        \letintwo{\var{s}}{\mbnew} \\
                        \send{\var{f}}{Put}{5}; \cdots
                    \el
                }
                { \one }
            }
        }
        {
            \tseq
            { \var{f} : \retty{\mbrecvone} }
            {
                {
                    \bl
                        \spawn{\mkwd{emptyFuture}(\var{f})}; \\
                        \letintwo{\var{s}}{\mbnew} \\
                        \send{\var{f}}{Put}{5}; \cdots
                    \el
                }
            }
            { \one }
        }
    }
    {
        \tseq
        { \cdot }
        {
            {
                \bl
                \letintwo{\var{f}}{\mbnew} \\
                \spawn{\mkwd{emptyFuture}(\var{f})}; \\
                \letintwo{\var{s}}{\mbnew} \\
                \send{\var{f}}{Put}{5}; \;
                \send{\var{f}}{Get}{\var{s}}; \\
                \guardone{\var{s}} \\
                \quad \receivethree{Reply}{\var{r}}{\var{s}} \\
                \qquad \sugarfree{\var{s}};\\
                \qquad \mkwd{print}(\mkwd{intToString}(\var{r})) \\
                \}
                \el
            }
        }
        { \one }
    }
\end{mathpar}
}

Since we let-bind \var{f} to $\mbnew$, $\var{f}$ must have type
$\retty{\mbrecvone}$. Definition $\mkwd{emptyFuture}$ requires an argument of
type $\retty{\tyrecvp{\patconcattwo{\msgtag{Put}}{\many{\msgtag{Get}}}}}$; since
the function application appears in the body of the $\calcwd{spawn}$ we can mask
the usage annotation to $\usable$, and  use environment subtyping to
rewrite the type of \var{f} to $\usablety{\tyrecvp{(\msgtag{Put} \patconcat
        \msgtag{Get})
\patconcat \mbone}}$.
This then balances out with the use of \var{f} in
$\derivd_2$, completing the derivation.

\begin{figure}[t]
    {\small
  \header{Runtime syntax}

\[
  \begin{array}{lrcl}
    \text{Runtime names} & a \\
    \text{Values} & \vala, \valb & ::= & \cdots \midspace a \\
    \text{Names} & \nma, \nmb, \nmc & ::= & x \midspace a \\
    \text{Frames} & \sigma & ::= & \fframe{x}{M} \\
    \text{Frame stacks} & \framestack & ::= & \emptystack \midspace \frameconcat{\mvframe}{\framestack}\\
    \text{Guard contexts} & \ctx{G} & ::= & \seq{G_1} \cdot [~] \cdot \seq{G_2} \\
    \text{Configurations} & \config{C}, \config{D} & ::= &
      \thread{\tma}{\framestack} \midspace \sentmsg{a}{m}{\seq{V}}
                          \midspace \config{C} \parallel \config{D} \midspace \res{a} \config{C} \\
    \text{Type environments} & \tyenv & ::= & \cdots \midspace \tyenv, a : \usagety{\mbtya} \\
    \text{Runtime type environments} & \plainenv & ::= & \cdot \midspace \plainenv, \nma : \pretya
  \end{array}
\]

\headersig{Reduction rules}{$\config{C} \cevalp \config{D}$}

\[
    \begin{array}{lrcl}
        \textsc{E-Let} &
        \thread{\letin{x}{M}{N}}{\framestack} & \ceval &
        \thread{M}{\frameconcat{\fframe{x}{N}}{\framestack}}\\
\textsc{E-Return} &
        \thread{V}{\frameconcat{\fframe{x}{M}}{\framestack}} &
        \ceval & \thread{M \{ V / x \}}{\framestack} \\
\textsc{E-App} & \thread{\fnapp{f}{\seq{V}}}{\framestack} & \ceval &
        \thread{\tma \{ \seq{V} / \seq{x} \}}{\framestack} \\
                       & & &
        \quad (\text{if } \prog(f) = \fndef{f}{\seq{x: \tya}}{\tyb}{\tma}) \\
\textsc{E-New} & \thread{\mbnew}{\framestack} & \ceval &
        \resconf{a}{\thread{a}{\framestack}} \qquad (a \text{ is fresh}) \\
\textsc{E-Send} & \thread{\send{a}{m}{\seq{V}}}{\framestack} & \ceval &
        \thread{()}{\framestack} \parallel \sentmsg{a}{m}{\seq{V}} \\
\textsc{E-Spawn} & \thread{\spawn{M}}{\framestack} & \ceval &
        \thread{()}{\framestack} \parallel \thread{M}{\emptystack}
        \\
        \textsc{E-Free} & (\nu a)(\thread{\guard{a}{\ctx{G}[\free{M}]}}{\framestack}) & \ceval &
        \thread{M}{\framestack} 
        \\
            \multicolumn{4}{l}{
            \textsc{E-Recv} \:\:
                \thread
                    {\guard{a}{\ctx{G}[\receive{m}{\seq{x}}{y}{M}]}}
                    {\framestack}
                \parallel \sentmsg{a}{m}{\seq{V}} \: \ceval \:
                \thread
                    {M \{ \seq{V} / \seq{x}, a / y \}}
                    {\framestack}
                }
            \end{array}
    \]

       \begin{mathpar}
           \inferrule
           [E-Nu]
           {
               \config{C} \ceval \config{D}
           }
           {
               \res{a}{\config{C}} \ceval \res{a}{\config{D}}
           }

           \inferrule
           [E-Par]
           {
               \config{C} \ceval \config{C}'
           }
           { \config{C} \parallel \config{D} \ceval \config{C}' \parallel
           \config{D} }

           \inferrule
           [E-Struct]
           {
               \config{C} \equiv \config{C}' \\
               \config{C}' \ceval \config{D}' \\
               \config{D}' \equiv \config{D}
           }
           { \config{C} \ceval \config{D} }
       \end{mathpar}

    \headersig{Structural congruence}{$\config{C} \equiv \config{D}$}
    \begin{mathpar}
        \config{C} \parallel \config{D} \equiv \config{D} \parallel \config{C}

        \config{C} \parallel (\config{D} \parallel \config{D}')
        \equiv
        (\config{C} \parallel \config{D}) \parallel \config{D}'

        \config{C} \parallel (\nu a) \config{D} \equiv
        (\nu a)(\config{C} \parallel \config{D}) \quad \text{if } a \not\in
        \fn{\config{C}}

        (\nu a)(\nu b)\config{C} \equiv (\nu b)(\nu a)\config{C}
    \end{mathpar}
    }
  \caption{\langname operational semantics}
  \label{fig:declarative:semantics}
\end{figure}

\subsection{Operational semantics}
Figure~\ref{fig:declarative:semantics} shows the runtime syntax and reduction
rules for $\langname$. To allow runtime names to occur in values and
computations during reduction, we extend values $V$ with runtime names $a$. We
also extend type environments with mappings $a : \usagety{\mbtya}$ and lift the
operations on type environments to these extended environments.

The concurrent semantics of the language is described as a nondeterministic
reduction relation on a language of \emph{configurations}, which resemble
processes
in the $\pi$-calculus. A thread $\thread{\tma}{\framestack}$ evaluates term
$\tma$ with frame stack $\framestack$ (discussed shortly).  Configuration
$\sentmsg{a}{m}{\seq{\vala}}$ is a message $\msg{m}{\seq{\vala}}$ sent to
mailbox $a$; name restriction $\res{a}{\config{C}}$ binds name $a$ in
$\config{C}$; and $\config{C} \parallel \config{D}$ denotes the parallel
composition of $\config{C}$ and $\config{D}$.

Structural congruence $\config{C} \equiv \config{D}$ is defined as the smallest
congruence relation satisfying the rules in Figure~\ref{fig:declarative:semantics} and
is standard, capturing the commutativity and associativity of parallel
compositions; scope extrusion; and the unordered nature of name restrictions. We
let $\fn{\config{C}}$ denote the set of free runtime names in configuration
$\config{C}$.

\paragraph*{Running a program.}
To run a program $(\sigs, \seq{D}, \tma)$, we place its body $\tma$
into a thread with an empty frame stack: $\thread{\tma}{\emptystack}$.

\paragraph*{Frame stacks.}
We use frame stacks~\citep{Pitts98, EnnalsSM04} rather than evaluation
contexts for technical convenience: specifically, whereas evaluation
contexts are defined directly on the structure of terms, the explicit definition
of frame stacks makes it more convenient to define runtime typing rules that
allow us to more easily reason about quasi-linearity during reduction
(see~\secref{sec:metatheory}). 
A frame $\fframe{x}{\tma}$ is a pair of a
variable $x$ and a continuation $M$, where $x$ binds the result of the current
computation in $M$. A frame stack is an ordered sequence of frames, where
$\emptystack$ denotes the empty stack.

\paragraph*{Reduction rules.}
Frame stacks are best demonstrated by the \textsc{E-Let} and \textsc{E-Return}
rules: intuitively, the $\letin{x}{\tma}{\tmb}$ construct evaluates $\tma$, binding
its result to $x$ in $\tmb$. To capture this behaviour, the \textsc{E-Let} rule
adds a fresh frame $\fframe{x}{\tmb}$ to the top of a frame stack, and evaluates
$\tma$. Conversely, \textsc{E-Return}
returns $\vala$ into the top frame on the stack: if the top frame is $\fframe{x}{\tma}$,
then we can evaluate the continuation $\tma$ with $\vala$ substituted for $x$.
Rule \textsc{E-App} evaluates the body of function $f$ with arguments
$\seq{\vala}$ substituted for the parameters $\seq{x}$.

Rule \textsc{E-New} creates a fresh mailbox name restriction and returns the
fresh name. Rule \textsc{E-Send} sends a message with
tag $\msgtag{m}$ and payloads $\seq{V}$ to a mailbox $a$: it evaluates to $()$
and creates a sent message configuration $\sentmsg{a}{m}{\seq{\vala}}$.
Rule \textsc{E-Spawn} spawns a computation as a fresh process, with an empty
frame stack.
Rule \textsc{E-Free} allows a name $a$ to be garbage collected if it is not
contained in any other thread, evaluating the continuation $\tma$ of the
$\calcwd{free}$ guard. Finally, rule \textsc{E-Recv} handles receiving a message
from a mailbox, binding the payload values to $\seq{x}$ and the updated mailbox
name to $y$ in continuation $\tma$.
The remaining rules are administrative.

\subsection{Metatheory}\label{sec:metatheory}

In this section we describe \langname's metatheory: specifically type preservation and
a progress result. Consequently we show that a well-typed \langname
process will never process any messages that are not supported by its mailbox
types (\emph{mailbox conformance}),
and that processes are free of self-deadlocks, unexpected messages, and
forgotten replies.
We achieve this by introducing a runtime type system to maintain inductive
invariants during evaluation, and proving type preservation and a progress
result.

\subsubsection{Runtime typing}
To prove metatheoretical properties about $\langname$ we introduce
a type system on configurations; this type system is used only for reasoning and
is not required for typechecking.

\paragraph*{Runtime type environments.}
The runtime typing rules make use of a type environment $\plainenv$ that maps
variables to types that \emph{do not} contain usage information. Usage
information is inherently only useful in constraining \emph{sequential} uses of
a mailbox variable, where guards are blocking, whereas it makes little sense to
constrain \emph{concurrent} usages of a variable.
Runtime type environment combination $\pcombtwo{\plainenv_1}{\plainenv_2}$ is
similar to usage-annotated type environment combination but with two differences: it
is \emph{commutative} to account for the unordered nature of parallel
threads, and type combination does not include usage information.

\begin{definition}[Environment combination ($\plainenv$)]
  Environment combination $\pcombtwo{\plainenv_1}{\plainenv_2}$ is the
  smallest partial commutative binary operator on type environments closed under
  the following rules:

  {\small
  \begin{mathpar}
      \inferrule
      { }
      { \pcombtwo{\cdot}{\cdot} = \cdot }

      \inferrule
      {
        \nma \not\in \dom{\plainenv_2} \\
        \pcombtwo{\plainenv_1}{\plainenv_2} = \plainenv
      }
      {
        \pcombtwo{(\plainenv_1, \nma {:} \pretya)}{\plainenv_2} = \plainenv, \nma {:} \pretya
      }

      \inferrule
      {
        \nma \not\in \dom{\plainenv_1} \\
        \pcombtwo{\plainenv_1}{\plainenv_2} = \plainenv
      }
      {
        \pcombtwo{\plainenv_1}{(\plainenv_2, \nma {:} \pretya)} =
        \plainenv, \nma {:} \pretya
      }

      \inferrule
      { \pcombtwo{\plainenv_1}{\plainenv_2} = \plainenv }
      {
        \pcombtwo{(\plainenv_1, \nma {:} \pretya)}{(\plainenv_2, \nma {:} \pretyb)}
        =
        \plainenv, \nma {:} (\mbcombtwo{\pretya}{\pretyb})
      }
  \end{mathpar}
    }
\end{definition}

We can derive a runtime environment from a type environment by erasing all usage
annotations.

\begin{definition}[Usage erasure]
    The \emph{usage erasure} operator is defined as follows:
\begin{mathpar}
    \strip{\basety} = \basety

    \strip{\usagety{\mbtya}} = \mbtya
\end{mathpar}
We extend the operator to type environments by applying erasure 
pointwise on types, \ie,
$\strip{\nma_1 : \tya_1, \ldots, \nma_n : \tya_n} = \nma_1: \strip{\tya_1}, \ldots,
\nma_n : \strip{\tya_n}$.
\end{definition}

\begin{figure}[t]
{\small
\headersig{Configuration typing}{$\cseq{\plainenv}{\config{C}}$}
\begin{mathpar}
    \inferrule
    [TP-Nu]
    { \cseq{\plainenv, a : \mbrecvone}{ \config{C} } }
    { \cseq{\plainenv}{\res{a} \config{C}} }

    \inferrule
    [TP-Par]
    {
        \cseq{\plainenv_1}{\config{C}} \\
        \cseq{\plainenv_2}{\config{D}}
    }
    { \cseq{\pcombtwo{\plainenv_1}{\plainenv_2}}{\config{C} \parallel \config{D}} }

    \inferrule
    [TP-Message]
    { (\vseq{\makeusable{\plainenv_i}}{V_i}{\tya_i})_{i \in 1..n} \\
      \seq{\tya} \subtype \makeusable{\siglookup{m}}
    }
    { \cseq{\plainenv_1 + \cdots + \plainenv_n, a : \tysend{\msgtag{m}}}{\sentmsg{a}{m}{\seq{V}} } }

    \inferrule
    [TP-Thread]
    {
        \plainenv = \strip{\scombtwo{\tyenv_1}{\tyenv_2}}  \\
        \tseq{\tyenv_1}{\tma}{\tya} \\
        \stackseq{\tyenv_2}{\tya}{\framestack}
    }
    { \cseq{\plainenv}{\thread{\tma}{\framestack}} }

    \inferrule
    [TP-Sub]
    { \plainenv \subtype \plainenv' \\
      \cseq{\plainenv'}{\config{C}}
    }
    { \cseq{\plainenv}{\config{C}} }
\end{mathpar}

\headersig{Frame stack typing}{$\stackseq{\tyenv}{\tya}{\framestack}$}
\begin{mathpar}
    \inferrule
    [TF-Empty]
    {
    }
    { \stackseq{\cdot}{\tya}{\emptystack} }
\quad
    \inferrule
    [TF-Frame]
    {
        \tseq{\tyenv_1, x : \tya}{M}{\tyb} \\
        \returnablepred{\tya} \\
        \stackseq{\tyenv_2}{\tyb}{\framestack}
    }
    { \stackseq
        {\scombtwo{\tyenv_1}{\tyenv_2}}
        {\tya}
        {\frameconcat{\fframe{x}{\tma}}{\framestack}}
    }
\quad
    \inferrule
    [TF-Sub]
    {
        \tyenv_1 \subtype \tyenv_2 \\
\stackseq{\tyenv_2}{\tya}{\framestack}
    }
    { \stackseq
        {\tyenv_1}
        {\tya}
        {\framestack}
    } 
\end{mathpar}
}

  \caption{\langname runtime typing}
  \label{fig:declarative:rt-typing}
\end{figure}

Disjoint combination on runtime type environments $\plainenv_1 + \plainenv_2$
(omitted) is defined analogously to disjoint combination on $\tyenv$.

\paragraph*{Runtime typing rules.}
Figure~\ref{fig:declarative:rt-typing} shows the runtime typing rules.
Rule \textsc{TP-Nu} types a name restriction if the name is of type
$\mbrecvone$; in turn this ensures that sends and receives on the mailbox
``balance out'' across threads. Rule \textsc{TP-Par} allows configurations $\config{C}$ and
$\config{D}$ to be composed in parallel if they are typable under combinable
runtime type environments.
Rule \textsc{TP-Message} types a message configuration
$\sentmsg{a}{m}{\seq{V}}$. 
\ADDED{Since contexts can only contain a single occurrence of a variable,
name $a$ of type $\tysend{\msgtag{m}}$ cannot appear in
$\plainenv_1, \ldots, \plainenv_n$ and thus cannot occur in any of the values
sent as a payload.
This property is ensured in the corresponding static rule for message sends
(\textsc{T-Send}) because the environments used to type the target of the
sends and the message payloads are combined using the disjoint environment
combination operator $\pluscombtwo{\tyenv_1}{\tyenv_2}$, and therefore cannot
share mailbox-typed variables.
}
Each payload value $\vala$ must be a subtype of the type defined by the message
signature, under the second-class lifting of a disjoint runtime type
environment.
Rule \textsc{TP-Sub} allows subtyping on runtime type environments; the
subtyping relation $\plainenv \subtype \plainenv'$ is analogous to subtyping on
$\tyenv$.

\paragraph*{Thread and frame stack typing.}
Rule \textsc{TP-Thread} types a thread, consisting of a currently-evaluating
term (typable under $\tyenv_1$) and a frame stack (typable under $\tyenv_2$).
Since the term and the stack frames execute sequentially, $\tyenv_1 \scomb
\tyenv_2$ must be defined. Because usage annotations are thread-local, the
runtime environment needed to type the thread is $\tyenv_1 \scomb \tyenv_2$ with
annotations erased.
\textsc{TP-Thread} makes use of the frame stack typing judgement
$\stackseq{\tyenv}{\tya}{\framestack}$ (inspired by~\citet{EnnalsSM04}), which
can be read ``under type environment $\tyenv$, given a value of type $\tya$,
frame stack $\framestack$ is well-typed''.
The empty frame stack is typable under the empty
environment given any type (\textsc{TF-Empty}).
Rule \textsc{TF-Frame} details the typing rule for a non-empty frame stack
$\frameconcat{\fframe{x}{\tma}}{\framestack}$, which is
well-typed if continuation $\tma$ has type $\tyb$, given a variable $x$ of
returnable type $\tya$. The remainder of the stack must then be well-typed given $\tyb$.
We combine the environments used for typing the head term and the remainder of
the stack using $\scomb$ as we wish to account for sequential uses of a mailbox;
for example, in the term $\send{x}{m}{\vala}; \send{x}{n}{\valb}$, $x$ would
have type $\usablety{\tysendp{\patconcattwo{\msgtag{m}}{\msgtag{n}}}}$.
Finally, rule \textsc{TF-Sub} allows the use of subtyping when typing frame stacks.

\subsubsection{Preservation}

We can now state some metatheoretical results.  We give outlines of the salient
proofs here but full proofs can be found in
the extended version~\citep{extended}.

Typability is preserved by reduction; the proof is nontrivial since we must do
extensive reasoning about environment combination.

To begin with, we need to tame some of the complexity introduced by environment
subtyping, since environment subtyping allows a notion of weakening.  Read
top-down, \textsc{T-Sub} allows us to use environment subtyping to add a
variable with mailbox type $\usablety{\mbsendone}$ or base type $\basety$, or replace a
type with its subtype.
It is therefore useful to introduce a definition referring to environments
that can be populated by repeated uses of \textsc{T-Sub} and which might not be
used by a term.  We call these environments \emph{cruft}.

\begin{definition}[Cruft]
    A type environment $\tyenv$ is \emph{cruft} if $\tyenv \subtype \cdot$.
\end{definition}

It also helps to define a stricter version of environment subtyping that does not
permit weakening:

\begin{definition}[Strict environment subtyping]
An environment $\tyenv$ is a \emph{strict subtype environment} of an environment
$\tyenv'$, written $\strictsubtytwo{\tyenv}{\tyenv'}$, if
$\tyenv \subtype \tyenv'$ and $\dom{\tyenv} = \dom{\tyenv'}$.
\end{definition}

Let $\fv{-}$ denote the set of free variables in a term $\tma$ or frame stack
$\framestack$. We can now characterise when an environment is \emph{cruftless}
for a given term.

\begin{definition}[Cruftless]
We say that an environment $\tyenv$ is \emph{cruftless for a term} $\tma$ if
$\tseq{\tyenv}{\tma}{\tya}$ and $\dom{\tyenv} = \fv{\tma}$.
\end{definition}

The following lemma allows us to separate the type environment required for
typing the term from the cruft introduced by environment subtyping, and is used
extensively within the preservation proof.

\begin{lemma}\label{lem:disjoint-from-cruft}
    If $\tseq{\tyenv}{\tma}{\tya}$, then there exist $\tyenvb_1, \tyenvb_2,
    \tyenvb_3$ such
    that:

    \begin{itemize}
        \item $\tyenv = \tyenvb_1, \tyenvb_2$
        \item $\tseq{\tyenvb_3}{\tma}{\tya'}$
        \item $\tyenvb_1$ is cruftless for $\tma$, and $\tyenvb_1 \strictsubty \tyenvb_3$
        \item $\tya' \subtype \tya$
        \item $\cruftpred{\tyenvb_2}$
    \end{itemize}
\end{lemma}
\begin{proof}
    Follows from the definition of environment subtyping: read top-down, each
    application of environment subtyping will either add a variable with an
    unrestricted type, or alter the type of an existing variable.
\end{proof}

One of the most important lemmas used in the preservation proof uses
quasi-linear typing to show that if two environments $\tyenvb_1, \tyenvb_2 $ are
combined with a third environment $\tyenv$ using the $\scomb$ operator (\ie,
$\tyenvb_1, \tyenvb_2$ are used to type an evaluation frame), and all types in
environment $\tyenvb_1$ are returnable, then none of the
mailbox variables in $\tyenvb_1$ are present in $\tyenvb_2 \scomb \tyenv$.
This lemma is crucial for reasoning about nested evaluation contexts and follows
from the definition of usage combination.

\begin{lemma}\label{lem:returnable-non-aliased}
    If $\scombtwo{(\tyenvb_1, \tyenvb_2)}{\tyenv}$ is defined and
    $\returnablepred{\tyenvb_1}$,
    then $\scombtwo{(\tyenvb_1, \tyenvb_2)}{\tyenv} = \tyenvb_1 + (\scombtwo{\tyenvb_2}{\tyenv})$.
\end{lemma}
\begin{proof}
    Follows from the definition of usage combination: the $\scomb$ operation is
    not commutative for returnable mailbox types, so the returnable mailbox type
    must be the last occurrence of that name in the combination. For base types,
    the definitions of combination for $\scomb$ and $+$ coincide.
\end{proof}

The preservation theorem is interesting in that the type environment remains the
same before and after reduction, reflecting the observation that sends and
receives on a mailbox must eventually ``balance out''. This reasoning is
exemplified by the key \emph{balancing lemma} used in the proof of preservation,
showing the symmetry inherent in pattern inclusions:

\begin{lemma}[Balancing~\citep{dP18:mbc}]
    \label{lem:balancing}
    If $\patconcattwo{\msgtag{m}}{\patb} \subpatone \pata$
    and $\patb \not\subpatone \mbzero$, then
    $\subpat{\patb}{\pata \without \msgtag{m}}$.
\end{lemma}

We need several other lemmas to allow us to do equational reasoning on
environments.  For example, since we identify mailbox patterns up to
associativity and commutativity, we can straightforwardly show that environment
combination $\scombtwo{\tyenv_1}{\tyenv_2}$ is associative, and that runtime
environment combination $\pcombtwo{\plainenv_1}{\plainenv_2}$ is both
associative and commutative. 

\ADDED{
Recall that a type environment $\tyenv$ is reliable if all input mailbox
types $\usagety{\tyrecv{\pata}}$ in $\tyenv$ are reliable, specifically that each
$\usageann{\usage_1}{\tyrecv{\pata}} \not\subtype
\usageann{\usage_2}{\tyrecv{\mbzero}}$.
We extend this definition to runtime environments $\plainenv$.
}

\begin{restatable}[Preservation]{theorem}{preservation}
    If $\progseq{\prog}$, and
    $\cseq[\prog]{\plainenv}{\config{C}}$ with $\plainenv$ reliable, and
    $\config{C} \cevalpnopad \config{D}$, then $\cseq[\prog]{\plainenv}{\config{D}}$.
\end{restatable}
\begin{proof}
    By induction on the derivation of $\config{C} \ceval \config{D}$.
\end{proof}

Preservation implies mailbox conformance: the property that a configuration will
never evaluate a $\sugarfail{\vala}$ construct. To
state mailbox conformance, it is useful to define the notion of a
\emph{configuration context} $\config{H} ::= (\nu a)\ctx{H}
        \midspace \ctx{H} \parallel \config{C} \midspace
        \thread{[~]}{\framestack}$, which allows us to focus on a single
thread.

\begin{corollary}[Mailbox conformance]
    If $\progseq{\prog}$ and $\cseq[\prog]{\plainenv}{\config{C}}$ with $\plainenv$ reliable, then $\config{C}
    \mathop{\not{\cevalnopad^{*}}} \ctx{H}[\sugarfail{\vala}]$.
\end{corollary}

This corollary follows from preservation because,
in order to evaluate $\sugarfail{\vala}$, we would need a
mailbox with type $\retty{\tyrecv{\mbzero}}$ in the environment. This would
contradict the premise that $\plainenv$ is reliable.

\subsubsection{Progress}
To prove a progress result for $\langname$, we begin with some auxiliary definitions.

\begin{definition}[Message set]
    A \emph{message set} $\mathcal{M}$ is a configuration of the form:
    $\sentmsgg{a_1}{\msgtag{m}_1}{\seq{\vala_1}}
        \parallel \cdots \parallel
        \sentmsgg{a_n}{\msgtag{m}_n}{\seq{\vala_n}}$.
We say that a message set $\msgset$ \emph{contains a message} $\msgtag{m}$
\emph{for} $a$ if $\msgset \equiv \sentmsg{a}{m}{\seq{\vala}} \parallel
\msgset'$ for some $\msgset'$.
\end{definition}

Next, we define \emph{canonical forms}, which give us a global view of a
configuration.

\begin{definition}[Canonical form]
    A configuration $\config{C}$ is in \emph{canonical form} if it is of the
    form:
    \[
        (\nu a_1) \cdots (\nu a_m)(\thread{\tma_1}{\framestack_1} \parallel
        \cdots \parallel \thread{\tma_n}{\framestack_n} \parallel \msgset)
    \]
\end{definition}

Every configuration can be written in canonical form; the result follows from
repeated application of the structural congruence rules. As a result we can
reason about any arbitrary configuration by rewriting it in canonical form.

\begin{proposition}[Canonical forms]
    \ADDED{For every configuration $\config{C}$}, there exists some $\config{D}$
    such that $\config{C} \equiv \config{D}$ and $\config{D}$ is in canonical
    form.
\end{proposition}

We need two further definitions to characterise a thread that is blocked while
waiting for a message to arrive. 

\begin{definition}[Waiting]
We say that a term $\tma$ is \emph{waiting on mailbox $a$} if $\tma$ can be
written $\guard{a}{\seq{G}}$ for some guards $\seq{G}$.

We say that a term $\tma$ is \emph{waiting on mailbox $a$ for a message with
tag} $\msgtag{m}$,
written $\waiting{\tma}{a}{\msgtag{m}}$,
if
$\tma$ can be written
$
    \guard{a}{\mathcal{G}[\receive{m}{\seq{x}}{y}{\tmb}]}
$ for some guard context $\ctx{G}$.
We say that $\tma$ is a \emph{waiting term} if there exists some $a$ such that
$\tma$ is waiting on mailbox $a$.
\end{definition}

\begin{definition}[Guard clauses]
    The \emph{guard clauses} of a waiting term $\tma =
    \guard{\vala}{\seq{\gd}}$ are the guards $\seq{G}$. 
\end{definition}

With these definitions in hand, we can state the notion of progress enjoyed by
\langname.

We begin by showing functional reduction, \ie, that threads can always reduce
up to communication and concurrency constructs.

\begin{lemma}[Progress (functional reduction)]\label{lem:thread-progress}
    If $\cseq[\prog]{\plainenv}{\thread{\tma}{\framestack}}$ with $\plainenv$
    reliable, then either:
    \begin{itemize}
        \item $\tma$ is a value and $\framestack = \emptystack$; or
        \item there exist some $\tma', \framestack'$ such that $\thread{\tma}{\framestack} \ceval
            \thread{\tma'}{\framestack'}$; or
            \item $\tma$ is a communication and concurrency construct, \ie, $\mbnew$, or
        $\spawn{\tma}$, or \\$\send{\vala}{m}{\seq{\valb}}$, or $\guard{\vala}{\seq{\gd}}$.
    \end{itemize}
\end{lemma}
\begin{proof}
    By induction on the derivation of $\cseq{\plainenv}{\thread{\tma}}{\framestack}$
    and inspection of the reduction rules.
\end{proof}

We can then use canonical forms to characterise a progress
result: either a configuration can reduce,
or each constituent thread has either reduced to a value, or is waiting for
a message that has not yet been sent by a different thread.

\begin{restatable}[Partial progress]{theorem}{progress}\label{thm:weak-progress}
    Suppose $\progseq{\prog}$ and $\cseq[\prog]{\cdot}{\config{C}}$ where $\config{C}$ is in canonical form:
    \[
\config{C} = (\nu a_1) \cdots (\nu a_m)(\thread{\tma_1}{\framestack_1} \parallel
    \cdots \parallel \thread{\tma_n}{\framestack_n} \parallel \msgset)
    \]

    Then either there exists some $\config{C}'$ such that
    $\config{C} \ceval \config{C}'$, or
    for each $\tma_i$, either:
    \begin{itemize}
        \item $\tma_i$ is a value and $\framestack_i = \epsilon$; or
        \item $\tma_i$ is waiting on some mailbox $a$ with guard clauses
            $\seq{G_i}$,
            and for all
            $\msgtag{m}_j$ such that
            $\waiting{\tma_i}{a}{\msgtag{m}_j}$,
            message set $\msgset$ does not contain a message
            $\sentmsgg{a}{\msgtag{m}_j}{\seq{\vala}}$,
            and $a \not\in \fn{\seq{\gd_i}} \cup \fn{\framestack_i}$.
    \end{itemize}
\end{restatable}
\begin{proof}
    Functional reduction enjoys progress~(Lemma~\ref{lem:thread-progress}), and the
    constructs $\mbnew$, $\spawn{\tma}$, and $\send{a}{m}{\seq{V}}$ can all
    always reduce by \textsc{E-New}, \textsc{E-Spawn}, or \textsc{E-Send}.
    Therefore, the body of an irreducible thread
    $\thread{\tma_k}{\framestack_k}$ must be waiting: \ie, it must be of the
    form $\thread{\guard{a}{\seq{\gd}}}{\framestack_k}$ for some name
    $a$ and guards $\seq{\gd}$. 

    There are four cases to consider:

    \begin{enumerate}
        \item That there exists some $\msgtag{m}$ such that
            $\waiting{\tma_k}{a}{\msgtag{m}}$ and there exists some sent
            message $\sentmsg{a}{\msgtag{m}}{\vala}$ in the message set
            $\msgset$. In this case, we can reduce by \textsc{E-Recv}.
        \item That there exist no messages for $a$ in $\msgset$, and $a$ does
            not occur free in $\seq{\gd}$, $\framestack_k$, or any other thread.
            In this case by \textsc{TP-Nu}, $a$ must have some type
            $\usagety{\mbtya}$ such
            that $\retty{\mbrecvone} \subtype \usagety{\mbtya}$ and as such
            $\seq{\gd}$ must include a $\calcwd{free}$ guard. We can therefore
            reduce by \textsc{E-Free}.
        \item That $a$ occurs free in $\seq{\gd}$ or $\framestack_k$. This is
            impossible because the subject of a guard must be returnable, and
            therefore by Lemma~\ref{lem:returnable-non-aliased} cannot occur in
            the guards or frame stack.
        \item That there exist no messages for $a$ in $\msgset$ but $a$ occurs
            free in some other waiting thread, indicating a cyclic
            inter-process dependency.  This satisfies the second clause of the
            theorem statement.\qedhere
    \end{enumerate}
\end{proof}

\paragraph*{Consequences of progress.}
The progress theorem shows how \langname can rule out self-deadlocks, forgotten
replies, and unexpected messages.

Specifically, a self-deadlock occurs when a process is waiting on a name $a$
where the message is sent in a later part of the same thread: in our case,
either the guard body or the frame stack. This is ruled out by the second case
of the progress theorem.

A forgotten reply occurs when a process is waiting for a message that will never
be sent. In this case, we would have a process that is waiting on a name $a$ but
no other thread has a reference to send the message. This is evidenced by the
second case of the theorem statement and more specifically case (2) of the
proof: the only way a thread can be waiting on $a$ while no other threads have a
reference to $a$ is if the guard expression has a $\calcwd{free}$ guard, in
which case the configuration would be able to free the mailbox.

An unexpected message occurs when a process receives a message that is not
accounted for by any of its $\calcwd{receive}$ clauses: in a language like
Erlang this would manifest itself as a process terminating while still containing
messages in its mailbox. Again, this cannot occur: consider the case where all
processes are terminated. By \textsc{TP-Nu} it would be impossible for there to
be a message in the message set without a corresponding input reference.

\paragraph*{Deadlocks.}

Although \langname rules out self-deadlocks, it does not rule out
inter-process deadlocks. For example, the following (correct) processes encode
a request-response pattern:

{\small
    \begin{minipage}{0.6\textwidth}
\[
    \bl
    \fndefthree{\mkwd{requester}}{\var{self}:
    \tyrecv{\msgtag{Response}}, \var{other}: \tysend{\msgtag{Request}}}{\tyunit} \\
        \quad \send{\var{other}}{Request}{}; \\
        \quad \guardanntwo{\var{self}}{\msgtag{Response}} \\
        \qquad \receive{Response}{}{\var{self}}{\sugarfree{\var{self}}} \\
        \quad \} \\
    \} \\ \\
\fndefthree{\mkwd{responder}}{\var{self}:
    \tyrecv{\msgtag{Request}}, \var{other}: \tysend{\msgtag{Response}}}{\tyunit} \\
        \quad \guardanntwo{\var{self}}{\msgtag{Request}} \\
        \qquad \receivethree{Request}{}{\var{self}} \\
        \qqquad \send{\var{other}}{Response}{}; \\
        \qqquad \sugarfree{\var{self}} \\
        \quad \} \\
    \} \\
    \el
\]
\end{minipage}
\hfill
\begin{minipage}{0.35\textwidth}
    \[
        \bl
    \fndefthree{\mkwd{main}}{}{\tyunit} \\
        \quad \letintwo{\var{mb1}}{\mbnew} \\
        \quad \letintwo{\var{mb2}}{\mbnew} \\
        \quad \spawn{\mkwd{requester}(\var{mb1}, \var{mb2})}; \\
        \quad \spawn{\mkwd{responder}(\var{mb2}, \var{mb1})} \\
    \}
    \el
    \]
\end{minipage}
} \\

However, if we were to modify the $\mkwd{requester}$ process to send the request
only after receiving the response, which would result in a deadlock, the program
would still be accepted by our type system because we do not have any static way
of ruling out inter-process deadlocks:

{\small
\[
    \bl
    \fndefthree{\mkwd{badRequester}}{\var{self}:
    \tyrecv{\msgtag{Response}}, \var{other}: \tysend{\msgtag{Request}}}{\tyunit} \\
        \quad \guardanntwo{\var{self}}{\msgtag{Response}} \\
        \qquad \receivethree{Response}{}{\var{self}} \\
        \qqquad \send{\var{other}}{Request}{}; \: \sugarfree{\var{self}} \\
        \quad \} \\
    \}
    \el
\]
}

Although the mailbox calculus can rule out deadlocks using a dependency
graph~\citep{dP18:mbc}, dependency graphs are difficult to integrate with the
richer structure of a programming language (see~\secref{sec:dgs}). We look forward to
investigating inter-process deadlock detection in future work.
 \section{Algorithmic typing}\label{sec:algorithmic}
Writing a typechecker based on \langname's declarative
typing rules is challenging due to nondeterministic context splits, environment
subtyping, and pattern inclusion. \emph{MC}$^2$~\citep{mc2} is a
typechecker for the mailbox calculus, based on a typechecker for
concurrent object usage protocols~\citep{Padovani18}. The \emph{MC}$^2$ type
system has, however, not been formalised, and does not scale naturally to a
programming language.  We adopt several ideas from \emph{MC}$^2$, especially
algorithmic type combination, and adapt the approach for a programming language.

\begin{figure}[t]

{\small
    \[
    \begin{array}{lrcl}
        \text{Pattern variables} & \patvara, \patvarb \\
\text{Mailbox patterns} & \ppata, \ppatb & ::= & \mbzero \midspace \mbone \midspace
           \algomsg{m} \midspace \patplustwo{\ppata}{\ppatb} \midspace
           \patconcattwo{\ppata}{\ppatb} \midspace \many{\ppata} \midspace
           \patvara \\
\text{Mailbox types} & \pmbty & ::= & \tysend{\ppata}  \midspace \tyrecv{\ppata} \\
\text{Types} & \ppretya, \ppretyb & ::= & \basety \midspace \pmbty \\
\text{Usage-annotated types} & \ptya, \ptyb & ::= & \basety \midspace \usagety{\pmbty} \\
\text{Type environments} & \penv & ::= & \cdot \midspace \penv, x : \ptya \\
\text{Constraints} & \phi & ::= & \subpatconstr{\ppata}{\ppatb} \\
        \text{Constraint sets} & \constrs \\
    \end{array}
\]

\[
    \ADDED{
        \begin{array}{lrcl}
        \text{Open signatures} & \psigs & ::= & \seq{\msgtag{m} \mapsto \seq{\ppretya}} \\
        \text{Annotated definitions} & \pdef & ::= & \fndef{f}{\seq{x : \ptya}}{\ptyb}{\ptma} \\
                \text{Annotated programs} & \pprog & ::= &
                \progdef{\psigs}{\seq{\pdef}}{\ptma} \\
        \text{Annotated computations} & \ptma, \ptmb & ::=  & \vala \midspace
          \letin[\tya]{x}{\ptma}{\ptmb} \midspace \fnapp{f}{\seq{V}} \\
                       &         & \mid & \spawn{\ptma} \midspace \mbnew
                       \midspace \send{\vala}{m}{\seq{\valb}} \midspace
                       \guardann{\vala}{\pata}{\seq{\pgd}} \midspace
                       \sugarfail{\vala} \\
\text{Annotated guards} & \pgd & ::= & \free{\ptma} \midspace
          \receive{m}{\seq{x}}{y}{\ptma}
    \end{array}
    }
    \]
}
    \caption{\langname syntax extended for algorithmic typing}
    \label{fig:extended-syntax}
\end{figure}
\paragraph*{Type system overview.}
Our algorithmic type system takes a
\emph{co-contextual} approach~\citep{ErdwegBKKM15}: rather than taking a type environment
as an \emph{input} to the typechecking algorithm, we produce a type environment
as an \emph{output}. The intuition is that, read bottom-up,
\emph{splitting} an environment into two sub-environments is harder than
\emph{merging} two environments inferred from subexpressions.
We also generate \emph{pattern inclusion constraints} to be
solved later.

Bidirectional type systems~\citep{PierceT00, DunfieldK21} split typing rules into
two classes: those that \emph{synthesise} a type $\tya$ for a term $\tma$
($\tradsynth{\tyenv}{\tma}{\tya}$), and those that
\emph{check} that a term $\tma$ has type $\tya$
($\tradchk{\tyenv}{\tma}{\tya}$). Bidirectional type systems
are syntax-directed and amenable to implementation.

We use a co-contextual variant of bidirectional typing first introduced
by~\citet{Zeilberger18:bbt}. The main twist is the variable rule, which becomes a
\emph{checking} rule and records the given variable-type mapping in the
inferred environment.

\subsection{Algorithmic type system}
\label{sec:AlgorithmicTypeSystem}

\paragraph*{Extended syntax and annotation.}
A key difference in comparison to the declarative type system is the addition of
\emph{pattern variables} $\patvara$, which act as placeholders for parts of 
patterns and are generated during typechecking.
We can then generate and solve \emph{inclusion constraints} $\constr$
on patterns.  Figure~\ref{fig:extended-syntax} shows the extended syntax used in
the algorithmic system.

Our algorithmic type syntax uses a type annotation on
$\calcwd{let}$-expressions, used when establishing completeness with respect to
the declarative system.  As we will see, in the vast majority of cases this
annotation is unnecessary, and such type annotations are not required by our
implementation.
The algorithmic type system also requires annotations on
$\calcwd{guard}$
expressions; these \emph{are} required by our implementation and we will discuss
the necessity of these annotations when describing the algorithmic typing rules.
Annotated computations $\ptma, \ptmb$ 
replace the $\letin{x}{\tma}{\tmb}$ construct from the declarative system with
an annotated $\letin[\tya]{x}{\ptma}{\ptmb}$ construct, 
and replace the $\guard{\vala}{\seq{\gd}}$
construct from the declarative system with a
$\guardann{\vala}{\pata}{\seq{\pgd}}$ expression that involves guarding on
mailbox $\vala$ and asserting that it has pattern $\pata$. Annotated guards
$\pgd$ are modified to include annotated expressions in their bodies.

We also introduce \emph{open signatures} $\psigs$ that allow message payload
types to contain pattern variables; \emph{annotated definitions} $\pdef$ that
allow function arguments and return types to contain pattern variables and where
the function body is an annotated computation; and extend a program to include
an open signature, annotated definitions, and an annotated body.

\paragraph*{Constraints.}
An important challenge for the algorithmic type system is determining whether one
pattern is \emph{included} within another: for example, $\msgtag{m} \subpatone
\many{\msgtag{m}}$.
Given that patterns may contain pattern variables, we may need to defer
inclusion checking until more pattern variables are known, so we
introduce inclusion constraints $\subpatconstr{\ppata}{\ppatb}$ to require
that pattern $\ppata$ is included in pattern $\ppatb$.

\begin{figure}[t]
{\small
\headersig{Unrestrictedness}{$\unrseq{\ptya}{\constrs}$}
\begin{mathpar}
\inferrule
    { }
    { \unrseq{\basety}{\emptyset} }

\inferrule
    { }
    { \unrseq{\usagety{\tysend{\ppata}}}{\set{\subpatconstr{\mbone}{\ppata}}} }
\end{mathpar}

\headertwo
{Subtyping}
    {\framebox{$\usage_1 \subtype \usage_2$}
     \framebox{$\subtyseq{\ptya}{\ptyb}{\constrs}$}
}
\begin{mathpar}
    \inferrule
    { }
    { \subtypetwo{\usage}{\usage} }
\quad\:
    \inferrule
    { }
    { \subtypetwo{\bcirc}{\wcirc} }
\quad\:
\inferrule
    { }
    { \subtyseq{\basety}{\basety}{\emptyset} }
\quad\:
\inferrule
    { \usage_1 \subtype \usage_2}
    { \subtyseq
        {\usageann{\usage_1}{\pmbty}}
        {\usageann{\usage_2}{\pmbty}}
        {\emptyset}
    }
\quad\:
    \inferrule
    {
        \subtypetwo{\usage_1}{\usage_2}
    }
    {
        \subtyseq
            {\usageann{\usage_1}{\tysend{\ppata}} }
            {\usageann{\usage_2}{\tysend{\ppatb}} }
            { \set{\subpatconstr{\ppatb}{\ppata}} }
    }
\quad\:
    \inferrule
    {
        \subtypetwo{\usage_1}{\usage_2}
    }
    {
        \subtyseq
            {\usageann{\usage_1}{\tyrecv{\ppata}}}
            {\usageann{\usage_2}{\tyrecv{\ppatb}}}
            { \set{\subpatconstr{\ppata}{\ppatb}} }
    }
\end{mathpar}

\headertwo
    {Sequential merge}
    {\framebox{$\joinseq{\pmbty_1}{\pmbty_2}{\pmbty}{\constrs}$}~\framebox{$\joinseq{\ptya_1}{\ptya_2}{\ptyb}{\constrs}$}}
\begin{mathpar}
\inferrule
    { }
    { \joinseq
        {\tysend{\ppata}}
        {\tysend{\ppatb}}
        {\tysendp{\patconcattwo{\ppata}{\ppatb}}}
        {\emptyset}
    }

\inferrule
    {
        \patvara {\text{ fresh}}
    }
    {
        \joinseq
            { \tysend{\ppata} }
            { \tyrecv{\ppatb} }
            { \tyrecv{\patvara} }
            { \{ \subpatconstr{(\patconcattwo{\ppata}{\patvara})}{\ppatb} \} }
    }

\inferrule
    {
        \patvara {\text{ fresh}}
    }
    {
        \joinseq
            {\tyrecv{\ppata}}
            {\tysend{\ppatb}}
            {\tyrecv{\patvara}}
            { \{ \subpatconstr{(\patconcattwo{\ppatb}{\patvara})}{\ppata} \} }
    }

\inferrule
    {
        \joinseq{\pmbty_1}{\pmbty_2}{\pmbty}{\constrs}
    }
    {
        \joinseq
            {\usageann{\usage_1}{\pmbty_1}}
            {\usageann{\usage_2}{\pmbty_2}}
            {\usageann{\usagecombtwo{\usage_1}{\usage_2}}{\pmbty}}
            { \constrs }
    }

\inferrule
    {
}
    { \joinseq
        {\basety}
        {\basety}
        {\basety}
        {\emptyset}
    }
\end{mathpar}

\headertwo
    {Branching merge}
    {\framebox{$\mergeseq{\pmbty_1}{\pmbty_2}{\pmbty}{\constrs}$}
     \framebox{$\mergeseq{\ptya_1}{\ptya_2}{\ptyb}{\constrs}$}}
\begin{mathpar}
\inferrule
    { }
    {
        \mergeseq
          { \tysend{\ppata} }
          { \tysend{\ppatb} }
          { \tysendp{\patplustwo{\ppata}{\ppatb}} }
          { \emptyset}
    }

\inferrule
    {
        \patvara \text{ fresh}
    }
    {
        \mergeseq
        { \tyrecv{\ppata} }
        { \tyrecv{\ppatb} }
        { \tyrecv{\patvara} }
        { \set
            {
                \subpatconstr{\patvara}{\ppata},
                \subpatconstr{\patvara}{\ppatb}
            }
        }
    }

    \inferrule
    {
        \mergeseq{\pmbty_1}{\pmbty_2}{\pmbty}{\constrs}
    }
    {
        \mergeseq
        { \usageann{\usage_1}{\pmbty_1} }
        { \usageann{\usage_2}{\pmbty_2}}
        { \usageann{\minusage{\usage_1}{\usage_2}}{\pmbty}}
        { \constrs }
    }

\inferrule
    { }
    { \mergeseq
        {\basety}
        {\basety}
        {\basety}
        {\emptyset}
    }
\end{mathpar}
}
\caption{\langname algorithmic type operations}
\label{fig:algo-type-operations}
\end{figure}

Figure~\ref{fig:algo-type-operations} shows the algorithmic type operators.

\paragraph*{Unrestrictedness and subtyping.}
The algorithmic unrestrictedness operation $\unrseq{\ptya}{\constrs}$ states that
$\ptya$ is unrestricted subject to constraints $\constrs$, and the definition
reflects the corresponding definitions in the declarative system.
Algorithmic subtyping is similar: a base type is a subtype of itself,
and we check that one mailbox type is a subtype of another by checking that the
capabilities match, and generating a contravariant constraint for a send type,
and a covariant constraint for a receive type.

\paragraph*{Algorithmic sequential merge.}
Declarative mailbox typing relies  on
the subtyping rule to manipulate types into a form where
they can be combined with the type combination operators, \eg,
$\mbcombtwo{\tysend{\pata}}{\tyrecvp{\pata \patconcat \patb}} =
\tyrecv{\patb}$. The algorithmic type system cannot apply the same technique as it does
not know, \emph{a priori}, the form of each pattern. Instead, the
\emph{algorithmic sequential merge} operation allows the combination of two mailbox types irrespective
of their syntactic form. Combining two send types is the same as in the
declarative system, but combining a send type with a receive type (and vice
versa) is more interesting: say we wish to combine $\tysend{\ppata}$ and
$\tyrecv{\ppatb}$. In this case, we generate a fresh pattern variable
$\patvara$; the result is $\tyrecv{\patvara}$ along with the constraint that
$\subpatconstr{(\ppata \patconcat \patvara)}{\ppatb}$, namely that the send
pattern composed with the fresh pattern variable is included in the pattern
$\ppatb$.

As an example, applying the sequential merge to types $\tysend{\msgtag{m}}$ and
$\tyrecvp{\patconcattwo{\msgtag{n}}{\msgtag{m}}}$ produces an input mailbox
type $\tyrecv{\patvara}$ and a constraint $\subpatconstr{(\msgtag{m} \patconcat
\patvara)}{(\patconcattwo{\msgtag{n}}{\msgtag{m}})}$, for which a valid solution
is $\patvara \mapsto \msgtag{n}$, and hence gives rise to the expected combined
type $\tyrecv{\msgtag{n}}$.

\paragraph*{Algorithmic branching merge.}
In the declarative type system, branching control flow requires that each branch
is typable under the same type environment (using the \textsc{T-Sub} rule).
The algorithmic type system instead generates constraints that ensure that each
type is used consistently across branches using the \emph{algorithmic branching merge} operation
$\mergeseq{\ptya_1}{\ptya_2}{\ptyb}{\constrs}$.
Two base types are merged if they are identical.
In the case of mailbox types, the function takes the
\ADDED{minimum usage annotation, \ie,
$\minusage{\returnable}{\usable} = \returnable$}. It ensures that when
merging two output capabilities the patterns are combined using
pattern disjunction. Conversely merging two input capabilities generates a
new pattern variable that must be included in both merged patterns.

\begin{figure}[t]
    
    {\small
    \headersig{Environment sequential merge}{$\joinseq{\penv_1}{\penv_2}{\penv}{\constrs}$}
    \begin{mathpar}
    \inferrule
    { }
    { \joinseq{\cdot}{\cdot}{\cdot}{\emptyset} }

    \inferrule
    {
        x \not\in \dom{\penv_2} \\
        \joinseq{\penv_1}{\penv_2}{\penv}{\constrs}
    }
    { \joinseq{\penv_1, x {:} \ptya}{\penv_2}{\penv, x {:} \ptya}{\constrs}  }

    \inferrule
    {
        x \not\in \dom{\penv_1} \\
        \joinseq{\penv_1}{\penv_2}{\penv}{\constrs}
    }
    { \joinseq{\penv_1}{\penv_2, x {:} \ptya}{\penv, x {:} \ptya}{\constrs}  }

    \inferrule
    {
        \joinseq{\ptya_1}{\ptya_2}{\ptyb}{\constrs_1} \\
        \joinseq{\penv_1}{\penv_2}{\penv}{\constrs_2}
    }
    { \joinseq
        {\penv_1, x {:} \ptya_1}
        {\penv_2, x {:} \ptya_2}
        {\penv, x {:} \ptyb}{\constrs_1 \cup \constrs_2}
    }
    \end{mathpar}

\headersig{Environment branching merge}{$\mergeseq{\penv_1}{\penv_2}{\penv}{\constrs}$}
    \begin{mathpar}
    \inferrule
    { }
    { \mergeseq{\cdot}{\cdot}{\cdot}{\emptyset} }

    \inferrule
    {
        x \not\in \dom{\penv_2} \\
        \mergeseq{\penv_1}{\penv_2}{\penv}{\constrs}
    }
    {
        \mergeseq
        { \penv_1, x {:} \basety }
        { \penv_2 }
        { \penv, x {:} \basety }
        { \constrs }
    }

    \inferrule
    {
        x \not\in \dom{\penv_2} \\
        \mergeseq{\penv_1}{\penv_2}{\penv}{\constrs}
    }
    {
        \mergeseq
        { \penv_1, x {:} \usagety{\tysend{\ppata}} }
        { \penv_2 }
        { \penv, x {:} \usagety{\tysendp{\ppata \patplus \mbone}} }
        { \constrs }
    }

    \inferrule
    {
        x \not\in \dom{\penv_1} \\
        \mergeseq{\penv_1}{\penv_2}{\penv}{\constrs}
    }
    {
        \mergeseq
        { \penv_1 }
        { \penv_2, x {:} \basety }
        { \penv, x {:} \basety }
        { \constrs  }
    }

    \inferrule
    {
        x \not\in \dom{\penv_1} \\
        \mergeseq{\penv_1}{\penv_2}{\penv}{\constrs}
    }
    {
        \mergeseq
        { \penv_1 }
        { \penv_2, x {:} \usagety{\tysend{\ppata}} }
        { \penv, x {:} \usagety{\tysendp{\ppata \patplus \mbone}} }
        { \constrs }
    }

    \inferrule
    {
        \mergeseq{\ptya_1}{\ptya_2}{\ptyb}{\constrs_1} \\
        \mergeseq{\penv_1}{\penv_2}{\penv}{\constrs_2}
    }
    {
        \mergeseq
        { \penv_1, x {:} \ptya_1 }
        { \penv_2, x {:} \ptya_2 }
        { \penv, x {:} \ptyb }
        { \constrs_1 \cup \constrs_2 }
    }
    \end{mathpar}

    \headersig{Disjoint combination}{$\combineseq{\penv_1}{\penv_2}{\penv}{\constrs}$}
    \begin{mathpar}
        \inferrule
        { }
        { \combineseq{\cdot}{\cdot}{\cdot}{\emptyset} }
~~~
        \inferrule
        {
            x \not\in \dom{\penv_2} \\\\
            \combineseq{\penv_1}{\penv_2}{\penv}{\constrs}
        }
        { \combineseq{\penv_1, x {:} \ptya}{\penv_2}{\penv, x {:} \ptya}{\constrs}}
~~~
        \inferrule
        {
            x \not\in \dom{\penv_1} \\\\
            \combineseq{\penv_1}{\penv_2}{\penv}{\constrs}
        }
        { \combineseq{\penv_1}{\penv_2, x {:} \ptya}{\penv, x {:} \ptya}{\constrs}}
~~~
        \inferrule
        {
            \combineseq{\penv_1}{\penv_2}{\penv}{\constrs} \\
        }
        { \combineseq
            {\penv_1, x {:} \basety}
            {\penv_2, x {:} \basety}
            {\penv, x {:} \basety}
            {\constrs}
        }
    \end{mathpar}
    }
    \caption{Algorithmic environment combination}
    \label{fig:algorithmic:env-comb}
\end{figure}

\paragraph*{Algorithmic environment combination.}
Figure~\ref{fig:algorithmic:env-comb} shows how the algorithmic type combination
operators can be extended to type environments.

The environment sequential merge operator
$\joinseq{\penv_1}{\penv_2}{\penv}{\constrs}$
concatenates $\penv_1$ and $\penv_2$, computing the sequential merge of any
types for overlapping variables, and produces constraints $\constrs$.
The environment branching merge operator
$\mergeseq{\penv_1}{\penv_2}{\penv}{\constrs}$
computes the algorithmic branching merge of $\penv_1$ and $\penv_2$.
If a variable is contained in both environments, then the
merged type is used in the output environment.
If the variable is only contained in one of the environments, then the result
depends on the type.
If the variable is only contained in one of the environments and has type
$\usagety{\tysend{\ppata}}$, then its type is changed to
$\usagety{\tysendp{\ppata \patplus \mbone}}$ to denote the fact that it may not
be used. Note that this only applies to mailbox types with an \emph{output}
capability since mailbox types with an \emph{input} capability must be treated
linearly.

Disjoint environment combination combines two environments; if a variable is
used in both environments then it must have an identical base type.

\begin{figure}[t]
{\small
    \headertwo
    {Constraint generation for programs and definitions}
    {
        \framebox{$\algoprogseq{\pprog}{\constrs}$}~
        \framebox{$\algodefseq[\pprog]{\pdef}{\constrs}$}
    }
    \begin{mathpar}
        \inferrule
        {
            \pprog = \progdef{\psigs}{\seq{\pdef}}{\ptma} \\\\
            (\algodefseq{\pdef_i}{\constrs_i})_{i \in 1..n} \\
            \chkseq{\ptma}{\one}{\cdot}{\constrs}
        }
        { \algoprogseq{\pprog}{\constrs \cup \constrs_1 \cup \cdots \cup \constrs_n} }

        \inferrule
        {
            \chkseq{\ptma}{\ptyb}{\penv}{\constrs_1} \\\\
            \checkenvseq{\penv}{\seq{x}}{\seq{\ptya}}{\constrs_2} \\
            \penv \envwithout \seq{x} = \cdot
        }
        { \algodefseq[]{\fndef{f}{\seq{x : \ptya}}{\ptyb}{\ptma}}{ \constrs_1
        \cup \constrs_2} }
    \end{mathpar}

    \headertwo
        {Constraint generation (synthesis)}
        {
            \framebox{$\synthseq[\pprog]{\ptma}{\ptya}{\penv}{\constrs}$}
        }
\begin{mathpar}
    \inferrule
    [TS-Const]
    { c \text{ has base type } \basety }
    { \synthseq{c}{\basety}{\cdot}{\emptyset} }

    \inferrule
    [TS-New]
    { }
    {
        \synthseq{\mbnew}{\retty{\mbrecvone}}{\cdot}{\emptyset}
    }

    \inferrule
    [TS-Spawn]
    {
        \chkseq{\ptma}{\one}{\penv}{\constrs}
    }
    { \synthseq{\spawn{\ptma}}{\one}{\makeusable{\penv}}{\constrs} }

    \inferrule
    [TS-Send]
    {
        \pprog(\msgtag{m}) = \seq{\ppretya} \\
        \chkseq{\vala}{\usablety{\tysend{\algomsg{m}}}}{\penv'}{\constrs} \\
        (\chkseq{\valb_i}{\makeusable{\ppretya_i}}{\penv'_i}{\constrs'_i})_{i
        \in 1..n} \\
        \combineseqmany{\penv' + \penv'_1 + \cdots + \penv'_n}{\penv}{\constrs''}
    }
    { \synthseq
        {\send{\vala}{m}{\seq{\valb}}}
        {\one}
        {\penv}
        {\constrs \cup \constrs'_1 \cup \cdots \cup \constrs'_n \cup \constrs''}
    }

    \inferrule
    [TS-App]
    {
        \pproglookup{f} = \fndef{f}{\seq{x : \ptya}}{\ptyb}{\ptma} \\
        (\chkseq{\vala_i}{\ptya_i}{\penv_i}{\constrs_i})_{i \in 1..n} \\
        \combineseqmany{\penv_1 + \cdots + \penv_n}{\penv}{\constrs}
    }
    { \synthseq
        { \fnapp{f}{\vala_1, \ldots, \vala_n} }
        {\ptyb}
        {\penv}
        {\constrs \cup \constrs_1 \cup \cdots \cup \constrs_n}
    }
\end{mathpar}

\headertwo
    {Constraint generation (checking)}
    {
        \framebox{$\chkseq[\pprog]{\ptma}{\ptya}{\penv}{\constrs}$}
    }
\begin{mathpar}
    \inferrule
    [TC-Var]
    { }
    { \chkseq{x}{\ptya}{x : \ptya}{\emptyset} }

    \inferrule
    [TC-Let]
    {
        \chkseq{\ptma}{\tya}{\penv_1}{\constrs_1} \\
        \chkseq{\ptmb}{\ptya}{\penv_2}{\constrs_2} \\
        \returnablepred{\tya} \\\\
        \checkenvseq{\penv_2}{x}{\tya}{\constrs_3} \\
        \joinseq{\penv_1}{(\penv_2 \envwithout x)}{\penv}{\constrs_4}
    }
    {
        \chkseq
            {\letin[\tya]{x}{\ptma}{\ptmb}}
            { \ptya }
            { \penv }
            { \constrs_1 \cup \cdots \cup \constrs_4 }
    }

    \inferrule
    [TC-Guard]
    {
        \chkgseq
            {\pata}
            {\seq{\pgd}}
            {\ptya}
            {\penv_1}
            {\constrs_1}
            {\patb} \\\\
        \chkseq{\vala}{\retty{\tyrecv{\patb}}}{\penv_2}{\constrs_2} \\
        \combineseq{\penv_1}{\penv_2}{\penv}{\constrs_3}
    }
    { \chkseq
        {\guardann{\vala}{\pata}{\seq{\pgd}}}
        {\ptya}
        {\penv}
        {\constrs_1 \cup \constrs_2 \cup \constrs_3 \cup
        \set{\subpatconstr{\pata}{\patb} }}
    }

    \inferrule
    [TC-Fail]
    {
        \chkseq{\vala}{\retty{\tyrecv{\mbzero}}}{\penv}{\constrs}
    }
    { \chkseq{\sugarfail{\vala}}{\ptya}{\penv}{\constrs} }

    \inferrule
    [TC-Sub]
    { \synthseq{\ptma}{\ptya}{\penv}{\constrs_1} \\
      \subtyseq{\ptya}{\ptyb}{\constrs_2}
    }
    { \chkseq{\ptma}{\ptyb}{\penv}{\constrs_1 \cup \constrs_2} }
\end{mathpar}

\headertwo
    {Environment lookup}
    {\framebox{$\checkenvseq{\penv}{x\vphantom{\seq{x}}}{\ptya}{\constrs}$}
     ~
     \framebox{$\checkenvseq{\penv}{\seq{x}}{\seq{\ptya}}{\constrs}$}
    }
    \begin{mathpar}
        \inferrule
        { x \not\in \dom{\penv} \\ \unrseq{\ptya}{\constrs} }
        { \checkenvseq{\penv}{x}{\ptya}{\constrs} }
\quad\:\:
        \inferrule
        { \subtyseq{\ptyb}{\ptya}{\constrs} }
        { \checkenvseq{(\penv, x : \ptya)}{x}{\ptyb}{\constrs} }
\quad\:\:
        \inferrule
        {
            (\checkenvseq{\penv}{x_i}{\ptya_i}{\constrs_i})_{i \in 1..n}
        }
        { \checkenvseq
            {\penv}
            {\seq{x}}
            {\seq{\ptya}}
            {\constrs_1 \cup \cdots \cup \constrs_n} }
    \end{mathpar}
}

\caption{\langname algorithmic typing (programs, definitions, and terms)}
\label{fig:algo-typing-terms}
\end{figure}

Figure~\ref{fig:algo-typing-terms} shows the \langname algorithmic
typing of programs, definitions and terms.  The key idea is to remain in
checking mode for as long as possible, in order to propagate type information to
the variable rule and construct a type environment.
We write
$\penv \envwithout \seq{x}$ for $\{ y : \ptya \mathop{\mid} y : \ptya \in \penv
    \wedge y \not\in
\seq{x} \}$.

\paragraph*{Synthesis.}
Our synthesis judgement has the form
$\synthseq[\pprog]{\ptma}{\ptya}{\penv}{\constrs}$, which can be read
``synthesise type $\ptya$ for term $\ptma$ under program $\pprog$, producing type
environment $\penv$ and constraints $\constrs$''.
Here, $\ptma$ and $\pprog$ are inputs of the judgement, whereas $\ptya$, $\penv$,
and $\constrs$ are outputs.
The checking judgement $\chkseq[\pprog]{\ptma}{\ptya}{\penv}{\constrs}$
can be read
``check that term $\ptma$ has type $\ptya$ under program $\pprog$, producing type
environment $\penv$ and constraints $\constrs$''.
Here, $\ptma$, $\pprog$, and $\ptya$ are inputs of the judgement, whereas $\penv$
and $\constrs$ are outputs.
As in the declarative system we omit the $\pprog$ annotation in the rules for readability.

Rule \textsc{TS-Const} assigns a known base type to a constant, and rule
\textsc{TS-New} synthesises a type $\retty{\tyrecv{\mbone}}$ (analogous to
\textsc{T-New}); both rules produce an empty environment and constraint set.
Rule \textsc{TS-Spawn} checks that the given computation $\ptma$ has the unit type,
synthesises type $\one$, and infers a type environment $\penv$ and constraint
set $\constrs$. Like \textsc{T-Spawn} in the declarative system, the usage
annotations in the produced environment are masked as second-class since
usage restrictions are process-local.

Message sending $\send{\vala}{m}{\seq{\valb}}$ is a side-effecting
operation, and so we synthesise type $\tyunit$. Rule \textsc{TS-Send}
first looks up the payload types $\seq{\ppretya}$ in the signature,
and \emph{checks} that message target $\vala$ has mailbox type
$\usablety{\tysend{\msgtag{m}}}$. In performing this check, the type
system will produce environment $\penv'$ that contains an entry
mapping the variable in $\vala$ to the desired mailbox type
$\usablety{\tysend{\msgtag{m}}}$.
Next, the algorithm checks each payload value against the
payload type described by the signature. The resulting environment is
the algorithmic disjoint combination of the environments produced by
checking each payload, and the resulting constraint set is the union of all
generated constraints.

Function application is similar: rule \textsc{TS-App} looks up the
type signature for function $f$ and checks that all arguments have the
expected types. The resulting environment is again the disjoint
combination of the environments, and the constraint set is the union
of all generated constraints.

\paragraph*{Checking.}

Rule \textsc{TC-Var} \emph{checks} that a variable $x$ has type $\ptya$,
producing a type environment $x : \ptya$.
The \textsc{TC-Let} rule checks that a let binding
$\letin[\tya]{x}{\ptma}{\ptmb}$ has type $\ptya$: first, we check that $\ptma$
has returnable type $\tya$, producing environment $\penv_1$ and constraints
$\constrs_1$.  Next we check that the body $\ptmb$ has type $\ptya$, producing
environment $\penv_2$ and constraints $\constrs_2$.
The next step is to check whether the type of the variable inferred in
$\penv_2$ corresponds with the annotation. The $\mkwd{check}$ meta-function
ensures that if $x$ is not contained within $\penv_2$, then the type of $x$ is
unrestricted; and conversely if $x$ \emph{is} contained within $\penv_2$, then the
annotation is a subtype of the inferred type as the annotation is a \emph{lower
bound} on what the body can expect of $x$.

Rule \textsc{TC-Guard} checks that a guard expression
$\guardann{\vala}{\pata}{\seq{\pgd}}$ has return type $\ptya$. First, the rule
checks that the guard sequence $\seq{\pgd}$ has type $\ptya$, producing 
environment $\penv_1$, constraint set $\constrs_1$, and pattern $\patb$ in
pattern normal form
\ADDED{(recall that a pattern $\pata$ is in
    pattern normal form if it is a sum
of $\mbone$ and $\msgtag{m} \patconcat \patb$ patterns, where $\patb$
is equivalent to the pattern residual of $\pata$ with respect to $\msgtag{m}$)}.
Next, the rule checks that the mailbox name $\vala$ has
type $\retty{\tyrecv{\patb}}$, producing environment $\penv_2$ and constraint
set $\constrs_2$. Finally, the rule calculates the disjoint combination
of $\penv_1$
and $\penv_2$, producing final environment $\penv$ and constraints $\constrs_3$.
The resulting constraint set is the union of the constraint sets produced by the
premises, along with a constraint that the annotation $\pata$
is a subpattern of the pattern $\patb$ produced by the guards.

Rule \textsc{TC-Fail} allows $\sugarfail{\vala}$ to be checked at
any type, provided that $\vala$ has type $\retty{\mbrecvzero}$.

Finally, rule \textsc{TC-Sub} states that if a term $\ptma$ is synthesisable with
type $\ptya$, where $\ptya$ is a subtype of $\ptyb$, then $\ptma$ is checkable
with type $\ptyb$. The resulting environment is that produced by synthesising
the type for $\ptma$, and the resulting constraint set is the union of the
synthesis and subtyping constraints.

\paragraph*{Unannotated $\calcwd{let}$ expressions.}
    Although our algorithmic core calculus uses an annotation on $\calcwd{let}$
    expressions, this is unnecessary if the let-bound variable is
    used in the continuation $\ptmb$, or $\ptma$ has a synthesisable type.
Specifically, \textsc{TC-LetNoAnn1} allows us to check the type of the
    continuation and inspect the produced environment for the type of $x$, which
    can be used to check $\ptma$.
Similarly, \textsc{TC-LetNoAnn2} allows us to type a
    $\calcwd{let}$-binding where $x$ is \emph{not} used in the continuation,
    as long as the type of $\ptma$ is synthesisable and unrestricted.

    {\small
    \begin{mathpar}
        \inferrule
        [TC-LetNoAnn1]
        {
            \chkseq{\ptmb}{\ptyb}{\penv_1, x : \ptya}{\constrs_1} \\
            \chkseq{\ptma}{\makereturnable{\ptya}}{\penv_2}{\constrs_2} \\
            \joinseq{\penv_2}{\penv_1}{\penv}{\constrs_3}
        }
        {
            \chkseq
                {\letin{x}{\ptma}{\ptmb}}
                { \ptyb }
                { \penv }
                { \constrs_1 \cup \constrs_2 \cup \constrs_3 }
        }

        \inferrule
        [TC-LetNoAnn2]
        {
            \chkseq{\ptmb}{\ptyb}{\penv_1}{\constrs_1} \\
            x \not\in \dom{\penv_1} \\
            \synthseq{\ptma}{\ptya}{\penv_2}{\constrs_2} \\\\
            \returnablepred{\ptya} \\
            \unrseq{\ptya}{\constrs_3} \\
            \joinseq{\penv_2}{\penv_1}{\penv}{\constrs_4}
        }
        {
            \chkseq
                {\letin{x}{\ptma}{\ptmb}}
                { \ptyb }
                { \penv }
                { \constrs_1 \cup \cdots \cup \constrs_4 }
        }
\end{mathpar}
    }

    We use the explicitly typed representation in the core language for
    simplicity and uniformity; however, the implementation follows the above
    approach to avoid needless annotations.

\paragraph*{Guards.}

\begin{figure}
{\small
\headertwo{Constraint generation for guards}{
    \framebox{$
        \chkgseq
            [\pprog]
            {\pata}
            {\seq{\pgd}}
            {\ptya}
            {\penv}
            {\constrs}
            {\patb}$}~
    \framebox{$
        \chkgseq
            [\pprog]
            {\pata}
            {\pgd\vphantom{\seq{\pgd}}}
            {\ptya}
            {\penv}
            {\constrs}
            {\patb}$}}

\begin{mathpar}
    \inferrule
    [TCG-Guards]
    { (\chkgseq
        {\pata}
        {\pgd_i}
        {\ptya}
        {\penv_i}
        {\constrs_i}
        {\patb_i})_{i \in 1..n} \\\\
      \patb = {\patb_1 \patplus \cdots \patplus \patb_n}\\
      \manymergeseq
        {\penv_1}
        {\penv_n}
        {\penv}
        {\constrs}
    }
    { \chkgseq
        {\pata}
        {\seq{\pgd}}
        {\ptya}
        {\penv}
        {\constrs \cup \constrs_1 \cup \cdots \cup \constrs_n}
        { \patb }
    }

    \inferrule
    [TCG-Free]
    { \chkseq{\ptma}{\ptya}{\penv}{\constrs} }
    { \chkgseq{\pata}{\free{\ptma}}{\ptya}{\penv}{\constrs}{\mbone} }

    \inferrule
    [TCG-Recv]
    {
        \chkseq{\ptma}{\ptya}{\penv', y : \retty{\tyrecv{\ppata}}}{\constrs_1} \\
        \pprog(\msgtag{m}) = {\seq{\ppretya}} \\
        \penv = \penv' \envwithout \seq{x} \\
        \mbbase{\seq{\ppretya}} \vee \mbbase{\penv} \\
        \checkenvseq{\penv'}{\seq{x}}{\seq{\makeusable{\ppretya}}}{\constrs_2}
    }
    { \chkgseq
        {\pata}
        {\receive{m}{\seq{x}}{y}{\ptma}}
        {\ptya}
        {\penv}
        {\constrs_1 \cup \constrs_2 \cup
            \set{\subpatconstr{\pata \without \msgtag{m}}{\ppata}}
        }
        {\algomsg{m} \patconcat (\pata \without \msgtag{m})}
    }
\end{mathpar}
}

    \caption{\langname algorithmic typing (guards)}
    \label{fig:algo-typing-guards}
\end{figure}

Figure~\ref{fig:algo-typing-guards} shows the typing rules for
guards; the judgement
$\chkgseq{\pata}{\pgd}{\ptya}{\penv}{\constrs}{\patb}$ can be read
``Check that guard $\pgd$ has type $\ptya$,
producing environment $\penv$, constraints $\constrs$, and
closed pattern literal $\patb$ in pattern normal form with respect to $\pata$''.
Rule \textsc{TCG-Guards} types a guard sequence, producing the algorithmic
branching merge of all environments and the sum of all produced patterns. Rule
\textsc{TCG-Free} checks that guard $\free{\ptma}$ has type $\ptya$ by checking
that $\ptma$ has type $\ptya$; the guard produces pattern $\mbone$.

Finally, rule \textsc{TCG-Recv} checks that a receive guard
$\receive{m}{\seq{x}}{y}{\ptma}$ has type $\ptya$. First, the rule checks that
$\ptma$ has type $\ptya$, producing environment $\penv', y :
\retty{\tyrecv{\ppata}}$ and constraint set $\constrs_1$;  since a
mailbox type with input capability is linear, it \emph{must} be present in the
inferred environment. Next, the rule checks that the inferred types for $\seq{x}$
in $\penv'$ are compatible with the payloads for $\msgtag{m}$ declared in the
signature, producing constraint set $\constrs_2$.
As with the declarative rule, to rule out unsafe aliasing, either the
payloads or the inferred environment must consist only of base types. The resulting
environment is $\penv$ (\ie, the inferred environment without the mailbox
variable or any payloads).
The resulting constraint set is the union of $\constrs_1$ and $\constrs_2$ along
with an additional constraint that ensures $\pata \without \msgtag{m}$ is
included in  $\ppata$, allowing us to produce the closed PNF literal
$\msgtag{m} \patconcat (\pata \without \msgtag{m})$.
The annotation $\pata$ on guard expressions is necessary because the
pattern residual, used in the final constraint, is only defined on
\emph{closed} patterns.

\subsection{Metatheory}

We can now establish that the algorithmic type system is sound and complete with
respect to the declarative type system. We begin by introducing pattern
substitutions and solutions.

A \emph{pattern substitution} $\patsubst$ is a mapping from pattern variables $\patvara$ to
(fully defined) patterns $\pata$: applying $\patsubst$ to a pattern $\ppata$
replaces each occurrence of a pattern variable $\patvara$ with $\aps{\patvara}$.
We extend application of pattern substitutions to types and environments.
We write $\pv{\ppata}$ for the set of pattern variables in a pattern and extend it to
types and environments.

\begin{definition}[Pattern solution]
    A pattern substitution $\patsubst$ is a \emph{pattern solution} for a
    constraint set $\constrs$ (or \emph{solves} $\constrs$) if
    $\pv{\constrs} \subseteq \dom{\patsubst}$ and
    for each $\subpatconstr{\ppata}{\ppatb} \in \ADDED{\constrs}$, we have
    that $\subpat{\apppatsubst{\ppata}}{\apppatsubst{\ppatb}}$.
A solution $\patsubst$ is a \emph{usable} solution if its range does not
    contain any pattern equivalent to $\mbzero$.
\end{definition}

It is useful to define the notion of a \emph{covering solution} to characterise
a solution that resolves all pattern variables present in an algorithmic typing
derivation.

\begin{definition}[Covering solution]
We say that a pattern substitution $\patsubst$ is a \emph{covering solution} for a
derivation $\synthseq[\pprog]{\ptma}{\ptya}{\penv}{\constrs}$ or
$\chkseq[\pprog]{\ptma}{\ptya}{\penv}{\constrs}$ if given
$\algoprogseq{\pprog}{\constrs'}$, it is the case that $\patsubst$ is a usable
solution for $\constrs \cup \constrs'$ such that $\pv{\ptya} \cup \pv{\pprog}
\subseteq \dom{\patsubst}$.
\end{definition}

\subsubsection{Algorithmic soundness}\label{sec:algo:alg-soundness}

If a term is well typed in the algorithmic system then, given a covering
solution, the term is also well typed in the declarative system. Proving this
result involves establishing several auxiliary results on the soundness of
the various type operators and type combination operators.
Full proof details can be found in the extended version.

\paragraph*{Properties of pattern variables and solutions.}
The first auxiliary result states that a solution for a set of constraints is
also a solution for a subset of those constraints.

\begin{lemma}
    \label{lem:pat-subset}
    If $\patsubst$ is a solution for a constraint set
    $\constrs_1 \cup \constrs_2$,
    then $\patsubst$ is a solution for $\constrs_1$.
\end{lemma}
\begin{proof}
Since $\patsubst$ is a solution for $\constrs_1 \cup \constrs_2$,
it follows that $\pv{\constrs_1 \cup \constrs_2} \subseteq \dom{\patsubst}$.
The result follows from the fact that $\pv{\constrs_1} \subseteq
\pv{\constrs_1 \cup \constrs_2} \subseteq \dom{\patsubst}$.
\end{proof}

We also need to reason about the provenance of pattern variables that appear in
an inferred environment. Specifically, any pattern variable that appears in an
inferred environment must either occur in the type of an expression, in the
program, or in the constraint set. 

\begin{lemma}\label{lem:penv-constrs-subset}
    If $\synthseq[\pprog]{\ptma}{\ptya}{\penv}{\constrs}$ or
       $\chkseq[\pprog]{\ptma}{\ptya}{\penv}{\constrs}$,
       then $\pv{\penv} \subseteq \pv{\ptya} \cup \pv{\pprog} \cup \pv{\constrs}$.
\end{lemma}
\begin{proof}
    By mutual induction on the two derivations, noting that whenever a pattern
    variable is introduced fresh, it is always added to the constraint set.
\end{proof}

\paragraph*{Properties of type operations.}

Next, we need to show the relation between algorithmic and declarative versions
of the various type operations. 
Given a usable solution of a constraint set, we can show the soundness of
algorithmic subtyping.

\begin{lemma}
    \label{lem:subtype-soundness}
    If $\subtyseq{\ptya}{\ptyb}{\constrs}$ and
    $\patsubst$ is a usable solution of $\constrs$ with
    $\pv{\ptya} \cup \pv{\ptyb} \subseteq \dom{\patsubst}$,
    then $\subtypetwo{\apppatsubst{\ptya}}{\apppatsubst{\ptyb}}$.
\end{lemma}
\begin{proof}
    By case analysis on the derivation of $\subtyseq{\ptya}{\ptyb}{\constrs}$.
    Base types are trivial, and the property follows for mailbox types from the
    definition of a usable solution.
\end{proof}

We can also show the soundness of the algorithmic unrestrictedness operation.

\begin{lemma}
    \label{lem:unr-soundness}
    If $\unrseq{\ptya}{\constrs}$ and $\patsubst$ is a usable solution of
    $\constrs$ with $\pv{\ptya} \subseteq \dom{\patsubst}$, then
    there exists some $\tya$ such that $\un{\tya}$ and $\apppatsubst{\ptya}
    \subtype \tya$.
\end{lemma}
\begin{proof}
    By case analysis on the derivation of $\unrseq{\ptya}{\constrs}$, noting
    that cases are undefined for linear types, and that the result follows
    immediately for base types.
The only interesting case is
    $\unrseq{\usagety{\tysend{\ppata}}}{\set{\subpatconstr{\mbone}{\ppata}}}$;
    since $\patsubst$ is a usable solution, we have that $\mbone \subpatone
    \apppatsubst{\ppata}$. Since $\returnable \subtype \usable$ we can therefore show that
    $\usagety{\tysend{(\aps{\ppata})}} \subtype \usablety{\mbsendone}$ where
    $\un{\usablety{\mbsendone}}$ as required.
\end{proof}

\paragraph*{Properties of type combination operations.}

Finally we need to show the soundness of the merge operators, which in both
cases follows by case analysis on the respective derivations.

\begin{restatable}[Soundness of algorithmic sequential merge]{lemma}{scombsol}
    \label{lem:scomb-sol}
    If $\joinseq{\ptya_1}{\ptya_2}{\ptyb}{\constrs}$
    and $\patsubst$ is a usable solution of $\constrs$
    such that $\pv{\ptya_1} \cup \pv{\ptya_2} \subseteq \dom{\patsubst}$,
    then there exist
    $\tya \subtype \apppatsubst{\ptya_1}$ and
    $\tyb \subtype \apppatsubst{\ptya_2}$ such that
    $\scombtwo{\tya}{\tyb} = \apppatsubst{\ptyb}$.
\end{restatable}

\begin{restatable}[Soundness of algorithmic branching merge]{lemma}{mergesol}
    \label{lem:merge-soundness}
    If
    $\mergeseq{\ptya_1}{\ptya_2}{\ptyb}{\constrs}$
    and $\patsubst$ is a usable solution of $\constrs$
    such that $\pv{\ptya_1} \cup \pv{\ptya_2} \subseteq \dom{\patsubst}$, then
    $\apppatsubst{\ptyb} \subtype \apppatsubst{\ptya_1}$
    and
    $\apppatsubst{\ptyb} \subtype \apppatsubst{\ptya_2}$.
\end{restatable}

\ADDED{
To relate annotated terms with unannotated terms in the declarative system, we
define an \emph{erasure} operator $\erase{\ptma} = \tma$
on annotated computations that removes annotations on let binders and guards.
The erasure operator is defined by the homomorphic extension of the
following rules over computations and guards:

{\small
\[
    \begin{array}{rcl}
        \erase{\letin[\tya]{x}{\ptma}{\ptmb}} & =  &
    \letin{x}{\erase{\ptma}}{\erase{\ptmb}}
    \\
        \erase{\guardann{\vala}{\pata}{\seq{\pgd}}} & = &
    \guard{\vala}{\seq{\erase{\pgd}}}
    \el
\]
}

We also extend the erasure operator to definitions and programs:

{\small
    \[
    \begin{array}{rcl}
        \erase{\fndef{f}{\seq{x : \ptya}}{\ptyb}{\ptma}} & =  &
    \fndef{f}{\seq{x : \ptya}}{\ptyb}{\erase{\ptma}} \\
\erase{\progdef{\psigs}{\seq{\pdef}}{\ptma}} & = &
    \progdef{\psigs}{\erase{\seq{\pdef}}}{\erase{\ptma}}
    \end{array}
\]
}
}

With these auxiliary results defined, we can show that the algorithmic type
system is sound with respect to the declarative type system, meaning that our
algorithmic type system will never accept ill-typed terms.

\begin{restatable}[Algorithmic soundness]{theorem}{algosoundness}\label{thm:algo-soundness}
\hfill
\begin{itemize}[leftmargin=*]
        \item If $\patsubst$ is a covering solution for
            $\synthseq[\pprog]{\ptma}{\ptya}{\penv}{\constrs}$, then
           $\tseq
               [\ADDED{\erase{\aps{\pprog}}}]
               {\apppatsubst{\penv}}
               {\ADDED{\erase{\ptma}}}
               {\apppatsubst{\ptya}}$.
       \item If $\patsubst$ is a covering solution for
           $\chkseq[\pprog]{\ptma}{\ptya}{\penv}{\constrs}$, then
           $\tseq
               [\ADDED{\erase{\aps{\pprog}}}]
               {\apppatsubst{\penv}}
               {\ADDED{\erase{\ptma}}}
               {\apppatsubst{\ptya}}$.
\end{itemize}
\end{restatable}

The result follows from a generalised version of algorithmic soundness on
all of the typing judgements (for values, computations, guards, and guard
sequences), which is established by mutual induction.

\subsubsection{Algorithmic completeness}\label{sec:algo:alg-completeness}

We also obtain a completeness result, but only for the checking direction.
This is because the type system \emph{requires} type information to construct a
type environment.
In practice the lack of a completeness result for synthesis is unproblematic
since all functions have return type annotations, and therefore the only terms
typable in the declarative system but unsynthesisable are top-level terms
containing free variables.
\ADDED{Recall that by the definition in~\secref{sec:declarative}, program
$\prog$ is \emph{closed}, \ie, no definitions or message payloads contain
pattern variables.}

Again, the proof of algorithmic completeness requires several auxiliary lemmas.
Full proofs can be found in the extended version.

\paragraph*{Closed and satisfiable constraint sets.}
We first define \emph{closed} and \emph{satisfiable} constraint sets.

\begin{definition}[Closed and satisfiable constraint sets]
    A constraint set $\constrs$ is \emph{closed} if $\pv{\constrs} = \emptyset$.
A closed constraint set $\constrs$ is \emph{satisfiable} if the empty
    solution is a solution for $\constrs$
    (\ie, $\constrs = (\subpatconstr{\pata_i}{\patb_i})_{i \in 1..n}$
    and $(\pata_i \subpatone \patb_i)_{i \in 1..n}$).
\end{definition}

\paragraph*{Checkability of values and synthesisable terms.}

Values are checkable, without creating any additional constraints.

\begin{restatable}{lemma}{valcompleteness}
    \label{lem:val-completeness}
    If $\tseq{\tyenv}{\vala}{\tya}$, then there exists some
    $\tyenv'$ such that $\tyenv \subtype \tyenv'$ and
    $\chkseq{\vala}{\tya}{\tyenv'}{\emptyset}$.
\end{restatable}
\begin{proof}
    By induction on the derivation of $\tseq{\tyenv}{\vala}{\tya}$.
    \textsc{T-Var} and \textsc{T-Const} follow immediately, and \textsc{T-Sub}
    follows from the IH and the transitivity of subtyping.
\end{proof}

Next, any synthesisable term is checkable with the same type, without needing to
introduce any additional constraints.

\begin{restatable}{lemma}{synthchk}
    \label{lem:synth-to-chk}
    If $\synthseq{\ptma}{\ptya}{\penv}{\constrs}$, then
    $\chkseq{\ptma}{\ptya}{\penv}{\constrs}$.
\end{restatable}
\begin{proof}
    Follows from the definition of \textsc{TC-Sub}, noting that the subtyping
    constraint is instantiated as $\subtyseq{\ptya}{\ptya}{\emptyset}$.
\end{proof}

\paragraph*{Properties of type operations.}
As with algorithmic soundness, we need various lemmas that show the
relationship between the algorithmic and declarative versions of the operations
on types and environments. We begin by showing three useful lemmas relating to
subtyping.

First, algorithmic subtyping on closed types is transitive.

\begin{restatable}[Transitivity of algorithmic subtyping]{lemma}{algosubtypetrans}
    \label{lem:subtype-solution}
    If $\subtyseq{\tya}{\tya'}{\constrs}$ where $\constrs$ is satisfiable,
    and $\tya' \subtype \tyb$, then
    $\subtyseq{\tya}{\tyb}{\constrs'}$
    and $\constrs'$ is satisfiable.
\end{restatable}
\begin{proof}
    By case analysis on the derivation of $\subtyseq{\tya}{\tya'}{\constrs}$ and
    the transitivity of pattern inclusion.
\end{proof}

Second, it is useful to show that if a term is checkable at some type $\tya$,
then it is also checkable at some supertype $\tyb$.

\begin{lemma}[Checkability at a supertype]\label{lem:supertype-checkability}

    If $\chkseq[\prog]{\ptma}{\tya}{\penv}{\constrs}$ where
    $\patsubst$ is a usable solution of $\constrs$,
    and $\tya \subtype \tyb$,
then there exist $\penv', \constrs'$ such that
    $\chkseq[\prog]{\ptma}{\tyb}{\penv'}{\constrs'}$
    where $\patsubst$ is a usable
    solution of $\constrs'$ and
    $\aps[\patsubst]{\penv} \strictsubty \aps[\patsubst]{\penv'}$.
\end{lemma}
\begin{proof}
    Follows from a more general result established by mutual induction,
    making essential use of Lemma~\ref{lem:subtype-solution}.
\end{proof}

The completeness of subtyping follows by case analysis on the derivation of 
$\aps{\ptya} \subtype \aps{\ptyb}$ and the definition of pattern inclusion.

\begin{restatable}[Completeness of subtyping]{lemma}{subtycomplete}
    \label{lem:subtya-ptya}
    Given a pattern substitution $\patsubst$ such that
    $\aps{\ptya} \subtype \aps{\ptyb}$, it follows that
    $\subtyseq{\ptya}{\ptyb}{\constrs}$
    and $\patsubst$ is a usable solution of $\constrs$.
\end{restatable}

The completeness of the $\mkwd{check}$ meta-function follows from case analysis
on whether the given variable is contained within an environment, with appeal to
Lemma~\ref{lem:subtya-ptya}.

\begin{restatable}[Completeness of \mkwd{check}]{lemma}{checkcomplete}\label{lem:checkfn-subtype}
    Given a pattern substitution $\patsubst$ such that
    $\aps{\penv, x : \ptya} \subtype \aps{\penv'}$, it follows that
    $\checkenvseq{\penv'}{x}{\ptya}{\constrs}$ and $\patsubst$ is a usable
    solution of $\constrs$.
\end{restatable}

\paragraph*{Properties of type combination operations.}

The final set of lemmas concentrates on the completeness of the type combination
operators. Both proofs follow by case analysis on the respective declarative
derivations and appeal to the underlying pattern semantics.

\begin{restatable}[Completeness of algorithmic sequential merge]{lemma}{joincomplete}
\label{lem:join-completeness-types}
If $\tya_1 \scomb \tya_2  = \tyb$
where
$\tya_1 \subtype \aps[\patsubst_1]{\ptya_1}$ and
$\tya_2 \subtype \aps[\patsubst_2]{\ptya_2}$
for pattern substitutions $\patsubst_1, \patsubst_2$ such that
$\dom{\patsubst_1} \cap \dom{\patsubst_2} = \emptyset$,
then there exist $\ptyb, \constrs$ such that
$\joinseq{\ptya_1}{\ptya_2}{\ptyb}{\constrs}$,
and there exists a usable solution $\patsubst \supseteq \patsubst_1 \cup
\patsubst_2$ of $\constrs$
such that $\tyb \subtype \aps{\ptyb}$.
\end{restatable}

\begin{restatable}[Completeness of algorithmic branching merge]{lemma}{mergecomplete}
\label{lem:merge-completeness-types}
If $\tya \subtype \aps[\patsubst_1]{\ptya_1}$ and
    $\tya \subtype \aps[\patsubst_2]{\ptya_2}$ for
    pattern substitutions $\patsubst_1, \patsubst_2$ such that
    $\dom{\patsubst_1} \cap \dom{\patsubst_2} = \emptyset$,
then there exist $\ptyb, \constrs$ such that
    $\mergeseq{\ptya_1}{\ptya_2}{\ptyb}{\constrs}$
    and there exists a usable solution
    $\patsubst \supseteq \patsubst_1 \cup \patsubst_2$ of $\constrs$
    such that $\tya \subtype \aps{\ptyb}$.
\end{restatable}

\paragraph*{Algorithmic completeness.}

With these intermediate results in hand, we can state the completeness of the
algorithmic type system with respect to the declarative type system.

\ADDED{
To relate unannotated terms that are typable in the declarative system to
annotated terms required for the algorithmic type system, we introduce
type-directed annotation rules. The interesting rules are those for
$\calcwd{let}$-bindings and guard expressions, which use information from the
declarative type system to add appropriate annotations.
The remaining rules (detailed in the extended version) are defined similarly.

{\small
\headersig{Type-directed annotation}{$\anntmseq{\tyenv}{\tma}{\tya}{\ptma}$}
\begin{mathpar}
    \inferrule
    {
        \anntmseq{\tyenv_1}{\tma}{\tya}{\ptma} \\
        \returnablepred{\tya} \\
        \anntmseq{\tyenv_2, x : \tya}{\tmb}{\tyb}{\ptmb}
    }
    { \anntmseq
        {\scombtwo{\tyenv_1}{\tyenv_2}}
        {\letin{x}{\tma}{\tmb}}
        {\tyb}
        {\letin[\tya]{x}{\ptma}{\ptmb}}
    }

    \inferrule
    {
      \tseq{\tyenv_1}{\vala}{\retty{\tyrecv{\pata}}} \\
      \anngseq{\pata}{\tyenv_2}{\seq{\gd}}{\tya}{\seq{\pgd}} \\
      \pnf{\pata}
    }
    { \anntmseq{{\tyenv_1} + {\tyenv_2}}
        {\guard{\vala}{\seq{\gd}}}
        {\tya}
    {\guardann{\vala}{\pata}{\seq{\pgd}}} }
\end{mathpar}
}
}

Finally, we can state our algorithmic completeness result.

\begin{restatable}
    [Algorithmic completeness]
    {theorem}
    {algocompleteness}
    \label{thm:algo-completeness}
    If $\progseq{\prog}$ where
        \ADDED{$\anntmseq[\prog]{\tyenv}{\tma}{\tya}{\ptma}$},
        then there exist some $\penv, \constrs$
        and usable solution $\patsubst$ of $\constrs$ such that
        $\chkseq[\prog]{\ptma}{\tya}{\penv}{\constrs}$ where $\tyenv \subtype \aps{\penv}$.
\end{restatable}

The proof in the extended version is again by mutual induction using a
generalised statement showing completeness of typing values, computations,
guards, and guard sequences.

Although the completeness result is aided by the explicit annotations
on let bindings,
an unannotated \calcwd{let} binding $\letin{x}{\ptma}{\ptmb}$ is also
typable by the algorithmic type system if either $x$ occurs free in $\ptmb$, or
the type of $\ptma$ is synthesisable. In practice this encompasses both base
types and linear usages of mailbox types, covering the vast majority of use
cases.

\subsection{Constraint solving}\label{sec:constraint-solving}
Constraint solving is covered in depth by~\citet{Padovani18}, and is not a
contribution of this work. However, as an informal overview, constraint solving
proceeds as follows:

\begin{description}
    \item[Identify and group bounds]
        A \emph{pattern bound} is of the form $\subpatconstr{\ppata}{\patvara}$,
        \ie, a constraint whose right-hand side is a pattern variable.
        We first group all pattern bounds using pattern disjunction,
        \eg, a constraint set
        $\set{\subpatconstr{\ppata}{\patvara}, \subpatconstr{\ppatb}{\patvara}}$
        would result in the constraint
        $\subpatconstr{\ppata \patplus \ppatb}{\patvara}$.
\item[Calculate closed-form solutions]
        \citet{HopkinsK99} define a closed-form solution for a set of pattern
        bounds $(\subpatconstr{\ppata_i}{\patvara_i})_{i \in 1..n}$:
        there exists a closed-form solution $\ppatb_i$ for each $\ppata_i$ such
        that $\patvara_i \not\in \pv{\ppatb_i}$.
We can then substitute each closed pattern through the system to
        eliminate all pattern variables in the remaining constraints and obtain a
        system of closed inclusion constraints.
\item[Translate to Presburger formulae and check satisfiability]
        Finally, we translate the closed constraints into Presburger formulae.
        Commutative regular expressions, and therefore patterns, can be
        expressed as semilinear sets~\citep{Parikh66} that are precisely those
        defined by Presburger formulae~\citep{Ginsburg66}.
To determine whether each constraint holds, we use the 
        quantifier elimination pass and quantifier-free linear integer
        arithmetic solver supported by the Z3 tool~\citep{MouraB08}.
\end{description}
 \section{Extensions}\label{sec:extensions}
\ADDED{
In this section we show how we can extend the base language to include
additional language features, and describe how this impacts typechecking.
}

It is straightforward to extend \langname\ with product and sum types, and
building on these allows us to introduce list types, which are a stepping stone
towards extending \langname\ with general recursive data types in the future.

We then show how to extend \langname\ with first-class functions, and
\emph{interfaces} that describe the set of messages a mailbox is allowed to
receive, which in turn increase the precision of typechecking and allow
finer-grained alias control.
The latter two extensions require contextual typing information prior to
constraint generation.

\subsection{Data types}

\subsubsection{Product types}

\begin{figure}[t]
{\small
    \header{Additional syntax}
    \[
    \begin{array}{lrcl}
        \text{Usage-annotated types} & \tya, \tyb & ::= & \cdots \midspace \typair{\tya}{\tyb}
        \\
        \text{Values} & \vala, \valb & ::= & \cdots \midspace (\vala, \valb)
        \\
        \text{Computations} & \tma, \tmb & ::= & \cdots \midspace
        \letin{(x, y)}{\vala}{\tma}
        \\
        \text{Annotated computations} & \ptma, \ptmb & ::= & \cdots \midspace
        \letin[\typairp{\tya}{\tyb}]{(x, y)}{\vala}{\ptma}
    \end{array}
    \]
    \headertwo
        {Additional declarative typing rules}
        {\framebox{$\vseq{\tyenv}{\vala}{\tya}$}~
         \framebox{$\tseq{\tyenv}{\tma}{\tya}$}}
\begin{mathpar}
    \inferrule
    [T-Pair]
    {
        \vseq{\tyenv_1}{\vala}{\tya} \\
        \vseq{\tyenv_2}{\valb}{\tyb} \\\\
        \returnablepred{\tya} \\
        \returnablepred{\tyb}
    }
    { \tseq{\tyenv_1 + \tyenv_2}{(\vala, \valb)}{\typair{\tya}{\tyb}} }

    \inferrule
    [T-LetPair]
    {
        \vseq{\tyenv_1}{\vala}{\typair{\tya_1}{\tya_2}} \\
        \tseq{\tyenv_2, x : \tya_1, y : \tya_2}{\tma}{\tyb}
    }
    { \tseq
        {\tyenv_1 + \tyenv_2}
        {\letin{(x, y)}{\vala}{\tma}}
        {\tyb}
    }
\end{mathpar}

\headertwo
    {Additional declarative operations on types}
    {\framebox{$\un{\tya}$}~\framebox{$\returnablepred{\tya}$}~\framebox{$\tya \subtype \tyb$}~\framebox{$\tya \scomb \tyb$}}

\begin{mathpar}
    \inferrule
    { \un{\tya} \\ \un{\tyb} }
    { \un{\typair{\tya}{\tyb}}}

    \inferrule
    {}
    { \returnablepred{\typair{\tya}{\tyb}} }

    \inferrule
    { \tya_1 \subtype \tyb_1 \\ \tya_2 \subtype \tyb_2 }
    { \typairp{\tya_1}{\tya_2} \subtype
        \typairp{\tyb_1}{\tyb_2} }

    \inferrule
    {}
    { \typairp{\tya_1}{\tya_2} \scomb \typairp{\tyb_1}{\tyb_2} = 
        \typair{(\scombtwo{\tya_1}{\tyb_1})}{(\scombtwo{\tya_2}{\tyb_2})}
    }
\end{mathpar}

\headertwo
    {Additional algorithmic typing rules}
    {\framebox{$\synthseq[\pprog]{\ptma}{\ptya}{\penv}{\constrs}$}~
        \framebox{$\chkseq[\pprog]{\ptma}{\ptya}{\penv}{\constrs}$}}
\begin{mathpar}
    \inferrule
    [TC-Pair]
    {
        \chkseq{\vala}{\ptya}{\penv_1}{\constrs_1} \\
        \chkseq{\valb}{\ptyb}{\penv_2}{\constrs_2} \\\\
        \returnablepred{\ptya} \\
        \returnablepred{\ptyb} \\
        \combineseq{\penv_1}{\penv_2}{\penv}{\constrs_3}
    }
    { \chkseq
        {(\vala, \valb)}
        {\typair{\ptya}{\ptyb}}
        {\penv}
        {\constrs_1 \cup \constrs_2 \cup \constrs_3}
    }

    \inferrule
    [TC-LetPair]
    {
        \chkseq{\vala}{\typair{\tya}{\tyb}}{\penv_1}{\constrs_1} \\
        \chkseq
            {\ptma}
            {\ptya}
            {\penv_2}
            {\constrs_2} \\
            \checkenvseq{\penv_2}{x}{\tya}{\constrs_3} \\
            \checkenvseq{\penv_2}{y}{\tyb}{\constrs_4} \\
        \combineseq{\penv_1}{((\penv_2 - x) - y)}{\penv}{\constrs_5}
    }
    { \chkseq
        {\letin[(\typair{\tya}{\tyb})]{(x, y)}{\vala}{\ptma}}
        {\ptya}
        {\penv}
        {\constrs_1 \cup \cdots \cup \constrs_5}
    }

    \inferrule
    [TC-LetPairNoAnn]
    {
        \chkseq{\ptma}{\ptyb}{\penv_1, x : \ptya_1, y : \ptya_2}{\constrs_1} \\
        \chkseq{\vala}{\typair{\ptya_1}{\ptya_2}}{\penv_2}{\constrs_2} \\
        \combineseq{\penv_1}{\penv_2}{\penv}{\constrs_3}
    }
    { \chkseq
        {\letin{(x, y)}{\vala}{\ptma}}
        {\ptyb}
        {\penv}
        {\constrs_1 \cup \constrs_2 \cup \constrs_3}
    }
\end{mathpar}
}
\caption{Extension of \langname with product types}
\label{fig:extensions:products}
\end{figure}

An advantage of adding product types is that we can avoid nested
$\calcwd{guard}$ expressions, by allowing a $\calcwd{guard}$ expression to return a
pair of a received value and an updated mailbox name. Consider the following two
expressions: the term on the left receives two integers and returns their sum
using nested $\calcwd{guard}$ expressions, whereas the term on the right avoids
nesting by returning a pair of the received value and the updated mailbox. 

{
\begin{minipage}[t]{0.45\textwidth}
        {\small
\[
    \bl
    \\
    \guardanntwo{\var{mb}}{\msgtag{Arg} \patconcat \msgtag{Arg}} \\
    \quad \receivethree{Arg}{x}{\var{mb}'} \\
    \qquad
        \guardanntwo{\var{mb}'}{\msgtag{Arg}} \\
        \qquad \quad \receivethree{\msgtag{Arg}}{y}{\var{mb}''} \\
        \qquad \qquad \sugarfree{\var{mb}''}; \\
        \qquad \qquad x {+} y \\
        \qquad \} \\
    \}
    \el
\]
}
\end{minipage}
\hfill
\begin{minipage}[t]{0.45\textwidth}
        {\small
    \[
        \bl
        \letinone{(x, \var{mb}')} \\
        \quad \guardanntwo{\var{mb}}{\msgtag{Arg} \patconcat \msgtag{Arg}} \\
        \qquad \receivethree{\msgtag{Arg}}{x}{\var{mb}'} \\
        \qqquad {(x, \var{mb}')} \\
        \quad \} \; \calcwd{in} \\
        \guardanntwo{\var{mb}'}{\msgtag{Arg}} \\
        \quad \receivethree{\msgtag{Arg}}{y}{\var{mb}''} \\
        \qquad {\sugarfree{\var{mb}''}; \; x {+} y} \\
        \}
        \el
    \]
}
\end{minipage}
}

\ADDED{This pattern can avoid deeply nested guard blocks when a process wants to make
several receives in a row. Additionally, recall from~\secref{sec:declarative} that
$\calcwd{receive}$ clauses use a conservative check to rule out
communication-based aliasing, where either all payloads of a message must be
base types, or all variables free in the body of the $\calcwd{receive}$ clause
must be base types. Returning received values along with the continuation of the
mailbox is a useful tool to increase expressiveness in the presence of this
restriction.}

\paragraph*{Formalism.}

Figure~\ref{fig:extensions:products} shows how to extend \langname with product
types.
The rules for constructing and deconstructing pairs (\textsc{T-Pair} and
\textsc{T-LetPair}) are standard aside from the condition that the types of both
values used to construct the pair are returnable. If we were to lift this
restriction then we would be able to violate the quasi-linearity conditions, for
example by using pair construction and deconstruction to subvert the condition
that only the last lexical occurrence of a variable can be returnable. We prefer
\textsc{T-LetPair} to individual projection functions due to the possibility
that one of the pair components may be linear, though we can use the usual
syntactic sugar
(\eg, $\calcwd{fst}\:\vala \defeq \letin{(x, y)}{\vala}{x}$ for fresh $x, y$).
Since product types can only consist of returnable types,
they cannot be used to replace $n$-ary argument sequences in function
definitions and $\calcwd{receive}$ clauses in full generality.

As usual, subtyping for product types is covariant in the component types,
and a product type is unrestricted if both of its component types
are unrestricted. We also extend type combination to handle products through
pointwise combination of the product's component types. To see why this is
necessary, consider the following example:

{\small
\begin{minipage}{0.55\textwidth}
\[
    \bl
    \fndefthree
        {\mkwd{sendToBoth}}
        {\var{p}:
            \typairp{\tysend{\msgtag{M1}}}{\tysend{\msgtag{M2}}}}
        {\one} \\
    \quad \letintwo{(\var{mb}_1, \var{mb}_2)}{\var{p}} \\
    \quad \send{\var{mb}_1}{M1}{}; \\
    \quad \send{\var{mb}_2}{M2}{} \\
    \} \\[1ex]
    \fndefthree
        {\mkwd{receiveAndFree}}
        {\var{p}: \typairp
                        {\tyrecvp{\msgtag{M1} \patconcat \mbone}}
                        {\tyrecvp{\msgtag{M2} \patconcat \mbone}}}
        {\one} \\
    \quad \letintwo{(\var{mb}_1, \var{mb}_2)}{\var{p}} \\
    \quad \guardone{\var{mb}_1} \\
    \qquad \receive{M1}{}{\var{mb}_1}{\sugarfree{\var{mb}_1}} \\
    \quad \}; \\
    \quad \guardone{\var{mb}_2} \\
    \qquad \receive{M2}{}{\var{mb}_2}{\sugarfree{\var{mb}_2}} \\
    \quad \} \\
    \}
    \el
    \]
\end{minipage}
\hfill
\begin{minipage}{0.35\textwidth}
    \[
    \bl
    \letintwo{\var{mb}_1}{\mbnew} \\
    \letintwo{\var{mb}_2}{\mbnew} \\
    \letintwo{\var{p}}{(\var{mb}_1, \var{mb}_2)} \\
    \spawn{(\mkwd{sendToBoth}(\var{p}))}; \\
    \mkwd{receiveAndFree}(\var{p})
    \el
\]
\end{minipage}
}

Here we create two mailboxes $\var{mb}_1$ and $\var{mb}_2$ and package them into
a pair $p$. We spawn a process that calls $\mkwd{sendToBoth}$: this sends
$\msgtag{M1}$ to the first component of the pair and $\msgtag{M2}$ to the second
component of the pair, and therefore uses
$\var{p}$ at type $\typair{\tysend{\msgtag{M1}}}{\tysend{\msgtag{M2}}}$. The
main process subsequently calls $\mkwd{receiveAndFree}$, which uses $\var{p}$ at
type $\typairp
        {\tyrecvp{\msgtag{M1} \patconcat \mbone}}
        {\tyrecvp{\msgtag{M2} \patconcat \mbone}}$.
Component-wise type combination gives $\var{p}$ the combined type
of $\typairp{\mbrecvone}{\mbrecvone}$ as needed.

\begin{figure}[t]
    {\small
    \header{Additional syntax}
    \[
    \begin{array}{lrcl}
        \text{Usage-annotated types} & \tya, \tyb & ::= & \cdots \midspace \tysum{\tya}{\tyb}
        \\
        \text{Values} & \vala, \valb & ::= & \cdots \midspace \inl{\vala}
        \midspace \inr{\vala} \\
        \text{Computations} & \tma, \tmb & ::= & \cdots \midspace
        \caseofcore{\vala}{x}{\tma}{y}{\tmb}
        \\
        \text{Annotated computations} & \ptma, \ptmb & ::= & \cdots \midspace
        \caseofcore{\vala}{x : \tya_1}{\ptma}{y : \tya_2}{\ptmb}
    \end{array}
    \]
    \headertwo
        {Additional declarative typing rules}
        {\framebox{$\vseq{\tyenv}{\vala}{\tya}$}~\framebox{$\tseq{\tyenv}{\tma}{\tya}$}}
\begin{mathpar}
    \inferrule
    [T-Inl]
    { \vseq{\tyenv}{\vala}{\tya} \\
      \returnablepred{\tya} \\
      \returnablepred{\tyb}
    }
    { \vseq{\tyenv}{\inl{\vala}}{\tysum{\tya}{\tyb}} }

    \inferrule
    [T-Inr]
    { \vseq{\tyenv}{\vala}{\tyb} \\
      \returnablepred{\tya} \\
      \returnablepred{\tyb}
    }
    { \vseq{\tyenv}{\inr{\vala}}{\tysum{\tya}{\tyb}} }

    \inferrule
    [T-Case]
    {
        \tseq{\tyenv_1 }{\vala}{\tysum{\tya_1}{\tya_2}} \\
        \tseq{\tyenv_2, x : \tya_1}{\tma}{\tyb} \\
        \tseq{\tyenv_2, y : \tya_2}{\tmb}{\tyb}
    }
    { \tseq{\tyenv_1 + \tyenv_2}{\caseofcore{\vala}{x}{\tma}{y}{\tmb}}{\tyb} }
\end{mathpar}

\headertwo
    {Additional declarative operations on types}
    {\framebox{$\un{\tya}$}~\framebox{$\returnablepred{\tya}$}~\framebox{$\tya \subtype \tyb$}~\framebox{$\tya \scomb \tyb$}}

\begin{mathpar}
    \inferrule
    { \un{\tya} \\ \un{\tyb} }
    { \un{\tysum{\tya}{\tyb}}}

    \inferrule
    {}
    { \returnablepred{\tysum{\tya}{\tyb}} }

    \inferrule
    { \tya_1 \subtype \tyb_1 \\ \tya_2 \subtype \tyb_2 }
    { \tysump{\tya_1}{\tya_2} \subtype
        \tysump{\tyb_1}{\tyb_2} }

    \inferrule
    {}
    { \tysump{\tya_1}{\tya_2} \scomb \tysump{\tyb_1}{\tyb_2} = 
        \tysum{(\scombtwo{\tya_1}{\tyb_1})}{(\scombtwo{\tya_2}{\tyb_2})}
    }
\end{mathpar}

\headertwo
    {Additional algorithmic typing rules}
    {\framebox{$\synthseq[\pprog]{\ptma}{\ptya}{\penv}{\constrs}$}~\framebox{$\chkseq[\pprog]{\ptma}{\ptya}{\penv}{\constrs}$}}
\begin{mathpar}
    \inferrule
    [TC-Inl]
    { \chkseq{\vala}{\ptya}{\penv}{\constrs} \\\\
        \returnablepred{\ptya} \\
        \returnablepred{\ptyb}
    }
    { \chkseq{\inl{\vala}}{\tysum{\ptya}{\ptyb}}{\penv}{\constrs} }

    \inferrule
    [TC-Inr]
    { \chkseq{\vala}{\ptyb}{\penv}{\constrs}\\\\
        \returnablepred{\ptya} \\
        \returnablepred{\ptyb}
    }
    { \chkseq{\inr{\vala}}{\tysum{\ptya}{\ptyb}}{\penv}{\constrs} }

    \inferrule
    [TC-Case]
    {
        \chkseq{\vala}{\tysum{\tya_1}{\tya_2}}{\penv_1}{\constrs_1} \\
        \chkseq{\ptma}{\ptya}{\penv_2}{\constrs_2} \\
        \chkseq{\ptmb}{\ptya}{\penv_3}{\constrs_3} \\
        \checkenvseq{\penv_2}{x}{\tya_1}{\constrs_4} \\
        \checkenvseq{\penv_3}{y}{\tya_2}{\constrs_5} \\
        \mergeseq
            {(\envwithouttwo{\penv_2}{x})}
            {(\envwithouttwo{\penv_3}{y})}
            {\penv_4}
            {\constrs_6} \\
        \combineseq{\penv_1}{\penv_4}{\penv}{\constrs_7}
    }
    { \chkseq
        {\caseof
            {\vala}
            {\inl{x}: \tya_1 \mapsto \ptma; \inr{y}: \tya_2 \mapsto \ptmb}
        }
        {\ptya}
        { \penv }
        { \constrs_1 \cup \cdots \cup \constrs_7 }
    }

    \inferrule
    [TC-CaseNoAnn]
    {
        \chkseq{\ptma}{\ptyb}{\penv_1, x : \ptya_1}{\constrs_1} \\
        \chkseq{\ptmb}{\ptyb}{\penv_2, y : \ptya_2}{\constrs_2} \\\\
        \chkseq{\vala}{\ptya_1 + \ptya_2}{\penv_3}{\constrs_3} \\
        \mergeseq{\penv_1}{\penv_2}{\penv_4}{\constrs_4} \\
        \combineseq{\penv_3}{\penv_4}{\penv}{\constrs_5}
    }
    { \chkseq
        {\caseof{\vala}{\inl{x} \mapsto \ptma; \inr{y} \mapsto \ptmb}}
        {\ptyb}
        {\penv}
        {\constrs_1 \cup \cdots \cup \constrs_5}
    }
\end{mathpar}
}
\caption{Extension of \langname with sum types}
\label{fig:extensions:sums}
\end{figure}

As for the algorithmic rules,
pair construction (\textsc{TC-Pair}) checks that both components have the given
types, and that both given types are returnable.
Environment combination and constraints are handled as usual.
Deconstructing the pair in general requires an annotation (\textsc{TC-LetPair});
as with the let rule, we check that the pair has the given annotation and that
the types inferred in the environment of the continuation are consistent with
the annotation.  If both components are used within the continuation
then we can omit the annotation (\textsc{TC-LetPairNoAnn}): the rule first
checks that the continuation has the given type, and inspects the resulting
environment to construct the product type used for checking $\vala$.
We omit the algorithmic type operations, which are the usual adaptations of the
declarative operations.

\subsubsection{Sum types}

It is also useful to include sum types in order to express multiple ways of
constructing data. The main principles are the same as for product types.

\paragraph*{Formalism.}
Figure~\ref{fig:extensions:sums} shows how to extend \langname with sum types,
which largely follows the development for product types.
Again, the declarative rules are unremarkable except for the requirement that
sum components must be returnable in the introduction rules. Subtyping
and type combination are defined component-wise, and a sum type is unrestricted
if both of its components are unrestricted.
\ADDED{As for the algorithmic rules}, sum injections are checking cases; similar
to the product rules we must ensure that the constituent types are both
returnable. We also have two separate rules for case expressions that allow
annotations to be elided if both $x$ and $y$ are used within continuations
$\tma$ and $\tmb$ respectively.

\begin{figure}[t]
{\small
    \header{Additional syntax}
    \[
    \begin{array}{lrcl}
        \text{Usage-annotated types} & \tya, \tyb & ::= & \cdots \midspace \tylist{\tya}
        \\
        \text{Values} & \vala, \valb & ::= & \cdots \midspace \nil \midspace \cons{\vala}{\valb}
        \\
        \text{Computations} & \tma, \tmb & ::= & \cdots \midspace
        \caselofcoredecl{\vala}{\tma}{y}{\var{ys}}{\tmb}
        \\
        \text{Annotated computations} & \ptma, \ptmb & ::= & \cdots \midspace
        \caselofcore{\vala}{\tylist{\tya}}{\ptma}{y}{\var{ys}}{\ptmb}
    \end{array}
    \]
    \headertwo
        {Additional declarative typing rules}
        {\framebox{$\vseq{\tyenv}{\vala}{\tya}$}~
         \framebox{$\tseq{\tyenv}{\tma}{\tya}$}}
\begin{mathpar}
    \inferrule
    [T-Nil]
    {
        \returnablepred{\tya}
    }
    { \tseq{\cdot}{\nil}{\tylist{\tya}} }

    \inferrule
    [T-Cons]
    {
        \vseq{\tyenv_1}{\vala}{\tya} \\
        \vseq{\tyenv_2}{\valb}{\tylist{\tya}} \\
        \returnablepred{\tya}
    }
    { \tseq{\tyenv_1 + \tyenv_2}{\cons{\vala}{\valb}}{\tylist{\tya}} }

    \inferrule
    [T-CaseL]
    {
        \tseq{\tyenv_1 }{\vala}{\tylist{\tya}} \\
        \tseq{\tyenv_2}{\tma}{\tyb} \\
        \tseq{\tyenv_2, y : \tya, \var{ys} : \tylist{\tya}}{\tmb}{\tyb}
    }
    { \tseq
        {\tyenv_1 + \tyenv_2}
        {\caselofcoredecl{\vala}{\tma}{y}{\var{ys}}{\tmb}}
        {\tyb}
    }
\end{mathpar}

\headertwo
    {Additional declarative operations on types}
    {\framebox{$\un{\tya}$}~\framebox{$\returnablepred{\tya}$}~\framebox{$\tya \subtype \tyb$}~\framebox{$\tya \scomb \tyb$}}

\begin{mathpar}
    \inferrule
    { \un{\tya} }
    { \un{\tylist{\tya}} }

    \inferrule
    {}
    { \returnablepred{\tylist{\tya}} }

    \inferrule
    { \tya \subtype \tyb }
    { \tylist{\tya} \subtype \tylist{\tyb}}

    \inferrule
    {}
    { (\scombtwo{\tylist{\tya}}{\tylist{\tyb}})
        = \tylist{\scombtwo{\tya}{\tyb}}    }
\end{mathpar}

\headertwo
    {Additional algorithmic typing rules}
    {\framebox{$\synthseq[\pprog]{\ptma}{\ptya}{\penv}{\constrs}$}~\framebox{$\chkseq[\pprog]{\ptma}{\ptya}{\penv}{\constrs}$}}
\begin{mathpar}
\inferrule
[TC-Nil]
{
    \returnablepred{\ptya}
}
{ \chkseq
    {\nil}
    {\tylist{\ptya}}
    { \cdot }
    {\emptyset}
}

\inferrule
[TC-Cons]
{
    \chkseq{\vala}{\ptya}{\penv_1}{\constrs_1} \\
    \chkseq{\valb}{\tylist{\ptya}}{\penv_2}{\constrs_2} \\\\
    \returnablepred{\ptya} \\
    \combineseq{\penv_1}{\penv_2}{\penv}{\constrs_3}
}
{ \chkseq
    {\cons{\vala}{\valb}}
    {\tylist{\ptya}}
    {\penv}
    {\constrs_1 \cup \constrs_2 \cup \constrs_3}
}

\inferrule
[TC-CaseL]
{
    \chkseq{\vala}{\tylist{\tya}}{\penv_1}{\constrs_1} \\
    \chkseq{\ptma}{\ptya}{\penv_2}{\constrs_2} \\
    \chkseq{\ptmb}{\ptya}{\penv_3}{\constrs_3} \\
    \checkenvseq{\penv_3}{y}{\tya}{\constrs_4} \\
    \checkenvseq{\penv_3}{\var{ys}}{\tylist{\tya}}{\constrs_5} \\
    \mergeseq
        {\penv_2}
        {(\envwithouttwo{(\envwithouttwo{\penv_3}{y})}{\var{ys}})}
        {\penv_4}
        {\constrs_6} \\
    \combineseq{\penv_1}{\penv_4}{\penv}{\constrs_7}
}
{ \chkseq
    {\caselof
        {\vala}
        {\tylist{\tya}}
        {\nil \mapsto \ptma; \cons{y}{\var{ys}} \mapsto \ptmb}
    }
    {\ptya}
    { \penv }
    { \constrs_1 \cup \cdots \cup \constrs_7 }
}

\inferrule
[TC-CaseLNoAnn]
{
    \chkseq{\ptma}{\ptyb}{\penv_1}{\constrs_1} \\
    \chkseq{\ptmb}{\ptyb}{\penv_2, y : \ptya, \var{ys} : \tylist{\ptya}}{\constrs_2} \\
    \chkseq{\vala}{\tylist{\ptya}}{\penv_3}{\constrs_3} \\
    \mergeseq{\penv_1}{\penv_2}{\penv_4}{\constrs_4} \\
    \combineseq{\penv_3}{\penv_4}{\penv}{\constrs_5}
}
{ \chkseq
    {\caselnotype
        {\vala}
        {\nil \mapsto \ptma; \cons{y}{\var{ys}} \mapsto \ptmb}
    }
    {\ptyb}
    { \penv }
    { \constrs_1 \cup \cdots \cup \constrs_5 }
}
\end{mathpar}
    }
\caption{Extension of \langname with list types}
\label{fig:extensions:lists}
\end{figure}

\subsubsection{List types}\label{sec:extensions:lists}

We can apply a similar approach to extend \langname to support
inductively defined lists.

Lists are useful as they allow us to encode patterns
such as broadcasting a message to a number of clients. For example, the
following code (adapted from the K-Fork benchmark described
in~\secref{sec:implementation:examples}) broadcasts a request to a list of actor
references:

{\small
    \[
    \bl
        \fndefthree
            {\var{broadcast}}
            {\var{actorMbs}: \tylist{\tysend{\msgtag{Request}}}}
            {\tyunit} \\
        \quad \caseltwo{\var{actorMbs}}{\tylist{\tysend{\msgtag{Request}}}}{\tyunit} \\
        \qquad \nil \mapsto () \\
        \qquad \cons{\var{mb}}{\var{mbs}} \mapsto
        \send{\var{mb}}{Request}{};
        \var{broadcast}(\var{mbs})
        \\
        \quad \}
        \\
        \}
    \el
\]
}

\paragraph*{Formalism.}
Figure~\ref{fig:extensions:lists} shows how to extend \langname with list types,
which follows the development for product and sum types presented above. As one
might expect, constructing a list is similar to constructing a product after
injecting into a sum, while pattern matching on a list is similar to case
matching on a sum followed by deconstructing a product.

The declarative rules are unremarkable aside from the type of values in the
list needing to be returnable in the introduction rules. Since sums and lists
are currently the only types in \langname that can be pattern matched against,
we use different syntax for each: $\calcwd{case}$ and $\calcwd{caseL}$ to avoid
overloading. A future extension introducing general recursive types would aim to
unify these and allow for general pattern matching against the constructors of
any data type, but this extension is outside the scope of the current work.
Note that we need the $\returnablepred{\tya}$ condition on
\emph{both} \textsc{T-Cons} \emph{and} \textsc{T-Nil}, rather than just
\textsc{T-Nil}, to avoid issues with subtyping.

In the algorithmic rules, rule \textsc{TC-Nil} requires that the type of list
elements is returnable. In \textsc{TC-Cons}, environment combination and
constraints are handled as in the analogous product rule.  Case expressions,
like sums, have two separate rules allowing the annotation to be elided if both
$y$ and $\var{ys}$ are used in the continuation $\tmb$.

\ADDED{
\subsection{Towards more liberal data types}

We have required that the data contained within each of the described data types
must be returnable. This ensures that we cannot deconstruct a data type
containing a second-class name and unpackage it later, thus breaking the lexical
scoping requirements of quasi-linearity.  However, this approach can be
restrictive.

A potential solution is to allow the construction of data types that may contain
names with second-class types, but only to allow them to be deconstructed in a
situation where unsafe aliasing cannot occur: namely where the continuation
$\tma$ of a $\letin{(x, y)}{\vala}{\tma}$ construct or the relevant
continuations of a $\calcwd{case}$ construct do not close over any mailbox types
(or, as we will see in~\secref{sec:extensions:interfaces}, do not close over any
mailbox types that might lead to aliasing). We can formalise the declarative
rules for product types as follows; the corresponding rules for sums and lists
are similar.

{\small
\begin{mathpar}
    \inferrule
    {
        \tseq{\tyenv_1}{\vala}{\tya}
        \\
        \tseq{\tyenv_2}{\valb}{\tyb}
    }
    { \tseq{\tyenv_1 + \tyenv_2}{(\vala, \valb)}{\typair{\tya}{\tyb}} }

    \inferrule
    {
        \tseq{\tyenv_1}{\vala}{\typair{\tya_1}{\tya_2}} \\
        \tseq{\tyenv_2, x : \tya_1, y : \tya_2}{\tma}{\tyb} \\\\
        (\returnablepred{\tya_1} \wedge \returnablepred{\tya_2}) 
        \vee \mbbase{\tyenv_2}
    }
    { \tseq{\tyenv_1 + \tyenv_2}{\letin{(x, y)}{\vala}{\tma}}{\tyb}}
\end{mathpar}
}

This approach has the advantage that it rules out problematic cases that
introduce unsafe aliasing, where we have two static names for the same
underlying runtime name. Consider the following example (assuming $x$ is bound
elsewhere):

{\small
\[
    \bl
    \letintwo{\var{mb}}{\mbnew} \\
    \letintwo{\var{pair}}{(\var{mb}, 1)} \\
    \letintwo{(a, b)}{\var{pair}} \\
    \send{x}{m}{a}; \\
    \guardone{\var{mb}} \\
    \quad \receive{m}{\var{y}}{\var{mb}}{\sugarfree{\var{mb}}} \\
    \}
    \el
\]
}

This code snippet packages $\var{mb}$ (with type $\usablety{\mbsendone}$)
into a pair, then uses pair deconstruction to alias the first component of the
pair to $a$, and would be ruled out as mailbox names $x$ and $\var{mb}$ are free
in the continuation of the pair deconstruction construct.

However, the approach is not sound according to our declarative rules due to
quasi-linearity, and does not rule out self-deadlocks. For example, the
following code would be well-typed even though it introduces a self-deadlock:

{\small
\[
    \bl
        \letintwo{\var{mb}}{\mbnew} \\
        \letintwo{\var{pair}}{(\var{mb}, 1)} \\
        \guardone{\var{mb}} \\
        \quad \receive{m}{}{\var{mb}}{\sugarfree{\var{mb}}} \\
        \}; \\
        \letintwo{(x, y)}{\var{pair}} \\
        \send{x}{m}{}
    \el
\]
}

We have implemented this approach in our typechecker as an opt-in (unsound)
feature, but we expect finer-grained alias analysis techniques to be an
important area of future work.
None of the examples presented in the paper or our evaluation rely on this more
liberal treatment of data types.
}

\ADDED{
    \subsection{First-class functions}\label{sec:extensions:lambdas}
}

\begin{figure}[t]
    {\small
    \ADDED{
\header{Additional syntax}
\[
    \begin{array}{lrcl}
        \text{Linearity annotations} & \linann & ::= & \linear \midspace
        \unrestricted \\
\text{Modified usage-annotated types} & \tya, \tyb & ::= &
            \cdots \midspace \tyannfun{\seq{\tya}}{\tyb}
            \\
        \text{Modified values} & \vala, \valb & ::= &
        \cdots \midspace
            \flaglambda{\seq{x : \tya}}{\tyb}{\tma} \\
        \text{Modified computations} & \tma, \tmb & ::= &
        \cdots \midspace
        \fnapp{\vala}{\seq{\valb}} \\
    \end{array}
\]

\headersig{Additional declarative typing rules}{$\tseq{\tyenv}{\tma}{\tya}$}
\begin{mathpar}
    \inferrule
    [T-LinLambda]
    {
        \tseq{\tyenv, \seq{x : \tya}}{\tma}{\tyb}
        \\
        \returnablepred{\tyenv}
    }
    { \tseq
        {\tyenv}
        {\linlambda{\seq{x : \tya}}{\tyb}{\tma}}
        {\tylinfun{\seq{\tya}}{\tyb}}
    }

    \inferrule
    [T-UnLambda]
    {
        \tseq{\tyenv, \seq{x : \tya}}{\tma}{\tyb}
        \\
        \returnablepred{\tyenv} \\ \un{\tyenv}
    }
    { \tseq
        {\tyenv}
        {\unlambda{\seq{x : \tya}}{\tyb}{\tma}}
        {\tyunrfun{\seq{\tya}}{\tyb}}
    }

    \inferrule
    [T-AppLambda]
    {
        \tseq{\tyenv}{\vala}{\tyannfun{\seq{\tya}}{\tyb}} \\
        (\tseq{\tyenv_i}{\valb_i}{\tya_i})_{i \in 1..n}
    }
    {
        \tseq
            {\tyenv + \tyenv_1 + \cdots + \tyenv_n}
            {\fnapp{\vala}{\seq{\valb}}}
            {\tyb}
    }
\end{mathpar}

\headertwo
    {Additional declarative operations on types}
    {\framebox{$\un{\tya}$}~\framebox{$\returnablepred{\tya}$}~\framebox{$\tya \subtype \tyb$}~\framebox{$\tya \scomb \tyb$}}

\begin{mathpar}
    \inferrule
    {}
    { \un{\tyunrfun{\seq{\tya}}{\tyb}} }

    \inferrule
    {}
    { \returnablepred{\tyannfun{\seq{\tya}}{\tyb}} }

    \inferrule
    { \seq{\tya_2} \subtype \seq{\tya_1} \\ \tyb_1 \subtype \tyb_2 }
    { (\tyannfun{\seq{\tya_1}}{\tyb_1}) \subtype
        (\tyannfun{\seq{\tya_2}}{\tyb_2}) }

    \inferrule
    {}
    { \scombtwo
        {(\tyunrfun{\seq{\tya}}{\tyb})}
        {(\tyunrfun{\seq{\tya}}{\tyb})} = \tyunrfun{\seq{\tya}}{\tyb} }
\end{mathpar}
\headersig{Reduction rule}{$\config{C} \cevalp \config{D}$}
\[
        \thread{(\flaglambda{\seq{x : \tya}}{\tyb}{\tma})(\seq{\vala})}{\framestack}
        \ceval
        \thread{\subst{\tma}{\seq{\vala}}{\seq{x}}}{\framestack}
\]
}
}
\caption{Additional syntax, typing rules, and reduction rules for first-class
functions}
\label{fig:extensions:lambdas}
\end{figure}

\ADDED{
We can extend \langname\ with first-class functions,
but this requires care with quasi-linearity and requires additional typing
information during typechecking. We will discuss quasi-linearity now but defer
discussion of typechecking to~\secref{sec:extensions:typechecking-contextual}.
}

\paragraph*{Quasi-linearity.}
Consider the following expression \ADDED{(where 
the $\linear$ annotation on the function abstraction
denotes a \emph{linear} function abstraction that must be applied precisely once)}:

{\small
\[
    \bl
        \letintwo{\var{mb}}{\mbnew} \\
        \letintwo{f}{(\linlambda{}{\one}{\send{\var{mb}}{m}{}})} \\
        \guardanntwo{\var{mb}}{\msgtag{m}} \\
        \quad \receive{m}{}{\var{mb}}{\sugarfree{\var{mb}}} \\
        \}; \\
        \fnapp{f}{}
    \el
\]
}

Here we bind $f$ to a function that sends message $\msgtag{m}$ to mailbox
$\var{mb}$; note that it is used lexically before the $\calcwd{guard}$, which
aligns with type combination. However, after reducing the expression (assuming
that $a$ is chosen as a runtime name), we obtain the following term:

{\small
\[
    \bl
        \guardanntwo{\var{a}}{\msgtag{m}} \\
        \quad \receive{m}{}{\var{mb}}{\sugarfree{\var{mb}}} \\
        \}; \\
        \fnapp{(\linlambda{}{\one}{\send{\var{a}}{m}{}})}{}
    \el
\]
}

After substituting the function body for $f$ we now have a second-class use
\emph{after} the first-class use, violating the ordering of returnable and
second-class usages.

\ADDED{
\paragraph*{Formalism.}
Figure~\ref{fig:extensions:lambdas} shows the additional syntax, declarative
typing rules, and reduction rule for extending \langname\ with first-class
functions. We defer discussion of algorithmic typing rules to~\secref{sec:extensions:typechecking-contextual}.
We extend values with fully annotated, $n$-ary anonymous functions
$\flaglambda{\seq{x : \tya}}{\tyb}{\tma}$, where $\linann$ is a
linearity annotation that specifies whether a function is linear or unrestricted.
We include $n$-ary functions rather than using currying because anonymous
functions may only close over returnable values, to ensure they do not violate
the quasi-linearity conditions on lexical scoping once applied, and therefore
unary functions would be less expressive.
We also extend computations with $n$-ary function application
$\fnapp{\vala}{\seq{\valb}}$. 

Rule \textsc{T-LinLambda} types a linear function, \ie, a function that must be
applied precisely once.
A linear function may, for example, close over returnable mailbox output
capabilities.
The rule is similar to the regular function abstraction rule, but binds
multiple parameters and requires that the function body closes over only
variables with returnable types.
Rule \textsc{T-UnLambda} types an unrestricted function and is similar, but
requires that the function closes over only variables with unrestricted types.

The $\un{-}$ predicate holds only for unrestricted function types, and since
functions must close over returnable environments, all
function types are returnable. We use the usual subtyping rule for functions
where function types are contravariant in their argument types and covariant in
their result types; we write $\seq{\tya} \subtype \seq{\tyb}$ to denote that
$\tya_i \subtype \tyb_i$ for all $i \in 1..n$.
The sequential combination of
two unrestricted function types does not affect the argument or result types.

The reduction rule for function application is the standard $\beta$-reduction
rule adapted for frame stacks.

\paragraph*{Metatheory.}
The addition of first-class functions does not violate the metatheoretical
properties of the system: preservation and self-deadlock-freedom are maintained
because we restrict $\lambda$-abstractions to close over only returnable
variables and thus applying a function cannot break the invariants on
combining usage annotations.
}

\ADDED{
    \subsection{Mailbox interfaces}\label{sec:extensions:interfaces}
}

In the core \langname language, a global signature maps
message tags to payload types. While technically convenient, this can be
inflexible.
First, distinct entities may wish to use the same message tags with different
payload types. For example, a client may send a $\msgtag{Login}$ message
containing credentials to a server, which may then send a $\msgtag{Login}$
message containing the credentials and a timestamp to a session management
server.
Second, we need a syntactic check on a $\calcwd{receive}$ guard
to avoid aliasing introduced by communication, as outlined
in~\secref{sec:whatstheissue}: either the received payloads or free variables in
the guard body must be base types.
Consider the following expression:

{\small
    \[
    \bl
    \fndefthree{\mkwd{waitAndSend}}{\var{mb}:
    \tyrecv{(\patconcattwo{\msgtag{M1}}{\msgtag{M2}}})}{\one} \\
    \quad \guardanntwo{\var{mb}}{\msgtag{M1} \patconcat \msgtag{M2}} \\
    \qquad \receivethree{M1}{\var{x}}{\var{mb}'} \\
    \qquad
        {
        \bl
            \guardanntwo{\var{mb}'}{\msgtag{M2}} \\
            \quad \receivethree{M2}{\var{y}}{\var{mb}''} \\
            \qquad \send{\var{x}}{Go}{}; \\
            \qquad
            \send{\var{y}}{Go}{}; \\
            \qquad \sugarfree{\var{mb}''} \\
            \}
        \el
        }\\
    \quad \} \\
    \}
    \el
    \]
}

The \mkwd{waitAndSend} definition waits for messages $\msgtag{M1}$ and
$\msgtag{M2}$ from mailbox $\var{mb}$. The messages carry mailbox names $x$ and
$y$ respectively.  After receiving both messages, the function sends a
$\msgtag{Go}$ message to both $x$ and $y$ before freeing $\var{mb}$.
This safe code is \emph{not} typable in the calculus without interfaces, because
the mailbox variable $x$ occurs free in the second $\calcwd{receive}$ guard.

\ADDED{
To address this issue we can associate each mailbox with an \emph{interface}
$\iface$ that maps tags to payload types, and allows us to syntactically
distinguish different kinds of mailbox (\eg, a future and its client).
}

Since a name cannot have two interfaces at once, we can loosen our syntactic
check on $\calcwd{receive}$ guards to require only that the \emph{interfaces} of
mailbox names in the payloads and free variables differ, as typing guarantees
that they will refer to different mailboxes. We could therefore type the
$\mkwd{waitAndSend}$ example above, as long as we statically know $x$ and $y$
have different interfaces, and therefore must be different mailboxes.

\begin{figure}[t]
    {\small
    \ADDED{
        \header{Modified syntax}
    \[
    \begin{array}{lrcl}
        \text{Interface names} & \iface \\
        \text{Interfaces} & \ifacemv & ::= & \seq{\msgtag{m} \mapsto \seq{\pretya}} \\
\text{Interface mapping} & \mathcal{I} & ::= & \seq{\iface \mapsto
            \ifacemv} \\
        \text{Modified programs} & \prog & ::= &
            \progdef{\proginterfaces}{\seq{D}}{\tma} \\
\text{Modified mailbox types} & \mbtya, \mbtyb & ::= & \tysendiface{\pata} \midspace
            \tyrecviface{\pata} \\
\text{Modified computations} & \tma, \tmb & ::= & \cdots \midspace \mbnewiface
    \end{array}
    \]

    \headersig{Modified typing rules for computations}{$\tseq{\tyenv}{\tma}{\tya}$}
    \begin{mathpar}
       \inferrule
       [TI-New]
       { }
       { \tseq{\cdot}{\mbnewiface}{\retty{\tyrecviface{\mbone}}} }

       \inferrule
        [TI-Guard]
        {
          \tseq{\tyenv_1}{\vala}{\retty{\tyrecviface{\pata}}} \\
          \gseq{\iface; \pata}{\tyenv_2}{\seq{G}}{\tya} \\
          \pnf{\pata}
        }
        { \tseq{{\tyenv_1} + {\tyenv_2}}{\guard{\vala}{\seq{G}}}{\tya} }

      \inferrule
      [TI-Fail]
      { \tseq{\tyenv}{\vala}{\retty{\tyrecviface{\mbzero}}} }
      { \tseq{\tyenv}{\sugarfail{\vala}}{\tya} }
    \end{mathpar}

    \headertwo
    {Typing rules for guards}
    {
        \framebox{$\gseq[\prog]{\iface; \pata}{\tyenv}{\seq{\gd}}{\tya}$}
        ~
        \framebox{$\gseq[\prog]{\iface; \pata}{\tyenv}{\gd\vphantom{\seq{\gd}}}{\tya}$}
    }
    \begin{mathpar}
        \inferrule
        [TGI-GuardSeq]
        {
            (\gseq{\iface; \pata_i}{\tyenv}{G_i}{\tya})_{i \in 1..n}
        }
        {
            \gseq
                {\iface; \pata_1 \patplus \cdots \patplus \pata_n}
                {\tyenv}
                {\seq{G}}
                {\tya}
        }

        \inferrule
        [TGI-Free]
        {
            \tseq{\tyenv}{\tma}{\tya}
        }
        { \gseq{\iface; \mbone}{\tyenv}{\free{\tma}}{\tya} }

        \inferrule
        [TGI-Recv]
        {
          \iface(\msgtag{m}) = \seq{\pretya} \\
          \interfaces{\seq{\pretya}} \cap \interfaces{\tyenv} = \emptyset \\
          \tseq
            {\tyenv,
                y : \usageann{\returnable}{\tyrecviface{\pata}},
                \seq{x} : \seq{\makeusable{\pretya}}}
            {\tma}
            {\tyb}
        }
        {
          \gseq{\iface; \algomsg{m} \patconcat
          E}{\tyenv}{\receive{m}{\seq{x}\!}{\!y}{\tma}}{\tyb}
        }
    \end{mathpar}
}
}
\caption{Extension of \langname to support mailbox interfaces}
\label{fig:extensions:interfaces}
\end{figure}

\ADDED{
\paragraph*{Formalism.}
Figure~\ref{fig:extensions:interfaces} shows the extensions to \langname to
support mailbox interfaces.
We modify the definition of programs $\progdef{\sig}{\seq{D}}{\tma}$ (where
$\sig$ is a mapping from message tags to sequences of payload types) to
$\progdef{\proginterfaces}{\seq{D}}{\tma}$, where $\proginterfaces$ is a mapping
from \emph{interface names} $\iface$ to interfaces $\ifacemv$.
Like signatures in the unextended core calculus, interfaces $\ifacemv$ map
message tags to sequences of payload types.
Given some program $\prog = \progdef{\proginterfaces}{\seq{D}}{\tma}$ and
interface name $\iface$, we write $\iface(\msgtag{m})$ as syntactic sugar for
$\proginterfaces(\iface)(\msgtag{m})$.
Note that interfaces can be mutually recursive.

We extend mailbox types $\tysendiface{\pata}$ and $\tyrecviface{\pata}$
to include their interface name.  The main alteration to the term syntax is to
require an annotation on the $\calcwd{new}$ construct to specify the interface
of the newly created mailbox.

We update the judgement form for guards to include the interface name of the
current mailbox. Rule \textsc{TI-New} includes the specified interface in the
mailbox type, and rule \textsc{TI-Guard} uses the mailbox type's interface name
when typing the guards.
Instead of using the program signature to determine message types, rule
\textsc{TGI-Recv} looks up the message tag in the given interface, and also
includes a more liberal check that only requires that the sets of interfaces of
mailbox types contained in the payload type 
and the interfaces of any mailbox types used in typing $\tma$ are disjoint.

We also require that combination of two mailbox types, and mailbox
subtyping, are only defined if the two types have the same interface.
}

\subsection{Using contextual type information: \ADDED{typechecking} first-class
functions and mailbox interfaces}\label{sec:extensions:typechecking-contextual}
\ADDED{
As described in~\secref{sec:algorithmic}, \langname's typechecker uses a
\emph{co-contextual} approach in order to generate the pattern inclusion
constraints required for algorithmic typechecking.
While this suffices for \langname without extensions and \langname with the
data type extensions, typechecking first-class functions and mailbox interfaces
requires \emph{contextual} type information \emph{before} the constraint
generation pass.

\subsubsection{Issues with typechecking first-class functions and mailbox interfaces}
}

\paragraph*{Typechecking first-class functions.}
Consider typechecking the application of a first-class function (here
$\unrestricted$ signifies that the function is unrestricted):
\[
(\unlambda{x : \tyint}{\tyint}{x})(5)
\]
The annotated $\lambda$ expression allows us to synthesise a type and
use a rule similar to \textsc{TS-App}.
Unfortunately, the lack of contextual type information
means that the approach fails as soon as we stray from applying function
literals, for example:
\[
    \letin{f}{(\unlambda{x : \tyint}{\tyint}{x})}{f(5)}
\]
This is because we \emph{do not} have information about the type of $f$ when
attempting to type $f(5)$.
A typical backwards bidirectional typing approach requires \emph{synthesising}
function argument types, but this is too inflexible in our setting as each
mailbox name argument would need a type annotation.

\paragraph*{Typechecking mailbox interfaces.}
\ADDED{
The complexity with typing mailbox interfaces comes with recording the interface
associated with each mailbox name. In \langname without extensions, the required
type information can be gained from context (\ie, through the message tag) and
the payload types are globally available through the program's message signature
$\sigs$.

For example, when typing a message send $\send{a}{m}{5}$, the base \langname
type system would look up message tag $\msgtag{m}$ in the program's signature
and see that the message had payload type $\tyint$, before checking that $5$ had
the corresponding type $\tyint$ and $a$ had type $\tysend{\msgtag{m}}$.
However, in a system with interfaces, we need knowledge of $a$'s interface in
order to look up the payload types. The checking judgement also requires
knowledge of $a$'s interface in addition to the expected pattern.
}

\paragraph*{Typechecking strategy.}
We implement the above extensions via a contextual type-directed elaboration
pass. We can annotate variables with an annotation that is useful when typing
function applications. As for interfaces, users specify an interface when
creating a mailbox ($\mbnewiface$); our pass then annotates sends, guards, and
fail constructs with interface information (\ie,
$\sendann{\vala}{\iface}{m}{\seq{\valb}}$ and 
    $\guardifaceann{\vala}{\pata}{\seq{\gd}}$ and $\sugarfailiface{\vala}$)
    for use in constraint generation.

\begin{figure}[t]
    {\small
    \header{Modified syntax}
    \[
        \begin{array}{lrcl}
            \text{Interface names} & \iface \\
                \text{Interfaces} & \ifacemv & ::= & \seq{\msgtag{m} \mapsto \seq{\ppretya}} \\
\text{Modified programs} & \prog & ::= &
            \progdef{\proginterfaces}{\seq{\pdef}}{\ptma} \\
\text{Modified mailbox types} & \pmbty & ::= & \tysendiface{\ppata} \midspace
                \tyrecviface{\ppata} \\
\text{Modified types} & \ppretya, \ppretyb & ::= &
                \basety \midspace \pmbty \midspace \tyannfun{\seq{\ptya}}{\ptyb}
            \\
            \text{Modified usage-annotated types} & \ptya, \ptyb & ::= &
            \basety \midspace \usageann{\usage}{\pmbty} \midspace \tyannfun{\seq{\ptya}}{\ptyb}
            \\
            \text{Pre-types} & \contya, \contyb & ::= & \basety \midspace \conmb
                \midspace \tyannfun{\seq{\ptya}}{\ptyb} \\
            \ADDED{\text{Modified annotated values}} & \ADDED{\pvala, \pvalb} & ::= & 
            \ADDED{\cdots \midspace \varann{x}{\shade{\contya}} \midspace
                \flaglambda{\seq{x : \ptya}}{\ptyb}{\ptma}}
            \\
            \text{Modified annotated computations} & \ptma, \ptmb & ::= & \cdots \midspace
            \fnapp{\pvala}{\seq{\pvalb}} \midspace
            \mbnewiface
            \midspace
            \sendann{\pvala}{\shade{\iface}}{m}{\seq{\pvalb}}  \\
             & &  \midspace &
            \guardifaceann[\shade{\iface}]{\pvala}{\pata}{\seq{\pgd}}
                \midspace \sugarfailiface[\shade{\iface}]{\pvala} \\
                                         
        \end{array}
    \]
\headersig{Promotion from pre-types to types}{$\promote{\contya} = \ppretya$}
    \begin{mathpar}
        \promote{\basety} = \basety

        \promote{\tyannfun{\seq{\ptya}}{\ptyb}} = \tyannfun{\seq{\ptya}}{\ptyb}

        \promote{\conmb} \text{ undefined}
    \end{mathpar}

    \headersig{Demotion from types to pre-types}{$\demote{\ppretya} = \contya$}
    \begin{mathpar}
        \demote{\basety} = \basety

        \demote{\tyannfun{\seq{\ptya}}{\ptyb}} = \tyannfun{\seq{\ptya}}{\ptyb}

        \demote{\tysendiface{\ppata}} = \conmb

        \demote{\tyrecviface{\ppata}} = \conmb
    \end{mathpar}
    }
    \caption{Extension of \langname to support \ADDED{typechecking of} first-class functions and interfaces (1)}
    \label{fig:extensions:contextual-1}
\end{figure}

\subsubsection{Extended syntax}
Figure~\ref{fig:extensions:contextual-1} shows the additional syntax required to
add first-class functions and interfaces.

\paragraph*{Types and pre-types.}
\emph{Pre-types} $\contya, \contyb$ are similar to types $\ptya$ but type
mailbox variables differently. Unlike the types we have seen so far, pre-types
can be inferred and checked algorithmically using an entirely standard
\emph{contextual} approach. The key difference between pre-types and types is
that mailbox names have type $\conmb$ and therefore do not contain either a
capability or a pattern. The main benefit of using pre-types is that they allow
us to propagate mailbox interface information, and to annotate
(non-mailbox-typed) variables with type information that can be used in
constraint generation (\eg, when typing functions). Pre-types that are not
mailbox types can be \emph{promoted} to a full type using the promotion operator
$\promote{\contya}$, and all types can be \emph{demoted} into a pre-type using
the demotion operator $\demote{\ppretya}$.

We modify types to include \emph{interface-annotated}
mailbox types, as well as $n$-ary function types
$\tyannfun{\seq{\ptya}}{\ptyb}$.

\ADDED{
\paragraph*{Values and computations.}

Since first-class functions can refer to computations, we add an additional
syntactic class $\pvala$ of annotated values (thus allowing first-class
functions to contain function bodies that have annotated $\calcwd{guard}$
expressions used for typechecking). We also extend variables $x$ with a pre-type
annotation.

\paragraph*{Mailbox terms.}
We extend computations so that a user specifies an interface $\iface$ when creating a
mailbox ($\mbnewiface$). We also augment send and guard expressions
with the interface of the mailbox they operate on.
Only the annotation on $\mbnew$ must be specified by a user: annotations on
variables, send expressions, fail expressions, and guard expressions (shaded)
are instead added by type-directed elaboration.
}

\begin{figure}[t]
{\small
    \headertwo
    {Type-directed elaboration}
    {\framebox{$\elabsynth{\contyenv}{\ptma}{\contya}{\ptmb}$}~\framebox{$\elabchk{\contyenv}{\ptma}{\contya}{\ptmb}$}}
\begin{mathpar}
    \inferrule
    [Elab-Var]
    { x : \contya \in \contyenv }
    { \elabsynth{\contyenv}{x}{\contya}{\varann{x}{\contya}} }

    \inferrule
    [Elab-New]
    { }
    { \elabsynth{\contyenv}{\mbnewiface}{\conmb}{\mbnewiface}  }

    \inferrule
    [Elab-Send]
    {
        \elabsynth{\contyenv}{\pvala}{\conmb}{\pvala'} \\
        I(\msgtag{m}) = \seq{\ppretya} \\\\
        (\elabchk{\contyenv}{\pvalb_i}{\demote{\ppretya_i}}{\pvalb_i'})_{i \in 1..n}
    }
    {
        \elabsynth
            {\contyenv}
            {\send{\pvala}{m}{\seq{\pvalb}}}
            {\one}
            {\sendann{\pvala'}{\iface}{m}{\seq{\pvalb'}}}
    }

    \inferrule
    [Elab-Guard]
    {
        \elabsynth{\contyenv}{\pvala}{\conmb}{\pvala'} \\\\
        (\elabgdsynth{\iface}{\contyenv}{\pgd_i}{\contya}{\pgd_i'})_{i \in 1..n}
    }
    { \elabsynth
        {\contyenv}
        {\guardann{\pvala}{\pata}{\seq{\pgd}}}
        {\contya}
        {\guardifaceann{\pvala'}{\pata}{\seq{\pgd'}}}
    }

    \inferrule
    [Elab-Fail]
    { \elabsynth
        {\contyenv}
        {\pvala}
        {\conmb}
        {\pvala'}
    }
    { \elabchk
        {\contyenv}
        {\sugarfail{\pvala}}
        {\contya}
        {\sugarfailiface{\pvala'}}
    }
\end{mathpar}
}
\caption{Type-directed elaboration (selected rules)}
\label{fig:extensions:elaboration}
\end{figure}
\subsubsection{Type-directed elaboration}
\ADDED{We propagate pre-type annotations} via a
contextual type-directed elaboration phase
(Figure~\ref{fig:extensions:elaboration}).
Pre-type environments $\contyenv$ map variables to pre-types. We use the
judgements $\elabsynth[\pprog]{\contyenv}{\ptma}{\contya}{\ptma'}$
(read ``under pre-type environment
$\contyenv$ and in the context of program $\pprog$,
synthesise pre-type $\contya$ for computation $\ptma$, and
produce elaborated term $\ptma'$''), and
$\elabchk[\pprog]{\contyenv}{\ptma}{\contya}{\ptma'}$
(read ``under pre-type environment
$\contyenv$ and in the context of program $\pprog$,
check that computation $\ptma$ has pre-type $\contya$, and
produce elaborated term $\ptma'$'').
We use analogous judgements for values and guards.
The rules for each judgement follow the usual bidirectional typing rules for the
simply typed $\lambda$-calculus, so we concentrate on the pertinent elaboration
rules.

Rule \textsc{Elab-Var} annotates a variable with its pre-type, and rule
\textsc{Elab-New} synthesises type $\conmb$, where $I$ corresponds to the
user-specified interface.
Rule \textsc{Elab-Send} states that if pre-type $\conmb$
can be synthesised for the target mailbox, then the rule looks up the payload
types for the message in interface $\iface$ and checks all payloads against the
(demoted) payload types. The result is a send expression annotated with the
interface $\iface$.

Rule \textsc{Elab-Guard} synthesises pre-type $\conmb$ for the given
mailbox, and then checks that every guard clause synthesises the same pre-type
$\contya$.
The interface is passed to the guard synthesis judgement so that the payload
types can be retrieved when typing a $\calcwd{receive}$ guard. Again, the result
is the guard annotated with interface $\iface$.

Finally, rule \textsc{Elab-Fail} is a checking rule for $\sugarfail{\vala}$: it
synthesises pre-type $\conmb$ for $\vala$ and annotates the
$\calcwd{fail}$ construct with interface $\iface$.

\begin{figure}[t]
    {\small
    \headertwo
    {Modified constraint generation rules}
    {   \framebox{\vphantom{$\Gamma$}$\synthseq[\pprog]{\ptma}{\ptya}{\penv}{\constrs}$}~
        \framebox{$\chkseq[\pprog]{\ptma}{\ptya}{\penv}{\constrs}$}~
        \fbox{$\chkgseqcon[\pprog]{\iface}{\pata}{\pgd}{\ptya}{\penv}{\constrs}{\patb}$} }
\begin{mathpar}
        \inferrule
        [TS-LinLam]
        {
            \chkseq{\ptma}{\ptyb}{\penv'}{\constrs_1} \\\\
            \checkenvseq{\penv'}{\seq{x}}{\seq{\ptya}}{\constrs_2} \\
            \penv = \makereturnable{\penv' \envwithout \seq{x}}
        }
        {
            \synthseq
                {\linlambda{\seq{x : \ptya}}{\ptyb}{\ptma}}
                {\tylinfun{\seq{\ptya}}{\ptyb}}
                {\penv}
                {\constrs_1 \cup \constrs_2}
        }
~~~~
        \inferrule
        [TS-UnLam]
        {
            \chkseq{\ptma}{\ptyb}{\penv'}{\constrs_1} \\
            \checkenvseq{\penv'}{\seq{x}}{\seq{\ptya}}{\constrs_2} \\\\
            \penv = \makereturnable{\penv' \envwithout \seq{x}} \\
            \unrseq{\penv}{\constrs_3}
        }
        {
            \synthseq
                {\unlambda{\seq{x : \ptya}}{\ptyb}{\ptma}}
                {\tyunrfun{\seq{\ptya}}{\ptyb}}
                {\penv}
                {\constrs_1 \cup \constrs_2 \cup \constrs_3}
        }

        \inferrule
        [TS-Var]
        {
            \promote{\contya} = \ptya
        }
        {
            \synthseq
                {\varann{x}{\contya}}
                {\ptya}
                {x : \ptya}
                {\emptyset}
        }

        \inferrule
        [TS-FnApp]
        {
            \synthseq
                {\pvala}
                {\tyannfun{\seq{\ptya}}{\ptyb}}
                {\penv_V}
                {\constrs_V}
            \\\\
            (\chkseq{\pvalb_i}{\ptya_i}{\penv_i}{\constrs_i})_{i \in 1..n} \\
            \combineseqmany{\penv_V + \penv_1 + \cdots + \penv_n}{\penv}{\constrs}
        }
        { \synthseq
            { \fnapp{\pvala}{\seq{\pvalb}} }
            {\ptyb}
            {\penv}
            {\constrs \cup \constrs_V \cup \constrs_1 \cup \cdots \cup \constrs_n}
        }

        \inferrule
        [TS-Send]
        {
            \iface(\msgtag{m}) = \seq{\ppretya} \\
            \chkseq{\pvala}{\usablety{\tysendiface{\msgtag{m}}}}{\penv_V}{\constrs_V}
            \\\\
            (\chkseq{\pvalb_i}{\makeusable{\ppretya_i}}{\penv_i}{\constrs'_i})_{i \in 1..n} \\
            \combineseqmany{\penv_V + \penv_1 + \cdots + \penv_n}{\penv}{\constrs}
        }
        { \synthseq
            {\sendann{\pvala}{\iface}{m}{\seq{\pvalb}}}
            {\one}
            {\penv}
            { \constrs \cup \constrs_V \cup \constrs'_1 \cup \cdots \cup \constrs'_n}
        }

        \inferrule
        [TS-New]
        { }
        { \synthseq{\mbnewiface}{\retty{\tyrecviface{\mbone}}}{\cdot}{\emptyset} }

        \inferrule
        [TC-Guard]
        {
            \chkgseqcon
                {\iface}
                {\pata}
                {\seq{\pgd}}
                {\ptya}
                {\penv_1}
                {\constrs_1}
                {\patb} \\\\
            \chkseq{\pvala}{\retty{\tyrecviface{\patb}}}{\penv_2}{\constrs_2} \\
            \combineseq{\penv_1}{\penv_2}{\penv}{\constrs_3}
        }
        { \chkseq
            {\guardifaceann{\pvala}{\pata}{\seq{\pgd}}}
            {\ptya}
            {\penv}
            {\constrs_1 \cup \constrs_2 \cup \constrs_3 \cup
            \set{\subpatconstr{\pata}{\patb} }}
        }

        \inferrule
        [TC-Fail]
        {
            \chkseq{\pvala}{\retty{\tyrecviface{\mbzero}}}{\penv}{\constrs}
        }
        { \chkseq{\sugarfailiface{\pvala}}{\ptya}{\penv}{\constrs} }

        \inferrule
        [TCG-Recv]
        {
            \chkseq{\ptma}{\ptya}{\penv', y : \retty{\tyrecviface{\ppata}}}{\constrs_1} \\
            I(\msgtag{m}) = {\seq{\ppretya}} \\
            \penv = \penv' \envwithout \seq{x} \\\\
            \interfaces{\seq{\ppretya}} \cap \interfaces{\penv} = \emptyset \\
            \checkenvseq{\penv'}{\seq{x}}{\seq{\makeusable{\ppretya}}}{\constrs_2}
        }
        { \chkgseqcon
            {\iface}
            {\pata}
            {\receive{m}{\seq{x}}{y}{\ptma}}
            {\ptya}
            {\penv}
            {\constrs_1 \cup \constrs_2 \cup
                \set{\subpatconstr{\pata \without \msgtag{m}}{\ppata}}
            }
            {\algomsg{m} \patconcat (\pata \without \msgtag{m})}
        }
    \end{mathpar}
}
\caption{Extension of \langname to support \ADDED{typechecking of} first-class
functions and interfaces (2)}
\label{fig:extensions:contextual-2}
\end{figure}

\subsubsection{Constraint generation rules}

Figure~\ref{fig:extensions:contextual-2} shows the constraint generation rules
for the extended calculus.
We require three new rules for first-class functions: rule \textsc{TS-LinLam} types a
linear anonymous function by checking that the body has the given result type,
and the inferred environment uses variables consistently with the parameter
annotations: the rule synthesises a type consistent with the annotation.
Further, we require that the inferred environment only closes over
variables with returnable types. Rule \textsc{TS-UnLam} is similar, but
additionally requires that the inferred environment is unrestricted. Rule
\textsc{TS-FnApp} synthesises a type for the function (made possible using
either the type annotation on the function abstraction, or the annotation on the
function variable); it then checks that the arguments and results have the
correct types.

As for the rules that support interfaces, rule \textsc{TS-Send} is similar but
looks up the types according to the interface rather than the global signature,
and checks that the target mailbox has the given interface. Rule \textsc{TS-New}
synthesises a mailbox type with the user-supplied interface. Finally, we modify
the shape of the guard typing judgement to record the interface of the mailbox
being guarded upon, and use this to look up the desired payload types in
\textsc{TCG-Recv}.

\paragraph*{Example.}

To illustrate this approach, consider our earlier troublesome example:
\[
\letin{f}{(\unlambda{x : \tyint}{\tyint}{x})}{f(5)}
\]
This example cannot be typed purely co-contextually since we do not have the
required type information for $f$ when typing the function application.

Let $\contyenv = f : \tyunrfun{\tyint}{\tyint}$.
The type-directed elaboration phase results in the following derivation:

{\footnotesize
\begin{mathpar}
    \inferrule*
    {
        \inferrule*
        {
            \inferrule*
            {
                \inferrule*
                { }
                {
                    \elabsynth
                        { x{:}~\tyint }
                        { x }
                        { \tyint }
                        { \varann{x}{\tyint} }
                }
            }
            { \elabchk
                { x{:}~\tyint }
                { x }
                { \tyint }
                { \varann{x}{\tyint} }
            }
        }
        {
            \elabsynth
                {\cdot}
                {\unlambda{x{:}~\tyint}{\tyint}{x}}
                { \tyunrfun{\tyint}{\tyint} }
                {\unlambda{x{:}~\tyint}{\tyint}{\varann{x}{\tyint}}}
        }
        ~~
        \inferrule*
        {
            \inferrule*
            { }
            {
                \elabsynth
                { \contyenv }
                { f }
                { \tyunrfun{\tyint}{\tyint} }
                { \varann{f}{\tyunrfun{\tyint}{\tyint}}}
            }
            ~~
            \inferrule*
            { }
            { \elabchk{ \contyenv }{5}{\tyint}{5} }
        }
        {
            \elabsynth
            { \contyenv }
                { \fnapp{f}{5} }
                { \tyint }
                { \fnapp{\varann{f}{\tyunrfun{\tyint}{\tyint}}}{5} }
        }
    }
    { \elabsynth
        {\cdot}
        {\letin{f}{(\unlambda{x{:}~\tyint}{\tyint}{x})}{f(5)}}
        { \tyint }
        {\letin{f}{(\unlambda{x{:}~\tyint}{\tyint}{\varann{x}{\tyint}})}{\varann{f}{\tyunrfun{\tyint}{\tyint}}(5)}}
    }
\end{mathpar}
}

Finally, we can type the expression with the modified constraint generation
rules. We omit the straightforward environment
operations in the premises of the rules for simplicity.
Let $\derivd$ be the following subderivation:

{\footnotesize
    \begin{mathpar}
        \inferrule*
        {
            \inferrule*
            {
                \inferrule*
                { }
                {
                    \synthseq
                    { \varann{f}{\tyunrfun{\tyint}{\tyint}} }
                    { \tyunrfun{\tyint}{\tyint} }
                    { f : \tyunrfun{\tyint}{\tyint} }
                    { \emptyset }
                }
                \\
                \inferrule*
                {
                    \inferrule*
                    { }
                    { \synthseq
                        { 5 }
                        { \tyint }
                        { \cdot }
                        { \emptyset }
                    }
                }
                {
                    \chkseq
                    { 5 }
                    { \tyint }
                    { \cdot }
                    { \emptyset }
                }
}
            {
                \synthseq
                { \fnapp{\varann{f}{\tyunrfun{\tyint}{\tyint}}}{5} }
                { \tyint }
                { f : \tyunrfun{\tyint}{\tyint} }
                { \emptyset }
            }
        }
        {
            \chkseq
            { \fnapp{\varann{f}{\tyunrfun{\tyint}{\tyint}}}{5} }
            { \tyint }
            { f : \tyunrfun{\tyint}{\tyint} }
            { \emptyset }
        }
    \end{mathpar}
}

Then, we can construct the whole derivation using \textsc{TC-LetNoAnn1}:

{\footnotesize
    \begin{mathpar}
    \inferrule*
    {
        \derivd
        \\
        \inferrule*
        {
            \inferrule*
            {
                \inferrule*
                {
                }
                {
                \chkseq
                    {\varann{x}{\tyint}}
                    { \tyint }
                    {x : \tyint}
                    {\emptyset}
                }
            }
            {
                \synthseq
                { \unlambda{x : \tyint}{\tyint}{\varann{x}{\tyint}} }
                { \tyunrfun{\tyint}{\tyint}}
                { \cdot }
                { \emptyset }
            }
        }
        {
            \chkseq
            { \unlambda{x : \tyint}{\tyint}{\varann{x}{\tyint}} }
            { \tyunrfun{\tyint}{\tyint}}
            { \cdot }
            { \emptyset }
        }
    }
    { \chkseq
        { \letin{f}{(\unlambda{x : \tyint}{\tyint}{\varann{x}{\tyint}})}{\varann{f}{\tyunrfun{\tyint}{\tyint}}(5)} }
        { \tyint }
        { \cdot }
        { \emptyset }
    }
    \end{mathpar}
}

\section{Implementation and evaluation}\label{sec:implementation}

We outline the implementation of a mailbox type checker written in OCaml,
and evidence the expressiveness of \langname\ via a selection of representative example
programs taken from the literature.
We first show that using quasi-linear typing in place  of
dependency graphs (cf.~\secref{sec:quasi-linear-typing}) does not prevent
\langname\ from expressing \emph{all} of the examples in~\citet{dP18:mbc}.
The Savina benchmarks~\citep{ImamS14a:savina} capture typical
concurrent communication patterns and are used both to compare actor
languages and to demonstrate expressiveness; we show that \langname\ (once extended with sums, products, and lists) can express
all of the 11 Savina expressiveness benchmarks used by~\citet{NeykovaY16}.
This selection confirms that our language can express real-world
scenarios that arise in concurrent and distributed computing. The full
Savina benchmark suite contains 30 programs; we base our choice of Savina
programs on the selection implemented by~\citet{NeykovaY16} in order to
demonstrate that mailbox types are as expressive as multiparty session types
for actor systems, at least within the context of this set of examples.

Finally, we \ADDED{describe the \emph{Sleeping Barber} example in detail}, and
show a case study provided by an industrial partner that develops control
software for factories.

\begin{figure}[t]
    \centering
    \includegraphics[width=\textwidth]{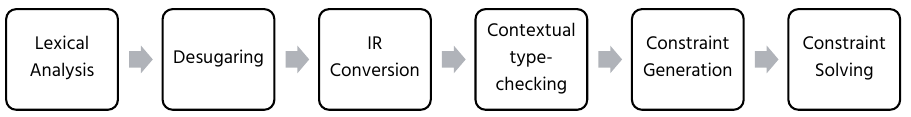}
    \caption{\langname\ typechecking pipeline}
    \label{fig:implementation-pipeline}
\end{figure}

\subsection{Implementation overview}\label{sec:implementation_overview}

\noindent
\langname\ programs consist of \emph{interface definitions} that detail the
messages and payload types supported by a mailbox; a series of \emph{function
definitions}; and finally a \emph{program body} that serves as an entry point.
Programs are typechecked in the six-stage pipeline outlined in
\Cref{fig:implementation-pipeline}:

\begin{description}
     \item[Lexical analysis] Performs standard lexing and parsing using
         the OCaml Menhir library.
     \item[Desugaring]
         \sloppypar
         Desugars $\sugarfree{\vala}$ to
         $\guardann{\vala}{\mbone}{\free{()}}$ and adds omitted pattern variables.
     \item[IR conversion]
         Transforms the surface language (supporting nested expressions) to our
         explicitly sequenced intermediate representation.
     \item[Contextual typechecking]
         Performs a (standard) bidirectional typing
         pass to propagate contextual type information
         \secrefp{sec:extensions:typechecking-contextual}.
     \item[Constraint generation]
         Implements the algorithmic type system
         from~\secref{sec:algorithmic}, and generates a set of pattern inclusion
         constraints.
     \item[Constraint solving]
         Applies the constraint solving approach detailed
         in~\secref{sec:constraint-solving}, and invokes Z3~\citep{MouraB08} to
         determine whether the generated constraints are
         satisfiable.
 \end{description}

The \langname\ typechecker operates in two \emph{modes} that determine how receive guards are typechecked.
\emph{Strict} mode uses the lightweight syntactic checks outlined in
\secref{sec:declarative} and \secref{sec:algorithmic}, whereas  \emph{interface}
mode uses interface type information
(\secref{sec:extensions:typechecking-contextual}) to relax these checks. This
means that every \langname\ program accepted in strict mode is also
accepted in interface mode.

\subsection{Expressiveness and typechecking time}\label{sec:implementation:examples}

\begin{table}[t]
    \caption{Typechecking concurrent actor examples in \langname}
    \label{tbl:examples-table}
    {\footnotesize
    \setlength\tabcolsep{4pt}
    \begin{tabular}{cc>{\raggedright\arraybackslash}p{6.6cm}@{}ccc}
      \toprule
      \textbf{\commentpound} & \textbf{Name} & \textbf{Description} &
      \textbf{Strict} & \textbf{SLOC} & \textbf{Time} (ms) \\
      \midrule
      \multicolumn{6}{l}{\itshape Original mailbox calculus models taken from~\citet{dP18:mbc}}\\
      \midrule
      1\label[id]{id:lock} & Lock & Concurrent lock modelling mutual exclusion & \yes\ & $35$ & $29.4$\\
      2\label[id]{id:future} & Future & Future variable written once and read multiple times & \yes\ & $32$ & $25.0$\\
      3\label[id]{id:account} & Account & Concurrent accounts exchanging debits and credits & \yes\ & $38$ & $20.0$\\
      4\label[id]{id:accountf} & AccountF & Concurrent accounts with debits effected via futures & \yes\ & $65$ & $37.1$\\
      5\label[id]{id:master_worker} & Master-Worker & Master-worker parallel network & \yes\ & $70$ & $29.8$\\
      6\label[id]{id:session_types} & Session Types & Session-typed communicating actors using one arbiter & \no\ & $96$ & $85.3$\\
      \midrule
      \multicolumn{6}{l}{\itshape Selected micro-benchmarks adapted from~\citet{ImamS14a:savina}, based on~\citet{NeykovaY16}}\\
      \midrule
      7 & Ping Pong & Process pair exchanging $k$ ping and pong messages & \yes\ & $47$ & $26.9$\\
      8 & Thread Ring & Ring network relaying one token with counter $k$ & \no\ & $76$ & $40.6$\\
      9 & Counter& One actor sending messages to a second that sums the count, $k$ & \no\ & $56$ & $30.0$\\
      10 & K-Fork & Fork-join where a central actor delegates $k$ requests & \yes\ & $41$ & $8.0$\\
      11 & Fibonacci & Server delegating terms $k-1$ and $k-2$ to parallel actors & \yes\ & $43$ & $27.1$\\
      12 & Big & Peer-to-peer actors randomly exchanging $k$ messages & \no\ & $108$ & $64.8$\\
      13 & Philosopher & Dining philosophers problem & \no\ & $90$ & $73.8$\\
      14 & Smokers & Central arbiter allocating $k$ messages to actors & \no\ & $82$ & $40.7$\\
      15 & Log Map & Computes $x_{k + 1} = r{\cdot}x_k(1 - x_k)$ using parallel actors & \no\ & $103$ & $63.2$\\
      16 & Transaction & Request-reply communication initiated by a teller actor & \no\ & $96$ & $51.2$\\
      17 & Barber & Multiple customers who interact with one `sleeping' barber & \no\ & $100$ & $77.5$\\
\bottomrule
    \end{tabular}
    }
    \smallskip
    \vspace{-1.5em}
  \end{table}

\Cref{tbl:examples-table} lists the
examples
implemented in \langname. Examples 1--6 are the mailbox calculus
examples from \citet[Ex.~1--3, and Sec.~4.1--4.3]{dP18:mbc}.
Examples 7--17 are the suite of Savina benchmarks~\citep[Table 1,
No.~1--4, 6, 7, 12--16]{ImamS14a:savina} used by~\citet{NeykovaY16}.
The table indicates whether a \langname\ program can be checked in strict mode,
and therefore also interface mode (denoted by \yes),
or whether it requires interface mode to typecheck (denoted by \no).
As a simple measure of their size, we report the number of Source Lines of
Code (SLOC) for each example.
We report the mean typechecking time, excluding phases 1--3 of the pipeline.
Measurements are made on a MacBook M1 Pro with 8GB of memory, running macOS 15.4
and OCaml 5.2. To minimise variability we report the mean time from 1000 repetitions.
The number of repetitions  was determined empirically by calculating the
coefficient of variation (CV)~\citep{Devore12}, \ie, the ratio of the standard
deviation to the mean, $\textrm{CV}=\sigma/\bar{x}$, for different repetitions
until an adequately low value ($<\!10\%$) was obtained.

\subsubsection{Benchmarks}

\Cref{tbl:examples-table} shows that all but
one of the mailbox calculus examples from~\citet{dP18:mbc}
can be checked in strict mode.
The Savina examples
capture typical concurrent programming patterns, namely, master-worker (K-Fork,
Fibonacci, Log Map), client-server (Ping Pong, Counter), and peer-to-peer (Big),
and common network topologies such as star (Philosopher, Smokers, Transaction)
and ring (Thread Ring).
Most of these programs (8, 9, and 12--17) require contextual type information to
typecheck.

The new list extension~\secrefp{sec:extensions:lists} allows us to
implement many examples more idiomatically: specifically
we can express examples that use fixed collections
(examples 8, 10, and 12--16) as lists. The original
implementations
in~\citet{DBLP:journals/pacmpl/FowlerASGT23} emulated the collections
using parameters.
The examples reveal the benefits of mailbox typing.
Runtime checks, such as
manual error handling
(\secref{sec:mailbox_types}), are unnecessary since errors (\eg, unexpected
messages) are \emph{statically} ruled out
by the type system.
Mailbox types also have an edge over session typing tools for actor systems,
\eg,~\citep{NeykovaY16,TaboneF22}, where developers typically specify protocols
in external tools and write code to accommodate the session typing framework.
In contrast, mailbox typing does not require actor programs to be rewritten to
accommodate the type system.

Despite supporting a permissive many-sender communication model, checking
mailbox types does not incur high typechecking runtimes (see
\Cref{tbl:examples-table}).
The aim of benchmarking typechecking time is to show that mailbox typechecking is not
prohibitively expensive, rather than
to claim comparative results.
Comparisons with other implementations of (non-mailbox-typed versions of) the
benchmarks written in other languages are unlikely to strengthen our results as
the benchmark source code would be different, and we would be measuring,
\eg, Java's entire type system implementation rather than the essence of the
typechecking algorithm.

Nevertheless, typechecking times for all of these benchmarks are under 100ms. The benchmark that takes the longest to
typecheck is the \emph{Session Types} benchmark, which has 12 different messages
that can be exchanged along with 12 different guard expressions. The benchmark
that has the smallest typechecking time is \emph{K-Fork}, which has only one type
of message and a single guard expression, and therefore a much simpler
communication structure. This would suggest that, as would be expected,
constraint solving takes the most time during typechecking.

\subsubsection{Worked example: Sleeping Barber}

To illustrate how \langname is used in practice, this section describes the
development of the \emph{Sleeping Barber} benchmark in greater detail.
The Sleeping Barber problem is a classic synchronisation problem,
originally specified by~\citet{Dijkstra2002}, and one of the Savina
benchmarks:

\begin{itemize}
    \item A barber is working in a barber shop with a waiting room.
    \item When the barber is ready for the next customer, they check the
        waiting room.
    \begin{itemize}
        \item If the waiting room has waiting customers, then the barber 
            calls the next customer in for their haircut. Once the barber is
            finished cutting the customer's hair, they  check the waiting
            room again.
        \item If the waiting room is empty, then the barber will sleep.
    \end{itemize}

    \item When a customer enters the barber shop, they check to see if the
        waiting room has space. If there is space, then the customer will wait;
if there is no space, then the customer will leave.
If the barber is asleep, then the customer will need to wake the barber.
\end{itemize}

\paragraph*{\langname syntax.}
In this section we use \langname syntax as typechecked by the implementation.
In contrast to the formal calculi, implemented \langname programs are typeset in
\texttt{teletype}. The syntax expected by the typechecker closely resembles the
core calculus, with some changes:

\begin{itemize}
    \item An interface-annotated mailbox type is written with the interface preceding
        the capability, and the pattern after the capability.
        For example, $\tyrecviface[I]{\msgtag{Msg}}$ is written in \langname source as
        \lstinline+I?Msg+.
\item Since \langname's typechecker allows a degree of pattern
        inference, it is also possible to omit the message pattern in type
        annotations: for example, \lstinline+I?+ is shorthand for
        $\tyrecviface[I]{\patvara}$, where $\patvara$ is a fresh pattern
        variable.
\item Patterns are written using ASCII regular expression syntax. Specifically,
        composition $\patconcattwo{\msgtag{M}}{\msgtag{N}}$ is written as
        \lstinline+M.N+;
        choice $\patplustwo{\msgtag{M}}{\msgtag{N}}$ is written as
        \lstinline|M + N|; and replication $\many{\msgtag{M}}$ is written as
        \lstinline+M*+. The unreliable ($\mbzero$) and empty ($\mbone$) mailbox
        patterns are written as \lstinline+0+ and \lstinline+1+ respectively.
\end{itemize}

\paragraph*{Interfaces.}

The Savina benchmarks implement an (untyped) actor-based version of the problem
by modelling the barber, customers, and waiting room as individual actors.
We
take a similar approach in \langname, and begin by defining interfaces for the
three types of mailbox.
We also need a fourth interface, \lstinline+WaitingRoomBuffer+, to model
customers waiting in the waiting room.

The \lstinline+WaitingRoom+ will receive three types of message:

\begin{lstlisting}
interface WaitingRoom { Enter(Customer!), Next(Barber!), Sleeping(Barber!) }
\end{lstlisting}

The \lstinline+Enter+ message is received from a customer when they enter the
barber shop, and contains an output reference to the customer's mailbox.
The \lstinline+Next+ message is received from the barber to signify
that they are ready to service the next customer, and the \lstinline+Sleeping+
message is received from the barber to signify that the barber has gone to
sleep. Both of the latter messages contain an updated output reference to the
barber's mailbox.

The \lstinline+WaitingRoomBuffer+ mailbox models the customers who are currently
in the waiting room. A mailbox with interface \lstinline+WaitingRoomBuffer+ can
receive a single kind of message, \lstinline+WaitingCustomer+, that contains a reference
to a \lstinline+Customer+ mailbox.

\begin{lstlisting}
interface WaitingRoomBuffer { WaitingCustomer(Customer!) }
\end{lstlisting}

The \lstinline+Barber+ will also receive three types of message:

\begin{lstlisting}
interface Barber { Wake(WaitingRoom!), CustomerReady(Customer!, WaitingRoom!),
    RoomEmpty(WaitingRoom!) }
\end{lstlisting}
\ADDED{
The \lstinline+Wake+ message is received from an empty waiting room when a
customer has entered and the barber is asleep. The \lstinline+CustomerReady+ and
\lstinline+RoomEmpty+ messages are received from the waiting room in response to
a \lstinline+Next+ message sent by the barber; the former notifies the barber
of the next customer, and the latter notifies the barber that the room is empty
and that they can go to sleep. The \lstinline+CustomerReady+ message includes a
reference that the barber can use to communicate with the customer, and both
messages include an updated \lstinline+WaitingRoom+ reference.

Finally, the \lstinline+Customer+ can receive four types of message:
}
\begin{lstlisting}
interface Customer { Full(), Wait(), Start(), Done() }
\end{lstlisting}
\ADDED{
After entering the waiting room, the customer will either receive a
\lstinline+Full+ or \lstinline+Wait+ message from the waiting room to state
that the customer should leave or wait in the waiting room respectively. When
the barber is ready to cut the customer's hair, the barber will send the
customer a \lstinline+Start+ message, and when the barber is finished with the
haircut, the barber will send a \lstinline+Done+ message.

\paragraph*{Customer.}
The implementation of the customer is fairly straightforward. A customer must
send an \lstinline+Enter+ message to the waiting room, and then wait for a
response:
}

\begin{lstlisting}
def customer(self: Customer?, waitingRoom: WaitingRoom!): Unit {
    waitingRoom ! Enter(self);
    guard self : Full + (Wait.Start.Done) {
        receive Full() from self ->
            print("Room is full. Oh well, best go somewhere else");
            free(self)
        receive Wait() from self ->
            print("Waiting");
            waitingCustomer(self)
    }
}
\end{lstlisting}

\ADDED{
There are two possible responses from the waiting room: either a
\lstinline+Full+ message to say that the waiting room is full (at which point
there are no possible interactions and the only thing to do is to free the
mailbox), or a \lstinline+Wait+ message. If the
latter, the mailbox types \emph{also} guarantee that the customer will need to
handle a \lstinline+Start+ message (when the barber begins their haircut), and a
\lstinline+Done+ message (when the barber has finished their haircut).

The \lstinline+waitingCustomer+ function processes the \lstinline+Start+ and
\lstinline+Done+ messages and frees the customer's mailbox after the haircut is
finished:
}

\begin{lstlisting}
def waitingCustomer(self: Customer?): Unit {
    guard self : Start.Done {
        receive Start() from self ->
            print("Barber is starting my haircut");
            guard self : Done {
                receive Done() from self ->
                    print("Haircut finished!");
                    free(self)
            }
    }
}
\end{lstlisting}

\ADDED{
\paragraph*{Waiting room.}

To obtain precise mailbox types, we model the waiting room as two
mutually recursive functions: one for when the barber is asleep, and one for
when there are waiting customers.
}

\begin{lstlisting}
def waitingRoomSleepingBarber(self: WaitingRoom?, barber: Barber!,
        capacity: Int): Unit {
    guard self : Enter* {
        free -> ()
        receive Enter(customer) from self ->
            let buffer = new[WaitingRoomBuffer] in
            buffer ! WaitingCustomer(customer);
            customer ! Wait();
            barber ! Wake(self);
            waitingRoom(self, buffer, 1, capacity)
    }
}
\end{lstlisting}

\ADDED{
The \lstinline+waitingRoomSleepingBarber+ function takes an input mailbox
reference to its own mailbox, an output mailbox reference to the barber, and an
integer denoting the capacity of the waiting room. In this state, the barber is
asleep and cannot send any more messages until they are awake, and so the
waiting room can only receive \lstinline+Enter+ messages (or free itself if no
more \lstinline+Enter+ messages can be sent).
When the waiting room receives an \lstinline+Enter+ message, it creates a new
mailbox \lstinline+buffer+ that is used to model the queue of customers waiting
for the barber; stores the request in the queue by sending the buffer a 
\lstinline+WaitingCustomer+ message containing an output reference to the
customer's mailbox; sends the customer a \lstinline+Wait+ message; sends the
barber a \lstinline+Wake+ message to wake them up; and then transitions to the
non-empty state.

In the non-empty state, the waiting room's mailbox can also receive many
\lstinline+Enter+ messages from customers, but must additionally receive a
single \lstinline+Next+ message from the barber:
}

\begin{lstlisting}
def waitingRoom(self: WaitingRoom?, buffer: WaitingRoomBuffer?,
                numCustomers: Int, capacity: Int): Unit {
    guard self : Enter* . Next {  
        receive Enter(customer) from self -> ...
        receive Next(barber) from self -> ...
    }
}
\end{lstlisting}

\ADDED{
Processing an \lstinline+Enter+ message is similar to before: if the number of
customers exceeds the capacity of the waiting room, then the waiting room will
respond with a \lstinline+Full+ message; otherwise, the waiting room will queue
the request and send the customer a \lstinline+Wait+ message:
}

\begin{lstlisting}
receive Enter(customer) from self ->
    if (numCustomers > capacity) {
        customer ! Full();
        waitingRoom(self, buffer, numCustomers, capacity)
    } else {
        buffer ! WaitingCustomer(customer);
        customer ! Wait();
        waitingRoom(self, buffer, numCustomers + 1, capacity)
    }
\end{lstlisting}

\ADDED{
To process a \lstinline+Next+ message, the waiting room will inspect the
\lstinline+buffer+, which may contain zero or more \lstinline+WaitingCustomer+
messages:
}

\begin{lstlisting}
receive Next(barber) from self ->
    guard buffer : WaitingCustomer* {
        receive WaitingCustomer(customer) from buffer ->
            barber ! CustomerReady(customer, self);
            waitingRoom(self, buffer, numCustomers - 1, capacity)
        free ->
            # In this case there are no more customers.
            # Notify the barber that they can sleep; move to sleeping state
            barber ! RoomEmpty(self);
            guard self : Enter* . Sleeping {
                # Receive notification that barber is asleep
                receive Sleeping(barber) from self ->
                    waitingRoomSleepingBarber(self, barber, capacity)
            }
    }
\end{lstlisting}

\ADDED{
When processing a \lstinline+WaitingCustomer+ message, the waiting
room responds to the barber with a \lstinline+CustomerReady+ message containing
the output reference with which to communicate with the customer, and
recursively invokes the \lstinline+waitingRoom+ function with a decremented
customer count.
If the \lstinline+free+ guard is triggered, we know
that there are no pending \lstinline+WaitingCustomer+ messages and therefore
that the waiting room is empty. So the function notifies the barber
that the room is empty, awaits a \lstinline+Sleeping+ notification, and
transitions to the \lstinline+waitingRoomSleepingBarber+ state.
}

\paragraph*{Remark.}
\ADDED{
    We use a mailbox, rather than a list, for storing the contents of the
    waiting room to avoid the potential for unsafe aliasing. 
    Consider a signature of the \lstinline+waitingRoom+ function that
    maintains the waiting room as a list:
}

\begin{lstlisting}
def waitingRoom(self: WaitingRoom?, buffer: List(Customer!),
                numCustomers: Int, capacity: Int): Unit
\end{lstlisting}

\ADDED{
    Here we would also need to modify the \lstinline+receive+ clause for
    the \lstinline+Enter+ message to add the customer to the waiting room:
}

\begin{lstlisting}
receive Enter(customer) from self ->
    if (numCustomers > capacity) {
        customer ! Full();
        waitingRoom(self, buffer, numCustomers, capacity)
    } else {
        customer ! Wait();
        waitingRoom(self, customer :: buffer, numCustomers + 1, capacity)
    }
\end{lstlisting}

\ADDED{
However there is no guarantee that a reference for the
customer does not already exist in the buffer, and therefore deconstructing the
list could introduce unsafe aliasing.
Moreover we can only safely store \emph{returnable} values in a
buffer, whereas the \lstinline+customer+ reference must be treated as
second-class.  Both issues are avoided by using a mailbox to model the
waiting room, since \langname's type system allows us to safely reason about one
customer at a time.

It is also possible (if slightly less elegant) to model the Sleeping Barber
problem without a separate buffer mailbox by using self-messages.
}

\ADDED{
\paragraph*{Barber.}
Finally, the barber process begins in the
\lstinline+sleepingBarber+ state, where they have only a reference to a
\lstinline+Barber+ mailbox. The \lstinline+self+ mailbox has pattern
\lstinline|Wake + 1| indicating that the barber is able to respond to
a \lstinline+Wake+ message (when a customer enters the waiting room), or
to free itself if no customers are ever spawned. When the barber
receives a \lstinline+Wake+ message, the process sends a \lstinline+Next+
message to the waiting room provided in the
message. Once awoken, the barber calls the \lstinline+barber+ function that
models an awake barber.
}

\begin{lstlisting}
def sleepingBarber(self: Barber?): Unit {
    guard self : Wake + 1 {
        free -> ()
        receive Wake(room) from self ->
            room ! Next(self);
            barber(self)
    }
}
\end{lstlisting}

\ADDED{
The \lstinline+barber+ function again takes a reference to the barber's mailbox.
However this time we know that the waiting room must respond with either a
\lstinline+RoomEmpty+ message if there are no waiting customers, at which point
the barber can go back to sleep, or a \lstinline+CustomerReady+ message if
another customer was waiting.
}

\begin{lstlisting}
def barber(self: Barber?): Unit {
    guard self : RoomEmpty + CustomerReady {
        receive RoomEmpty(room) from self ->
            print("Room empty; going to sleep");
            room ! Sleeping(self);
            sleepingBarber(self)
        receive CustomerReady(customer, room) from self ->
            customer ! Start();
            print("Cutting hair");
            print("Finished cutting hair; notifying customer and waiting room");
            customer ! Done();
            room ! Next(self);
            barber(self)
    }
}
\end{lstlisting}

\ADDED{
If the room is empty, the barber will send a \lstinline+Sleeping+
message to the waiting room before calling the \lstinline+sleepingBarber+
function, which models the barber falling asleep. When the barber
receives a \lstinline+CustomerReady+ message, which contains a reference to the
customer and the room, the barber will send a \lstinline+Start+ message to the
customer at the start of their haircut and a \lstinline+Done+ message at the end
of their haircut, before notifying the waiting room by sending a
\lstinline+Next+ message.
}

\ADDED{
\paragraph*{Discussion.}

Perhaps surprisingly, many of the messages exchanged between participants
contain references to the sender, even though a reference may already be in
scope. For example, a mailbox with the \lstinline+Barber+ interface can receive
\lstinline+Wake+, \lstinline+CustomerReady+, and \lstinline+RoomEmpty+ messages
from the \lstinline+WaitingRoom+, and all contain a \lstinline+WaitingRoom+
reference with which to respond.

As there is only one waiting room it is common practice in some  actor languages, like Erlang, to spawn the barber process with a reference to it, simplifying the \lstinline+Barber+ interface:

}
\begin{lstlisting}
interface Barber { Wake(), CustomerReady(Customer!), RoomEmpty() }
def barber(self: Barber?, room: WaitingRoom!): Unit { ... }
def sleepingBarber(self: Barber?, room: WaitingRoom!): Unit { ... }
\end{lstlisting}
The drawback of this approach is a loss of typing
  precision. With the previous approach we know statically that the
  barber will \emph{only have a reference to the waiting room when
  processing a request from the room} and this allows us to specify
  two precise mailbox types for the two states of the waiting room.
Specifically, when the barber is awake, the 
\lstinline+self+ mailbox in the
\lstinline+waitingRoom+ function has type
\lstinline+WaitingRoom?(Enter* . Next)+,
meaning that its mailbox can contain many \lstinline+Enter+ messages from
customers, but will also eventually contain a \emph{single} \lstinline+Next+
message from the barber when they are ready for the next customer.
When the barber is asleep, the barber will \emph{not} have a reference to the
waiting room, and thus we know statically that the waiting room \emph{cannot}
contain a \lstinline+Next+ message; we can therefore give the \lstinline+self+
mailbox the type \lstinline+WaitingRoom?Enter*+ in the 
\lstinline+waitingRoomSleepingBarber+ function.

If we were instead to keep the \lstinline+room+ variable in scope across both
functions, we would need to use the less precise mailbox type
\lstinline+WaitingRoom?(Enter* . Next* . Sleeping*)+ in both states of the waiting room, since
there are no guarantees about whether the barber is asleep or awake, and the
mailbox type no longer specifies that the barber will only ever send a single
\lstinline+Next+ or \lstinline+Sleeping+ message at a time.

The idiom of sending a mailbox reference even when it could be in
scope may seem contrary to common practice in untyped languages like
Erlang. It does, however, closely mirror the style employed by Akka's
typed references~\cite{akka-ref}, where messages often include an
actor reference that allows a process to respond with a message of a
different type. We posit that the practice of re-sending actor
references with a different type is therefore more of a typed actor
idiom than an idiom specific to mailbox typing.

\subsubsection{Factory case study}\label{sec:case_study_robots}

Finally we describe a real-world use case provided by \emph{Actyx
AG}\footnote{\url{https://www.actyx.com}}, who develop control software for factories.  The use case
captures a scenario where multiple robots on a factory floor acquire parts from
a warehouse that provides access through a single door.
Robots negotiate with the door to gain entry into the warehouse and obtain the part they require.
The behaviour of the three entities, \emph{Robot}, \emph{Door}, and
\emph{Warehouse}, is shown in~\Cref{fig:factory-seqdiag}.
Typechecking our case study relies on contextual type information
(see \secref{sec:extensions:typechecking-contextual}), and takes 89.6ms.

\begin{figure}[t]
\vspace{-1em}
    \centering
    \includegraphics[scale=0.46, trim={0 1.2cm 0 0},clip]{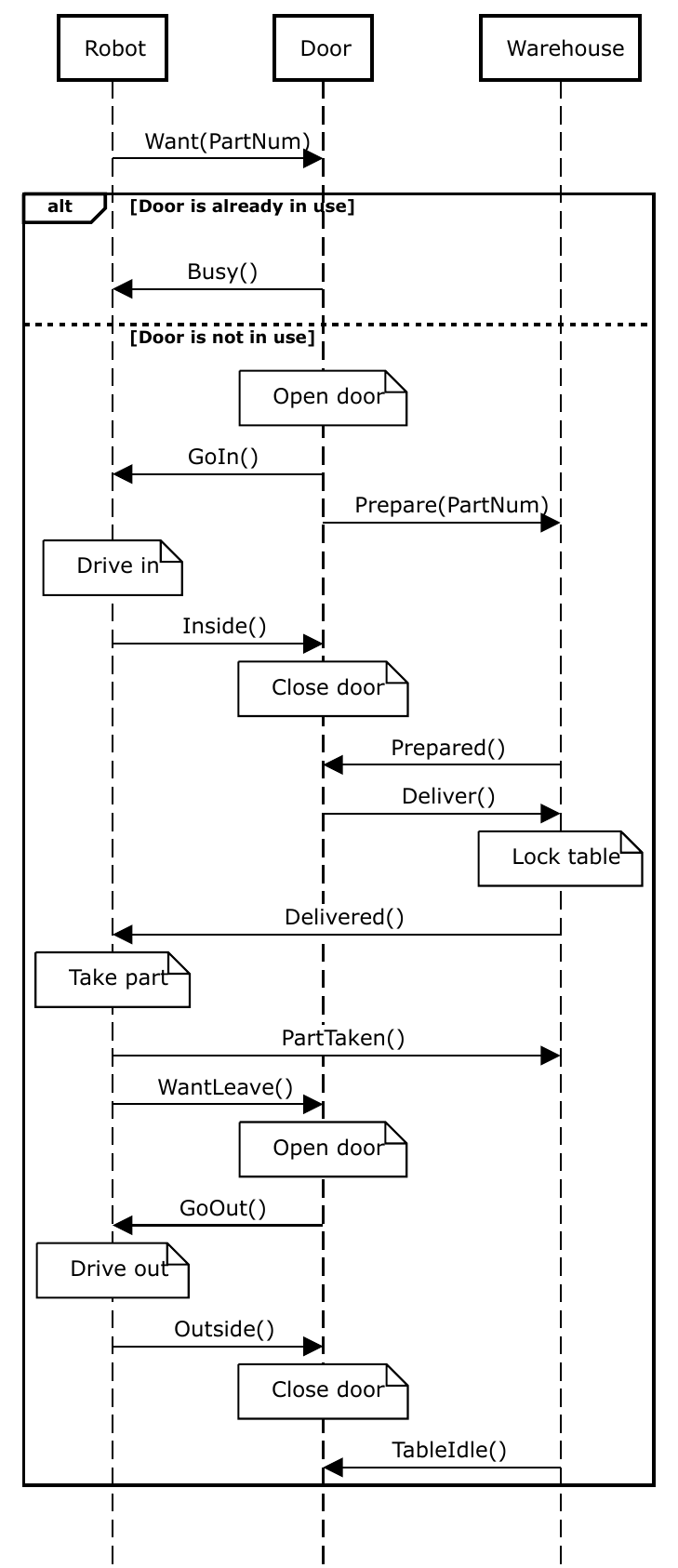}
    \caption{Factory use case}
    \label{fig:factory-seqdiag}
\end{figure}

\paragraph*{Interfaces.}
The messages accepted by the \lstinline+Robot+, \lstinline+Door+, and
\lstinline+Warehouse+ are defined by the following interfaces.

\noindent
  \begin{lstlisting}[language=Pat, emph={self}, firstnumber=1]
interface Robot { (*\label[ln]{ln:begin_interfaces}*)
  GoIn(Door!), GoOut(Door!), Busy(), Delivered(Warehouse!, Door!)
}

interface Door {
  Want(Int, Robot!), Inside(Robot!), Outside(),
  WantLeave(Robot!), Prepared(Warehouse!),
  TableIdle(Warehouse!)
}

interface Warehouse {
  Prepare(Int, Door!), Deliver(Robot!, Door!), PartTaken()
} (*\label[ln]{ln:end_interfaces}*)
\end{lstlisting}

\paragraph*{Robot.}
Each \lstinline+Robot+ is initially \lstinline+idle+ and issues a
\lstinline+Want+ message to the \lstinline+Door+ to obtain access to the
\lstinline+Warehouse+ (line~\ref{ln:door_want}).
The \lstinline+Door+ replies either with the message \lstinline+Busy+, in which case the
\lstinline+Robot+ terminates (lines
\ref{ln:robot_terminate_start}\crefrangeconjunction\ref{ln:robot_terminate_end}),
or \lstinline+GoIn+, to which the \lstinline+Robot+ replies with an
\lstinline+Inside+ message before transitioning to the \lstinline+working+ state
(lines
\ref{ln:robot_working_start}\crefrangeconjunction\ref{ln:robot_working_end}).
When in the \lstinline+working+ state, the \lstinline+Robot+ expects one \lstinline+Delivered+
message to inform the \lstinline+Robot+ that the part is delivered by the
\lstinline+Warehouse+, as asserted by the guard on
line~\ref{ln:robot_guard_delivered}.
The recipient \lstinline+Robot+ replies to the \lstinline+Warehouse+
with the \lstinline+PartTaken+ message, and notifies the
\lstinline+Door+ that it wants to exit by sending a \lstinline+WantLeave+
message on lines
\ref{ln:robot_delivered_start}\crefrangeconjunction\ref{ln:robot_delivered_end}.
It then awaits a \lstinline+GoOut+ message and finalises its negotiation with the
\lstinline+Door+ through an \lstinline+Outside+ message.

\begin{lrbox}{\listingbox}
  \begin{minipage}[b]{0.55\textwidth}
    \begin{lstlisting}[language=Pat, emph={self}, firstnumber=14]
def idle(self: Robot?, door: Door!): Unit {
  door ! Want(0, self); (*\label[ln]{ln:door_want}*)
  guard self: (Busy + GoIn) {
    receive Busy() from self (*$\shortrightarrow$*) (*\label[ln]{ln:robot_terminate_start}*)
      free(self) (*\label[ln]{ln:robot_terminate_end}*)
    receive GoIn(door) from self (*$\shortrightarrow$*)  (*\label[ln]{ln:robot_working_start}*)
      door ! Inside(self);
      working(self) (*\label[ln]{ln:robot_working_end}*)
  }
}
    \end{lstlisting}
  \end{minipage}
\end{lrbox}
\noindent
\raisebox{0.5em}{\usebox{\listingbox}}
\hfill
\begin{minipage}[b]{.43\textwidth}
  \begin{lstlisting}[language=Pat, emph={self}, firstnumber=last]
def working(self: Robot?): Unit {
  let self = guard self: Delivered { (*\label[ln]{ln:robot_guard_delivered}*)
    receive Delivered(wh, door)
        from self (*$\shortrightarrow$*)
      wh ! PartTaken(); (*\label[ln]{ln:robot_delivered_start}*)
      door ! WantLeave(self); self (*\label[ln]{ln:robot_delivered_end}*)
  } in guard self: GoOut {
    receive GoOut(door) from self (*$\shortrightarrow$*)
      door ! Outside();
      free(self)
  }
}
  \end{lstlisting}
\end{minipage}

\paragraph*{Door.}
The \lstinline+Door+ accepts zero or more \lstinline+Want+ messages, replying to each with
\lstinline+Busy+ or \lstinline+GoIn+.
In the latter case, the \lstinline+Door+ informs the \lstinline+Warehouse+ of an
inbound \lstinline+Robot+ by sending it a \lstinline+Prepare+ message, and
then transitions to the \lstinline+busy+ state (lines
\ref{ln:door_go_in_start}\crefrangeconjunction\ref{ln:door_go_in_end}).
Both \lstinline+GoIn+ and \lstinline+Prepare+ include an updated self-reference
to ensure precise types.
The \lstinline+free+ guard on line~\ref{ln:door_guard_free} handles the case
where no \lstinline+Robot+s are present, \ie, no \lstinline+Want+ messages are
received by the \lstinline+Door+.

\begin{lstlisting}[language=Pat, emph={self}, firstnumber=36]
def clear(self: Door?, wh: Warehouse!): Unit {
  guard self: Want* {
    free (*$\shortrightarrow$*) () (*\label[ln]{ln:door_guard_free}*)
    receive Want(part, robot) from self (*$\shortrightarrow$*) (*\label[ln]{ln:door_go_in_start}*)
      robot ! GoIn(self);
      wh ! Prepare(part, self);
      busy(self) (*\label[ln]{ln:door_go_in_end}*)
  }
}
\end{lstlisting}

When \lstinline+busy+, the \lstinline+Door+ mailbox potentially contains an
\lstinline+Inside+ message from the admitted \lstinline+Robot+, a \lstinline+Prepared+
message from the \lstinline+Warehouse+,
and \lstinline+Want+ messages sent by other \lstinline+Robot+s requesting access (line~\ref{ln:door_busy_guard}).
These \lstinline+Want+ messages are answered with \lstinline+Busy+, as lines \ref{ln:door_answer_busy_start}\crefrangeconjunction\ref{ln:door_answer_busy_end} show.
Once the \lstinline+Door+ receives the \lstinline+Inside+ message, it awaits a
\lstinline+Prepared+ message issued by the \lstinline+Warehouse+, before notifying the
latter that the \lstinline+Robot+ is collecting its part via \lstinline+Deliver+ (lines
\ref{ln:door_inside_deliver_start}\crefrangeconjunction\ref{ln:door_inside_deliver_end}).

\begin{lstlisting}[language=Pat, emph={self}, firstnumber=45]
def busy(self: Door?): Unit {
  guard self: Inside.Prepared.Want* { (*\label[ln]{ln:door_busy_guard}*)
    receive Want(partNum, robot) from self (*$\shortrightarrow$*) (*\label[ln]{ln:door_answer_busy_start}*)
      robot ! Busy();
      busy(self) (*\label[ln]{ln:door_answer_busy_end}*)
    receive Inside(robot) from self (*$\shortrightarrow$*) (*\label[ln]{ln:door_inside_deliver_start}*)
      guard self: Prepared.Want* {
        receive Prepared(wh) from self (*$\shortrightarrow$*)
          wh ! Deliver(robot, self); (*\label[ln]{ln:door_inside_deliver_end}*)
          guard self: WantLeave.TableIdle.Want* {
            receive WantLeave(robot) from self (*$\shortrightarrow$*) (*\label[ln]{ln:door_handle_want_leave_start}*)
              robot ! GoOut(self); (*\label[ln]{ln:door_handle_want_leave_end}*)
              ready(self)
            }
        }
  }
}
\end{lstlisting}

Eventually, the \lstinline+Robot+ requests to exit the \lstinline+Warehouse+
by sending \lstinline+WantLeave+ to the \lstinline+Door+, which handles it on lines~\ref{ln:door_handle_want_leave_start}\crefrangeconjunction\ref{ln:door_handle_want_leave_end}.
The \lstinline+Door+ transitions to the \lstinline+ready+ state, whereupon it
confirms that the \lstinline+Robot+ has exited and that the
\lstinline+Warehouse+ is available; these interactions are captured by the
\lstinline+Outside+ and \lstinline+TableIdle+ messages respectively (lines
\ref{ln:door_outside_tableidle_start}\crefrangeconjunction\ref{ln:door_outside_tableidle_end}).
Finally, the \lstinline+Door+ transitions back to \lstinline+clear+ on
lines~\ref{ln:door_transition_clear_1} and~\ref{ln:door_transition_clear_2},
ready to service other \lstinline+Robot+s.

\begin{lstlisting}[language=Pat, emph={self}, firstnumber=62]
def ready(self: Door?): Unit {
  guard self: Outside.TableIdle.Want* {
    # Handle messages Outside and TableIdle in
    # any order and clear door.
    receive Outside() from self (*$\shortrightarrow$*) (*\label[ln]{ln:door_outside_tableidle_start}*)
      guard self: TableIdle.Want* {
          receive TableIdle(wh) from self (*$\shortrightarrow$*)
            clear(self, wh) (*\label{ln:door_transition_clear_1}*)
      }
    receive TableIdle(wh) from self (*$\shortrightarrow$*)
      guard self: Outside.Want* {
        receive Outside() from self (*$\shortrightarrow$*)
          clear(self, wh) (*\label{ln:door_transition_clear_2}*)
      } (*\label[ln]{ln:door_outside_tableidle_end}*)
  }
}
\end{lstlisting}

\paragraph*{Warehouse.}
The \lstinline+Warehouse+ in its \lstinline+empty+ state expects a
\lstinline+Prepare+ message (if there are \lstinline+Robot+s in the system), or
none (if no \lstinline+Robot+ requests access), \ie, the guard \lstinline:Prepare + 1: on
line~\ref{ln:warehouse_prepare_or_not}.
When a part is requested, the \lstinline+Warehouse+ transitions to the \lstinline+engaged+
state and awaits a \lstinline+Deliver+ message from the \lstinline+Door+, notifying the
\lstinline+Robot+ collecting the part via a \lstinline+Delivered+ message (lines
\ref{ln:warehouse_collect_part_start}\crefrangeconjunction\ref{ln:warehouse_collect_part_end}),
and then transitions to the \lstinline+given+ state (lines
\ref{ln:warehouse_given_part_start}\crefrangeconjunction\ref{ln:warehouse_given_part_end}).
The \lstinline+Robot+ acknowledges the delivery by sending
\lstinline+PartTaken+, as required by the guard on line~\ref{ln:warehouse_part_taken}.
To conclude its interaction with the \lstinline+Door+, the \lstinline+Warehouse+
sends \lstinline+TableIdle+ before transitioning back to the \lstinline+empty+
state.

\begin{lrbox}{\listingbox}
  \begin{minipage}[b]{.47\textwidth}
    \begin{lstlisting}[language=Pat, emph={self}, firstnumber=78]
def empty(self: Warehouse?): Unit {
  guard self: Prepare + 1 { (*\label[ln]{ln:warehouse_prepare_or_not}*)
    free (*$\shortrightarrow$*) ()
    receive Prepare(partNum, door)
        from self (*$\shortrightarrow$*)
      door ! Prepared(self);
      engaged(self)
  }
}

def engaged(self: Warehouse?): Unit { (*\label[ln]{ln:warehouse_collect_part_start}*)
  guard self: Deliver {
    receive Deliver(robot, door)
        from self (*$\shortrightarrow$*)
      robot ! Delivered(self, door);
      given(self, door)
  }
} (*\label[ln]{ln:warehouse_collect_part_end}*)
    \end{lstlisting}
  \end{minipage}
\end{lrbox}

\noindent
\raisebox{0em}{\usebox{\listingbox}}
\hfill
\begin{minipage}[b]{.47\textwidth}
  \begin{lstlisting}[language=Pat, emph={self}, firstnumber=last]
def given(self: Warehouse?, (*\label{ln:warehouse_given_part_start}*) 
          door: Door!): Unit { 
  guard self : PartTaken { (*\label[ln]{ln:warehouse_part_taken}*)
    receive PartTaken() from self (*$\shortrightarrow$*)
      door ! TableIdle(self);
      empty(self)
  }
} (*\label{ln:warehouse_given_part_end}*)

# Launcher function.
def main(): Unit {
  # (*\color{Gray}{$n$}*) Robot mailboxes.
  let robot(*$_i$*) = new[Robot] in
  let door = new[Door] in
  let wh = new[Warehouse] in
  # Door.
  spawn { clear(door, wh) }; (*\label[ln]{ln:main_spawn_start}*)
  # (*\color{Gray}{$n$}*) Robots.
  spawn { idle(robot(*$_i$*), door) };
  # Warehouse.
  spawn { empty(wh) } (*\label[ln]{ln:main_spawn_end}*)
}
  \end{lstlisting}
\end{minipage}

The \lstinline+main()+ function creates $n$ \lstinline+Robot+ mailboxes,
together with a \lstinline+Door+ and \lstinline+Warehouse+ mailbox, spawning the
respective processes on lines
\ref{ln:main_spawn_start}\crefrangeconjunction\ref{ln:main_spawn_end}.
 \section{Related work}
\label{sec:related}

\paragraph*{Behaviourally typed actors.}
The asymmetric nature of mailboxes makes developing
behavioural type systems for actor languages challenging.
\citet{MostrousV11} investigate session typing for Core Erlang, using selective
message reception and unique references to encode session-typed channels.
\citet{TaboneF21,TaboneF22} develop a tool that statically checks
Elixir~\citep{Juric19} actors against binary session types to prove session
fidelity.
\citet{NeykovaY16} propose a programming model for dynamically checking actor
communication against multiparty session types~\citep{HondaYC16},
implemented in Erlang by~\citet{Fowler16}.
\citet{NeykovaY17} show how causality information in global types
can support efficient recovery strategies.
\citet{HFDG21} use multiparty session types with explicit connection
actions~\citep{HuY17} to give strong guarantees about actors that support runtime
adaptation, but an actor can only participate in one session at a time.
\citet{FowlerH26} introduce a language design that allows session-typed actor
communication by enforcing multiparty session typing using a flow-sensitive
effect system, and their language allows actors to be involved in multiple
sessions by using ideas from event-driven programming.

Session types are helpful when there are clear, structured communication flows
between a fixed set of participants. Session types also provide a convenient
top-down development methodology, whereas mailbox types need to be added to
individual components in a more bottom-up fashion.
However, the big disadvantage of using session types over mailbox types is that
session types are specified using point-to-point interactions, and this requires
either designing applications with session types from the beginning, or
rewriting existing applications to use a session-typed communication style.
In contrast, our mailbox typing approach naturally fits actor-style programming
without needing to restructure the code to group participants into sessions.

\citet{BagherzadehR17} define a type system for active objects~\citep{BoerCJ07}
which can rule out message races; this work targets an imperative calculus and
is not validated via an implementation.
\citet{KamburjanDC16} apply session-based reasoning to a core active object
calculus where types encode remote calls and future resolutions;
communication correctness is ensured by static checks against session
automata~\citep{BolligHLM13}.

Mailbox types are inspired by behavioural type systems~\citep{CrafaP17} for the \emph{objective join calculus}~\citep{FournetG96}.
The technique can be implemented in Java using code generation via
\emph{matching automata}~\citep{GerboP19}, and dependency graphs can rule out
deadlocks~\citep{Padovani18DF}, but the authors do not consider a programming
language design.
\citet{ScalasYB19} define a behavioural type system for Scala actors.  Types
are written in a domain-specific language, and type-level model checking
determines safety and liveness properties.
Their system focuses on the behaviour
of a process, rather than the state of the mailbox.

\paragraph*{Session-typed functional languages.}
Session types~\citep{DBLP:conf/concur/Honda93, HondaVK98} were originally
considered in the setting of process calculi. \citet{NeubauerT04} were the first to
design a Haskell embedding of session types, whereas \citet{VasconcelosGR06}
were the first to design a core functional language with session types, though their
approach had an imperative-style interface and required complex alias checking.
\citet{GayV10} integrated session types in a functional language based on the
linear $\lambda$-calculus, and their approach has been adopted by several other
works (\eg,~\citep{Wadler14, LindleyM15, AlmeidaMTV22}).
Linear types are insufficient for mailbox typing since we require
\emph{multiple} uses of a mailbox name as a sender; we believe our
use of quasi-linearity for behavioural typing is novel, and we conjecture that it could
be used to support other paradigms (\eg, publish-subscribe) that require
non-linear variable use.

\paragraph*{Co-contextual typing.}
Co-contextual typing~\citep{ErdwegBKKM15} was originally introduced to support
efficient incremental typechecking, and has also been used to support
intrinsically typed compilation~\citep{RouvoetKV21}.
\citet{Padovani14} uses a co-contextual type algorithm for the linear
$\pi$-calculus with sums, products, and recursive types; and~\citet{CicconeP22}
use it when analysing fair termination properties.
\emph{Backwards bidirectional typing}~\citep{Zeilberger18:bbt} is a co-contextual
formulation of bidirectional typing, and to the best of our knowledge
we are the first to use it in a language implementation.
Co-contextual typing has parallels with the co-de Bruijn nameless variable
representation~\citep{McBride18}, where subterms are annotated with the
variables they contain.

\paragraph*{Safety via static analysis.}
\citet{ChristakisS11} implement a static analyser for Erlang that detects errors
such as receiving from an empty mailbox, payload mismatches, redundant patterns,
and orphan messages.
All of these issues can be detected with mailbox types, which also allow us
to specify the mailbox state.
\citet{Harrison18} implements an approach incorporating both typechecking and
static analysis to detect errors such as orphan messages and redundant patterns.

 \section{Conclusion and future work}\label{sec:conclusion}
Concurrent and distributed applications can harbour subtle and insidious bugs,
including protocol violations and deadlocks.
Behavioural types support \emph{correct-by-construction} communication-centric
software, but are difficult to apply to actor languages.
We have developed the \emph{first} language design incorporating \emph{mailbox
types}, which characterise communication in actor languages and can statically
detect communication errors that would otherwise be difficult to locate and fix.
The many-sender, single-receiver nature of actor communication makes
the integration of mailbox types in programming languages challenging.
We have addressed these challenges through a novel use of quasi-linear types and
have formalised and implemented an algorithmic type system based on backwards
bidirectional typing \secrefp{sec:algorithmic}, proving it to be sound and
complete with respect to the declarative type system \secrefp{sec:declarative}.
\langname can encode common communication patterns (\eg, master-worker), and 
we have shown that it can also express a real-world case study based on factory
automation.

\paragraph*{Ongoing and future work.}
Mailbox typing is a young field and there are many areas that are ripe for
exploration. 

We are currently investigating using mailbox types to verify
communication behaviour in mainstream actor languages such as
Erlang. In \langname a process may have multiple mailboxes and
\emph{explicitly} creates and destroys each mailbox. In contrast,
mainstream actor languages \emph{implicitly} create a single
\emph{monolithic} mailbox that holds messages from all
protocols. Our approach overlays multiple virtual mailboxes on a
monolithic mailbox, and annotates the code to indicate what messages
are expected by $\calcwd{receive}$s and when a mailbox should be created or
reused.

An important area for future work is better inference:
both at the level of mailbox patterns in order to allow developers to elide
annotations on $\calcwd{guard}$ expressions, and at the level of types
to support more interesting type system features (\eg, polymorphism or
set-theoretic typing).
For mailbox types to be adopted in practice, it is also important to consider 
how mailbox types can be adapted to handle failure. 

\ADDED{Another avenue for future work is finer-grained deadlock and alias
detection. Quasi-linearity provides some guarantees, but since we cannot easily
adopt the dependency graph formalism introduced by~\citet{dP18:mbc}, we
cannot guarantee inter-process deadlock-freedom.  Approaches such
as~\emph{priorities}~\cite{Kobayashi06, Padovani14:priorities, KokkeD23} may
prove a useful starting point, but it is not yet clear how to adapt these to the
many-sender, single-receiver communication model.}

This paper has concentrated on the design and implementation of a
\emph{typechecker} for \langname. We are currently working on 
implementing an interpreter for \langname's semantics, which is nontrivial due
to features like the $\calcwd{free}$ guard. Our approach uses 
an adaptation of the Perceus reference counting algorithm~\cite{ReinkingXML21}.

\subsubsection*{Acknowledgements}
    We are deeply grateful to the JFP reviewers for their extremely thorough
    reading of the paper and their insightful suggestions, which we believe have
    significantly improved the final version. We are also grateful to the
    ICFP'23 reviewers and Artifact Evaluation Committee for their helpful
    comments on a previous version of this paper.  Thanks also to our STARDUST
    colleagues for many interesting discussions; to Roland Kuhn for discussion of
    the case study; and to Edgard Schiebelbein for initial work mechanising
    \langname~\cite{SchiebelbeinBF26} that resulted in several significant
    technical improvements.
    This work was supported by EPSRC Grant EP/T014628/1 (STARDUST).

\subsubsection*{Declaration of competing interests}
The authors have no competing interests to declare.

\bibliographystyle{ACM-Reference-Format}
\bibliography{mb-typingJFP}

\clearpage
\appendix
\section{Proofs}
\subsection{Proofs for \S\ref{sec:declarative} (Preservation)}\label{ap:preservation}

\subsubsection{Auxiliary Definitions and Lemmas}

We begin with some further auxiliary definitions and lemmas.

We extend $\returnablepred{-}$ to typing environments, writing
$\returnablepred{\tyenv}$ if $\returnablepred{\tya}$ for each $x : \tya \in
\tyenv$.
Similarly, we write $\irrelevantpred{\tya}$ if $\tya$ is irrelevant (\ie, it is
a mailbox type $\usagety{\tysend{\pata}} \subtype \usagety{\mbsendone}$), and
extend this to environments.

We write $\fv{\tma}$ to return the free variables of a term.

It helps to have an inversion lemma on values:

\begin{lemma}\label{lem:val-inversion}
    If $\vseq{\tyenv}{\vala}{\tya}$, then either:
    \begin{itemize}
        \item $\vala = c$ and $\tya = \basety$ for some constant $c$ and base type
            $\basety$ with $\cruftpred{\tyenv}$; or
        \item $\vala = x$ and $\tya = \usagety{\mbtya}$ for some name $x$, usage
            annotation $\usage$, and mailbox type $\mbtya$ such that
            $\tyenv = \tyenv', x : \usageann{\usage'}{\mbtya'}$ and
            $\usageann{\usage'}{\mbtya'} \subtype \usagety{\mbtya}$ and
            $\vseq{\tyenv', x : \usageann{\usage'}{\mbtya'}}{x}{\usagety{\mbtya}}$
            with $\cruftpred{\tyenv'}$.
    \end{itemize}
\end{lemma}
\begin{proof}
    By case analysis on the derivation of $\vseq{\tyenv}{\vala}{\tya}$.
\end{proof}

\begin{lemma}[$\subpatone$ is a precongruence]\label{lem:subpat-precongruence}
    The pattern inclusion relation $\subpatone$ is a precongruence:
    \begin{description}
        \item[Reflexivity] $\pata \subpatone \pata$.
        \item[Transitivity] If $\pata_1 \subpatone \pata_2$ and
            $\pata_2 \subpatone \pata_3$, then $\pata_1 \subpatone \pata_3$. 
        \item[Compatibility wrt.\ $\patconcat$]
            If $\pata_1 \subpatone \patb_1$ and $\pata_2 \subpatone \patb_2$,
            then
            $(\pata_1 \patconcat \pata_2) \subpatone (\patb_1 \patconcat \patb_2)$.
        \item[Compatibility wrt.\ $\patplus$]
            If $\pata_1 \subpatone \patb_1$ and $\pata_2 \subpatone \patb_2$,
            then
            $(\pata_1 \patplus \pata_2) \subpatone (\patb_1 \patplus \patb_2)$.
    \end{description}
\end{lemma}
\begin{proof}
    \begin{itemize}
        \item \textbf{Reflexivity}: Follows since $\sem{\pata} \subseteq
            \sem{\pata}$.
        \item \textbf{Transitivity}:
            We have that $\sem{\pata_1} \subseteq \sem{\pata_2}$ and
            $\sem{\pata_2} \subseteq \sem{\pata_3}$, and the result follows by
            the transitivity of $\subseteq$.
        \item \textbf{Compatibility wrt.\ $\patconcat$}:
            We have that $\sem{\pata_1} \subseteq \sem{\patb_1}$ and
            $\sem{\pata_2} \subseteq \sem{\patb_2}$.
By the definition of pattern semantics we have that
            $\sem{\pata_1 \patconcat \pata_2} =
                \set{\mseta \uplus \msetb \mid \mseta \in \sem{\pata_1}, \msetb
                \in \sem{\pata_2}}$.
                
                Take an arbitrary multiset $\mseta' \in \sem{\pata_1}$ and an
                arbitrary multiset $\msetb' \in \sem{\pata_2}$.
                Because $\sem{\pata_1} \subseteq \sem{\patb_1}$ we know that
                $\mseta' \in \sem{\patb_1}$ and similarly that
                $\msetb' \in \sem{\patb_2}$.
Consequently $(\mseta' \uplus \msetb') \in 
                \set{\mseta \uplus \msetb \mid \mseta \in \sem{\patb_1},
                     \msetb \in \sem{\patb_2}}$
                and thus
                $(\pata_1 \patconcat \pata_2)
                    \subpatone
                 (\patb_1 \patconcat \patb_2)$
                as required.
        \item \textbf{Compatibility wrt.\ $\patplus$}:
            We have that $\sem{\pata_1} \subseteq \sem{\patb_1}$ and
            $\sem{\pata_2} \subseteq \sem{\patb_2}$.
By the definition of pattern semantics we have that
            $\sem{\pata_1 \patplus \pata_2} = \sem{\pata_1} \cup \sem{\pata_2}$.
Since $\sem{\pata_1} \subseteq \sem{\patb_1}$ and
            $\sem{\patb_2} \subseteq \sem{\patb_2}$, it follows that
            $(\sem{\pata_1} \cup \sem{\pata_2}) \subseteq
             (\sem{\patb_1} \cup \sem{\patb_2})$
            and therefore 
            $(\pata_1 \patplus \pata_2) \subpatone (\patb_1 \patplus \patb_2)$
            as required.
    \end{itemize}
\end{proof}

The substitution lemma is only defined on disjoint environments: we should not
be substituting a name into a term where it is already free. This is ensured by
distinguishing between returnable and second-class usages of a variable: if a
variable is returnable, then we know it cannot be used within the term into
which it is being substituted. If a variable is second-class, then there will be
no applicable reduction rules which result in substitution.

\begin{lemma}[Substitution]\label{lem:substitution}
    If:
    \begin{itemize}
        \item $\tseq{\tyenv_1, x : \tya}{M}{\tyb}$
        \item $\vseq{\tyenv_2}{V}{\tya'}$
        \item $\subtypetwo{\tya'}{\tya}$
        \item $\tyenv_1 + \tyenv_2$ is defined
    \end{itemize}

    then $\tseq{\tyenv_1 + \tyenv_2}{M \{ V / x \}}{\tyb}$.
\end{lemma}
\begin{proof}
    By induction on the derivation of $\tseq{\tyenv_1, x : \tya}{M}{\tyb}$.

    Most of the cases are standard.  \textsc{T-Let} is more intricate so we
    prove it explicitly:

    \begin{proofcase}{T-Let}
        The typing rule for \textsc{T-Let} is:

    \begin{mathpar}
        \inferrule
        {
            \tseq{\tyenvb_1}{\tma}{\tya_M} \\
            \returnablepred{\tya_M} \\
            \tseq{\tyenvb_2, y : \tya_M}{\tmb}{\tyb}
        }
        { \tseq{\scombtwo{\tyenvb_1}{\tyenvb_2}}{\letin{y}{\tma}{\tmb}}{\tyb} }
    \end{mathpar}

    There are several subcases:
    \begin{itemize}
        \item $x \in \dom{\tyenvb_1}$ and $x \not\in \dom{\tyenvb_2}$
        \item $x \in \dom{\tyenvb_2}$ and $x \not\in \dom{\tyenvb_1}$
        \item $x \in \dom{\tyenvb_1}$ and $x \in \dom{\tyenvb_2}$
    \end{itemize}

    The first two cases are straightforward so we concentrate on the final case.

    Assumption:

    \begin{mathpar}
        \inferrule
        {
            \tseq{\tyenvb_1, x : \tya_1}{\tma}{\tya_M} \\
            \returnablepred{\tya_M} \\
            \tseq{\tyenvb_2, x : \tya_2, y : \tya_M}{\tmb}{\tyb}
        }
        { \tseq
            {\scombtwo{\tyenvb_1}{\tyenvb_2}, x : (\scombtwo{\tya_1}{\tya_2})}
            {\letin{y}{\tma}{\tmb}}
            {\tyb}
        }
    \end{mathpar}

    We can also assume that $\vseq{\tyenv}{\vala}{\tya'}$ and
    $\tya' \subtype \scombtwo{\tya_1}{\tya_2}$.

    By Lemma~\ref{lem:val-inversion} we have that either $\vala = c$ for some
    constant $c$, or $\vala = z$ for some mailbox name $z$ with mailbox type
    $\usagety{\mbtya}$ such that $\vseq{\tyenv', z : \usageann{\usage'}{\mbtya'}}{z}{\usagety{\mbtya}}$ and where
    $\usageann{\usage'}{\mbtya'} \subtype \usagety{\mbtya}$ and $\cruftpred{\tyenv'}$.
The case where $\vala = c$ is uninteresting so we concentrate on the latter
    case.

    Since $\tya' = \usagety{\mbtya}$ and $\tya' \subtype (\scombtwo{\tya_1}{\tya_2})$
    it must be the case that $\tya_1$ and $\tya_2$ are mailbox types, \ie,
    there exist usages $\usage_1, \usage_2$ and mailbox types $\mbtyb_1, \mbtyb_2$ such that
    $\scombtwo{\tya_1}{\tya_2} = \scombtwo{\usageann{\usage_1}{\mbtyb_1}}{\usageann{\usage_2}{\mbtyb_2}}$ and we can
    refine our initial derivation:

    \begin{mathpar}
        \inferrule
        {
            \tseq{\tyenvb_1, x : \usageann{\usage_1}{\mbtyb_1}}{\tma}{\tya_M} \\
            \tseq{\tyenvb_2, x : \usageann{\usage_2}{\mbtyb_2}, y : \tya_M}{\tmb}{\tyb}
        }
        { \tseq{\scombtwo{\tyenvb_1}{\tyenvb_2}, x :
        (\scombtwo{\usageann{\usage_1}{\mbtyb_1}}{\usageann{\usage_2}{\mbtyb_2}})}{\letin{y}{\tma}{\tmb}}{\tyb} }
    \end{mathpar}

    We also know that $\usageann{\usage'}{\mbtya'} \subtype \usagety{\mbtya} \subtype (\scombtwo{\usageann{\usage_1}{\mbtyb_1}}{\usageann{\usage_2}{\mbtyb_2}})$.

    We can therefore construct:
    \begin{itemize}
        \item $\vseq{z : \usageann{\usage_1}{\mbtyb_1}}{z}{\usageann{\usage_1}{\mbtyb_1}}$
        \item $\vseq{z : \usageann{\usage_2}{\mbtyb_2}}{z}{\usageann{\usage_2}{\mbtyb_2}}$
    \end{itemize} 

    We are now in a position to use the IH:
    \begin{itemize}
        \item Since
            $\tseq{\tyenvb_1, x : \usageann{\usage_1}{\mbtyb_1}}{\tma}{\tya_M}$
            and $\usageann{\usage_1}{\mbtyb_1} \subtype \usageann{\usage_1}{\mbtyb_1}$
            and $\vseq{z : \usageann{\usage_1}{\mbtyb_1}}{z}{\usageann{\usage_1}{\mbtyb_1}}$,
            by the IH we have that
            $\tseq{\tyenvb_1 + z : \usageann{\usage_1}{\mbtyb_1}}{\subst{\tma}{z}{x}}{\tya_M}$
            and by the definition of $+$ we have that
            $\tseq{\tyenvb_1, z : \usageann{\usage_1}{\mbtyb_1}}{\subst{\tma}{z}{x}}{\tya_M}$
        \item Since
            $\tseq{\tyenvb_2, x : \usageann{\usage_2}{\mbtyb_2}, y : \tya_M}{\tmb}{\tyb}$
            and $\usageann{\usage_2}{\mbtyb_2} \subtype \usageann{\usage_2}{\mbtyb_2}$
            and $\vseq{z : \usageann{\usage_2}{\mbtyb_2}}{z}{\usageann{\usage_2}{\mbtyb_2}}$,
            by the IH we have that
            $\tseq
                {(\tyenvb_2, y : \tya_M) + (z : \usageann{\usage_2}{\mbtyb_2})}
                {\subst{\tmb}{z}{x}}
                {\tyb}$
            and by the definition of $+$ we have that
            $\tseq
                {\tyenvb_2, y : \tya_M, z : \usageann{\usage_2}{\mbtyb_2}}
                {\subst{\tmb}{z}{x}}
                {\tyb}$.
    \end{itemize}

    Since $\cruftpred{\tyenv'}$ and
    $\usageann{\usage'}{\mbtya'} \subtype (\scombtwo{\usageann{\usage_1}{\mbtyb_1}}{\usageann{\usage_2}{\mbtyb_2}})$, we can recompose using
    an additional application of \textsc{T-Sub}:

    \begin{mathpar}
        \inferrule
        {
            \inferrule
            {
                \tseq{\tyenvb_1, z : \usageann{\usage_1}{\mbtyb_1}}{\subst{\tma}{z}{x}}{\tya_M} \\
                \returnablepred{\tya_M} \\
                \tseq
                    {\tyenvb_2, y : \tya_M, z : \usageann{\usage_2}{\mbtyb_2}}
                    {\subst{\tmb}{z}{x}}
                    {\tyb}
            }
            { \tseq
                {(\scombtwo{\tyenvb_1}{\tyenvb_2}),
                   z : (\scombtwo{\usageann{\usage_1}{\mbtyb_1}}{\usageann{\usage_2}{\mbtyb_2}})}
                {\letin{y}{\subst{\tma}{z}{x}}{\subst{\tmb}{z}{x}}}
                {\tyb}
            }
        }
        {
            \tseq
                {\scombtwo{\tyenvb_1}{\tyenvb_2} + (\tyenv', z : \usageann{\usage'}{\mbtya'})}
                {\letin{y}{\subst{\tma}{z}{x}}{\subst{\tmb}{z}{x}}}
                { \tyb }
        }
    \end{mathpar}
    as required.

\end{proofcase}
\end{proof}

\begin{lemma}[Subtyping preserves reliablility / usability~\citep{dP18:mbc}]
    \label{lem:subtype-usability}
    If $\subtypetwo{\tya}{\tyb}$, then:

    \begin{enumerate}
        \item $\tya$ reliable implies $\tyb$ reliable
        \item $\tyb$ usable implies $\tya$ usable
    \end{enumerate}
\end{lemma}

\begin{corollary}
    \label{lem:subenv-env-usability}
    If $\subtypetwo{\tyenv_1}{\tyenv_2}$ then:
    \begin{enumerate}
        \item $\tyenv_1$ reliable implies $\tyenv_2$ reliable
        \item $\tyenv_2$ usable implies $\tyenv_1$ usable
    \end{enumerate}
\end{corollary}

\begin{lemma}\label{lem:returnable-subtype}
    If $\tya \subtype \tyb$ and $\returnablepred{\tyb}$, then $\returnablepred{\tya}$
\end{lemma}
\begin{proof}
    Follows from the fact that $\returnable \subtype \usable$.
\end{proof}

\begin{corollary}\label{cor:returnable-subtype-env}
    If $\tyenv_1 \strictsubty \tyenv_2$ and
    $\returnablepred{\tyenv_2}$, then
    $\returnablepred{\tyenv_1}$.
\end{corollary}

\begin{lemma}\label{lem:returnable-val-env}
If $\vseq{\tyenv}{\vala}{\tya}$ where $\returnablepred{\tya}$ and $\tyenv$ is
cruftless for $\vala$, then $\returnablepred{\tyenv}$.
\end{lemma}
\begin{proof}
    By case analysis on the derivation of $\vseq{\tyenv}{\vala}{\tya}$.
\end{proof}

\begin{lemma}\label{lem:scomb-plus-defined}
    If $\scombtwo{\tyenv_1}{\tyenv_2}$ is defined, with $\tyenv_1$ and
    $\tyenv_2$ sharing only variables of base type, then
    $\pluscombtwo{\tyenv_1}{\tyenv_2}$ is defined.
\end{lemma}
\begin{proof}
    Immediate from the definitions.
\end{proof}

\begin{lemma}[$\scomb$ is associative]\label{lem:scomb-assoc}
      $\tya_1 \scomb (\tya_2 \scomb \tya_3) = (\tya_1 \scomb \tya_2) \scomb \tya_3$
\end{lemma}
\begin{proof}
    Follows from the fact that usage combination is associative, and that we
    identify patterns up to commutativity and associativity.
\end{proof}

Extending to usage-aware type environments, we get the following corollary:

\begin{corollary}\label{lem:scomb-env-assoc}
      $\tyenv_1 \scomb (\tyenv_2 \scomb \tyenv_3) = (\tyenv_1 \scomb \tyenv_2) \scomb \tyenv_3$
\end{corollary}

The same result holds for runtime type environments and $\pcomb$:

\begin{lemma}[$\pcomb$ is associative]\label{lem:pcomb-env-assoc}
      $\plainenv_1 \pcomb (\plainenv_2 \pcomb \plainenv_3) = (\plainenv_1 \pcomb \plainenv_2) \pcomb \plainenv_3$
\end{lemma}
\begin{proof}
    Follows the same reasoning as for $\scomb$.
\end{proof}

\begin{lemma}\label{lem:pcomb-env-comm}
    The $\pcomb$ operator is commutative:
    $\plainenv_1 \pcomb \plainenv_2 = \plainenv_2 \pcomb \plainenv_1$.
\end{lemma}
\begin{proof}
    Follows from the fact that $\mbcomb$ is commutative.
\end{proof}

\begin{lemma}\label{lem:plus-scomb-env}
    If $\tyenv_1 + (\tyenv_2 \scomb \tyenv_3)$ is defined, then $\tyenv_1 + (\tyenv_2 \scomb \tyenv_3)= (\tyenv_1 + \tyenv_2) \scomb \tyenv_3$.
\end{lemma}
\begin{proof}
    Follows directly from the definitions.
\end{proof}

\begin{lemma}\label{lem:disjoint-scomb}
    If $\tyenv_1, \tyenv_2 = \tyenv$, then $\tyenv_1 \scomb \tyenv_2 = \tyenv$
\end{lemma}
\begin{proof}
    Follows from the definition of $\scomb$ given that $\tyenv_1$ and
    $\tyenv_2$ are disjoint.
\end{proof}

\begin{lemma}\label{lem:scomb-to-pcomb}
    If $\tyenv_1 \scomb \tyenv_2 = \tyenv$, then $\strip{\tyenv_1} \pcomb
    \strip{\tyenv_2} = \strip{\tyenv}$.
\end{lemma}
\begin{proof}
    Follows directly from the definitions, since $\pcomb$ is more liberal than $\scomb$.
\end{proof}

\begin{lemma}\label{lem:pcomb-to-scomb}
    If 
    $\strip{\tyenv} \pcomb \plainenv$ is defined, then
    $\strip{\tyenv} \pcomb \plainenv = \strip{\makeusable{\plainenv} \scomb \tyenv}$.
\end{lemma}
\begin{proof}
    For each $x$ such that $x : \pretya \in \strip{\tyenv}$ and $x : \pretyb \in
    \plainenv$, since
    $\strip{\tyenv} \pcomb \plainenv$ is defined, we have that $\pretya \mbcomb
    \pretyb$ is defined. The result then follows from the definition of
    $\scomb$, noting that all types in $\makeusable{\plainenv}$ are second-class and
    therefore combinable with any other usage.
\end{proof}

Because of the use of environment subtyping in both the configuration and term
typing judgements, it is useful to be able to re-associate type combination in
the presence of subtyping.

\begin{lemma}\label{lem:ty-subty-assoc}
    If $\pretya = \scombtwo{\pretya_1}{\pretya_2}$ where $\pretya_1 \subtype
    (\scombtwo{\pretyb_1}{\pretyb_2})$, then there exist $\pretyb'_1 \subtype \pretyb_1$ and 
    $\pretyb'_2 \subtype \pretyb_2$ and $\pretya'_2 \subtype \pretya_2$ such that
    $\pretya \subtype \pretyb'_1 \scomb (\pretyb'_2 \scomb \pretya'_2)$.
\end{lemma}
\begin{proof}
    By case analysis on the derivation of $\scombtwo{\pretya_1}{\pretya_2}$.

    \begin{proofcase*}{$\pretya = \scombtwo{\basety}{\basety}$}
        In this case, by the definitions of combinations and subtyping it must
        be the case that $\pretyb_1, \pretyb_2, \pretya_2$ are all $\basety$ and thus
        trivially $\basety \subtype \basety \scomb (\basety \scomb \basety)$ as required.
    \end{proofcase*}

    \begin{proofcase*}{$\pretya = 
        \scombtwo
            {\tysend{\pata_1}}
            {\tysend{\pata_2}}$}

        By the definition of $\scomb$ we have that
        $\pretya = \tysendp{\pata_1 \patconcat \pata_2}$.

        In this case we have that
        $\tysend{\pata_1} \subtype \scombtwo{\pretyb_1}{\pretyb_2}$.

        By the definition of subtyping it must be the case that
        $\scombtwo{\pretyb_1}{\pretyb_2}$ is an output mailbox type, and by the
        definition of $\scomb$ it must be the case that $\tysend{\pata_1}
        \subtype \scombtwo{\tysend{\patb_1}}{\tysend{\patb_2}}$ for patterns
        $\patb_1, \patb_2$.

        Therefore, $\tysend{\pata_1} \subtype
        \tysendp{\patconcattwo{\patb_1}{\patb_2}}$ and by the definition of
        subtyping, $\patconcattwo{\patb_1}{\patb_2} \subpatone \pata_1$.

        Through the definition of $\scomb$ we can show:
\begin{itemize}
            \item $\scombtwo{\tysend{\patb_2}}{\tysend{\pata_2}} =
                \tysendp{\patconcattwo{\patb_2}{\pata_2}}$
            \item $\tysend{\patb_1} \scomb
                (\scombtwo{\tysend{\patb_2}}{\tysend{\pata_2}}) =
                \tysendp{\patb_1 \patconcat {\patb_2} \patconcat {\pata_2}}$
        \end{itemize}

        Since
        $\patb_1 \patconcat \patb_2 \subpatone \pata_1$, by 
        Lemma~\ref{lem:subpat-precongruence}
        we can show that $\patb_1 \patconcat \patb_2 \patconcat \pata_2 \subpatone
        \pata_1 \patconcat \pata_2$.
        and therefore that $\tysendp{\pata_1 \patconcat \pata_2}
        \subtype \tysendp{\patb_1 \patconcat \patb_2 \patconcat \pata_2}$, as
        required.

    \end{proofcase*}

    \begin{proofcase*}{$\pretya = \tyrecvp{\patconcattwo{\pata_1}{\pata_2}} \scomb
        \tysend{\pata_1}$}

        By the definition of $\scomb$ we have that
        $\pretya = \tyrecv{\pata_2}$.

        In this case we have that
        $\tyrecvp{\patconcattwo{\pata_1}{\pata_2}} \subtype \pretyb_1 \scomb \pretyb_2$.

        By the definition of subtyping it must be the case that $\pretyb_1
        \scomb \pretyb_2 = \tyrecv{\patb}$ for some pattern $\patb$ where
        $\tyrecvp{\patconcattwo{\pata_1}{\pata_2}} \subtype \tyrecv{\patb}$
        and therefore that $\patconcattwo{\pata_1}{\pata_2} \subpatone \patb$.

        We therefore have two subcases (for some pattern $\patb'$):
        \begin{itemize}
            \item $\pretyb_1 =  \tysend{\patb'}$ and
                $\pretyb_2 = \tyrecvp{\patconcattwo{\patb'}{\patb}}$ 
            \item $\pretyb_1 = \tyrecvp{\patconcattwo{\patb'}{\patb}}$ and
                $\pretyb_2 = \tysend{\patb'}$
        \end{itemize}

        Both are similar so we show the first case.
        We know that $\patconcattwo{\pata_1}{\pata_2} \subpatone \patb$.

            Let
            $\pretyb'_2 = \tyrecvp{\pata_1 \patconcat \patb' \patconcat \pata_2}$.
            By Lemma~\ref{lem:subpat-precongruence}
            we have that
            $\pata_1 \patconcat \patb' \patconcat \pata_2
                \subpatone \patconcattwo{\patb'}{\patb}$
            and by the definition of
            subtyping we have that
            $\pretyb'_2 \subtype \tyrecvp{\patconcattwo{\patb'}{\patb}}$.

            Thus we can show:

            \[
                \begin{array}{l}
                    \tysend{\patb'} \scomb
                        (\tyrecvp{\pata_1 \patconcat \patb' 
                           \patconcat \pata_2} \scomb \tysend{\pata_1})
                    \\ 
                    \quad =
                    \\
                    \tysend{\patb'} \scomb \tyrecvp{\patb' \patconcat \pata_2}
                    \\
                    \quad =
                    \\
                    \tyrecv{\pata_2}
                \end{array}
            \]

            Here we have that:
            \begin{itemize}
                \item $\pretyb'_1 = \tysend{\patb'} = \pretyb_1$
                \item $\pretyb'_2 = \tyrecvp{\pata_1 \patconcat \patb'
                    \patconcat \pata_2} \subtype \tyrecvp{\patb' \patconcat
                    \patb} = \pretyb_2$
                \item $\pretya'_2 = \tysend{\pata_1} = \pretya_2$
                \item $\pretya = \tysend{\pata_2}$
            \end{itemize}
            as required.
    \end{proofcase*}

    \begin{proofcase*}{$\pretya =
            \tysend{\pata_1} \scomb
            \tyrecvp{\patconcattwo{\pata_1}{\pata_2}}$}

    In this case we have that $\tysend{\pata_1} \subtype \pretyb_1 \scomb
    \pretyb_2 $
    and by the definitions of $\subtype$ and $\scomb$ it follows that
    $\pretyb_1, \pretyb_2 = \tysend{\patb_1}, \tysend{\patb_2}$ for some
    $\patb_1, \patb_2$ and thus that
    $\tysend{\pata_1} \subtype \tysendp{\patb_1 \scomb \patb_2}$.

    By the definition of $\scomb$ it follows that
    $\tysend{\pata_1} \subtype \tysendp{\patb_1 \patconcat \patb_2}$.

    By the definition of subtyping it follows that
    $\patb_1 \patconcat \patb_2 \subpatone \pata_1$.

    Let $\pretyb'_2 = \tyrecvp{\patb_2 \patconcat \patb_1 \patconcat \pata_2}$.
    By Lemma~\ref{lem:subpat-precongruence},
    $\patb_2 \patconcat \patb_1 \patconcat \pata_2 \subpatone \pata_1 \patconcat
    \pata_2$ and therefore 
    $\tyrecvp{\patb_2 \patconcat \patb_1 \patconcat \pata_2} \subtype
    \tyrecvp{\pata_1 \patconcat \pata_2}$.
    
    Thus we can show:

    \[
        \begin{array}{l}
            \tysend{\patb_1} \scomb (\tysend{\patb_2} \scomb \tyrecvp{\patb_2
            \patconcat \patb_1 \patconcat \pata_2})
            \\
            \quad =
            \\
            \tysend{\patb_1} \scomb \tyrecvp{\patb_1 \patconcat \pata_2}
            \\
            \quad =
            \\
            \tyrecv{\pata_2}
        \end{array}
    \]

    where:
    \begin{itemize}
        \item $\pretyb'_1 = \tysend{\patb_1} = \pretyb_1$
        \item $\pretyb'_2 = \tysend{\patb_2} = \pretyb_2$
        \item $\pretya'_2 = \tyrecvp{\patb_2 \patconcat \patb_1 \patconcat
            \pata_2} \subtype \tyrecvp{\pata_1 \patconcat \pata_2}= \pretya_2$
    \end{itemize}
    as required.
    \end{proofcase*}
\end{proof}

\begin{corollary}\label{cor:env-subtype-assoc-r}
    If $\tyenv = \scombtwo{\tyenv_1}{\tyenv_2}$ where $\tyenv_1 \subtype
    (\scombtwo{\tyenvb_1}{\tyenvb_2})$, then there exist $\tyenvb'_1 \subtype \tyenvb_1$ and 
    $\tyenvb'_2 \subtype \tyenvb_2$ and $\tyenv'_2 \subtype \tyenv_2$ such that
    $\tyenv \subtype \tyenvb'_1 \scomb (\tyenvb'_2 \scomb \tyenv'_2)$.
\end{corollary}

We can also re-associate to the left. The proof follows similar reasoning to
Corollary~\ref{cor:env-subtype-assoc-r}.

\begin{lemma}\label{lem:env-subtype-assoc-l}
    If $\tyenv = \scombtwo{\tyenv_1}{\tyenv_2}$ where $\tyenv_2 \subtype
    (\scombtwo{\tyenvb_1}{\tyenvb_2})$, then there exist
    $\tyenvb'_1 \subtype \tyenvb_1$ and 
    $\tyenvb'_2 \subtype \tyenvb_2$ and $\tyenv'_1 \subtype \tyenv_1$ such that
    $\tyenv \subtype (\tyenv'_1 \scomb \tyenvb'_1) \scomb \tyenvb'_2$.
\end{lemma}

\begin{lemma}\label{lem:cruft-usable-typable}
    If $\tseq{\tyenv}{\tma}{\tya}$ where $\cruftpred{\tyenv}$ and
    $\fv{\tma} \cap \tyenv = \emptyset$, then
    $\tseq{\makeusable{\tyenv}}{\tma}{\tya}$.
\end{lemma}
\begin{proof}
    Since $\fv{\tma} = \emptyset$, $\tseq{\cdot}{\tma}{\tya}$.
    The result follows by repeated applications of weakening and subtyping to
    add each cruft type, \emph{without} using usage subtyping.
\end{proof}

\begin{lemma}\label{lem:cruft-comb-typing}
    If $\tseq{\tyenv}{\tma}{\tya}$ and $\cruftpred{\tyenvb}$ and
    $\scombtwo{\tyenvb}{\tyenv}$ is defined, then
    $\tseq{\tyenvb \scomb \tyenv}{\tma}{\tya}$.
\end{lemma}
\begin{proof}
    It suffices to consider the case where $\tyenv = x : \usagety{\mbtya}$ and
    $\tyenvb = x : \tyb$ and $\tma = x$ for some type $\tyb$.

    We proceed by case analysis on $\mbtya$.

    \begin{proofcase*}{$\mbtya = \tyrecv{\pata_{\var{ty}}}$}
        Since $\scombtwo{\tyenvb}{\tyenv}$ is defined it must be the case that
        $\pata_{\var{ty}} = \patconcattwo{\pata}{\patb}$ and
        $\tyenv = x : \usagety{\tyrecvp{\patconcattwo{\pata}{\patb}}}$
        and $\tyenvb = x : \usablety{\tysend{\pata}}$.

        In this case:

        \begin{mathpar}
            \inferrule
            {
                \tseq
                { x : \usageann{\usage'}{\tyrecvp{\patconcattwo{\pata}{\patb}}} }
                { x }
                { \usageann{\usage'}{\tyrecvp{\patconcattwo{\pata}{\patb}}} }
            }
            { \tseq
                {x : \usagety{\tyrecvp{\patconcattwo{\pata}{\patb}}}}
                { x }
                { \usageann{\usage''}{\tyrecv{\patc}} }
            }
        \end{mathpar}

        with:
        \begin{itemize}
            \item $\tyrecv{\patconcattwo{\pata}{\patb}} \subtype
                \tyrecv{\patc}$ and therefore $\subpat{\patconcattwo{\pata}{\patb}}{\patc}$
            \item $\usage \subtype \usage'$ and $\usage' \subtype \usage''$
        \end{itemize}

        We must show 
        $
        \tseq
                {x : \usagety{\tyrecv{\patb}}}
                { x }
                { \usageann{\usage''}{\tyrecv{\patc}}}
        $.

        Since $\tyenvb$ is cruft we have that $\tysend{\pata} \subtype
        \mbsendone$ and thus $\mbone \subpatone \pata$. Since
        $\mbone \subpatone \pata$
        it follows that $\pata \tyequiv \mbone \patplus \pata'$ for
        some $\pata' \subpatone \pata$.

        Since $\pata \tyequiv \mbone \patplus \pata'$, by
        Lemma~\ref{lem:subpat-precongruence} and the axioms of commutative
        regular expressions,
        \[
            \begin{array}{l}
                \patconcattwo{\pata}{\patb}
                \\
                \quad \tyequiv
                \\
                (\mbone \patplus \pata') \patconcat \patb
                \\
                \quad \tyequiv
                \\
                \patb \patplus (\pata' \patconcat \patb)
            \end{array}
        \]

        From the definition of pattern semantics we have that 
        $\patb \subpatone \patb \patplus (\pata' \patconcat \patb)$
        and thus
        $\patb \subpatone (\pata \patconcat \patb)$.

        Thus we can construct:
        \begin{mathpar}
            \inferrule
            {
                \tseq
                { x : \usageann{\usage'}{\tyrecvp{\patconcattwo{\pata}{\patb}}} }
                { x }
                { \usageann{\usage'}{\tyrecvp{\patconcattwo{\pata}{\patb}}} }
            }
            { \tseq
                {x : \usagety{\tyrecv{\patb}}}
                { x }
                { \usageann{\usage''}{\tyrecv{\patc}} }
            }
        \end{mathpar}

        as required.

    \end{proofcase*}

    \begin{proofcase*}{$\mbtya = \tysend{\patb}$}
        Assumption:

        \begin{mathpar}
            \inferrule
            {
                \tseq
                { x : \usageann{\usage'}{\tysend{\patb}} }
                { x }
                { \usageann{\usage'}{\tysend{\patb}} }
            }
            { \tseq
                {x : \usagety{\tysend{\patb}}}
                { x }
                { \usageann{\usage''}{\tysend{\patc}} }
            }
        \end{mathpar}

        with $\usage \subtype \usage'$ and
        $\tysend{\patb} \subtype \tysend{\patc}$ and therefore $\patc \subpatone \patb$.

        Since $\tyenvb \scomb \tyenv$ is defined it must be the case that
        $\tyenvb = \usablety{\tysend{\pata}}$ where
        $\tysend{\pata} \subtype \mbsendone$ and therefore that
        $\mbone \subpatone \pata$.

        We need to show that 
        \[
        \tseq
                {x : \usagety{\tysendp{\pata \patconcat \patb}}}
                { x }
                { \usageann{\usage''}{\tysend{\patc}} }
            \]

        Since $\mbone \subpatone \pata$ we have that
        $\pata \tyequiv \mbone \patplus \pata'$ for some pattern $\pata' \subpatone \pata$.

        Thus by Lemma~\ref{lem:subpat-precongruence} and equational reasoning we can show:
        \[
            \bl
            \pata \patconcat \patb
            \\
            \quad \tyequiv
            \\
            (\mbone \patplus \pata') \patconcat \patb
            \\
            \quad \tyequiv
            \\
            (\patb \patconcat \mbone) \patplus (\pata' \patconcat \patb) \\
            \quad \tyequiv
            \\
            \patb \patplus (\pata' \patconcat \patb)
            \el
        \]

        and thus we have that $(\pata \patconcat \patb) \tyequiv (\patb \patplus (\pata' \patconcat \patb)) $
and therefore $\patb \subpatone (\pata \patconcat \patb)$.

        It follows that $\patc \subpatone \patb \subpatone (\patconcattwo{\pata}{\patb})$, so
        $\tysendp{\patconcattwo{\pata}{\patb}} \subtype \tysend{\patb} \subtype \tysend{\patc}$ and we can construct:

        \begin{mathpar}
            \inferrule*
            {
                \inferrule*
                {
                    \tseq
                    {x : \usageann{\usage'}{\tysendp{\pata \patconcat \patb}}}
                    { x }
                    { \usageann{\usage'}{\tysendp{\pata \patconcat \patb}} }
                }
                {
                    \tseq
                    {x : \usageann{\usage}{\tysendp{\pata \patconcat \patb}}}
                    { x }
                    { \usageann{\usage''}{\tysend{\patb}} }
                }
            }
            {
            \tseq
                {x : \usagety{\tysend{\pata \patconcat \patb}}}
                { x }
                { \usageann{\usage''}{\tysend{\patc}} }
            }
        \end{mathpar}
        as required.
    \end{proofcase*}
\end{proof}

We can derive an analogous result for combining with cruft on the right-hand
side of an environment.

\begin{lemma}\label{lem:cruft-comb-typing-r}
    If $\tseq{\tyenv}{\tma}{\tya}$ and $\cruftpred{\tyenvb}$ and
    $\scombtwo{\tyenv}{\tyenvb}$ is defined, then
    $\tseq{\tyenv \scomb \tyenvb}{\tma}{\tya}$.
\end{lemma}
\begin{proof}
    Analogous to the proof of Lemma~\ref{lem:cruft-comb-typing}.
\end{proof}

\begin{lemma}\label{lem:env-subtype-plus}
    If $\tyenv \subtype \tyenv_1 + \tyenv_2$ then
    $\tyenv = \tyenv'_1 + \tyenv'_2 + \tyenvb$ for some
    $\tyenv'_1, \tyenv'_2, \tyenvb$ such that
    $\tyenv'_1 \strictsubty \tyenv_1$ and
    $\tyenv'_2 \strictsubty \tyenv_2$ and
    $\cruftpred{\tyenvb}$.
\end{lemma}
\begin{proof}
    Follows from the definition of environment subtyping.
\end{proof}

\begin{lemma}\label{lem:scomb-inside-plus}
    If $\tyenv \scomb (\tyenvb_1 + \tyenvb_2)$ is defined and
    $\tyenv + \tyenvb_2$ is defined, then
    $\tyenv \scomb (\tyenvb_1 + \tyenvb_2) =
    (\tyenv \scomb \tyenvb_1) + \tyenvb_2$.
\end{lemma}
\begin{proof}
    Since $\tyenv + \tyenvb_2$ is defined, the two environments can only overlap
    on base types. Thus the result follows from the definitions of $\scomb$ and
    $+$.
\end{proof}

\begin{corollary}\label{cor:cruft-inside-plus}
    If $(\tyenv_1 \scomb \tyenvb) + \tyenv_2$ is defined and
    $\tseq{\tyenv_1 + \tyenv_2}{\tma}{\tya}$
    and $\cruftpred{\tyenvb}$
    then $\tseq{(\tyenv_1 \scomb \tyenvb) + \tyenv_2}{\tma}{\tya}$.
\end{corollary}
\begin{proof}
    Follows from Lemma~\ref{lem:cruft-comb-typing-r} and the fact that
    $(\tyenv_1 \scomb \tyenvb) + \tyenv_2$ is defined, since
    $\tyenv_1 \scomb \tyenvb$ and $\tyenv_2$ only overlap on base types.
\end{proof}

\subsubsection{Preservation proof}

\begin{lemma}[Preservation (Equivalence)]\label{lem:equiv-pres}
    If $\cseq{\plainenv}{\config{C}}$ and $\config{C} \equiv \config{D}$, then
    $\cseq{\plainenv}{\config{D}}$.
\end{lemma}
\begin{proof}
    By induction on the derivation of $\config{C} \equiv \config{D}$, relying on
    Lemmas~\ref{lem:pcomb-env-assoc} and~\ref{lem:pcomb-env-comm} and \textsc{T-Sub}.
\end{proof}

\preservation*
\begin{proof}
    By induction on the derivation of $\cseq{\plainenv}{\config{C}}$.

    \begin{proofcase}{E-Let}

            Assumption:

            \begin{mathpar}
                \inferrule*
                {
                    \inferrule*
                    {
                        \plainenv' = \strip{\scombtwo{\tyenv_1}{\tyenv_2}} \\
                        \inferrule*
                        {
                            \inferrule*
                            {
                                \scombtwo{\tyenvb_1}{\tyenvb_2} = \tyenv'_1 \\
                                \tseq{\tyenvb_1}{\tma}{\tya} \\\\
                                \returnablepred{\tya} \\
                                \tseq{\tyenvb_2, x : \tya}{\tmb}{\tyb'}
                            }
                            { \tseq{\tyenv'_1}{\letin{x}{\tma}{\tmb}}{\tyb'} }
                        }
                        {
                            \tseq{\tyenv_1}{\letin{x}{\tma}{\tmb}}{\tyb}
                        }
                        \\
\inferrule
                        {}
                        { \stackseq{\tyenv_2}{\tyb}{\framestack}}
                    }
                    { \cseq{\plainenv'}{\thread{\letin{x}{\tma}{\tmb}}{\framestack}} }
                }
                { \cseq{\plainenv}{\thread{\letin{x}{\tma}{\tmb}}{\framestack}} }
            \end{mathpar}

            where
            \begin{itemize}
                \item $\plainenv \subtype \plainenv'$
                \item $\plainenv' = \strip{\scombtwo{\tyenv_1}{\tyenv_2}}$
                \item $\tyb' \subtype \tyb$
                \item $\tyenv'_1 = \scombtwo{\tyenvb_1}{\tyenv_2}$
                \item $\tyenv_1 \subtype \tyenv'_1$
            \end{itemize}

            By Corollary~\ref{cor:env-subtype-assoc-r}
            there exist $\tyenvb'_1 \subtype \tyenvb_1$ and 
            $\tyenvb'_2 \subtype \tyenvb_2$ and
            $\tyenv'_2 \subtype \tyenv_2$ such that
            $\plainenv' \subtype
            \strip{(\tyenvb'_1 \scomb (\tyenvb'_2 \scomb \tyenv'_2))}$.

            Recomposing:

            {\footnotesize
            \begin{mathpar}
                \inferrule*
                {
                    \inferrule*
                    {
                        \inferrule*
                        {
                            \plainenv'' =
                            \strip{\scombtwo{\tyenvb'_1}{(\tyenvb'_2 \scomb
                            \tyenv'_2)}}
                            \\
                            \inferrule*
                            { \tseq{\tyenvb_1}{\tma}{\tya} }
                            { \tseq{\tyenvb'_1}{\tma}{\tya} }
                            \\
\inferrule*
                            {
                                \inferrule*
                                { \tseq{\tyenvb_2, x : \tya}{\tmb}{\tyb'} }
                                { \tseq{\tyenvb'_2, x : \tya}{\tmb}{\tyb} }
                                \\
                                \returnablepred{\tya}
                                \\
                                \inferrule*
                                { \stackseq{\tyenv_2}{\tyb}{\framestack} }
                                { \stackseq{\tyenv'_2}{\tyb}{\framestack} }
                            }
                            { \stackseq
                                {(\scombtwo{\tyenvb'_2}{\tyenv'_2})}
                                {\tya}
                                {\frameconcat{\fframe{x}{\tmb}}{\framestack}}
                            }
                        }
                        {
                            \cseq
                                {\plainenv''}
                                {\thread{\tma}{\frameconcat{\fframe{x}{\tmb}}{\framestack}}}
                        }
                    }
                    {
                        \cseq
                            {\plainenv'}
                            {\thread{\tma}{\frameconcat{\fframe{x}{\tmb}}{\framestack}}}
                    }
                }
                { \cseq{\plainenv}{\thread{\letin[\pretya]{x}{\tma}{\tmb}}{\framestack}} }
            \end{mathpar}
        }

            where $\plainenv' \subtype \plainenv''$ as required.

\end{proofcase}

    \begin{proofcase}{E-Return}
        Assumption:

            \begin{mathpar}
                \inferrule
                {
                        \plainenv = \strip{\scombtwo{\tyenv_1}{\tyenv_2}} 
                        \\
                        \tseq{\tyenv_1}{V}{\tya}
                    \\
                    \inferrule
                    {
                        \inferrule
                        {
                            \tyenv'_2 = \scombtwo{\tyenv_3}{\tyenv_4} \\
                            \returnablepred{\tya} \\\\
                            \tseq{\tyenv_3, x : \tya}{M}{\tyb} \\
                            \stackseq{\tyenv_4}{\tyb}{\framestack}
                        }
                        { \stackseq{\tyenv'_2}{\tya}{\frameconcat{\fframe{x}{M}}{\framestack} } }
                    }
                    {
                        \stackseq{\tyenv_2}{\tya}{\frameconcat{\fframe{x}{M}}{\framestack} }
                    }
                }
                {
                    \cseq
                        {\plainenv}
                        {
                            \thread
                                {\vala}
                                {\frameconcat{\fframe{x}{M}}{\framestack}}
                        }
                }
            \end{mathpar}

            By Lemma~\ref{lem:disjoint-from-cruft}, we have that there exist
            $\tyenvb_1, \tyenvb_2, \tyenvb_3$ such that:
 
            \begin{itemize}
                \item $\tyenv_1 = \tyenvb_1, \tyenvb_2$
                \item $\tseq{\tyenvb_3}{\vala}{\tya'}$
                \item $\tyenvb_1$ is cruftless for $\vala$, and $\tyenvb_1 \strictsubty \tyenvb_3$
                \item $\tya' \subtype \tya$
                \item $\cruftpred{\tyenvb_2}$
            \end{itemize}

            Refining our derivation:

            \begin{mathpar}
                \inferrule
                {
                        \plainenv =
                            \strip{\scombtwo{(\tyenvb_1, \tyenvb_2)}{\tyenv_2}}
                        \\
                        \tseq{\tyenvb_1, \tyenvb_2}{V}{\tya}
                        \\
                    \inferrule
                    {
                        \inferrule
                        {
                            \tyenv'_2 = \scombtwo{\tyenv_3}{\tyenv_4} \\
                            \returnablepred{\tya} \\\\
                            \tseq{\tyenv_3, x : \tya}{M}{\tyb} \\
                            \stackseq{\tyenv_4}{\tyb}{\framestack}
                        }
                        { \stackseq{\tyenv'_2}{\tya}{\frameconcat{\fframe{x}{M}}{\framestack} } }
                    }
                    {
                        \stackseq{\tyenv_2}{\tya}{\frameconcat{\fframe{x}{M}}{\framestack} }
                    }
                }
                {
                    \cseq
                        {\plainenv}
                        {
                            \thread
                                {\vala}
                                {\frameconcat{\fframe{x}{M}}{\framestack}}
                        }
                }
            \end{mathpar}

            By Lemma~\ref{lem:returnable-subtype}, $\returnablepred{\tya'}$, and
            by Lemma~\ref{lem:returnable-val-env}, $\returnablepred{\tyenvb_3}$.

            By Corollary~\ref{cor:returnable-subtype-env}, $\returnablepred{\tyenvb_1}$.

            Since
            $\plainenv = \strip{(\tyenvb_1, \tyenvb_2) \scomb \tyenv_2}$,
            by Lemma~\ref{lem:returnable-non-aliased},
            $\plainenv = \strip{\tyenvb_1 + (\tyenvb_2 \scomb \tyenv_2)}$.

            Since $\tyenv_2 \subtype (\tyenv_3 \scomb \tyenv_4)$, by
            Lemma~\ref{lem:env-subtype-assoc-l} and the definition of
            environment combination it follows that
            $\plainenv \subtype \strip{\tyenvb_1 +
                ((\tyenvb'_2 \scomb \tyenv'_3) \scomb \tyenv'_4)}$.

            By \textsc{T-Sub}, since $\tyenv'_3 \subtype \tyenv_3$, we have that
            $\tseq{\tyenv'_3, x : \tya}{\tma}{\tyb}$.

            Since $\cruftpred{\tyenvb_2}$ and $\tyenvb'_2 \subtype \tyenvb_2$ so
            $\cruftpred{\tyenvb'_2}$,
            by Lemma~\ref{lem:cruft-comb-typing} we have that
            $\tseq{(\tyenvb'_2 \scomb \tyenv'_3), x : \tya}{\tma}{\tyb}$.

            By Lemma~\ref{lem:substitution},
            $\tseq{\tyenvb_1 + (\tyenvb'_2 \scomb \tyenv'_3)}{\subst{\tma}{\vala}{x}}{\tyb}$.

            By Lemma~\ref{lem:plus-scomb-env},
            $
                \tyenvb_1 + ((\tyenvb'_2
                        \scomb \tyenv'_3) \scomb \tyenv'_4)
                =
                (\tyenvb_1 + (\tyenvb'_2
                        \scomb \tyenv'_3)) \scomb \tyenv'_4
            $.

            Thus, recomposing:

            \begin{mathpar}
                \inferrule*
                {
                    \inferrule*
                    {
                        \plainenv' =
                        \strip{(\tyenvb_1 + (\tyenvb'_2
                        \scomb \tyenv'_3)) \scomb \tyenv'_4}
                        \\
                        \tseq{\tyenvb_1 + (\tyenvb'_2 \scomb \tyenv'_3)}{\tma}{\tyb}
                        \\
                        \inferrule*
                        { \stackseq{\tyenv_4}{\tyb}{\framestack} }
                        { \stackseq{\tyenv'_4}{\tyb}{\framestack} }
                    }
                    {
                        \cseq{\plainenv'}{\thread{\subst{\tma}{\vala}{x}}{\framestack}}
                    }
                }
                { \cseq{\plainenv}{\thread{\subst{\tma}{\vala}{x}}{\framestack} } }
            \end{mathpar}
            as required.

\end{proofcase}

    \begin{proofcase}{E-App}

       Assumption:
       \begin{mathpar}
           \inferrule*
           {
               \plainenv = \strip{\tyenv_1 \scomb \tyenv_2} \\
\inferrule*
               {
                   \inferrule*
                   {
                   \prog(f) = \tyfun{\seq{\tya}}{\tyb'} \\
                   (\vseq{\tyenv'_i}{\vala_i}{\tya_i})_i
                   }
                   { \tseq{\tyenv'_1 + \cdots + \tyenv'_n}{\fnapp{f}{\seq{V}}}{\tyb'}} \\
               }
               { \tseq{\tyenv_1}{\fnapp{f}{\seq{V}}}{\tyb}} \\
\stackseq{\tyenv_2}{\tyb}{\framestack}
           }
           { \cseq{\plainenv}{\thread{\fnapp{f}{\seq{V}}}{\framestack}} }
       \end{mathpar}

       where $\tyenv_1 \subtype \tyenv'_1 + \cdots + \tyenv'_n$ and
       $\tyb' \subtype \tyb$

       Since we also assume $\progseq{\prog}$, we know by definition typing that:

       \begin{mathpar}
           \inferrule
           {
               \tseq[\prog]{\seq{x : \tya}}{\tma}{\tyb'}
           }
           { \defseq[\prog]{\fndef{f}{\seq{x : \tya}}{\tyb'}{\tma}} }
       \end{mathpar}

       By Lemma~\ref{lem:substitution} we have that
       $\tseq{\tyenv'_1 + \cdots + \tyenv'_n}{\tma \{ \seq{V} / \seq{x} \}}{\tyb'}$.

       Thus we can recompose:

       \begin{mathpar}
           \inferrule*
           {
               \plainenv = \strip{\tyenv_1 \scomb \tyenv_2} \\
\inferrule*
               { \tseq{\tyenv'_1 + \cdots + \tyenv'_n}{\tma \{ \seq{V} / \seq{x}\}}{\tyb'} }
               { \tseq{\tyenv_1}{\tma \{ \seq{V} / \seq{x}\}}{\tyb} } \\
\stackseq{\tyenv_2}{\tyb}{\framestack}
           }
           { \cseq{\plainenv}{\thread{\tma \{ \seq{V} / \seq{x}\}}{\framestack}} }
       \end{mathpar}

       as required.
    \end{proofcase}

    \begin{proofcase}{E-New}

        Assumption:

        \begin{mathpar}
            \inferrule*
            {
                \inferrule
                {
                    \plainenv' = \strip{\scombtwo{\tyenv_1}{\tyenv_2}} \\
                    \inferrule
                    {
                        \inferrule
                        { }
                        { \tseq{\cdot}{\mbnew}{\retty{\mbrecvone}} }
                    }
                    { \tseq{\tyenv_1}{\mbnew}{\retty{\mbrecvone}} }
                    \\
                    \stackseq{\tyenv_2}{\retty{\mbrecvone}}{\framestack}
                }
                { \cseq{\plainenv'}{\thread{\mbnew}{\framestack}} }
            }
            { \cseq{\plainenv}{\thread{\mbnew}{\framestack}} }
        \end{mathpar}

        where $\plainenv \subtype \plainenv'$ and $\cruftpred{\tyenv_1}$.

        Recomposing:
        \begin{mathpar}
            \inferrule*
            {
                \inferrule*
                {
                    \inferrule
                    {
                        \plainenv', a : \mbrecvone =
                        \strip{\scombtwo{\tyenv_1, a : \retty{\mbrecvone}}{\tyenv_2}} \\ 
                        \inferrule
                        {
                            \inferrule
                            { }
                            { \tseq{a : \retty{\mbrecvone}}{a}{\retty{\mbrecvone}} }
                        }
                        { \tseq{\tyenv_1, a : \retty{\mbrecvone}}{a}{\retty{\mbrecvone}} }
                        \\
                        \stackseq{\tyenv_2}{\retty{\mbrecvone}}{\framestack}
                    }
                    { \cseq{\plainenv', a : \mbrecvone}{\thread{a}{\framestack}} }
                }
                {
                    \cseq{\plainenv'}{(\nu a)(\thread{a}{\framestack})}
                }
            }
            { \cseq{\plainenv}{(\nu a)(\thread{a}{\framestack})} }
        \end{mathpar}

        as required.
    \end{proofcase}

    \begin{proofcase}{E-Send}

        \begin{mathpar}
            \inferrule
            {
                \inferrule*
                {
                    \plainenv = \strip{\tyenv_1 \scomb \tyenv_2}
\\
                    \tseq{\tyenv_1}{\send{a}{m}{\seq{V}}}{\one}
                    \\
\stackseq{\tyenv_2}{\one}{\framestack}
                }
                { \cseq{\plainenv'}{\thread{\send{a}{m}{\seq{V}}}{\framestack}} }
            }
            { \cseq{\plainenv}{\thread{\send{a}{m}{\seq{V}}}{\framestack}} }
        \end{mathpar}

        By Lemma~\ref{lem:disjoint-from-cruft} we have that:
        \begin{itemize}
            \item $\tyenv_1 = \tyenvb_1, \tyenvb_2$
            \item $\tseq{\tyenvb_3}{\send{a}{m}{\seq{V}}}{\one}$
            \item $\tyenvb_1$ is cruftless for $\send{a}{m}{\seq{V}}$
                and $\tyenvb_1\strictsubty \tyenvb_3$
            \item $\cruftpred{\tyenvb_2}$
        \end{itemize}

        Therefore we have that $\tyenvb_3 = \tyenvb'_3, a : \tysendm$ such
        that:

        \begin{mathpar}
            \inferrule*
            {
                \vseq{a : \usablety{\tysendm}}{a}{\tysendm} \\
                \vseq{\tyenvb'_3}{\seq{V}}{\seq{\tya}} \\
                \seq{\tya} \subtype \prog(\msgtag{m})
            }
            { \tseq{\tyenvb'_3, a : \usablety{\tysendm}}{\send{a}{m}{\seq{V}}}{\one} }
        \end{mathpar}

        \[
            \bl
            \plainenv' \\
            \quad = \text{(expanding)} \\
            \strip{\tyenv_1 \scomb \tyenv_2} \\
            \quad = \text{(expanding)} \\
            \strip{(\tyenvb_1, \tyenvb_2) \scomb \tyenv_2} \\
            \quad = \text{(expanding)} \\
            \strip{(\tyenvb'_1, a : \usablety{\tysendm}, \tyenvb_2) \scomb \tyenv_2} \\
            \quad = \text{(Lemma~\ref{lem:disjoint-scomb})} \\
            \strip{(\tyenvb'_1, a : \usablety{\tysendm} \scomb \tyenvb_2) \scomb \tyenv_2} \\
            \quad = \text{(Corollary~\ref{lem:scomb-env-assoc})}\\
            \strip{\tyenvb'_1, a : \usablety{\tysendm} \scomb (\tyenvb_2 \scomb \tyenv_2)} \\
            \quad = \text{(Lemma~\ref{lem:scomb-to-pcomb})} \\
            \strip{\tyenvb'_1, a : \usablety{\tysendm}} \pcomb \strip{(\tyenvb_2 \scomb \tyenv_2)} \\
            \quad = \text{($\pcomb$ is commutative)} \\
            \strip{\tyenvb_2 \scomb \tyenv_2} \pcomb \strip{\tyenvb'_1, a :
            \usablety{\tysendm}} \\
            \el
            \]

        Recomposing:

        {\footnotesize
        \begin{mathpar}
            \inferrule*
            {
                \inferrule*
                {
                    \inferrule*
                    {
                        \vseq{\tyenvb_2}{()}{\one}
                        \\
                        \stackseq{\tyenv_2}{\one}{\framestack}
                    }
                    { \cseq{\strip{\tyenvb_2 \scomb \tyenv_2}}{\thread{()}{\framestack}} }
                    \\
                    \inferrule*
                    {
                        \vseq{a : \tysendm}{a}{\tysendm} \\
                        \vseq{\makeusable{\tyenvb'_1}}{\seq{V}}{\seq{\tya}} \\
                        \seq{\tya} \subtype \makeusable{\prog(\msgtag{m})}
                    }
                    {
                        \cseq{\strip{\tyenvb'_1, a : \tysendm}}{\sentmsg{a}{m}{\seq{V}}}
                    }
                }
                {
                    \cseq
                        {
                            \strip{\tyenvb_2 \scomb \tyenv_2} \pcomb \strip{\tyenvb'_1, a : \tysendm}
                        }
                        {\thread{()}{\framestack}
                            \parallel
                         \sentmsg{a}{m}{\seq{V}}}
                }
            }
            { \cseq{\plainenv}{\thread{()}{\framestack}
                \parallel
              \sentmsg{a}{m}{\seq{V}}}
            }
        \end{mathpar}
        }

        as required.
    \end{proofcase}

    \begin{proofcase}{E-Spawn}

        Assumption:

        \begin{mathpar}
            \inferrule*
            {
                \inferrule
                {
                    \plainenv' = \strip{\scombtwo{\makeusable{\tyenv_1}}{\tyenv_2}} \\
                    \inferrule
                    {
                        \tseq{\tyenv_1}{M}{\one}
                    }
                    { \tseq{\makeusable{\tyenv_1}}{\spawn{M}}{\one} }
                    \\
                    \stackseq{\tyenv_2}{\one}{\framestack}
                }
                { \cseq{\plainenv'}{\thread{\spawn{M}}{\framestack}} }
            }
            { \cseq{\plainenv}{\thread{\spawn{M}}{\framestack}} }
        \end{mathpar}

        By Lemma~\ref{lem:disjoint-from-cruft}, there exist $\tyenvb_1,
        \tyenvb_2, \tyenvb_3$ such that:

        \begin{itemize}
            \item $\tyenv_1 = \tyenvb_1, \tyenvb_2$
            \item $\tseq{\tyenvb_3}{M}{\one}$
            \item $\tyenvb_1$ is cruftless for $\tma$ and $\tyenvb_1 \strictsubty \tyenvb_3$
            \item $\cruftpred{\tyenvb_2}$
        \end{itemize}

By Lemma~\ref{lem:cruft-usable-typable},
        since $\vseq{\tyenvb_2}{()}{\one}$ where $\cruftpred{\tyenvb_2}$,
        it follows that $\vseq{\makeusable{\tyenvb_2}}{()}{\one}$.

        Let $\derivd$ be the following derivation:

        {\small
        \begin{mathpar}
            \inferrule
            {
                \inferrule*
                { \tseq{\cdot}{()}{\one} }
                { \tseq{\makeusable{\tyenvb_2}}{()}{\one} }
                \\
                \stackseq{\tyenv_2}{\one}{\framestack}
            }
            {
                \cseq{\strip{\scombtwo{\makeusable{\tyenvb_2}}{\tyenv_2}}}{\thread{()}{\framestack}}
            }
        \end{mathpar}
        }

        We can then construct the full derivation:
        {\footnotesize
        \begin{mathpar}
            \inferrule*
            {
                \inferrule*
                {
                    \inferrule
                    {
                        \derivd
                        \\
                        \inferrule
                        {
                            \strip{\tyenvb_1}
                            = \scombtwo{\strip{\tyenvb_1}}{\cdot} \\
                            \inferrule
                                {
                                    \tseq{\tyenvb_3}{M}{\one}
                                }
                                { \tseq{\tyenvb_1}{M}{\one} }
                            \\
                            \inferrule
                            { }
                            { \stackseq{\cdot}{\one}{\emptystack} }
                        }
                        { \cseq{\strip{\tyenvb_1}}{\thread{M}{\emptystack}} }
                    }
                    { \cseq
                        {\strip{\scombtwo{\makeusable{\tyenvb_2}}{\tyenv_2}} \pcomb
                        \strip{\tyenvb_1}}
                        {\thread{()}{\framestack} \parallel \thread{M}{\emptystack}}
                    }
                }
                { \cseq{\plainenv'}{\thread{()}{\framestack} \parallel \thread{M}{\emptystack}} }
            }
            { \cseq{\plainenv}{\thread{()}{\framestack} \parallel \thread{M}{\emptystack}} }
        \end{mathpar}
            }

        Finally we now prove that
        $\plainenv' \subtype \strip{\makeusable{\tyenvb_2} \scomb \tyenv_2} \pcomb \strip{\tyenvb_1}$.

\begin{align*}
& \plainenv' \\
& \quad = \text{(expanding)} \\
& \strip{\makeusable{\tyenv_1} \scomb \tyenv_2} \\
& \quad = \text{(expanding)} \\
& \strip{(\makeusable{\tyenvb_1}, \makeusable{\tyenvb_2}) \scomb \tyenv_2} \\
& \quad = \text{(Lemma~\ref{lem:disjoint-scomb})} \\
& \strip{(\makeusable{\tyenvb_1} \scomb \makeusable{\tyenvb_2}) \scomb \tyenv_2} \\
& \quad = \text{(Corollary~\ref{lem:scomb-env-assoc})} \\
& \strip{\makeusable{\tyenvb_1} \scomb (\makeusable{\tyenvb_2} \scomb \tyenv_2)} \\
& \quad = \text{(Lemma~\ref{lem:scomb-to-pcomb})} \\
& \strip{\makeusable{\tyenvb_1}} \pcomb \strip{(\makeusable{\tyenvb_2} \scomb \tyenv_2)} \\
& \quad = \text{($\pcomb$ is commutative)} \\
& \strip{(\makeusable{\tyenvb_2} \scomb \tyenv_2)} \pcomb \strip{\makeusable{\tyenvb_1}} \\
& \quad = \text{($\strip{-}$ cancels $\makeusable{-}$)} \\
& \strip{(\makeusable{\tyenvb_2} \scomb \tyenv_2)} \pcomb \strip{\tyenvb_1}
\end{align*}

            as required.
    \end{proofcase}

    \begin{proofcase}{E-Free}

        Assumption (assuming WLOG that the $\calcwd{free}$ guard is the first
        guard in the sequence):

        {\small
            \begin{mathpar}
                \inferrule*
                {
                    \inferrule*
                    {
                        \inferrule*
                        {
                            \plainenv', a : \tyrecv{\patb_{\var{env}}} =
                            \strip{
                                \tyenv_1, a :
                                \retty{\tyrecv{\patb_{\var{env}}}} \scomb \tyenv_2} \\
\tseq{\tyenv_1, a :
                            \retty{\tyrecv{\patb_{\mathit{env}}}}}{\guard{a}{\free{\tma} \cdot \seq{G}}}{\tya}
                            \\
                            \stackseq{\tyenv_2}{\tya}{\framestack}
                        }
                        { \cseq{\plainenv', a :
                        \tyrecv{\patb_{\var{env}}}}{\thread{\guard{a}{\free{\tma} \cdot \seq{G}}}{\framestack}} }
                    }
                    { \cseq{\plainenv, a : \mbrecvone}{\thread{\guard{a}{\free{\tma} \cdot \seq{G}}}{\framestack}} }
                }
                { \cseq{\plainenv}{\resconf{a}{\thread{\guard{a}{\free{\tma} \cdot \seq{G}}}{\framestack}}} }
            \end{mathpar}
        }

        where $\plainenv \subtype \plainenv'$, and $\mbrecvone \subtype
        \tyrecv{\patb_{\var{env}}}$, and (by Lemma~\ref{lem:returnable-non-aliased}) $a \not\in
        \dom{\tyenv_2}$.

        Furthermore:

        {\small
            \begin{mathpar}
                \inferrule*
                {
                    \inferrule*
                    {
                        \tseq{\tyenvb_1, a : \retty{\tyrecv{\patb{'}_{\var{env}}}}}{a}{\tyrecvp{\mbone \patplus \patb'}} \\
                        \inferrule*
                        {
                            \inferrule*
                            { \tseq{\tyenvb_2}{\tma}{\tya'} }
                            { \gseq{\mbone}{\tyenvb_2}{\free{\tma}}{\tya'} } \\
                            \gseq{\patb'}{\tyenvb_2}{\seq{G}}{\tya'}
                        }
                        { \gseq{\mbone \patplus \patb'}{\tyenvb_2}{\free{\tma} \cdot \seq{G}}{\tya'} }
                    }
                    {
                        \tseq
                            {\tyenvb_1, a : \retty{\tyrecv{\patb{'}_{\var{env}}}} + \tyenvb_2}
                            {\guard{a}{\free{\tma} \cdot \seq{G}}}
                            {\tya'}
                    }
                }
                { \tseq
                    {\tyenv_1, a :
                        \retty{\tyrecv{\patb_{\var{env}}}}}
                    {\guard{a}{\free{\tma} \cdot \seq{G}}}
                    {\tya}
                }
            \end{mathpar}
        }

        where $\tya' \subtype \tya$ and $\tyenv_1, a : \retty{\tyrecv{{\patb_{\var{env}}}}} \subtype \tyenvb_1, a : \retty{\tyrecv{{\patb'_{\var{env}}}}}  + \tyenvb_2$
        and $\cruftpred{\tyenvb_1}$.
        Thus, $\tyenvb_1 + \tyenvb_2 \subtype
        \tyenvb_2$ and $\tseq{\tyenvb_1 + \tyenvb_2}{\tma}{\tya'}$. Furthermore,
        $\tyenv_1 \subtype \tyenvb_1 + \tyenvb_2$.

        Thus, recomposing:

        {\small
            \begin{mathpar}
                \inferrule*
                {
                    \inferrule*
                    {
                        \plainenv' = \strip{\tyenv_1 \scomb \tyenv_2}  \\
                        \inferrule*
                        {
                            \tseq{\tyenvb_1 + \tyenvb_2}{\tma}{\tya'}
                        }
                        { \tseq{\tyenv_1}{\tma}{\tya} }
                        \\
                        \stackseq{\tyenv_2}{\tya}{\framestack}
                    }
                    { \cseq{\plainenv'}{\thread{\tma}{\framestack}} }
                }
                { \cseq{\plainenv}{\thread{\tma}{\framestack}} }
            \end{mathpar}
        }

        as required.
    \end{proofcase}

    \begin{proofcase}{E-Recv}

        Assumption:

        {\small
        \begin{mathpar}
            \inferrule*
            {
                \inferrule*
                {
                    \inferrule*
                    {
                        \plainenv_1 = \strip{\scombtwo{\tyenv_1}{\tyenv_2}} \\
\deriv{D} \\
                        \stackseq{\tyenv_2}{\tya}{\framestack}
                    }
                    { \cseq
                        {\plainenv_1}
                        {
                            \thread
                                {\guard{a}{\ctx{G}[\receive{m}{\seq{x}}{y}{M}]}}
                                {\framestack}
                        }
                    }
                    \\
                    \inferrule*
                    {
                        \prog(\msgtag{m}) = \seq{\pretyb} \\
                        \tseq{\makeusable{\plainenv_2}}{\seq{V}}{\seq{\tyb}} \\\\
                        \seq{\tyb} \subtype \makeusable{\seq{\pretyb}}
                    }
                    { \cseq{\plainenv_2, a : \tysend{\msgtag{m}}}{\sentmsg{a}{m}{\seq{V}}} }
                }
                {
                    \cseq
                        { \pcombtwo{\plainenv_1}{(\plainenv_2, a : \tysend{\msgtag{m}})} }
                        {
                            \thread
                                {\guard{a}{\ctx{G}[\receive{m}{\seq{x}}{y}{M}]}}
                                {\framestack}
                            \parallel
                            \sentmsg{a}{m}{\seq{V}}
                        }
                }
            }
            {
                \cseq
                    { \plainenv }
                    {
                        \thread
                            {\guard{a}{\ctx{G}[\receive{m}{\seq{x}}{y}{M}]}}
                            {\framestack}
                        \parallel
                        \sentmsg{a}{m}{\seq{V}}
                    }
            }
        \end{mathpar}
    }

    where $\deriv{D}$ is the following derivation:

    {\small
    \begin{mathpar}
       \inferrule*
       {
           \inferrule*
           {
               \inferrule*
               { \vseq{a : \retty{\tyrecv{\pata}}}{a}{\retty{\tyrecv{\pata}}} }
               { \vseq{\tyenv_3}{a}{\retty{\tyrecv{\pata_{\mathit{ty}}}}} } \\
               \gseq
                 {\pata_{\mathit{ty}}}
                 {\tyenv_4}
                 {\ctx{G}[\receive{m}{\seq{x}}{y}{M}]}
                 {\tya'}
               \\
               {\bl
               \tyenv'_1 = \tyenv_3 + \tyenv_4 \\
               \pnf{\pata_{\mathit{ty}}}
               \el
               }
           }
           { \tseq{\tyenv'_1}{\guard{a}{\ctx{G}[\receive{m}{\seq{x}}{y}{M}]}}{\tya'} }
       }
       {
           \tseq{\tyenv_1}{\guard{a}{\ctx{G}[\receive{m}{\seq{x}}{y}{M}]}}{\tya}
       }
    \end{mathpar}
    }

    where $\tyenv_1 \subtype \tyenv'_1$ and $\tya' \subtype \tya$.

    By Lemma~\ref{lem:disjoint-from-cruft}, $\plainenv_2 = \plainenv'_2,
    \plainenv_{\var{cruft}}$ where $\cruftpred{\plainenv_{\var{cruft}}}$ and
    $\plainenv'_2$ is cruftless for $\seq{V}$.

    Again by Lemma~\ref{lem:disjoint-from-cruft},
    $\tyenv_3 = \tyenvb, a : \retty{\tyrecv{\pata_{\mathit{env}}}}$,
    where:
    \begin{itemize}
        \item $\tyrecv{\pata_{\mathit{env}}} \subtype \tyrecv{\pata}$
        \item $\tyrecv{\pata} \subtype \tyrecv{\pata_{\mathit{ty}}}$
        \item $\cruftpred{\tyenvb}$
    \end{itemize}
    and thus
    $\tyrecv{\pata_{\mathit{env}}} \subtype \tyrecv{\pata} \subtype
    \tyrecv{\pata_{\mathit{ty}}}$.

    By the definition of $\scomb$, we also know that $a \not\in \dom{\tyenv_4}$
    and therefore $\tyenv'_1 = (\tyenvb + \tyenv_4), a : \retty{\tyrecv{\pata_{\var{env}}}}$.
    Thus $\tyenv_1 = \tyenv''_1, a : \retty{\tyrecv{\pata_{\var{sub}}}}$ where
    $\tyenv''_1 \subtype (\tyenvb + \tyenv_4)$ and
    $\retty{\tyrecv{\pata_{\var{sub}}}} \subtype \retty{\pata_{\var{env}}}$.

    By Lemma~\ref{lem:env-subtype-plus}, $\tyenv''_1 = \tyenvb' + \tyenv'_4$
    where $\tyenvb' \subtype \tyenvb$ and $\cruftpred{\tyenvb'}$ and $\tyenv'_4
    \strictsubty \tyenv_4$.

    We can therefore refine our derivations. Let $\derivd'$ be the following
    derivation:

    {\footnotesize
    \begin{mathpar}
       \inferrule*
       {
           \inferrule*
           {
               \inferrule*
               { \vseq{a : \retty{\tyrecv{\pata}}}{a}{\retty{\tyrecv{\pata}}} }
               { \vseq{\tyenvb, a : \retty{\tyrecv{\pata_{\var{env}}}}}{a}{\retty{\tyrecv{\pata_{\mathit{ty}}}}} } \\
               \gseq
                 {\pata_{\mathit{ty}}}
                 {\tyenv_4}
                 {\ctx{G}[\receive{m}{\seq{x}}{y}{M}]}
                 {\tya'}
               \\
               \pnf{\pata_{\mathit{ty}}}
           }
           {
             \tseq
                {(\tyenvb + \tyenv_4), a : \retty{\tyrecv{\pata_{\var{env}}}}}
                {\guard{a}{\ctx{G}[\receive{m}{\seq{x}}{y}{M}]}}
                {\tya'}
           }
       }
       {
           \tseq
             {\tyenvb' + \tyenv'_4, a : \retty{\tyrecv{\pata_{\var{sub}}}}}
             {\guard{a}{\ctx{G}[\receive{m}{\seq{x}}{y}{M}]}}
             {\tya}
       }
    \end{mathpar}
    }

    By Lemma~\ref{lem:returnable-non-aliased} we know that
    \[
        \bl
            (\tyenvb' + \tyenv'_4, a : \retty{\tyrecv{\pata_{\var{sub}}}}) \scomb
                \tyenv_2 \\
            \quad = \\
            a : \retty{\tyrecv{\pata_{\var{sub}}}} + ((\tyenvb' + \tyenv'_4) \scomb
                \tyenv_2) \\
            \quad = \\
            ((\tyenvb' + \tyenv'_4) \scomb \tyenv_2),
            a : \retty{\tyrecv{\pata_{\var{sub}}}}
        \el
    \]

    By the definition of $\pcomb$, we know that
    $\pata_{\mathit{sub}} = \msgtag{m} \patconcat \pata_{\mathit{pat}}$ for some
    pattern $\pata_{\mathit{pat}}$,
    and therefore that
    $\plainenv_1 = \strip{((\tyenvb' + \tyenv'_4) \scomb \tyenv_2)},
      a : \tyrecvp{\msgtag{m} \patconcat \pata_{\var{pat}}}$.

    Our overall derivation is then:

    {\footnotesize
    \begin{mathpar}
            \inferrule*
            {
                \inferrule*
                {
                    \inferrule*
                    {
                        \plainenv' = \strip{\scombtwo{ \tyenvb' + \tyenv'_4, a :
                        \retty{\tyrecvp{\msgtag{m} \patconcat \pata_{\var{pat}}}} }{\tyenv_2}} \\
\derivd' \\
                        \stackseq{\tyenv_2}{\tya}{\framestack}
                    }
                    { \cseq
                        {\plainenv_1
                        }
                        {
                            \thread
                                {\guard{a}{\ctx{G}[\receive{m}{\seq{x}}{y}{M}]}}
                                {\framestack}
                        }
                    }
                    \\
                    \inferrule*
                    {
                        \prog(\msgtag{m}) = \seq{\pretyb} \\\\
                        \tseq{\makeusable{\plainenv'_2}, \makeusable{\plainenv_{\var{cruft}}}}{\seq{V}}{\seq{\tyb}} \\\\
                        \seq{\tyb} \subtype \makeusable{\seq{\pretyb}}
                    }
                    { \cseq{\plainenv'_2, \plainenv_{\var{cruft}}, a : \tysend{\msgtag{m}}}{\sentmsg{a}{m}{\seq{V}}} }
                }
                {
                    \cseq
                    { (\pcombtwo
                        {\plainenv_1}
                        {(\plainenv'_2, \plainenv_{\var{cruft}})})
                        , a : \tyrecv{\pata_{\var{pat}}}
                    }
                    {
                        \thread
                            {\guard{a}{\ctx{G}[\receive{m}{\seq{x}}{y}{M}]}}
                            {\framestack}
                        \parallel
                        \sentmsg{a}{m}{\seq{V}}
                    }
                }
            }
            {
                \cseq
                    { \plainenv }
                    {
                        \thread
                            {\guard{a}{\ctx{G}[\receive{m}{\seq{x}}{y}{M}]}}
                            {\framestack}
                        \parallel
                        \sentmsg{a}{m}{\seq{V}}
                    }
            }
    \end{mathpar}
    }

    Without loss of generality, let us consider the case where the
    $\calcwd{receive}$ is the first guard.
We can therefore write $\ctx{G}[\receive{m}{\seq{x}}{y}{M}]$ as
    $\receive{m}{\seq{x}}{y}{M} \cdot \seq{G}$ for some sequence $\seq{G}$.

    By \textsc{T-GuardSeq} and \textsc{TG-Recv}, and since
    $\pnf{\pata_{\mathit{ty}}}$, we have that $\pata_{\mathit{ty}} = \patb_1
    \patplus \cdots \patplus \patb_n$, where $\patb_1 = \msgtag{m} \patconcat
    \patb'$ and $\patb' \tyequiv \pata_{\mathit{ty}} \without \msgtag{m}$.

    Furthermore:

    \begin{mathpar}
        \inferrule
        {
            \prog(\msgtag{m}) = \seq{\pretyb} \\
\mbbase{\tyenv_4} \vee \mbbase{\seq{\makeusable{\pretyb}}}\\
\tseq
                {\tyenv_4, \seq{x} : \seq{\makeusable{\pretyb}},
                    y : \retty{\tyrecv{\patb'}}}
                {\tma}
                {\tya'}
        }
        { \gseq
            {(\msgtag{m} \patconcat \patb')}
            {\tyenv_4}
            {\receive{m}{\seq{x}}{y}{\tma}}
            {\tya'}
        }
    \end{mathpar} 

    By Lemma~\ref{lem:balancing}, we have that
    $\pata_{\mathit{pat}} \subpatone (\pata_{\var{ty}} \without \msgtag{m})$, and thus
    $\retty{\tyrecv{\pata_{\mathit{pat}}}} \subtype
    \retty{\tyrecvp{\pata_{\var{ty}} \without \msgtag{m}}} \subtype
    \retty{\tyrecv{\patb'}}$.

    Since either $\mbbase{\tyenv_4}$ or $\mbbase{\seq{\makeusable{\pretyb}}}$,
    we know that $\makeusable{\plainenv'_2} + \tyenv_4$ is defined.

    By Lemma~\ref{lem:substitution},
    $\tseq{\makeusable{\plainenv'_2} + \tyenv_4, a :
    \retty{\tyrecv{\patb'}}}{\tma \{ \seq{V} / \seq{x}, a / y \}}{\tya'}$
    and by \textsc{T-Sub} it follows that
    $\tseq{\makeusable{\plainenv'_2} + \tyenv'_4, a :
    \retty{\tyrecv{\pata_{\var{pat}}}}}{\tma \{ \seq{V} / \seq{x}, a / y \}}{\tya}$.

    By equational reasoning on environments:

        \begin{align*}
            & \plainenv_1 \pcomb (\plainenv_2, a : \tysend{\msgtag{m}}) \\
            & \quad = (\text{expanding}) \\
            & (\strip{((\tyenvb' + \tyenv'_4) \scomb \tyenv_2)},
                      a : \tyrecvp{\msgtag{m} \patconcat \pata_{\var{pat}}})
                      \pcomb (\plainenv'_2, \plainenv_{\var{cruft}}, a :
                      \tysend{\msgtag{m}}) \\
            & \quad = (\text{def.\ } \pcomb) \\
            & (\strip{((\tyenvb' + \tyenv'_4) \scomb \tyenv_2)}
                      \pcomb (\plainenv'_2, \plainenv_{\var{cruft}})), a : \tyrecv{\pata_{\var{pat}}}
                      \\
            &      \quad = (\text{Lemma~\ref{lem:pcomb-to-scomb}}) \\
            &      (\strip{\makeusable{(\plainenv'_2, \plainenv_{\var{cruft}})}
                  \scomb ((\tyenvb' + \tyenv'_4) \scomb \tyenv_2)}),
                  a : \tyrecv{\pata_{\var{pat}}}
                  \\
            & \quad = (\text{def.\ } \makeusable{-}) \\
            &    (\strip{(\makeusable{\plainenv'_2}, \makeusable{\plainenv_{\var{cruft}}})
                  \scomb ((\tyenvb' + \tyenv'_4) \scomb \tyenv_2)}),
                  a : \tyrecv{\pata_{\var{pat}}}
                  \\
            &      \quad = (\text{Lemma}~\ref{lem:scomb-assoc}) \\
            &    \strip{((\makeusable{\plainenv'_2}, \makeusable{\plainenv_{\var{cruft}}})
                  \scomb (\tyenvb' + \tyenv'_4)) \scomb \tyenv_2},
                  a : \tyrecv{\pata_{\var{pat}}}
                  \\
            & \quad = (\text{Lemma}~\ref{lem:scomb-inside-plus})
            \\
            & \strip{(\makeusable{\plainenv_{\var{cruft}}}
                  \scomb ((\makeusable{\plainenv'_2} \scomb \tyenvb') + \tyenv'_4)) \scomb \tyenv_2},
                  a : \tyrecv{\pata_{\var{pat}}}
            \\
            & \quad = (a \text{ disjoint})
            \\
            & \strip{(\makeusable{\plainenv_{\var{cruft}}}
                  \scomb ((\makeusable{\plainenv'_2} \scomb \tyenvb') +
              \tyenv'_4, a : \tyrecv{\pata_{\var{pat}}})) \scomb \tyenv_2} 
          \end{align*}

    We now need to show that typability of the guard body is maintained by cruft environments:
    by Corollary~\ref{cor:cruft-inside-plus} we have that
    $\tseq{(\makeusable{\plainenv'_2} \scomb \tyenvb') + \tyenv'_4, a :
    \retty{\tyrecv{\pata_{\var{pat}}}}}{\tma \{ \seq{V} / \seq{x}, a / y \}}{\tya'}$.
    
    By Lemma~\ref{lem:cruft-comb-typing} we have that 
    $\tseq{\makeusable{\plainenv_{\var{cruft}}} \scomb ((\makeusable{\plainenv'_2} \scomb \tyenvb') + \tyenv'_4, a :
    \retty{\tyrecv{\pata_{\var{pat}}}})}{\tma \{ \seq{V} / \seq{x}, a / y \}}{\tya'}$.

    Let $\plainenv' = \strip{(\makeusable{\plainenv_{\var{cruft}}}
                  \scomb ((\makeusable{\plainenv'_2} \scomb \tyenvb') +
              \tyenv'_4, a : \tyrecv{\pata_{\var{pat}}})) \scomb \tyenv_2}$
              (noting by equational reasoning that $\plainenv' = \plainenv_1
              \pcomb (\plainenv_2, a : \tysend{\msgtag{m}})$).

    Thus we can construct:
    {\footnotesize
    \begin{mathpar}
        \inferrule*
        {
            \inferrule*
            {
                \plainenv'
                =
                    \strip{(\makeusable{\plainenv_{\var{cruft}}}
                                      \scomb ((\makeusable{\plainenv'_2} \scomb \tyenvb') +
                      \tyenv'_4, a : \tyrecv{\pata_{\var{pat}}})) \scomb \tyenv_2} 
                \\
                \tseq
                    {\makeusable{\plainenv_{\var{cruft}}} \scomb 
                        ((\makeusable{\plainenv'_2} \scomb \tyenvb') + 
                        \tyenv'_4, a : \retty{\tyrecv{\pata_{\var{pat}}}})}
                    {\tma \{ \seq{V} / \seq{x}, a / y \}}
                    {\tya'}
                \\
                \stackseq{\tyenv_2}{\tya}{\framestack}
            }
            { \cseq
                {\plainenv' }
                {\thread{M \{ \seq{V} / \seq{x}, a / y \}}{\framestack}}
            }
        }
        {
            \cseq{\plainenv}{\thread{M \{ \seq{V} / \seq{x}, a / y \}}{\framestack}}
        }
    \end{mathpar}
    }

    as required.

\end{proofcase}

    \begin{proofcase}{E-Nu}
        Follows immediately from the induction hypothesis.
    \end{proofcase}

    \begin{proofcase}{E-Par}
        Follows immediately from the induction hypothesis.
    \end{proofcase}

    \begin{proofcase}{E-Struct}
        Follows immediately from Lemma~\ref{lem:equiv-pres} and the induction
        hypothesis.
    \end{proofcase}
\end{proof}

 \newpage
\subsection{Proofs for \S\ref{sec:algo:alg-soundness} (Algorithmic Soundness)}
\label{sec:soundness}

The pattern variables in an inferred environment must either occur in the type,
program, or constraint set.

Lemma~\ref{lem:subtype-soundness} shows the soundness of algorithmic subtyping.
As a direct corollary, we can show that constraints generated by equivalence
preserve subtyping in both directions.

\begin{corollary}
    \label{lem:equiv-soundness}
    If
    $\equivseq{\ptya}{\ptyb}{\constrs}$ and
    $\patsubst$ is a usable solution of $\constrs$ with $\pv{\ptyb} \subseteq
    \dom{\patsubst}$,
    then both
    $\subtypetwo{\apppatsubst{\ptya}}{\apppatsubst{\ptyb}}$ and
    $\subtypetwo{\apppatsubst{\ptyb}}{\apppatsubst{\ptya}}$.
\end{corollary}

If two environments are combinable, and we have a solution for the constraints
generated by their algorithmic combination, then their combination is defined.

\begin{lemma}
    \label{lem:combine-sol}
    If $\combineseq{\penv_1}{\penv_2}{\penv}{\constrs}$ and
    $\patsubst$ is a usable solution of $\constrs$
    where $\pv{\penv_1} \cup \pv{\penv_2} \subseteq \dom{\patsubst}$,
    there exist $\tyenv_1$ and $\tyenv_2$ such that
    $\tyenv_1 \subtype \apppatsubst{\penv_1}$ and
    $\tyenv_2 \subtype \apppatsubst{\penv_2}$ and
    $\tyenv_1 + \tyenv_2 = \apppatsubst{\penv}$.
\end{lemma}
\begin{proof}
    By induction on the derivation of
    $\combineseq{\penv_1}{\penv_2}{\penv}{\constrs}$.

    \begin{proofcase}{$\combineseq{\penv_1, x : \ptya}{\penv_2}{\penv}{\constrs}$}

        Assumption:
    \begin{mathpar}
        \inferrule
        {
            x \not\in \dom{\penv_2} \\
            \combineseq{\penv_1}{\penv_2}{\penv}{\constrs}
        }
        { \combineseq{\penv_1, x : \ptya}{\penv_2}{\penv, x : \ptya}{\constrs} }
    \end{mathpar}

    We also assume that $\patsubst$ is a usable solution for $\constrs$.

    By the IH, we have that there exists some $\tyenv \subtype
    \apppatsubst{\penv_2}$ such that
    $\apppatsubst{\penv_1} + \tyenv = \apppatsubst{\penv}$.

    Since $x \not\in \dom{\penv_2}$, by the definition of $+$ in the declarative
    setting, we have that
    \[
        \apppatsubst{\penv_1}, x : \apppatsubst{\ptya} + \apppatsubst{\penv_2}
        \subtype
        \apppatsubst{\penv}, x : \apppatsubst{\ptya}
    \]
    as required.
    \end{proofcase}

    \begin{proofcase}{$\combineseq{\penv_1}{\penv_2, x : \ptya}{\penv}{\constrs}$}
        Symmetric to the first case.
    \end{proofcase}

    \begin{proofcase}
        {$\combineseq{\penv_1, x : \ptya}{\penv_2, x : \ptyb}{\penv}{\constrs}$}

        Assumption:
        \begin{mathpar}
            \inferrule
            {
                \combineseq{\penv_1}{\penv_2}{\penv}{\constrs_1} \\
                \equivseq{\ptya}{\ptyb}{\constrs_2} \\\\
                \unrseq{\ptya}{\constrs_3} \\
                \unrseq{\ptyb}{\constrs_4}
            }
            { \combineseq
                {\penv_1, x : \ptya}
                {\penv_2, x : \ptyb}
                {\penv, x : \ptya}
                {\constrs_1 \cup \cdots \cup \constrs_4}
            }
        \end{mathpar}

        We also assume that $\patsubst$ is a usable solution for
        $\constrs_1 \cup \cdots \cup \constrs_4$.

        By the IH, there exists some $\tyenv$ such that $\tyenv \subtype
        \apppatsubst{\penv_2}$ and
        \[
            \apppatsubst{\penv_1} + \tyenv = \apppatsubst{\penv}
        \]

        By the definitions of $\equivsymb$ and $\mkwd{unr}(-)$, and knowing that
        $\patsubst$ is a usable solution for $\constrs_2 \cup
        \constrs_3 \cup \constrs_4$, we have that
        either $\ptya = \ptyb = \basety$ for some base type $\basety$ (in which case we can
        conclude with logic similar to the previous case),
        or $\ptya = \usablety{\tysend{\ppata}}$ and
        $\ptyb = \usablety{\tysend{\ppatb}}$ where
        $\subpat{\apppatsubst{\ppata}, \apppatsubst{\ppatb}}{\mbone}$.

        Since $\patsubst$ is usable, we know that
        $\apppatsubst{\ppata}, \apppatsubst{\ppatb} \not\subpatone \mbzero$.
        Therefore, we have that $\ptya, \ptyb \subtype
        \usablety{\tysend{\mbone}}$.

        We can then show that
        $\tyenv, x :
            \tyenv, x : \apppatsubst{\usablety{\tysend{\ptya}}} \subtype
            \apppatsubst{\penv_2}, x : \apppatsubst{\usablety{\tysend{\ptyb}}}$

        and further that
        $\apppatsubst{\penv_1}, x : \apppatsubst{\usablety{\tysend{\ptya}}} +
         \tyenv, x : \apppatsubst{\usablety{\tysend{\ptya}}} =
         \apppatsubst{\penv}, x :
         \apppatsubst{\usablety{\tysend{\ptya}}}
        $
        as required.
    \end{proofcase}

\end{proof}

We can generalise the previous result to an $n$-ary combination:

\begin{corollary}\label{cor:combine-many-sol}
If $\combineseqmany{\penv_1 + \cdots + \penv_n}{\penv}{\constrs}$
where $\patsubst$ is a usable solution for $\constrs$ such that
$\pv{\penv_1} \cup \cdots \cup \pv{\penv_n} \subseteq \dom{\patsubst}$,
then there exist $(\tyenv_i \subtype \penv_i)_i$ such that
$\tyenv_1 + \cdots + \tyenv_n = \aps{\penv}$.
\end{corollary}

We now turn our attention to the relation between the algorithmic join and type
combination operators.

\scombsol*
\begin{proof}
    By case analysis on the derivation of
    $\joinseq{\ptya_1}{\ptya_2}{\ptyb}{\constrs}$.

    \begin{proofcase}{
        $\joinseq
            {\usageann{\usage_1}{\tysend{\ppata}}}
            {\usageann{\usage_2}{\tysend{\ppatb}}}
            {\usageann{\usage_1 \scomb \usage_2}{\tysendp{\patconcattwo{\ppata}{\ppatb}}} }
            {\emptyset}$}

        We can immediately conclude with
            $\scombtwo
                {\usageann{\usage_1}{\tysendp{\apppatsubst{\ppata}}}}
                {\usageann{\usage_2}{\tysendp{\apppatsubst{\ppatb}}}}
              =
             \usageann
                {\usage_1 \scomb \usage_2}
                {\tysendp{\patconcattwo{\apppatsubst{\ppata}}{\apppatsubst{\ppatb}}}}
            $
         as required.
    \end{proofcase}

    \begin{proofcase}
    {
        $\joinseq
            {\usageann{\usage_1}{\tysend{\ppata}}}
            {\usageann{\usage_2}{\tyrecv{\ppatb}}}
            {\usageann{\scombtwo{\usage_1}{\usage_2}}{\tyrecvp{\patvara}}}
            {\subpatconstr{\patconcattwo{\ppata}{\patvara}}{\ppatb}}$
    }

    Since $\patsubst$ is a solution for
    $\subpatconstr{\patconcattwo{\ppata}{\patvara}}{\ppatb}$,
    we have that $\aps{\patconcattwo{\ppata}{\patvara}} \subpatone
    \aps{\ppatb}$.

    By expansion of $\aps{-}$, we have that
    $\patconcattwo{\aps{\ppata}}{\aps{\patvara}} \subpatone \aps{\ppatb}$.

    Since receive mailbox types are covariant in their patterns, we can show
    that
    $\tyrecvp{\patconcattwo{\aps{\ppata}}{\aps{\patvara}}}
        \subtype
     \tyrecv{\aps{\ppatb}}$

     and we can conclude that
     \[
         \scombtwo
            {\usageann{\usage_1}{\tysendp{\aps{\ppata}}}}
            {\usageann{\usage_2}{\tyrecvp{\patconcattwo{\aps{\ppata}}{\aps{\patvara}}}}}
            =
        \usageann{\scombtwo{\usage_1}{\usage_2}}{\tyrecv{\aps{\patvara}}}
     \]
     as required.
    \end{proofcase}

    \begin{proofcase}
    {
        $\joinseq
            {\tyrecv{\ppata}}
            {\tysend{\ppatb}}
            {\tyrecv{\patvara}}
            {\subpatconstr{\patconcattwo{\ppatb}{\patvara}}{\ppata}}$
    }

    Symmetric to the previous case.
    \end{proofcase}

    \begin{proofcase}
        {
            $
            \joinseq
            {\ptya}
            {\ptyb}
            {\ptya}
            {\constrs}
            $
        }

        Assumption: $\ptya$, $\ptyb$ are not mailbox types and
        $\equivseq{\ptya}{\ptyb}{\constrs}$.

        By Lemma~\ref{lem:subtype-soundness},
        $\tyequivtwo{\apppatsubst{\ptya}}{\apppatsubst{\ptyb}}$.
        Since neither type is a mailbox type we have that $\aps{\ptya} = \ptya =
        \ptyb = \aps{\ptyb} = \basety$ for some base type $\basety$, as
        required.
    \end{proofcase}
\end{proof}

We can extend this result to environments.

\begin{lemma}
    \label{lem:scomb-env-sol}
    If $\joinseq{\penv_1}{\penv_2}{\penv}{\constrs}$
    and $\patsubst$ is a usable solution of $\constrs$
    such that $\pv{\penv_1} \cup \pv{\penv_2} \subseteq \dom{\patsubst}$, then there exist
    $\tyenv_1 \subtype \apppatsubst{\penv_1}$
    and
    $\tyenv_2 \subtype \apppatsubst{\penv_2}$
    such that
    $\scombtwo{\tyenv_1}{\tyenv_2} = \apppatsubst{\penv}$.
\end{lemma}
\begin{proof}
    A direct consequence of Lemma~\ref{lem:scomb-sol}.
\end{proof}

\mergesol*
\begin{proof}
    By case analysis on the derivation of
    $\mergeseq{\ptya_1}{\ptya_2}{\ptyb}{\constrs}$.
The case for base types is straightforward.
    For two mailbox types $\usageann{\usage_1}{\mbtya}$ and
    $\usageann{\usage_2}{\mbtya}$, since $\returnable \subtype \usable$
    it is always the case that
    $\minusage{\usage_1}{\usage_2} \subtype \usage_1$ and
    $\minusage{\usage_1}{\usage_2} \subtype \usage_2$, so therefore it suffices
    to consider the non-usage-annotated merge
    $\mergeseq{\ppretya_1}{\ppretya_2}{\ppretyb}{\emptyset}$

    \begin{proofcase}{
        $
        \mergeseq
          { \tysend{\ppata} }
          { \tysend{\ppatb} }
          { \tysendp{\patplustwo{\ppata}{\ppatb}} }
          { \emptyset}
        $
        }

        By the definition of $\sem{-}$ we have that
        $\sem{\patplustwo{\apppatsubst{\ppata}}{\apppatsubst{\ppatb}}} =
            \sem{\apppatsubst{\ppata}} \cup \sem{\apppatsubst{\ppatb}}$ and
            therefore
            ${\sem{\apppatsubst{\ppata}}} \subseteq
                 {\sem{\patplustwo{\apppatsubst{\ppata}}{\apppatsubst{\ppatb}}}}$
            and
             $\sem{\apppatsubst{\ppatb}} \subseteq
                 \sem{\patplustwo{\apppatsubst{\ppata}}{\apppatsubst{\ppatb}}}$.
        By the definition of pattern inclusion it follows that
        ${\apppatsubst{\ppata}} \subpatone
                ({\patplustwo{\apppatsubst{\ppata}}{\apppatsubst{\ppatb}}})$ and
        $\apppatsubst{\ppatb} \subpatone
                 (\patplustwo{\apppatsubst{\ppata}}{\apppatsubst{\ppatb}})$.

        Since output mailbox types are contravariant in their
        patterns, it follows that both
        $\subtypetwo
                {\tysendp{\apppatsubst{\ppata} \patplus \apppatsubst{\ppatb}}}
                {\tysendp{\apppatsubst{\ppata}}}$ and
        $\subtypetwo
                {\tysendp{\apppatsubst{\ppata} \patplus \apppatsubst{\ppatb}}}
                {\tysendp{\apppatsubst{\ppatb}}}$
        as required.
    \end{proofcase}

    \begin{proofcase*}
        {
            $
                \mergeseq
                { \tyrecv{\ppata} }
                { \tyrecv{\ppatb} }
                { \tyrecv{\patvara} }
                { \set
                    {
                        \subpatconstr{\patvara}{\ppata},
                        \subpatconstr{\patvara}{\ppatb}
                    }
                }
            $
        (where $\patvara$ fresh).
        }

        Since $\patsubst$ is a usable solution, we know
        $\subpat{\apppatsubst{\patvara}}{\apppatsubst{\ppata}}$ and
        $\subpat{\apppatsubst{\patvara}}{\apppatsubst{\ppatb}}$.
Since input mailbox types are covariant in their pattern arguments, it
        follows that both
        $\tyrecvp{\apppatsubst{\patvara}} \subtype \tyrecvp{\apppatsubst{\ppata}}$
        and
        $\tyrecvp{\apppatsubst{\patvara}} \subtype \tyrecvp{\apppatsubst{\ppatb}}$
        as required.
    \end{proofcase*}
\end{proof}

\begin{lemma}
    \label{lem:merge-env-sol}
    If $\mergeseq{\penv_1}{\penv_2}{\penv}{\constrs}$ and
    $\patsubst$ is a usable solution of $\constrs$
    such that $\pv{\penv_1} \cup \pv{\penv_2} \subseteq \dom{\patsubst}$,
    then
    $\apppatsubst{\penv} \subtype \apppatsubst{\penv_1}$ and
    $\apppatsubst{\penv} \subtype \apppatsubst{\penv_2}$.
\end{lemma}
\begin{proof}
    By induction on the derivation of
    $\mergeseq{\penv_1}{\penv_2}{\penv}{\constrs}$ with appeal to
    Lemma~\ref{lem:merge-soundness}.
\end{proof}

We can apply environment subtyping to guards through by appeal to \textsc{T-Sub}.

\begin{lemma}\label{lem:guard-subtype}
    If $\gseq{\pata}{\tyenv}{\gd}{\tya}$ and $\tyenv' \subtype \tyenv$, then
    $\gseq{\pata}{\tyenv'}{\gd}{\tya}$.
\end{lemma}
\begin{proof}
    By case analysis on the derivation of
    $\gseq{\pata}{\tyenv}{\gd}{\tya}$ with appeal to \textsc{T-Sub} on guard
    bodies, noting that since we work up-to Barendregt's convention we can
    rename any variables bound in the body of a $\calcwd{receive}$ clause that
    also appear in $\tyenv'$.
\end{proof}

As a corollary, the same result holds for guard sequences.

\begin{corollary}\label{cor:guard-subtypes}
    If $\gseq{\pata}{\tyenv}{\seq{\gd}}{\tya}$ and $\tyenv' \subtype \tyenv$, then
    $\gseq{\pata}{\tyenv'}{\seq{\gd}}{\tya}$.
\end{corollary}

The next lemma allows us to reconcile the guard annotation with the sequence of
PNF literals produced by each guard rule.

\begin{lemma}\label{lem:equiv-pnf}
    If $(\pnflittwo{\pata}{\patb_i})_{i \in 1..n}$ for pattern $\pata$ and
    patterns $\patb_1, \ldots, \patb_n$,
    then $\pnf{\patb_1 \patplus \cdots \patplus \patb_n}$.
\end{lemma}
\begin{proof}
    Let $\patb = \patb_1 \patplus \cdots \patplus \patb_n$. 

    We need to show that
    $\pnftwo
        {\patb_1 \patplus \cdots \patplus \patb_n}
        {\patb_1 \patplus \cdots \patplus \patb_n}$.

    Pick an arbitrary $\patb_i$. Our goal is to show that
    $\pnflittwo{\patb}{\patb_i}$; since $\patb_i$ is picked arbitrarily, if
    $\pnflittwo{\patb}{\patb_i}$ then we can show by construction that
    $\pnftwo{\patb}{\patb_1 \patplus \cdots \patplus \patb_n}$
    and therefore that $\pnf{\patb}$.

    Since $\pnflittwo{\pata}{\patb_i}$ there are two cases:

    \begin{proofcase*}{$\patb_i = \mbone$}
        In this case, $\pnflittwo{\patb}{\mbone}$ as required.
    \end{proofcase*}

    \begin{proofcase*}{$\patb_i = \msgtag{m} \patconcat \patb'_i$}
        In this case we have:

        \begin{mathpar}
            \inferrule*
            { \pata \without \msgtag{m} \tyequiv \patb'_i }
            { \pnflittwo{\pata}{\msgtag{m} \patconcat \patb'_i} }
        \end{mathpar}

        Consider the subset of other pattern literals
        $\set{\msgtag{m} \patconcat \patb''_j}_{j \in 1..m}$
        with tag $\msgtag{m}$. 
For each of these we also have that
        $\pata \without \msgtag{m} \tyequiv \patb''_j$ and so it follows
        that $\patb'_i \tyequiv \patb''_j$ for all $j \in 1..m$.
        It follows by the definition of pattern semantics
        that $(\patb'_i \patplus \patb''_j) \tyequiv \patb'_i$.

        Conversely, consider the subset of other pattern literals
        $\set{\msgtag{n} \patconcat \patb''_k}_{k \in 1..p}$ such that
        $\msgtag{n} \ne \msgtag{m}$. In these cases, we have that
        $(\msgtag{n} \patconcat \patb''_k) \without \msgtag{m} = \mbzero$.
        Similarly, for all literals such that $\pnflittwo{\pata}{\mbone}$ it
        follows that $\mbone \without \msgtag{m} = \mbzero$ and therefore
        $\patb'_i \patplus \mbzero \tyequiv \patb'_i$.

        Consequently we have that $\patb \without \msgtag{m} \tyequiv \patb'_i$
        and we can show that:

        \begin{mathpar}
            \inferrule*
            { \patb \without \msgtag{m} \tyequiv \patb'_i }
            { \pnflittwo{\patb}{\msgtag{m} \patconcat \patb'_i} }
        \end{mathpar}

        as required.
    \end{proofcase*}
\end{proof}

Algorithmic soundness relies on the following generalised result:

\begin{lemma}[Algorithmic soundness (generalised)]
    \label{lem:algo-soundness-generalised}
    \hfill
    \begin{enumerate}
        \item If $\patsubst$ is a covering solution of
            $\synthseq[\pprog]{\ptma}{\ptya}{\penv}{\constrs}$ 
            then $\tseq
               [\aps{\erase{\pprog}}]
               {\apppatsubst{\penv}}
               {\erase{\ptma}}
               {\apppatsubst{\ptya}}$.
           \item If $\patsubst$ is a covering solution of
               $\chkseq[\pprog]{\ptma}{\ptya}{\penv}{\constrs}$ 
               then
           $\tseq
           [\aps{\erase{\pprog}}]
               {\apppatsubst{\penv}}
               {\erase{\ptma}}
               {\apppatsubst{\ptya}}$.
\item If $\patsubst$ is a covering solution of
               $\chkgseq[\pprog]{\pata}{\pgd}{\ptya}{\penv}{\constrs}{\patb}$
               then
           $\gseq
               [\apppatsubst{\erase{\pprog}}]
               {\patb}
               {\apppatsubst{\penv}}
               {\erase{\pgd}}
               {\apppatsubst{\ptya}}$ and
            $\pnflittwo{\pata}{\patb}$.
\item If $\patsubst$ is a covering solution of
               $\chkgseq[\pprog]{\pata}{\seq{\pgd}}{\ptya}{\penv}{\constrs}{\patb}$
               then
           $\gseq
               [\apppatsubst{\erase{\pprog}}]
               {\patb}
               {\apppatsubst{\penv}}
               {\erase{\pgd}}
               {\apppatsubst{\ptya}}$ and
            $\pnf{\patb}$.
    \end{enumerate}
\end{lemma}
\begin{proof}
    By mutual induction on all statements. 
    We know in all cases that the solution covers the pattern variables in
    the program, return type, and constraints. Therefore by
    Lemma~\ref{lem:penv-constrs-subset} we know that any produced environment
    will contain pattern variables contained in the solution. We make use of
    this fact implicitly throughout the proof.

    \paragraph*{Statement 1: Synthesis.}

    \begin{proofcase}{TS-Base}
        Assumption:

        \begin{mathpar}
            \inferrule
            { c \text{ has base type } D}
            { \synthseq{c}{D}{\cdot}{\emptyset} }
        \end{mathpar}

        By \textsc{T-Const}:

        \begin{mathpar}
            \inferrule
            { }
            { \tseq{\cdot}{c}{D} }
        \end{mathpar}

        noting that:
        \begin{itemize}
            \item $\un{\cdot}$
            \item $\apppatsubst{c} = c$
            \item $\apppatsubst{D} = D$
        \end{itemize}

        as required.
    \end{proofcase}
    \begin{proofcase}{TS-Unit}
        Similar to \textsc{TS-Base}.
    \end{proofcase}
    \begin{proofcase}{TS-New}
        Similar to \textsc{TS-Base}.
    \end{proofcase}
    \begin{proofcase}{TS-Spawn}
        Assumption:

        \begin{mathpar}
            \inferrule
            {
                \chkseq{\ptma}{\one}{\penv}{\constrs}
            }
            { \synthseq{\spawn{\ptma}}{\one}{\makeusable{\penv}}{\constrs} }
        \end{mathpar}

        Furthermore, we assume that $\patsubst$ is a usable solution for $\constrs$.

        By the IH (2),
        \[
            \tseq{\apppatsubst{\penv}}{\erase{\ptma}}{\one}
        \]

        Recomposing by \textsc{T-Spawn}:

        \begin{mathpar}
            \inferrule
            {
                \tseq{\apppatsubst{\penv}}{\erase{\ptma}}{\one}
            }
            {
                \tseq
                    {\makeusable{\apppatsubst{\penv}}}
                    {\spawn{(\erase{\ptma})}}
                    {\one}
            }
        \end{mathpar}

        as required.
    \end{proofcase}
    \begin{proofcase}{TS-Send}
        Assumption:
        \begin{mathpar}
            \inferrule
            {
                \proglookup{\msgtag{m}} = \seq{\ppretya} \\
                \chkseq{\vala}{\usablety{\tysend{\algomsg{m}}}}{\penv'}{\constrs} \\\\
                (\chkseq{\pvalb_i}{\makeusable{\ppretya_i}}{\penv'_i}{\constrs'_i})_{i \in 1..n} \\
                \combineseqmany{\penv' + \penv'_1 + \cdots + \penv'_n}{\penv}{\constrs''}
            }
            { \synthseq
                {\send{\vala}{m}{\seq{\pvalb}}}
                {\one}
                {\penv}
                {\constrs \cup \constrs'_1 \cup \cdots \cup \constrs'_n \cup \constrs''}
            }
        \end{mathpar}

        Also, we assume $\algoprogseq{\pprog}{\constrs_{\var{prog}}}$.

        Furthermore, we assume that $\patsubst$ is a solution for
        $\constrs_{\var{prog}} \cup \constrs \cup \constrs'_1 \cup \cdots \cup \constrs'_n \cup \constrs''$.
        By Lemma~\ref{lem:pat-subset}, we have that $\patsubst$ is also a
        solution for each constraint set individually.

        Thus, by the IH:

        \begin{itemize}
            \item
                $\tseq
                    { \apppatsubst{\penv'} }
                    { \vala }
                    { \usablety{\tysend{\msgtag{m}}} }$
            \item
                $\tseq
                    { \apppatsubst{\penv'_i} }
                    { \valb_i }
                    { \makeusable{\apppatsubst{\ppretya_i}}}$ for $i \in 1..n$
        \end{itemize}

        By Corollary~\ref{cor:combine-many-sol},
        there exist $ \tyenv' \subtype \aps{\penv'}$ and
        $\tyenv'_i \subtype \aps{\penv'_i}$ for $i \in 1..n$ such that
        $\tyenv' + \tyenv'_1 + \cdots + \tyenv'_n = \aps{\penv}$.
Therefore:

        \begin{mathpar}
            \inferrule
            {
              \proglookup[\aps{\erase{\pprog}}]{\msgtag{m}} = \seq{\aps{\ppretya}} \\
              \inferrule
              { \tseq{\aps{\penv'}}{\vala}{\usablety{\tysend{\algomsg{m}}}} }
              { \tseq{\tyenv'}{\vala}{\usablety{\tysend{\algomsg{m}}}} } \\
\inferrule
              { (\tseq{\aps{\penv'_i}}{\valb_i}{\aps{\ppretya_i}})_{i \in 1..n} }
              { (\tseq{\tyenv'_i}{\valb_i}{\aps{\ppretya_i}})_{i \in 1..n} } \\
            }
            { \tseq
                {\tyenv' + \tyenv'_1 + \cdots +  \tyenv'_n }
                {\send{\vala}{m}{\seq{\valb}}}
                {\one}
            }
        \end{mathpar}

        as required.
    \end{proofcase}
    \begin{proofcase}{TS-App}

        Assumption:
        \begin{mathpar}
            \inferrule
            [TS-App]
            {
                \pproglookup{f} = \fndef{f}{\seq{x : \ptya}}{\ptyb}{\ptma} \\\\
                (\chkseq{\vala_i}{\ptya_i}{\penv_i}{\constrs_i})_{i \in 1..n} \\
                \combineseqmany{\penv_1 + \cdots + \penv_n}{\penv}{\constrs}
            }
            { \synthseq
                { \fnapp{f}{\seq{\vala}} }
                {\ptyb}
                {\penv}
                {\constrs \cup \constrs_1 \cup \cdots \cup \constrs_n}
            }
        \end{mathpar}

        Also, we assume $\algoprogseq{\pprog}{\constrs_{\var{prog}}}$.

        We can also assume that there exists some $\patsubst$ which is a usable solution of
        $\constrs_{\var{prog}} \cup \constrs_1 \cup \cdots \cup \constrs_n$.

        By Lemma~\ref{lem:pat-subset}, we have that $\patsubst$ is a solution
        for all $\constrs_i$ individually.

        By the IH,
            $\tseq
                {\aps{\penv_i}}
                {\vala_i}
                {\aps{\ptya_i}}$
        for all $i$.

        By Corollary~\ref{cor:combine-many-sol}, there exist $(\tyenv_i \subtype
        \penv_i)_{i \in 1..n}$ such that $\tyenv_1 + \cdots + \tyenv_n =
        \aps{\penv}$.

        Thus by \textsc{T-Sub} and \textsc{T-App}:

        {\footnotesize
        \begin{mathpar}
            \inferrule
            {
                \erase{\aps{\pproglookup{f}}} = 
                    \fndef{f}{\seq{x : \aps{\ptya}}}{\aps{\ptyb}}{\erase{\ptma}} \\
\inferrule*
                {
                    (\vseq{\aps{\penv_i}}{\vala_i}{\aps{\ptya_i}})_{i \in
                    1..n}
                }
                { (\vseq{\tyenv_i}{\vala_i}{\aps{\ptya_i}})_{i \in
                    1..n} }
            }
            { \tseq{\tyenv_1 + \cdots + \tyenv_n}{\fnapp{f}{\seq{\vala}}}{\aps{\ptyb}} }
        \end{mathpar}
    }

        as required.
    \end{proofcase}

    \paragraph*{Statement 2: Checking.}
    \begin{proofcase}{TC-Var}
        Assumption:
        \begin{mathpar}
            \inferrule
            [TC-Var]
            { }
            { \chkseq{x}{\ptya}{x : \ptya}{\emptyset} }
        \end{mathpar}

        By \textsc{T-Var}:

        \begin{mathpar}
            \inferrule
            { }
            { \tseq
                {x : \apppatsubst{\ptya}}
                {x}
                {\apppatsubst{\ptya}}
            }
        \end{mathpar}
        as required.
    \end{proofcase}

    \begin{proofcase}{TC-Let}
        Assumption:
        \begin{mathpar}
        \inferrule
        {
            \chkseq{\ptma}{\tya}{\penv_1}{\constrs_1} \\
            \chkseq{\ptmb}{\ptya}{\penv_2}{\constrs_2} \\
            \returnablepred{\tya} \\\\
            \checkenvseq{\penv_2}{x}{\tya}{\constrs_3} \\
            \joinseq{\penv_1}{(\penv_2 \envwithout x)}{\penv}{\constrs_4}
        }
        {
            \chkseq
                {\letin[\tya]{x}{\ptma}{\ptmb}}
                { \ptya }
                { \penv }
                { \constrs_1 \cup \cdots \cup \constrs_4 }
        }
        \end{mathpar}

        We also assume that we have some usable solution $\patsubst$ for
        $\constrs_1 \cup \cdots \cup \constrs_4$, and by Lemma~\ref{lem:pat-subset},
        we know that $\patsubst$ is a usable solution for all $\constrs_i$
        individually.

        By the IH, we have that:
        \begin{itemize}
          \item $\tseq{\aps{\penv_1}}{\ptma}{\tya}$
          \item $\tseq{\aps{\penv_2}}{\ptmb}{\aps{\ptya}}$
        \end{itemize}

        By Lemma~\ref{lem:scomb-env-sol}, there exist some $\tyenv_1, \tyenv_2$ such that
        $\tyenv_1 \subtype \aps{\penv_1}$,
        $\tyenv_2 \subtype \aps{\penv_2 \envwithout x}$
        and $\tyenv_1 \scomb \tyenv_2 = \aps{\penv}$

        By the definition of $\mkwd{check}$, we have two subcases based on
        whether $x \in \dom{\penv_2}$:

        \begin{subcase}{$x \not\in \dom{\penv_2}$}
            In this case we have that
            $\unrseq{\tya}{\constrs}$.

            By Lemma~\ref{lem:unr-soundness}, we have that
            there exists some $\tyb$ such that
            $\tya \subtype \tyb$ and
            $\un{\tyb}$.

            Thus by \textsc{T-Let} and \textsc{T-Sub}:

            \begin{mathpar}
                \inferrule
                {
                    \inferrule*
                    { \tseq{\aps{\penv_1}}{\erase{\ptma}}{\tya} }
                    { \tseq{\tyenv_1}{\erase{\ptma}}{\tya} }
                        \\
                        \returnablepred{\tya}
                        \\
                    \inferrule*
                    {
                        \inferrule*
                        { \tseq{\aps{\penv_2}}{\erase{\ptmb}}{\aps{\ptya}} }
                        { \tseq{\aps{\penv_2}, x : \tyb}{\erase{\ptmb}}{\aps{\ptya}} }
                    }
                    {
                        \tseq
                            {\tyenv_2, x : \tya}
                            {\erase{\ptmb}}
                            {\aps{\ptya}}
                    }
                }
                {
                    \tseq
                        { \scombtwo{\tyenv_1}{\tyenv_2} }
                        { \letin{x}{\erase{\ptma}}{\erase{\ptmb}}}
                        { \aps{\ptya} }
                }
            \end{mathpar}

            as required.
        \end{subcase}
        \begin{subcase}{$x \in \dom{\penv_2}$}
            In this case, we have that
            $x : \ptyb \in \penv_2$ and
            $\subtyseq{\tya}{\ptyb}{\constrs_1}$.

            By Lemma~\ref{lem:subtype-soundness},
            $\tya \subtype \aps{\ptyb}$ and so
            $\tyenv_2, x : \tya \subtype \aps{\penv_2}$.

            Thus by \textsc{T-Let} and \textsc{T-Sub}:

            \begin{mathpar}
                \inferrule
                {
                    \inferrule*
                    { \tseq{\aps{\penv_1}}{\erase{\ptma}}{\tya} }
                    { \tseq{\tyenv_1}{\erase{\ptma}}{\tya} }
                    \\
                    \returnablepred{\tya}
                    \\
                    \inferrule*
                    {
                        \tseq
                            {\aps{\penv_2}}
                            {\erase{\ptmb}}
                            {\aps{\ptya}}
                    }
                    {
                        \tseq
                            {\tyenv_2, x : \tya}
                            {\erase{\ptmb}}
                            {\aps{\ptya}}
                    }
                }
                {
                    \tseq
                        { \aps{\penv} }
                        { \letin{x}{\erase{\ptma}}{\erase{\ptmb}}}
                        { \aps{\ptya} }
                }
            \end{mathpar}

            as required.
        \end{subcase}

    \end{proofcase}

    \begin{proofcase}{TC-Guard}

        Assumption:
        \begin{mathpar}
            \inferrule
                {
                    \chkgseq
                        {\pata}
                        {\seq{\pgd}}
                        {\ptya}
                        {\penv_1}
                        {\constrs_1}
                        {\patb} \\
                    \chkseq{\vala}{\retty{\tyrecv{\patb}}}{\penv_2}{\constrs_2} \\
                    \combineseq{\penv_1}{\penv_2}{\penv}{\constrs_3}
                }
                { \chkseq
                    {\guardann{\vala}{\pata}{\seq{G}}}
                    {\ptya}
                    {\penv}
                    {\constrs_1 \cup \constrs_2 \cup \constrs_3 \cup
                    \set{\subpatconstr{\pata}{\patb} }}
                }
        \end{mathpar}

        By the IH (4) we have that
        $\gseq{\patb}{\aps{\penv_1}}{\seq{\erase{\pgd}}}{\aps{\ptya}}$ with $\pnf{\patb}$.

        By the IH (2) we have that
        $\tseq{\aps{\penv_2}}{\erase{\vala}}{\aps{\retty{\tyrecv{\patb}}}}$, and by the definitions of $\erase{-}$ and $\aps{-}$ it follows that
        $\tseq{\aps{\penv_2}}{\vala}{\retty{\tyrecv{\patb}}}$.

        By Lemma~\ref{lem:combine-sol} there exist $\tyenv_1 \subtype \aps{\penv_1}$ and
        $\tyenv_2 \subtype \aps{\penv_2}$ such that
        $\tyenv_1 + \tyenv_2 = \aps{\penv}$. 

        By Corollary~\ref{cor:guard-subtypes},
        $\gseq{\patb}{\tyenv_1}{\seq{\erase{\pgd}}}{\aps{\ptya}}$.

        Recomposing by \textsc{T-Guard} and \textsc{T-Sub}:
        \begin{mathpar}
            \inferrule
            {
                \inferrule*
                {\tseq{\aps{\penv_2}}{\vala}{\retty{\tyrecv{\patb}}}}
                {\tseq{\tyenv_2}{\vala}{\retty{\tyrecv{\patb}}}}
                \\
                \gseq{\patb}{\tyenv_1}{\seq{\erase{\pgd}}}{\aps{\ptya}} \\
                \pnf{\patb}
            }
            { \tseq
                {\tyenv_1 + \tyenv_2}
                {\guard{\vala}{\seq{G}}}{\tya}
            }
        \end{mathpar}

        with $\tyenv_1 + \tyenv_2 = \aps{\penv}$ as required.
    \end{proofcase}
    \begin{proofcase}{TC-Fail}
        Assumption:
        \begin{mathpar}
            \inferrule
            {
                \chkseq{\vala}{\retty{\tyrecv{\mbzero}}}{\penv}{\constrs}
            }
            { \chkseq{\sugarfail{\vala}}{\ptya}{\penv}{\constrs} }
        \end{mathpar}

        By the IH (1), we have that
        $\tseq
            {\aps{\penv}}
            {\sugarfail{(\erase{\vala})}}
            {\aps{\retty{\tyrecv{\mbzero}}}}$.

        By the definition of $\erase{-}$ and application of substitutions,
        $\tseq{\aps{\penv}}{\sugarfail{\vala}}{\retty{\tyrecv{\mbzero}}}$.

        Thus by \textsc{T-Fail} we have that:

        \begin{mathpar}
            \inferrule
            { \tseq{\aps{\penv}}{\vala}{\retty{\tyrecv{\mbzero}}}}
            { \tseq{\aps{\penv}}{\sugarfail{\vala}}{\aps{\ptya}} }
        \end{mathpar}

        noting that the unconstrained type in \textsc{T-Fail} allows us to
        specialise the return type to $\aps{\ptya}$ as required.
    \end{proofcase}
    \begin{proofcase}{TC-Sub}
        Assumption:
        \begin{mathpar}
            \inferrule
            { \synthseq{\ptma}{\ptya}{\penv}{\constrs_1} \\
              \subtyseq{\ptya}{\ptyb}{\constrs_2}
            }
            { \chkseq{\ptma}{\ptyb}{\penv}{\constrs_1 \cup \constrs_2} }
        \end{mathpar}

        By the IH (statement 1),
        $\tseq
            {\apppatsubst{\penv}}
            {\erase{\ptma}}
            {\apppatsubst{\ptya}}$.

        By Lemma~\ref{lem:subtype-soundness},
        $\apppatsubst{\ptya} \subtype \apppatsubst{\ptyb}$.

        Therefore by \textsc{T-Sub}:

        \begin{mathpar}
            \inferrule
            {
                \tseq
                {\apppatsubst{\penv}}
                {\erase{\ptma}}
                {\apppatsubst{\ptya}}
            }
            {
                \tseq
                {\apppatsubst{\penv}}
                {\erase{\ptma}}
                {\apppatsubst{\ptyb}}
            }
        \end{mathpar}
        as required.
    \end{proofcase}

    \paragraph*{Statement 3: Guards.}

    \begin{proofcase}{TCG-Free}
        Assumption:
        \begin{mathpar}
            \inferrule
            { \chkseq{\ptma}{\ptya}{\penv}{\constrs} }
            { \chkgseq{\pata}{\free{\ptma}}{\ptya}{\penv}{\constrs}{\mbone} }
        \end{mathpar}

        By the IH (statement 2), we have that
        $\tseq{\aps{\penv}}{\erase{\ptma}}{\aps{\ptya}}$.

        Trivially, $\pnflittwo{\pata}{\mbone}$.

        Therefore, we can reconstruct by \textsc{TG-Free}:

        \begin{mathpar}
            \inferrule
            {
                \tseq{\aps{\penv}}{\erase{\ptma}}{\aps{\ptya}}
            }
            { \gseq{\mbone}{\aps{\penv}}{\free{\erase{\ptma}}}{\aps{\ptya}} }
        \end{mathpar}

        as required.
    \end{proofcase}
    \begin{proofcase}{TCG-Recv}
        Assumption:

        \begin{mathpar}
            \inferrule
            [TCG-Recv]
            {
                \chkseq{\ptma}{\ptyb}{\penv', y : \retty{\tyrecv{\ppatb}}}{\constrs_1} \\
                \proglookup{\msgtag{m}} = {\seq{\ppretya}} \\
                \penv = \penv' \envwithout \seq{x} \\
                \mbbase{\seq{\ppretya}} \vee \mbbase{\penv'} \\
                \checkenvseq{\penv'}{\seq{x}}{\seq{\makeusable{\ppretya}}}{\constrs_3}
            }
            { \chkgseq
                {\pata}
                {\receive{m}{\seq{x}}{y}{\ptma}}
                {\ptyb}
                {\penv}
                {\constrs_1 \cup \constrs_2 \cup \constrs_3 \cup
                    \set{\subpatconstr{\pata \without \msgtag{m}}{\ppatb}}
                }
                {\algomsg{m} \patconcat (\pata \without \msgtag{m})}
            }
        \end{mathpar}
        We also assume that we have some usable solution $\patsubst$
        for
        $
            \constrs_1 \cup \constrs_2 \cup \constrs_3 \cup
                    \set{\subpatconstr{\pata \without \msgtag{m}}{\ppatb}}
        $.

        As usual, by Lemma~\ref{lem:pat-subset} we can assume that $\patsubst$
        is a usable solution for all $\constrs_i$.

        By the IH,
            $\tseq
                { \aps{\penv'}, y : \retty{\tyrecv{\aps{\ppatb}}}}
                { \erase{\ptma} }
                { \aps{\ptyb} }$.

        Suppose
        $\penv' = \penv, x_1 : \ptya_1, \ldots, x_m : \ptya_m$ and
        $\seq{x} = x_1, \ldots, x_n$.

        Then by the definition of $\mkwd{check}$ we have that:
        \begin{itemize}
            \item $(\subtyseq{\ptya_i}{\makeusable{\ppretya_i}}{\constrs'_i})_{i
                \in 1..m}$
            \item $(\unrseq{\ptya_i}{\constrs'_i})_{i \in (m + 1)..n}$
        \end{itemize}

        Thus by Lemma~\ref{lem:subtype-soundness},
        $\makeusable{\aps{\ppretya_i}} \subtype \aps{\ptya_i}$ for each $i \in
        1..m$.

        By Lemma~\ref{lem:unr-soundness}, there exist $\{\tya_j\}_{j \in (m + 1) .. n}$
        such that $\aps{\ptya_j} \subtype  \tya_j $ and
        $\un{\tya_j}$.

        Thus it follows by the definition of environment subtyping that
        $\aps{\penv} \subtype \aps{\penv'}$.

        It follows from the fact that pattern substitution preserves type shape
        that if $\mbbase{\seq{\ppretya}} \vee \mbbase{\penv'}$, we have that
        $\mbbase{\seq{\aps{\ppretya}}} \vee \mbbase{\aps{\penv'}}$.

        Since $\patsubst$ is a usable solution of
        $\subpatconstr{\pata \without \msgtag{m}}{\ppatb}$ we know that
        $\pata \without \msgtag{m} \subpatone \aps{\ppatb}$ and therefore that
        $\tyrecvp{\pata \without \msgtag{m}} \subtype \tyrecvp{\aps{\ppatb}}$.

        It remains to be shown that $\pnflittwo{\pata}{\msgtag{m} \patconcat
        (\pata \without \msgtag{m})}$:

        \begin{mathpar}
            \inferrule
            {
               \msgtag{m} \patconcat (\pata \without \msgtag{m})
               \tyequiv
               \pata
            }
            { \pnflittwo{\pata}{\msgtag{m} \patconcat (\pata \without \msgtag{m})}}
        \end{mathpar}

The pattern residual and concatenation cancel, so the premise holds and
        therefore we can conclude that
        $\pnflittwo{\pata}{\msgtag{m} \patconcat (\pata \without \msgtag{m})}$.

        Finally, we can reconstruct using \textsc{TG-Recv}:

        {\footnotesize
        \begin{mathpar}
            \inferrule
            {
              \proglookup[\aps{\prog}]{\msgtag{m}} = \seq{\aps{\ppretya}} \\
              \mbbase{\seq{\aps{\ppretya}}} \vee \mbbase{\aps{\penv}} \\
              \inferrule*
              {
                \tseq
                    {\aps{\penv'}, y : \retty{\tyrecv{\aps{\ppatb}}}}
                    { \erase{\ptma} }
                    {\aps{\ptyb}}
              }
              {
                  \tseq
                    {\aps{\penv},
                        y : \retty{\tyrecvp{\pata \without \msgtag{m}}},
                        \seq{x} : \seq{\makeusable{\ppretya}}}
                    { \erase{\ptma} }
                    {\aps{\ptyb}}
              }
            }
            {
              \gseq
                {\msgtag{m} \patconcat (\pata \without \msgtag{m})}
                {\aps{\penv}}
                {\receive{m}{\seq{x}}{y}{\erase{\ptma}}}
                {\aps{\ptyb}}
            }
        \end{mathpar}
        }

        as required.
    \end{proofcase}

    \paragraph*{Statement 4: Guard sequences.}

    \begin{proofcase}{TCG-Guards}
    Assumption:
    \begin{mathpar}
    \inferrule
    { (\chkgseq
        {\pata}
        {\pgd_i}
        {\ptya}
        {\penv_i}
        {\constrs_i}
        {\patb_i})_{i \in 1..n} \\\\
      \patb = {\patb_1 \patplus \cdots \patplus \patb_n}\\
      \manymergeseq
        {\penv_1}
        {\penv_n}
        {\penv}
        {\constrs}
    }
    { \chkgseq
        {\pata}
        {\seq{\pgd}}
        {\ptya}
        {\penv}
        {\constrs \cup \constrs_1 \cup \cdots \cup \constrs_n}
        { \patb }
    }
    \end{mathpar}

    By the IH (3), we have that 
    $\gseq
        {\patb_i}
        {\aps{\penv_i}}
        {\erase{\pgd_i}}
        {\aps{\ptya}}$ and 
        $\pnflittwo{\pata}{\patb_i}$ for all $i \in 1..n$.

    Let $\patb = \patb_1 \patplus \cdots \patplus \patb_n$.
    By Lemma~\ref{lem:equiv-pnf}, $\pnf{\patb}$.

    By repeated uses of Lemma~\ref{lem:merge-env-sol} we have that
    $\aps{\penv} \subtype \aps{\penv_i}$ for all $i \in 1..n$.

    By Lemma~\ref{lem:guard-subtype}, 
    $\gseq
        {\patb_i}
        {\aps{\penv}}
        {\erase{\pgd_i}}
        {\aps{\ptya}}$ for all $i \in 1..n$.

    Thus by \textsc{TG-GuardSeq}:
    \begin{mathpar}
        \inferrule
        {
            (\gseq{\patb_i}{\aps{\penv}}{\erase{\pgd_i}}{\aps{\ptya}})_{i \in 1..n}
        }
        {
            \gseq
                {\patb_1 \patplus \cdots \patplus \patb_n}
                {\aps{\penv}}
                {\seq{\erase{\pgd}}}
                {\aps{\ptya}}
        }
    \end{mathpar}
    as required.

    \end{proofcase}

\end{proof}

\algosoundness*
\begin{proof}
    A direct consequence of Lemma~\ref{lem:algo-soundness-generalised}.
\end{proof}

 \newpage
\subsection{Proofs for \S\ref{sec:algo:alg-completeness} (Algorithmic Completeness)}
\label{sec:completeness}

Every $\tyenv$ is also a valid $\penv$ and every $\tya$ is a valid
$\ptya$. We will therefore allow ourselves to use $\tyenv$ and $\tya$ in
algorithmic type system derivations directly.

\subsubsection{Proofs of auxiliary properties}

The completeness of the check meta-function follows from the completeness of
subtyping.

\subtycomplete*
\begin{proof}
    By case analysis on $\aps{\ptya} \subtype \aps{\ptyb}$. Base cases follow
    straightforwardly so we concentrate on mailbox types.

In the case that
        $\aps{\usageann{\usage_1}{\tysend{\ppata}}} \subtype
        \aps{\usageann{\usage_2}{\tysend{\ppatb}}}$ we can assume that
        $\usage_1 \subtype \usage_2$ and $\aps{\ppatb} \subpatone \aps{\ppata}$.
Using algorithmic subtyping we can derive
        $
            \subtyseq
                {\usageann{\usage_1}{\tysend{\ppata}}}
                {\usageann{\usage_2}{\tysend{\ppatb}}}
                {\set{\subpatconstr{\ppatb}{\ppata}}}
        $
        and since $\aps{\ppatb} \subpatone \aps{\ppata}$ it follows that
        $\patsubst$ is a usable solution of $\subpatconstr{\ppatb}{\ppata}$ as
        required.

        In the case that
        $\aps{\usageann{\usage_1}{\tyrecv{\ppata}}} \subtype
        \aps{\usageann{\usage_2}{\tyrecv{\ppatb}}}$ we can assume that
        $\usage_1 \subtype \usage_2$ and $\aps{\ppata} \subpatone \aps{\ppatb}$.
        Using algorithmic subtyping we can derive
        $
        \subtyseq
                    {\usageann{\usage_1}{\tyrecv{\ppata}}}
                    {\usageann{\usage_2}{\tyrecv{\ppatb}}}
                    {\set{\subpatconstr{\ppata}{\ppatb}}}
        $
        Since $\aps{\ppata} \subpatone \aps{\ppatb}$ it follows that $\patsubst$
        is a usable solution of $\subpatconstr{\ppata}{\ppatb}$, as required.
\end{proof}

\joincomplete*
\begin{proof}
    We proceed by case analysis on the derivation of  $\scombtwo{\tya_1}{\tya_2} = \tyb$.
    Base types follow straightforwardly, so we concentrate on mailbox types.

    \begin{proofcase*}{
        $\tya_1 = \usageann{\usage_1}{\tysend{\pata_1}}$
        and $\tya_2 = \usageann{\usage_2}{\tysend{\pata_2}}$
    }
In this case we have that
    $
     \usageann{\usage_1}{\tysend{\pata_1}}
                \scomb
              \usageann{\usage_2}{\tysend{\pata_2}}
                =
              \usageann
                {\usage_1 \scomb \usage_2}
                {\tysendp{\patconcattwo{\pata_1}{\pata_2}}}
    $
    and therefore that
    $\scombtwo{\tysend{\pata_1}}{\tysend{\pata_2}} =
    \tysend{\patconcattwo{\pata_1}{\pata_2}}$.
Since $\tysend{\pata_i} \subtype \aps[\patsubst_i]{\ptya_i}$ (for $i \in
    1,2$), we have that $\ptya_i = \tysend{\ppata_i}$ with
    $\aps[\patsubst]{\ppata_i} \subpatone \pata_i$.
Since $\pv{\patsubst_1} \cap \pv{\patsubst_2} = \emptyset$, let $\patsubst =
    \patsubst_1 \cup \patsubst_2$.
Using the algorithmic join operator we can show
    $\joinseq
        {\usageann{\usage_1}{\tysend{\ppata_1}}}
        {\usageann{\usage_2}{\tysend{\ppata_2}}}
        {\usageann
            {\usage_1 \scomb \usage_2}
            {\tysendp{\patconcattwo{\ppata_1}{\ppata_2}}}}
        {\emptyset}$.
        Since $\aps{\ppata_i} \subpatone \pata_i$ it follows that $\aps{\ppata_1
        \patconcat \ppata_2} \subpatone \pata_1 \patconcat \pata_2$ and
        therefore that
$\usageann
            {\usage_1 \scomb \usage_2}
            {\tysendp{\pata_1 \patconcat \pata_2}}
        \subtype
         \usageann
            {\usage_1 \scomb \usage_2}
            {\aps{\tysendp{\ppata_1 \patconcat
        \ppata_2}}}$
        with $\patsubst \supseteq \patsubst_1 \cup \patsubst_2$ as a solution of
        $\emptyset$, as required.
    \end{proofcase*}

\begin{proofcase*}{$\tya_1 = \usageann{\usage_1}{\tysend{\pata}}$ and $\tya_2 =
        \usageann{\usage_2}{\tyrecvp{\pata \patconcat \patb}}$}
        In this case we have that
        $\scombtwo
                {\usageann{\usage_1}{\tysend{\pata}}}
                {\usageann{\usage_2}{\tyrecvp{\pata \patconcat \patb}}}
              =
              \usageann{\usage_1 \scomb \usage_2}{\tyrecv{\patb}}$
        and therefore that 
        $\scombtwo{\tysend{\pata}}{\tyrecvp{\pata \patconcat \patb}}  =
                \tyrecv{\patb}$.
For
        $\tysend{\pata} \subtype \aps[\patsubst_1]{\ptya_1}$
        and
        $\tyrecvp{\pata \patconcat \patb} \subtype \aps[\patsubst_2]{\ptya_2}$
        to hold,
        it must be the case that $\ptya_1 = \tysend{\ppata}$
        with $\aps[\patsubst_1]{\ppata} \subpatone \pata$, 
        and that $\ptya_2 = \tyrecv{\ppatb}$ with
        $(\pata \patconcat \patb) \subpatone \aps[\patsubst_2]{\ppatb}$.
Using the algorithmic type join operator, 
        we can show
        $
            \joinseq
                { \usageann{\usage_1}{\tysend{\ppata}} }
                { \usageann{\usage_2}{\tyrecv{\ppatb}} }
                { \usageann{\usage_1 \scomb \usage_2}{\tyrecv{\patvara}} }
                { \set{\subpatconstr{\ppata \patconcat \patvara}{\ppatb}}}$
        (for a fresh $\patvara$).
        Since $\pv{\patsubst_1} \cap \pv{\patsubst_2} = \emptyset$ 
        we can construct
        $\patsubst = \patsubst_1
        \cup \patsubst_2 \cup \patvara \mapsto \patb$.
        To show that $\patsubst$ is a solution it suffices to
        show that $(\aps{\ppata} \patconcat \patb) \subpatone \aps{\ppatb}$:
        by the pre-congruence and transitivity properties of $\incll[]$ we have that
        $
        (\aps{\ppata} \patconcat \patb)
            \subpatone
            (\pata \patconcat \patb)
            \subpatone
            \aps{\ppatb}
        $
        and as such $\aps{\tyrecv{\patvara}} = \tyrecv{\patb}$
        with $\patsubst \supset \patsubst_1 \cup \patsubst_2$ a solution for the
        constraint set, as required.
The case where 
        $\tya_1 =
        \usageann{\usage_1}{\tyrecvp{\pata \patconcat \patb}}$
        $\tya_2 = \usageann{\usage_2}{\tysend{\pata}}$ is symmetric.
        \end{proofcase*}
\end{proof}

\mergecomplete*
\begin{proof}
    By case analysis on the structure of $\tya$.
    Base types follow directly so we need only consider mailbox types.
    
    \begin{proofcase*}{$A = \usagety{\tysend{\pata}}$}
    By the definition of
    subtyping we have that
    $\usagety{\tysend{\pata}} \subtype
    \aps
        [\patsubst_1]
        {\usageann{\usage_1}{\tysend{\ppata}}}$ and
    $\usagety{\tysend{\pata}} \subtype
    \aps
        [\patsubst_2]
        {\usageann{\usage_2}{\tysend{\ppatb}}}$.
We first show that $\usage \subtype \minusage{\usage_1}{\usage_2}$.
    If $\usage = \usable$ then it must be the case that $\usage_1,
    \usage_2 = \usable$ and $\minusage{\usage_1}{\usage_2} = \usable$.
    If $\usage = \returnable$ then we have that $\usage \subtype
    \minusage{\usage_1}{\usage_2}$.
By the algorithmic merge operator,
    $
    \mergeseq
                    {\usageann{\usage_1}{\tysend{\ppata}}}
                    {\usageann{\usage_2}{\tysend{\ppatb}}}
                    { \usageann
                        {\minusage{\usage_1}{\usage_2}}
                        {\tysendp{\ppata \patplus \ppatb}}
                    }
                    {\emptyset}
    $, and since $\pv{\patsubst_1} \cap \pv{\patsubst_2} = \emptyset$,
    we can set $\patsubst = \patsubst_1 \cup \patsubst_2$ (trivially a
    solution of $\emptyset$).
It remains to be shown that $\usageann{\usage}{\tysend{\pata}} \subtype
        \usageann{\minusage{\usage_1}{\usage_2}}{\tysendp{\aps{\ppata \patplus
            \ppatb}}}$.
Applying the solution pointwise,
        $\aps{\tysendp{\ppata \patplus \ppatb}} = \tysendp{\aps{\ppata}
        \patplus \aps{\ppatb}}$.
Since $\tysend{\pata} \subtype \tysend{\aps{\ppata}}$ and
        $\tysend{\pata} \subtype \tysend{\aps{\ppatb}}$, it follows by
        subtyping that $\aps{\ppata} \subpatone \pata$ and
        $\aps{\ppatb} \subpatone \pata$.
By the definition of pattern semantics
        $\aps{\ppata} \patplus \aps{\ppatb} \subpatone \pata$ and therefore
        $\usageann{\usage}{\tysend{\pata}} \subtype
        \usageann{\minusage{\usage_1}{\usage_2}}{\tysendp{\aps{\ppata \patplus
        \ppatb}}}$ as required.
    \end{proofcase*}

    \begin{proofcase*}{$\tya = \usagety{\tyrecv{\pata}}$}
    By the definition of
    subtyping we have that
    $\usageann{\usage}{\tysend{\pata}} \subtype
        \aps[\patsubst_1]{\usageann{\usage_1}{\tysend{\ppata}}}$
    and
    $\usageann{\usage}{\tysend{\pata}} \subtype
        \aps[\patsubst_2]{\usageann{\usage_2}{\tysend{\ppatb}}}$.
    As before, $\usage \subtype \minusage{\usage_1}{\usage_2}$. 
    Using the algorithmic merge operation we can show
    $
        \mergeseq
                {\usageann{\usage_1}{\tyrecv{\ppata}}}
                {\usageann{\usage_2}{\tyrecv{\ppatb}}}
                {\usageann{\minusage{\usage_1}{\usage_2}}{\tyrecv{\patvara}}}
                {\set{\subpatconstr{\patvara}{\ppata},
                      \subpatconstr{\patvara}{\ppatb}}}
    $.
Since $\dom{\patsubst_1} \cap \dom{\patsubst_2} = \emptyset$, we can set
        $\patsubst = \patsubst_1 \cup \patsubst_2 \cup \set{\patvara \mapsto \pata}$.
To show that $\patsubst$ is a usable solution of the constraint set, it remains to be shown that
        $\pata \subpatone \aps{\ppata}$ and
        $\pata \subpatone \aps{\ppatb}$;
since $\tyrecv{\pata} \subtype \tyrecv{\aps[\patsubst_1]{\ppata}}$ it
        follows that $\pata \subpatone \aps[\patsubst_1]{\ppata}$ and likewise
        for $\aps[\patsubst_2]{\ppatb}$; since $\dom{\patsubst_1} \cap
        \dom{\patsubst_2} = \emptyset$ it follows that
        $\tyrecv{\pata} \subpatone \aps{\ppata}$ and likewise for $\ppatb$, as
        required.
    \end{proofcase*}
\end{proof}

\subsubsection{Useful auxiliary lemmas}

\valcompleteness*
\begin{proof}
    The proof is by induction on the derivation of $\tseq{\tyenv}{\vala}{\tya}$.

    \begin{proofcase}{T-Var}
        We assume that $\tseq{x : \tya}{x}{\tya}$.
        By \textsc{TC-Var} we can show $\chkseq{x}{\tya}{x : \tya}{\emptyset}$,
        as required.
    \end{proofcase}
    \begin{proofcase}{T-Const}
        We assume that $\tseq{\cdot}{c}{\basety}$, where $c$ has base type
        $\basety$. By \textsc{TS-Base}, we can show that
        $\synthseq{c}{\basety}{\cdot}{\emptyset}$.
        Finally, by \textsc{TC-Sub} (noting that
        $\subtyseq{\basety}{\basety}{\emptyset}$) we have that
        $\chkseq{c}{\basety}{\cdot}{\emptyset}$, as required.
    \end{proofcase}
    \begin{proofcase}{T-Sub}
        Assumption:
\begin{mathpar}
            \inferrule
            { \tyenv \subtype \tyenv' \\
              \tya \subtype \tyb \\
              \tseq{\tyenv'}{\vala}{\tya}
            }
            { \tseq{\tyenv}{\vala}{\tyb} }
        \end{mathpar}

        By the IH, there exists some $\tyenv''$ such that
        $\tyenv' \subtype \tyenv''$
        and
        $\chkseq{\tyenv''}{\vala}{\tya}{\emptyset}$.

        By the transitivity of subtyping, we have that
        $\tyenv \subtype \tyenv' \subtype \tyenv''$, as required.
    \end{proofcase}
\end{proof}

\subsubsection{Completeness of auxiliary definitions}

We now need to show completeness for all auxiliary judgements (\eg, subtyping,
environment combination).

\begin{lemma}[Completeness of environment join]\label{lem:join-completeness}
If:

\begin{itemize}
    \item $\tyenv_1 \scomb \tyenv_2 = \tyenv$,
    \item $\tyenv_1 \subtype \aps[\patsubst_1]{\penv_1}$,
    \item $\tyenv_2 \subtype \aps[\patsubst_2]{\penv_2}$; and
    \item $\pv{\patsubst_1} \cap \pv{\patsubst_2} = \emptyset$
\end{itemize}
    then there exist $\penv, \constrs$ such that
    $\joinseq{\penv_1}{\penv_2}{\penv}{\constrs}$,
    and there exists a usable solution $\patsubst \supseteq \patsubst_1 \cup
    \patsubst_2$ of $\constrs$
    such that $\tyenv \subtype \aps{\penv}$.
\end{lemma}
\begin{proof}
    By induction on the derivation of $\tyenv_1 \scomb \tyenv_2$, with appeal to
    Lemma~\ref{lem:join-completeness-types}.
\end{proof}

\begin{lemma}[Completeness of disjoint environment combination]\label{lem:combine-completeness}
If:

\begin{itemize}
    \item $\tyenv_1 + \tyenv_2 = \tyenv$,
    \item $\tyenv_1 \subtype \aps[\patsubst_1]{\penv_1}$,
    \item $\tyenv_2 \subtype \aps[\patsubst_2]{\penv_2}$; and
    \item $\pv{\patsubst_1} \cap \pv{\patsubst_2} = \emptyset$
\end{itemize}
    then there exist $\penv, \constrs$ such that
    $\combineseq{\penv_1}{\penv_2}{\penv}{\constrs}$,
    and there exists a usable solution $\patsubst \supseteq \patsubst_1 \cup
    \patsubst_2$ of $\constrs$
    such that $\tyenv \subtype \aps{\penv}$.
\end{lemma}
\begin{proof}
    By induction on the derivation of $\tyenv_1 + \tyenv_2 = \tyenv$.

    \begin{proofcase*}{$\tyenv_1 = \cdot$ and $\tyenv_2 = \cdot$}
        \begin{mathpar}
           \inferrule
            { }
            { \pluscombtwo{\cdot}{\cdot} = \cdot }
        \end{mathpar}

      By the definition of environment subtyping, the only environment that can
      be a supertype of the empty environment is $\cdot$. Therefore,
      we can immediately conclude with the corresponding base case in
      algorithmic type environment combination:

      \begin{mathpar}
          \inferrule
          { }
          { \combineseq{\cdot}{\cdot}{\cdot}{\emptyset} }
      \end{mathpar}
    \end{proofcase*}

    \begin{proofcase*}{$x \not\in \dom{\tyenv_2}$}
        Assumption:
        \begin{mathpar}
            \inferrule
            { x \not\in \dom{\tyenv_2} \\
              \pluscombtwo{\tyenv_1}{\tyenv_2} = \tyenv
            }
            { \pluscombtwo{\tyenv_1, x : \tya}{\tyenv_2} = \tyenv, x : \tya }
        \end{mathpar}

        where:
        \begin{itemize}
            \item $\tyenv_1, x : \tya \subtype \aps[\patsubst_1]{\penv_1}$
            \item $\tyenv_2 \subtype \aps[\patsubst_2]{\penv_2}$
            \item $\pv{\patsubst_1} \cap \pv{\patsubst_2} = \emptyset$
        \end{itemize}

        There are two subcases based on whether $x \in \dom{\penv_1}$.

        \begin{subcase}{$x \not\in \dom{\penv_1}$}
            In this case we have by the definition of environment subtyping that:
            \begin{mathpar}
                \inferrule*
                {
                    \inferrule*
                    { \un{\tya} }
                    { \tyenv, x : \tya \subtype \tyenv }
                    \\
                    \tyenv \subtype \aps[\patsubst_1]{\penv}
                }
                {
                    \tyenv, x : \tya \subtype \aps[\patsubst_1]{\penv}
                }
            \end{mathpar}

            By the IH we have that there exists some
            $\patsubst \supset \patsubst_1 \cup \patsubst_2$ such that
            $\combineseq{\penv_1}{\penv_2}{\penv}{\constrs}$ and there exists a
            usable solution $\patsubst \supseteq \patsubst_1 \cup \patsubst_2$
            of $\constrs$ such that
            $\tyenv \subtype \aps{\penv}$.

            Thus by environment subtyping we can show that
            $\tyenv, x : \tya \subtype \aps{\penv}$ as required.
        \end{subcase}
        \begin{subcase}{$x \in \dom{\penv_1}$}
            In this case, $\penv_1 = \penv'_1, x : \ptya$ with $\tya
            \subtype \aps[\patsubst_1]{\ptya}$.

            By the IH,
            $\combineseq{\penv'_1}{\penv_2}{\penv}{\constrs}$
            for some $\penv, \constrs$ and
            there exists some usable solution
            $\patsubst \supseteq \patsubst_1 \cup \patsubst_2$ of $\constrs$
            such that $\tyenv_1 + \tyenv_2 \subtype \aps{\penv}$.

            Since $\tya \subtype \aps[\patsubst_1]{\ptya}$ and $\patsubst_1
            \subseteq \patsubst$, it follows that $\tya \subtype \aps{\ptya}$.

            Therefore it follows that $\tyenv_1 + \tyenv_2, x : \tya \subtype
            \aps{\penv, x : \ptya}$ as required.
        \end{subcase}
    \end{proofcase*}

    \begin{proofcase*}{$x \not\in \dom{\tyenv_1}$}
       \begin{mathpar}
           \inferrule
           { x \not\in \dom{\tyenv_1} \\
             \pluscombtwo{\tyenv_1}{\tyenv_2} = \tyenv
           }
           { \pluscombtwo{\tyenv_1}{\tyenv_2, x : \tya} = \tyenv, x : \tya }
       \end{mathpar}

        Symmetric to the above case.
    \end{proofcase*}

    \begin{proofcase*}{$x \in \dom{\tyenv_1} \cap \dom{\tyenv_2}$}
        \begin{mathpar}
            \inferrule
            { \un{\tya} \\
              \pluscombtwo{\tyenv_1}{\tyenv_2} = \tyenv
            }
            { \pluscombtwo{\tyenv_1, x : \tya}{\tyenv_2, x : \tya} = \tyenv, x : \tya }
        \end{mathpar}

            In this case, we have that:
            \begin{itemize}
                \item $\penv_1, = \penv'_1, x : \ptyb_1$
                \item $\penv_2 = \penv'_2, x : \ptyb_2$
            \end{itemize}

            By the IH, there exist $\penv, \constrs$ such that
            $\combineseq
                {\penv'_1}
                {\penv'_2}
                {\penv}
                {\constrs}$
            and some usable solution $\patsubst \supseteq \patsubst_1 \cup
            \patsubst_2$ of $\constrs$
            such that $\tyenv \subtype \aps{\penv}$.

            By algorithmic environment combination we have:
            \begin{mathpar}
                \inferrule
                {
                    \combineseq{\penv'_1}{\penv'_2}{\penv}{\constrs_1} \\
                    {\equivseq{\ptyb_1}{\ptyb_2}{\constrs_2}} \\
                    {\unrseq{\ptyb_1}{\constrs_3}} \\
                    {\unrseq{\ptyb_2}{\constrs_4}}
                }
                { \combineseq
                    {\penv'_1, x : \ptyb_1}
                    {\penv'_2, x : \ptyb_2}
                    {\penv, x : \tyb_1 }
                    {\constrs_1 \cup \constrs_2 \cup \constrs_3 \cup \constrs_4}
                }
            \end{mathpar}

            From $\un{\tya}$, we have two subcases based on whether $\tya$ is a base type
            $\basety$, or a mailbox type $\usablety{\mbsendone}$.

            \begin{subcase}{$\tya = \basety$}
                In this case, by the definition of subtyping we have that
                $\tyb_1 = \tyb_2 = \basety$ and therefore:

                \begin{mathpar}
                    \inferrule
                    {
                        \combineseq{\penv'_1}{\penv'_2}{\penv}{\constrs} \\
                        {\equivseq{\basety}{\basety}{\emptyset}} \\
                        {\unrseq{\basety}{\emptyset}} \\
                        {\unrseq{\basety}{\emptyset}}
                    }
                    { \combineseq
                        {\penv'_1, x : \basety}
                        {\penv'_2, x : \basety}
                        {\penv, x : \basety }
                        {\constrs}
                    }
                \end{mathpar}

                with $\patsubst$ remaining a usable solution of $\constrs$.

                It follows that $\penv'_1, x : \basety + \penv'_2, x : \basety
                \subtype \aps{\penv}, x : \basety$, as required.
            \end{subcase}

            \begin{subcase}{$\tya = \usablety{\mbsendone}$}

                In this case, we have that
                $\tyb_1 = \usablety{\tysend{\ppatb_1}}$ and
                $\tyb_2 = \usablety{\tysend{\ppatb_2}}$.

                and:
                \begin{mathpar}
                    \inferrule
                    {
                        \combineseq{\penv'_1}{\penv'_2}{\penv}{\constrs} \\
                        {\equivseq
                            {\usablety{\tysend{\ppatb_1}}}
                            {\usablety{\tysend{\ppatb_2}}}
                            {\set
                                {\subpatconstr{\ppatb_1}{\ppatb_2},
                                 \subpatconstr{\ppatb_2}{\ppatb_1}}}} \\
                        {\unrseq{\ppatb_1}{\set{\subpatconstr{\ppatb_1}{\mbone}}}} \\
                        {\unrseq{\ppatb_2}{\set{\subpatconstr{\ppatb_2}{\mbone}}}}
                    }
                    { \combineseq
                        {\penv'_1, \usablety{\tysend{\ppatb_1}}}
                        {\penv'_2, \usablety{\tysend{\ppatb_2}}}
                        {\penv, x : \usablety{\tysend{\ppatb_1}} }
                        {\constrs \cup
                            \set{
                                \subpatconstr{\ppatb_1}{\ppatb_2},
                                \subpatconstr{\ppatb_2}{\ppatb_1},
                                \subpatconstr{\ppatb_1}{\mbone},
                                \subpatconstr{\ppatb_2}{\mbone}
                            }
                        }
                    }
                \end{mathpar}

                Let $\patsubst' = \patsubst[\ppatb_1 \mapsto \mbone, \ppatb_2
                \mapsto \mbone]$, which is now a usable solution for the
                additional constraints.

                Finally, we have that $\tyenv, x : \usablety{\mbsendone}
                \subtype \apppatsubst[\patsubst']{\penv, x :
                \usablety{\tysend{\ppatb_1}}}
                \subtype
                \apppatsubst
                    [\patsubst']
                    {\penv}, x : \usablety{\mbsendone}$, as required.
            \end{subcase}
    \end{proofcase*}
\end{proof}

\begin{lemma}\label{lem:merge-completeness}
If:
\begin{itemize}
    \item $\tyenv \subtype \aps[\patsubst_1]{\penv_1}$,
    \item $\tyenv \subtype \aps[\patsubst_2]{\penv_2}$; and
    \item $\pv{\patsubst_1} \cap \pv{\patsubst_2} = \emptyset$
\end{itemize}
    then there exist $\penv, \constrs$ such that
    $\mergeseq{\penv_1}{\penv_2}{\penv}{\constrs}$
    and there exists a usable solution
    $\patsubst \supseteq \patsubst_1 \cup \patsubst_2$ of $\constrs$
    such that $\tyenv \subtype \aps{\penv}$.
\end{lemma}
\begin{proof}
    By induction on the size of $\tyenv$ and inspection of $\penv_1$ and
    $\penv_2$; merging of types relies on Lemma~\ref{lem:merge-completeness-types}.
\end{proof}

\checkcomplete*
\begin{proof}
    We proceed by case analysis on whether $x \in \dom{\penv'}$.

    If $x \in \dom{\penv'}$ then the result follows directly from Lemma~\ref{lem:subtya-ptya}.

    If $x \not\in \dom{\penv'}$ then by the definition of environment subtyping it follows
    that $\aps{\ptya} \subtype \usablety{\tysend{\mbone}}$.

    By the definition of $\subtype$ it must be the case that
    $\aps{\ptya} = \usagety{\tysend{\aps{\patvara}}}$ for some $\patvara$ such that
    $\mbone \subpatone \aps{\patvara}$.

    In this case, by the definitions of $\mkwd{check}$ and $\mkwd{unr}$
    we can show that:

    \begin{mathpar}
        \inferrule
        { x \not\in \dom{\penv'} \\
          \unrseq
            {\usagety{\tysend{\patvara}}}
            {\set{\subpatconstr{\mbone}{\patvara}}}
        }
        { \checkenvseq
            {\penv'}
            {x}
            {\usagety{\tysend{\patvara}}}
            { \set{\subpatconstr{\mbone}{\patvara}}}
        }
    \end{mathpar}

    Since $\mbone \subpatone \aps{\patvara}$ it follows that $\patsubst$ is a usable solution of 
    $\checkenvseq
            {\penv'}
            {x}
            {\usagety{\tysend{\patvara}}}
            { \set{\subpatconstr{\mbone}{\patvara}}}$
    as required.
    \end{proof}

The $n$-ary version of Lemma~\ref{lem:checkfn-subtype} follows as a corollary:

\begin{corollary}[Completeness of n-ary check meta-function]\label{cor:checkfn-many-subtype}
    If $\aps{\penv, \seq{x} : \seq{\ptya}} \subtype \aps{\penv'}$ then
    $\checkenvseq{\penv'}{\seq{x}}{\seq{\ptya}}{\constrs}$ where $\patsubst$ is a usable
    solution of $\constrs$.
\end{corollary}

\subsubsection{Supertype checkability}

In order to show the completeness of \textsc{T-Sub}, we must show that if a
term is checkable at a subtype, then it is also checkable at a supertype.
To do this we require several intermediate results.

We firstly define \emph{closed} and \emph{satisfiable} constraint sets.

\begin{definition}[Closed constraint set]
    A constraint set $\constrs$ is \emph{closed} if $\pv{\constrs} = \emptyset$.
\end{definition}

\begin{definition}[Satisfiable constraint set]
    A closed constraint set $\constrs$ is \emph{satisfiable} if the empty
    solution is a solution for $\constrs$
    (\ie, $\constrs = (\subpatconstr{\pata_i}{\patb_i})_i$
    and $(\pata_i \subpatone \patb_i)_i$).
\end{definition}

If we have two types which do not contain pattern variables, algorithmic
subtyping does not introduce any pattern variables into the constraint set.

\begin{lemma}[Subtyping introduces no fresh variables]\label{lem:subtype-constraints-closed}

    If $\subtyseq{\tya}{\tyb}{\constrs}$, then $\pv{\constrs} = \emptyset$.
\end{lemma}
\begin{proof}
    A straightforward case analysis on the derivation of
    $\subtyseq{\ptya}{\ptyb}{\constrs}$.
\end{proof}

Next, if we have an algorithmic subtyping judgement which produces a satisfiable
constraint set, and a subtyping relation with a supertype, then we can show
that the algorithmic subtyping judgement instantiated with the supertype will
produce a satisfiable constraint set.

\algosubtypetrans*
\begin{proof}

   By case analysis on the derivation of $\subtyseq{\tya}{\tya'}{\constrs}$.

   Base types hold trivially, so we need only consider two cases:

   \begin{proofcase}{$\tysend{\pata} \subtype \tysend{\patb}$}
       Assumption:

       \begin{mathpar}
           \inferrule
           { \usage_1 \subtype \usage_2 }
           {
               \subtyseq
                   {\usageann{\usage_1}{\tysend{\pata}}}
                   {\usageann{\usage_2}{\tysend{\patb}}}
                   {\set{\subpatconstr{\patb}{\pata}}}
           }
       \end{mathpar}

       also we know that $\subpatconstr{\patb}{\pata}$ is satisfiable (therefore
       that $\subpat{\patb}{\pata}$), and
       $\tysend{\patb} \subtype \tyb$.

       By the definition of subtyping we have that $\tyb = \tysend{\patb'}$ for
       some pattern $\patb'$, and therefore that $\patb' \subpatone \patb$.

       By transitivity of pattern inclusion we have that
       $\patb' \subpatone \patb \subpatone \pata$ and therefore

       \begin{mathpar}
           \inferrule
           { \usage_1 \subtype \usage_2 }
           {
               \subtyseq
                   {\usageann{\usage_1}{\tysend{\pata}}}
                   {\usageann{\usage_2}{\tysend{\patb'}}}
                   {\set{\subpatconstr{\patb'}{\pata}}}
           }
       \end{mathpar}

       where $\subpatconstr{\patb'}{\pata}$ is satisfiable, as required.

   \end{proofcase}

   \begin{proofcase}{$\tyrecv{\pata} \subtype \tyrecv{\patb}$}

    Assumption:

       \begin{mathpar}
           \inferrule
           { \usage_1 \subtype \usage_2 }
           {
               \subtyseq
                   {\usageann{\usage_1}{\tyrecv{\pata}}}
                   {\usageann{\usage_2}{\tyrecv{\patb}}}
                   {\set{\subpatconstr{\pata}{\patb}}}
           }
       \end{mathpar}

       also we know that $\subpatconstr{\pata}{\patb}$ is satisfiable (therefore
       that $\subpat{\pata}{\patb}$), and
       $\tyrecv{\patb} \subtype \tyb$.

       By the definition of subtyping we have that $\tyb = \tyrecv{\patb'}$ for
       some pattern $\patb'$ and
       therefore that $\patb \subpatone \patb'$.

       Thus by transitivity of pattern inclusion we have that $\pata \subpatone
       \patb \subpatone \patb'$ and therefore that:

       \begin{mathpar}
           \inferrule
           { \usage_1 \subtype \usage_2 }
           {
               \subtyseq
                   {\usageann{\usage_1}{\tyrecv{\pata}}}
                   {\usageann{\usage_2}{\tyrecv{\patb'}}}
                   {\set{\subpatconstr{\pata}{\patb'}}}
           }
       \end{mathpar}

       where $\subpatconstr{\pata}{\patb'}$ is satisfiable, as required.

   \end{proofcase}
\end{proof}

We also need to show that environment joining respects subtyping, which we do by
firstly showing that type joining respects subtyping.

\begin{lemma}[Algorithmic type join respects subtyping]\label{lem:type-join-subtype}
    If:
    \begin{itemize}
        \item $\joinseq{\ptya_1}{\ptya_2}{\ptya}{\constrs}$
        \item $\patsubst$ is some usable solution of $\constrs$ such that
            $\aps{\ptya_2} \subtype \aps{\ptya_3}$  for some $\ptya_3$
    \end{itemize}
    then
    $\joinseq{\ptya_1}{\ptya_3}{\ptya'}{\constrs'}$ for some $\ptya'$,
    $\constrs'$ such that $\aps{\ptya} \subtype \aps{\ptya'}$ and $\patsubst$ is
    a usable solution of $\aps{\constrs'}$.
\end{lemma}
\begin{proof}

    Base type combination follows straightforwardly, so we have:

    \begin{mathpar}
        \inferrule*
        { \joinseq{\pmbty_1}{\pmbty_2}{\pmbty}{\constrs}}
        {
            \joinseq
                {\usageann{\usage_1}{\pmbty_1}}
                {\usageann{\usage_2}{\pmbty_2}}
                {\usageann{\usagecombtwo{\usage_1}{\usage_2}}{\pmbty}{\constrs}}
        }
    \end{mathpar}
    so it suffices to proceed by case analysis on the derivation of
    $\joinseq{\pmbty_1}{\pmbty_2}{\pmbty}{\constrs}$.

    \begin{proofcase}{$\pmbty_1 = \tysend{\ppata}$ and $\pmbty_2 = \tysend{\ppatb}$}
        Assumption:

        \begin{mathpar}
            \inferrule
            { }
            { \joinseq
                {\tysend{\ppata}}
                {\tysend{\ppatb}}
                {\tysendp{\patconcattwo{\ppata}{\ppatb}}}
                {\emptyset}
            }
        \end{mathpar}

        We also assume that $\aps{\tysend{\ppatb}} \subtype \aps{\ptya}$ for
        some $\ptya$, which by the definition of subtyping means that $\ptya =
        \tysend{\ppatb'}$ for some $\ppatb'$, where
        $\aps{\ppatb'} \subpatone \aps{\ppatb}$.

        It follows by the compositionality of pattern semantics that
        $\patconcattwo{\ppata}{\ppatb'} \subpatone
        \patconcattwo{\ppata}{\ppatb}$ and thus
        $\tysendp{\patconcattwo{\ppata}{\ppatb}} \subtype
        \tysendp{\patconcattwo{\ppata}{\ppatb'}}$, and we have that

        \begin{mathpar}
            \inferrule
            { }
            { \joinseq
                {\tysend{\ppata}}
                {\tysend{\ppatb'}}
                {\tysendp{\patconcattwo{\ppata}{\ppatb'}}}
                {\emptyset}
            }
        \end{mathpar}

        as required.
    \end{proofcase}

    \begin{proofcase}{$\pmbty_1 = \tysend{\ppata}$ and $\pmbty_2 = \tyrecv{\ppatb}$}
        Assumption:

        \begin{mathpar}
            \inferrule
                {
                    \patvara {\text{ fresh}}
                }
                {
                    \joinseq
                        { \tysend{\ppata} }
                        { \tyrecv{\ppatb} }
                        { \tyrecv{\patvara} }
                        { \{ \subpatconstr{(\patconcattwo{\ppata}{\patvara})}{\ppatb} \} }
                }
        \end{mathpar}

        By the assumptions we know that $\patsubst$ is a usable solution of
        $\constrs$ such that $\aps{\tyrecv{\ppatb}} \subtype \aps{\ptya}$ for
        some $\ptya$. By the definition of subtyping it must be the case that
        $\ptya = \tyrecv{\ppatb'}$ for some pattern $\ppatb'$.

        Since $\aps{\tyrecv{\ppatb}} \subtype \aps{\tyrecv{\ppatb'}}$ it follows
        that $\aps{\ppatb} \subpatone \aps{\ppatb'}$.

        Since $\patsubst$ is a usable solution of $\constrs$
        we have that
        $\aps{\patconcattwo{\ppata}{\patvara}} \subpatone \aps{\ppatb}$.

        Therefore by transitivity of subtyping we have that
        $\aps{\patconcattwo{\ppata}{\patvara}} \subpatone \aps{\ppatb'}$ and
        thus know that $\patsubst$ is a usable solution of
        $\{ \subpatconstr{(\patconcattwo{\ppata}{\patvara})}{\ppatb'} \}$.

        Recomposing:

        \begin{mathpar}
            \inferrule
                {
                    \patvara {\text{ fresh}}
                }
                {
                    \joinseq
                        { \tysend{\ppata} }
                        { \tyrecv{\ppatb'} }
                        { \tyrecv{\patvara} }
                        { \{ \subpatconstr{(\patconcattwo{\ppata}{\patvara})}{\ppatb'} \} }
                }
        \end{mathpar}
    \end{proofcase}

    \begin{proofcase}{$\pmbty_1 = \tyrecv{\ppata}$ and $\pmbty_2 = \tysend{\ppatb}$}
        Similar to the previous case.
    \end{proofcase}

\end{proof}

The desired result falls out as a corollary:

\begin{corollary}[Algorithmic environment join respects subtyping]\label{cor:env-join-subtype}
    If:
    \begin{itemize}
        \item $\joinseq{\penv_1}{\penv_2}{\penv}{\constrs}$
        \item $\patsubst$ is some usable solution of $\constrs$ such that
            $\aps{\penv_2} \subtype \aps{\penv_3}$  for some $\penv_3$
    \end{itemize}
    then
    $\joinseq{\penv_1}{\penv_3}{\penv'}{\constrs'}$ for some $\penv'$,
    $\constrs'$ such that $\aps{\penv} \subtype \aps{\penv'}$ and $\patsubst$ is
    a usable solution of $\aps{\constrs'}$.
\end{corollary}

Where environments are related by strict subtyping, we can obtain a stronger
result:

\begin{corollary}[Algorithmic environment join respects strict
    subtyping]\label{cor:env-join-strict-subtype}
    If:
    \begin{itemize}
        \item $\joinseq{\penv_1}{\penv_2}{\penv}{\constrs}$
        \item $\patsubst$ is some usable solution of $\constrs$ such that
            $\aps{\penv_2} \strictsubty \aps{\penv_3}$  for some $\penv_3$
    \end{itemize}
    then
    $\joinseq{\penv_1}{\penv_3}{\penv'}{\constrs'}$ for some $\penv'$,
    $\constrs'$ such that $\aps{\penv} \strictsubty \aps{\penv'}$ and $\patsubst$ is
    a usable solution of $\aps{\constrs'}$.
\end{corollary}

Finally we want to see that algorithmic environment combination respects
subtyping.

\begin{lemma}[Algorithmic combination respects subtyping]\label{lem:env-comb-subtype}
    If:
    \begin{itemize}
        \item $\combineseq{\penv_1}{\penv_2}{\penv}{\constrs}$
        \item $\patsubst$ is some usable solution of $\constrs$ such that
            $\aps{\penv_1} \subtype \aps{\penv_3}$  for some $\penv_3$
    \end{itemize}
    then
    $\combineseq{\penv_3}{\penv_2}{\penv'}{\constrs'}$ for some $\penv'$,
    $\constrs'$ such that $\aps{\penv} \subtype \aps{\penv'}$ and $\patsubst$ is
    a usable solution of $\aps{\constrs'}$.
\end{lemma}
\begin{proof}
    By induction on the derivation of
    $\combineseq{\penv_1}{\penv_2}{\penv}{\constrs}$.
\end{proof}

As a corollary, a stronger result holds for environments related by strict
subtyping:

\begin{corollary}[Algorithmic combination respects
    strict subtyping]\label{cor:env-comb-strict-subtype}
    If:
    \begin{itemize}
        \item $\combineseq{\penv_1}{\penv_2}{\penv}{\constrs}$
        \item $\patsubst$ is some usable solution of $\constrs$ such that
            $\aps{\penv_1} \strictsubty \aps{\penv_3}$  for some $\penv_3$
    \end{itemize}
    then
    $\combineseq{\penv_3}{\penv_2}{\penv'}{\constrs'}$ for some $\penv'$,
    $\constrs'$ such that $\aps{\penv} \strictsubty \aps{\penv'}$ and $\patsubst$ is
    a usable solution of $\aps{\constrs'}$.
\end{corollary}

Relying on the previous results, we can now show the supertype checkability
lemma.

\begin{lemma}[Supertype checkability]\label{lem:subtype-checking}\hfill

\begin{itemize}
    \item If:
        \begin{itemize}
            \item $\chkseq[\prog]{\ptma}{\tya}{\penv}{\constrs}$
            \item $\patsubst$ is a usable solution of $\constrs$
            \item $\tya \subtype \tyb$
        \end{itemize}
        then $\chkseq[\prog]{\ptma}{\tyb}{\penv'}{\constrs'}$,
        where
        $\patsubst$ is a usable solution of $\constrs'$ and
        $\aps{\penv} \strictsubty \aps{\penv'}$.
    \item
        If:
        \begin{itemize}
            \item $\chkgseq[\prog]{\pata}{\seq{\pgd}}{\tya}{\penv}{\constrs}{\patb}$
            \item $\patsubst$ is a usable solution of $\constrs$
            \item $\tya \subtype \tyb$
        \end{itemize}
        then $\chkgseq[\prog]{\pata}{\seq{\pgd}}{\tyb}{\penv'}{\constrs'}{\patb}$ where
        $\patsubst$ is a usable solution of $\constrs'$ and $\aps{\penv}
        \strictsubty \aps{\penv'}$.
    \item If:
        \begin{itemize}
            \item $\chkgseq[\prog]{\pata}{\pgd}{\tya}{\penv}{\constrs}{\patb}$
            \item $\patsubst$ is a usable solution of $\constrs$
            \item $\tya \subtype \tyb$
        \end{itemize}
        then $\chkgseq[\prog]{\pata}{\pgd}{\tyb}{\penv'}{\constrs'}{\patb}$ where
        $\patsubst$ is a usable solution of $\constrs'$ and $\aps{\penv}
        \strictsubty \aps{\penv'}$.
\end{itemize}
\end{lemma}
\begin{proof}
    By mutual induction on the three premises. We concentrate on proving premise
    1 in detail, and \textsc{TCG-Recv} for premise 3; premise 2 follows from
    premise 3, and the remaining guard cases are straightforward.

    By induction on the derivation of $\chkseq{\ptma}{\tya}{\penv}{\constrs}$.

    \begin{proofcase}{TC-Var}

        Assumption:

        \begin{mathpar}
            \inferrule
            { }
            { \chkseq{x}{\tya}{x : \tya}{\emptyset} }
        \end{mathpar}

        Now given that we have $\tya \subtype \tyb$, we can construct:

        \begin{mathpar}
            \inferrule
            { }
            { \chkseq{x}{\tyb}{x : \tyb}{\emptyset} }
        \end{mathpar}

        As $\constrs = \cdot$ it straightforwardly follows that $\patsubst$ is a
        usable solution, and since $\tya \subtype \tyb$ we have that $x : \tya
        \strictsubty x : \tyb$ as required.
    \end{proofcase}

    \begin{proofcase}{TC-Let}

        Assumption:
        \begin{mathpar}
            \inferrule
            {
                \chkseq{\ptma}{\tya_{\var{ann}}}{\penv_1}{\constrs_1} \\
                \chkseq{\ptmb}{\tya}{\penv_2}{\constrs_2} \\
                \returnablepred{\tya_{\var{ann}}} \\\\
                \checkenvseq{\penv_2}{x}{\tya_{\var{ann}}}{\constrs_3} \\
                \joinseq{\penv_1}{(\penv_2 \envwithout x)}{\penv}{\constrs_4}
            }
            {
                \chkseq
                    {\letin[\tya_{\var{ann}}]{x}{\ptma}{\ptmb}}
                    { \tya }
                    { \penv }
                    { \constrs_1 \cup \cdots \cup \constrs_4 }
            }
        \end{mathpar}

        By the IH we have that:

        \begin{itemize}
            \item $\chkseq{\ptmb}{\tyb}{\penv_3}{\constrs_5}$ for some $\penv_3$,
                $\constrs_5$
            \item $\aps{\penv_2} \strictsubty \aps{\penv_3}$
            \item $\patsubst$ is a usable solution of $\constrs_5$
        \end{itemize}

        By the definition of strict environment subtyping, it follows that
        $\aps{\penv_2 \envwithout x} \strictsubty \aps{\penv_3 \envwithout x}$.

        By Corollary~\ref{cor:env-join-strict-subtype} we have that
        $\joinseq{\penv_1}{(\penv_3 \envwithout x)}{\penv'}{\constrs_6}$, where
        $\aps{\penv} \strictsubty \aps{\penv'}$ and $\patsubst$ is a usable
        solution of $\constrs_6$.

        By Lemma~\ref{lem:checkfn-subtype}, we have that
        $\checkenvseq{\penv_3}{x}{\tya_{\var{ann}}}{\constrs_7}$
        where $\patsubst$ is a usable solution of $\constrs_7$.

        Therefore we can show that:

        \begin{mathpar}
            \inferrule
            {
                \chkseq{\ptma}{\tya_{\var{ann}}}{\penv_1}{\constrs_1} \\
                \chkseq{\ptmb}{\tyb}{\penv_3}{\constrs_5} \\
                \returnablepred{\tya_{\var{ann}}} \\\\
                \checkenvseq{\penv_3}{x}{\tya_{\var{ann}}}{\constrs_7} \\
                \joinseq{\penv_1}{(\penv_3 \envwithout x)}{\penv'}{\constrs_6}
            }
            {
                \chkseq
                    {\letin[\tya_{\var{ann}}]{x}{\ptma}{\ptmb}}
                    { \tyb }
                    { \penv' }
                    { \constrs_1 \cup \constrs_5 \cup \constrs_7 \cup \constrs_6}
            }
        \end{mathpar}

        where $\patsubst$ is a usable solution of
        $\constrs_1 \cup \constrs_5 \cup \constrs_7 \cup \constrs_6$
        and $\aps{\penv} \strictsubty \aps{\penv'}$,
        as required.
    \end{proofcase}

    \begin{proofcase}{TC-Guard}

        Assumption:
        \begin{mathpar}
            \inferrule
            {
                \chkgseq
                    {\pata}
                    {\seq{\pgd}}
                    {\tya}
                    {\penv_1}
                    {\constrs_1}
                    {\patb} \\\\
                \chkseq{\vala}{\retty{\tyrecv{\patb}}}{\penv_2}{\constrs_2} \\
                \combineseq{\penv_1}{\penv_2}{\penv}{\constrs_3}
            }
            { \chkseq
                {\guardann{\vala}{\pata}{\seq{\pgd}}}
                {\tya}
                {\penv}
                {\constrs_1 \cup \constrs_2 \cup \constrs_3 \cup
                \set{\subpatconstr{\pata}{\patb} }}
            }
        \end{mathpar}

        By the IH we have that
        $\chkgseq{\pata}{\seq{\pgd}}{\tyb}{\penv'_1}{\constrs'_1}{\patb}$
        with $\patsubst$ a usable solution of $\constrs'_1$
        and $\aps{\penv_1} \strictsubty \aps{\penv'_1}$.
Since the output pattern $\patb$ is unchanged, the premise
        $\chkseq{\vala}{\retty{\tyrecv{\patb}}}{\penv_2}{\constrs_2}$ and
        the constraint $\subpatconstr{\pata}{\patb}$ are unaffected.

        By Corollary~\ref{cor:env-comb-strict-subtype}
        $\combineseq{\penv'_1}{\penv_2}{\penv'}{\constrs'_3}$ with
        $\aps{\penv} \strictsubty \aps{\penv'}$ and $\patsubst$ a usable
        solution of $\constrs'_3$.

        Recomposing:

        \begin{mathpar}
            \inferrule
            {
                \chkgseq
                    {\pata}
                    {\seq{\pgd}}
                    {\tyb}
                    {\penv'_1}
                    {\constrs'_1}
                    {\patb} \\\\
                \chkseq{\vala}{\retty{\tyrecv{\patb}}}{\penv_2}{\constrs_2} \\
                \combineseq{\penv'_1}{\penv_2}{\penv'}{\constrs'_3}
            }
            { \chkseq
                {\guardann{\vala}{\pata}{\seq{\pgd}}}
                {\tyb}
                {\penv'}
                {\constrs'_1 \cup \constrs_2 \cup \constrs'_3 \cup
                \set{\subpatconstr{\pata}{\patb} }}
            }
        \end{mathpar}

        with $\patsubst$ a usable solution of $\constrs'_1 \cup \constrs_2 \cup
        \constrs'_3 \cup \set{\subpatconstr{\pata}{\patb}}$ and
        $\aps{\penv} \strictsubty \aps{\penv'}$
        as required.
    \end{proofcase}

    \begin{proofcase}{TC-Sub}

        Assumptions:

        \begin{mathpar}
            \inferrule
            { \synthseq{\ptma}{\tya}{\penv}{\constrs_1} \\
              \subtyseq{\tya}{\tya'}{\constrs_2}
            }
            { \chkseq{\ptma}{\tya'}{\penv}{\constrs_1 \cup \constrs_2} }
        \end{mathpar}

        and:

        \begin{itemize}
            \item $\patsubst$ is a usable solution of $\constrs_1 \cup
                \constrs_2$
            \item $\tya' \subtype \tyb$
        \end{itemize}

        Since $\tya$, $\tya'$, and $\tyb$ contain no pattern variables, by
        Lemma~\ref{lem:subtype-constraints-closed} we have that
        $\pv{\constrs_2} = \emptyset$ (however, since $\patsubst$ is a usable
        solution of $\constrs_1 \cup \constrs_2$, it follows that $\constrs_2$
        is satisfiable).

        By Lemma~\ref{lem:subtype-solution}, we have that
        $\subtyseq{\tya}{\tyb}{\constrs_3}$, where $\constrs_3$ is satisfiable.

        Since $\constrs_3$ is satisfiable and (again by
        Lemma~\ref{lem:subtype-constraints-closed}) $\pv{\constrs_3} =
        \emptyset$, it follows that $\patsubst$ is a usable solution of
        $\constrs_1 \cup \constrs_3$.

        Thus by \textsc{TC-Sub} we have that:

        \begin{mathpar}
            \inferrule
            { \synthseq{\ptma}{\tya}{\penv}{\constrs_1} \\
              \subtyseq{\tya}{\tyb}{\constrs_3}
            }
            { \chkseq{\ptma}{\tyb}{\penv}{\constrs_1 \cup \constrs_3} }
        \end{mathpar}

        where $\patsubst$ is a usable solution of $\constrs_1 \cup \constrs_3$,
        as required.
    \end{proofcase}

    \begin{proofcase}{TCG-Recv}

        Assumption:
        \begin{mathpar}
            \inferrule
            {
                \chkseq{\ptma}{\tya}{\penv', y : \retty{\tyrecv{\ppata}}}{\constrs_1} \\
                \prog(\msgtag{m}) = {\seq{\pretya}} \\
                \penv = \penv' \envwithout \seq{x} \\
                \mbbase{\seq{\pretya}} \vee \mbbase{\penv} \\
                \checkenvseq{\penv'}{\seq{x}}{\seq{\makeusable{\pretya}}}{\constrs_2}
            }
            { \chkgseq
                {\pata}
                {\receive{m}{\seq{x}}{y}{\ptma}}
                {\tya}
                {\penv}
                {\constrs_1 \cup \constrs_2 \cup
                    \set{\subpatconstr{\pata \without \msgtag{m}}{\ppata}}
                }
                {\algomsg{m} \patconcat (\pata \without \msgtag{m})}
            }
        \end{mathpar}

        Also we have that:
        \begin{itemize}
            \item $\patsubst$ is a usable solution of
                $\constrs_1 \cup \constrs_2 \cup \set{\subpatconstr{\pata
                \without \msgtag{m}}{\ppata}}$
            \item $\tya \subtype \tyb$
        \end{itemize}

        By the IH we have that
        $\chkseq
            {\ptma}
            {\tyb}
            {\penv'', y : \retty{\tyrecv{\ppatb}}}
            {\constrs'_1}
        $
        where $\aps{\penv', y : \retty{\tyrecv{\ppata}}} \strictsubty
        \aps{\penv'', y : \retty{\tyrecv{\ppatb}}}$ and
        where $\patsubst$ is a usable solution of $\constrs'_1$.

        By the definition of strict environment subtyping we have that
        $\aps{\tyrecv{\ppata}} \subtype \aps{\tyrecv{\ppatb}}$ and therefore
        $\aps{\ppata} \subpatone \aps{\ppatb}$.
Since $\patsubst$ is a solution of
        $\subpatconstr{\pata \without \msgtag{m}}{\ppata}$, by transitivity
        of pattern inclusion $\patsubst$ is also a solution of
        $\subpatconstr{\pata \without \msgtag{m}}{\ppatb}$.

        Let $\penv''' = \penv'' \envwithout \seq{x}$. It follows by the
        definition of strict environment subtyping that $\aps{\penv}
        \strictsubty \aps{\penv'''}$.

        Due to the definition of the subtyping relation it remains the case that
        $\mbbase{\seq{\pretya}} \vee \mbbase{\penv'''}$.

        By Corollary~\ref{cor:checkfn-many-subtype} we have that
        $\checkenvseq{\penv''}{\seq{x}}{\seq{\makeusable{\pretya}}}{\constrs'_2}$
        where $\patsubst$ is a usable solution of $\constrs'_2$.

        Recomposing:
        \begin{mathpar}
            \inferrule
            {
                \chkseq{\ptma}{\tyb}{\penv'', y :
                \retty{\tyrecv{\ppatb}}}{\constrs'_1} \\
                \prog(\msgtag{m}) = {\seq{\pretya}} \\
                \penv''' = \penv'' \envwithout \seq{x} \\
                \mbbase{\seq{\pretya}} \vee \mbbase{\penv'''} \\
                \checkenvseq{\penv''}{\seq{x}}{\seq{\makeusable{\pretya}}}{\constrs'_2}
            }
            { \chkgseq
                {\pata}
                {\receive{m}{\seq{x}}{y}{\ptma}}
                {\tyb}
                {\penv'''}
                {\constrs'_1 \cup \constrs'_2 \cup
                    \set{\subpatconstr{\pata \without \msgtag{m}}{\ppatb}}
                }
                {\algomsg{m} \patconcat (\pata \without \msgtag{m})}
            }
        \end{mathpar}

        where $\patsubst$ is a usable solution of
        $
        {\constrs'_1 \cup \constrs'_2 \cup
            \set{\subpatconstr{\pata \without \msgtag{m}}{\ppatb}}
        }
        $
        and $\aps{\penv} \strictsubty \aps{\penv'''}$, as required.
    \end{proofcase}
\end{proof}

\subsubsection{Freshness of type variables}

It is convenient to reason about fresh variables.

\begin{definition}[Created fresh]
    A pattern variable $\patvara$ is \emph{created fresh} in a derivation
    $\derivd$ if $\derivd$ contains a leaf with the premise ``$\patvara$
    fresh''.
\end{definition}

\begin{lemma}[Pattern variable freshness]\label{lem:patvar-fresh}
    If $\derivd = \chkseq[\prog]{\tma}{\tya}{\penv}{\constrs}$ or
       $\derivd = \synthseq[\prog]{\tma}{\tya}{\penv}{\constrs}$,
       then all pattern
       variables in $\pv{\penv} \cup \pv{\constrs}$ are created fresh in
       $\derivd$.
\end{lemma}
\begin{proof}
    By induction on the respective derivation, noting that since the signature
    and types are closed, pattern variables are only introduced through the type
    join and type merge operators, where they are created fresh.
\end{proof}

\subsubsection{Full details of annotation relation}

\ADDED{
{\small
    \headertwo
        {Annotation rules for programs and definitions}
        {\framebox{$\annprogseq{\prog}{\pprog}$}~\framebox{$\anndefseq{D}{\pdef}$}}
\begin{mathpar}
    \inferrule
    {
        \prog = \progdef{\sigs}{\seq{D}}{\tma} \\
        (\anndefseq[\prog]{D_i}{\pdef_i})_i \\
        \anntmseq[\prog]{\cdot}{\tma}{\tyunit}{\ptma}
    }
    { \annprogseq{\prog}{\progdef{\sigs}{\seq{\pdef}}{\ptma}} }

    \inferrule
    {
        \anntmseq[\prog]{\seq{x : \tya}}{\tma}{\tyb}{\ptma}
    }
    { \anndefseq[\prog]
        {\fndef{f}{\seq{x : \tya}}{\tyb}{\tma}}
        {\fndef{f}{\seq{x : \tya}}{\tyb}{\ptma}}
    }
\end{mathpar}

\headersig{Annotation rules for computations}{$\anntmseq{\prog}{\tyenv}{\tma}{\tya}{\ptma}$}
\begin{mathpar}
    \inferrule
    [T-Ann-Var]
    { }
    { \anntmseq{\hastypetwo{x}{\tya}}{x}{\tya}{x}  }

    \inferrule
    [T-Ann-Const]
    { c \text{ has base type } \basety }
    { \anntmseq{\cdot}{c}{\basety}{c} }

    \inferrule
    [T-Ann-App]
    {
        \prog(f) = \fndef{f}{\seq{x : \tya}}{\tyb}{\tma} \\
        (\tseq{\tyenv_i}{\vala_i}{\tya_i})_{i \in 1..n}
    }
    { \anntmseq{\tyenv_1 + \cdots + \tyenv_n}
        {\fnapp{f}{\vala_1, \ldots, \vala_n}}
        {\tyb}
        {\fnapp{f}{\vala_1, \ldots, \vala_n}} }

    \inferrule
    [T-Ann-Let]
    {
        \anntmseq{\tyenv_1}{\tma}{\tya}{\ptma} \\
        \returnablepred{\tya} \\
        \anntmseq{\tyenv_2, x : \tya}{\tmb}{\tyb}{\ptmb}
    }
    { \anntmseq
        {\scombtwo{\tyenv_1}{\tyenv_2}}
        {\letin{x}{\tma}{\tmb}}
        {\tyb}
        {\letin[\tya]{x}{\ptma}{\ptmb}}
    }

    \inferrule
    [T-Ann-Spawn]
    {
      \anntmseq{\tyenv}{\tma}{\one}{\ptma}
    }
    {
    \anntmseq{\makeusable{\tyenv}}
        {\spawn{\tma}}
        {\one}
        {\spawn{\ptma}} }

    \inferrule
    [T-Ann-New]
    { }
    { \anntmseq{\cdot}{\mbnew}{\retty{\mbrecvone}}{\mbnew} }

    \inferrule
    [T-Ann-Send]
    {
      \proglookup{\algomsg{m}} = \seq{\pretya} \\
      \tseq{\tyenv}{\vala}{\usablety{\tysend{\algomsg{m}}}} \\
      (\tseq{\tyenv'_i}{\valb_i}{\makeusable{\pretya_i}})_{i \in 1..n}
    }
    { \anntmseq
        {\tyenv + \tyenv'_1 + \cdots +  \tyenv'_n }
        {\send{\vala}{m}{\seq{\valb}}}
        {\one}
        {\send{\vala}{m}{\seq{\valb}}}
    }

    \inferrule
    [T-Ann-Guard]
    {
      \tseq{\tyenv_1}{\vala}{\retty{\tyrecv{\pata}}} \\
      \anngseq{\pata}{\tyenv_2}{\seq{\gd}}{\tya}{\seq{\pgd}} \\
      \pnf{\pata}
    }
    { \anntmseq{{\tyenv_1} + {\tyenv_2}}
        {\guard{\vala}{\seq{\gd}}}
        {\tya}
    {\guardann{\vala}{\pata}{\seq{\pgd}}} }

    \inferrule
    [T-Ann-Fail]
    {
        \tseq{\tyenv}{\vala}{\retty{\mbrecvzero}}
    }
    {
        \anntmseq
            {\tyenv}
            {\sugarfail{\vala}}
            {\tya}
            {\sugarfail{\vala}}
    }

    \inferrule
    [T-Ann-Sub]
    { \subtypetwo{\tyenv}{\tyenv'} \\
      \subtypetwo{\tya}{\tyb} \\
      \anntmseq{\tyenv'}{\tma}{\tya}{\ptma} }
    { \anntmseq{\tyenv}{\tma}{\tyb}{\ptma}  }
\end{mathpar}

\headertwo
    {Annotation rules for guards}
    {
        \framebox{$\anngseq[\prog]{\pata}{\tyenv}{\seq{\gd}}{\tya}{\pgd}$}
        ~
        \framebox{$\anngseq[\prog]{\pata}{\tyenv}{\gd\vphantom{\seq{\gd}}}{\tya}{\seq{\pgd}}$}
    }
\begin{mathpar}
    \inferrule
    [TG-Ann-GuardSeq]
    {
        (\anngseq{\pata_i}{\tyenv}{\gd_i}{\tya}{\pgd_i})_{i \in 1..n}
    }
    {
        \anngseq
            {\pata_1 \patplus \cdots \patplus \pata_n}
            {\tyenv}
            {\seq{\gd}}
            {\tya}
            {\seq{\pgd}}
    }

    \inferrule
    [TG-Ann-Free]
    {
        \anntmseq{\tyenv}{\tma}{\tya}{\ptma}
    }
    { \anngseq
        {\mbone}{\tyenv}{\free{\tma}}{\tya}
        {\free{\ptma}}
    }

    \inferrule
    [TG-Ann-Recv]
    {
      \prog(\msgtag{m}) = \seq{\pretya} \\
      \mbbase{\seq{\pretya}} \vee \mbbase{\tyenv} \\
      \anntmseq
        {\tyenv,
            y : \usageann{\returnable}{\tyrecv{E}},
            \seq{x} : \seq{\makeusable{\pretya}}}
        {\tma}
        {\tyb}
        {\ptma}
    }
    {
      \anngseq
        {\algomsg{m} \patconcat E}
        {\tyenv}
        {\receive{m}{\seq{x}\!}{\!y}{\tma}}
        {\tyb}
        {\receive{m}{\seq{x}\!}{\!y}{\ptma}}
    }
\end{mathpar}
}
}

\subsubsection{Completeness proof}

Finally, we can tie the above results together to show algorithmic completeness.

\algocompleteness*
\begin{proof}
    A direct consequence of Lemma~\ref{lem:algo-completeness-generalised}.
\end{proof}

\begin{lemma}[Algorithmic completeness (generalised)]
    \label{lem:algo-completeness-generalised}\hfill
\begin{itemize}
    \item If $\anntmseq[\prog]{\tyenv}{\tma}{\tya}{\ptma}$, then
            there exist some $\penv, \constrs$
            and usable solution $\patsubst$ of $\constrs$ such that
            $\chkseq{\ptma}{\tya}{\penv}{\constrs}$ where
            $\dom{\patsubst} = \pv{\penv} \cup \pv{\constrs}$ and
            $\tyenv \subtype \aps{\penv}$.
\item If $\anngseq[\prog]{\pata}{\tyenv}{\seq{\gd}}{\tya}{\seq{\pgd}}$ where
              $\pnf{\pata}$, then
              there exist some $\penv, \constrs$, usable
              solution $\patsubst$ of $\constrs$, and $\pata' \tyequiv \pata$ such that
         $\chkgseq{\pata}{\seq{\pgd}}{\tya}{\penv}{\constrs}{\pata'}$
              where $\dom{\patsubst} = \pv{\penv} \cup \pv{\constrs}$
              and $\tyenv \subtype \aps{\penv}$.
\item Given
            $\pata = \patb_1 \patplus \cdots \patplus \patb_n$ where $\pnf{\pata}$,
            if $\anngseq[\prog]{\patb_i}{\tyenv}{\gd}{\tya}{\pgd}$ for some $i \in 1..n$,
            then there exist some $\penv, \constrs$, usable
            solution $\patsubst$ of $\constrs$, and $\patb'_i \tyequiv \patb_i$ such that
            $\chkgseq{\pata}{\pgd}{\tya}{\penv}{\constrs}{\patb'_i}$
            where $\dom{\patsubst} = \pv{\penv} \cup \pv{\constrs}$
            and $\tyenv \subtype \aps{\penv}$.
    \end{itemize}
\end{lemma}
\begin{proof}
    By mutual induction on the three premises.

    \paragraph*{Premise 1.}

\begin{proofcase}{T-Ann-Var}
    Assumption:

    \begin{mathpar}
        \inferrule
        { }
        { \anntmseq{x : \tya}{x}{\tya}{x} }
    \end{mathpar}

    Recomposing via \textsc{TC-Var}:

    \begin{mathpar}
        \inferrule
        { }
        { \chkseq{x}{\tya}{x : \tya}{\emptyset} }
    \end{mathpar}

    with $\patsubst = \cdot$.

\end{proofcase}

\begin{proofcase}{T-Ann-Const}

    Assumption:

    \begin{mathpar}
        \inferrule
        { c \text{ has base type } \basety }
        { \anntmseq{\cdot}{c}{\basety}{c} }
    \end{mathpar}

    By \textsc{TS-Const}:

    \begin{mathpar}
        \inferrule
        { c \text{ has base type } \basety }
        { \synthseq{c}{\basety}{\cdot}{\emptyset} }
    \end{mathpar}

    By Lemma~\ref{lem:synth-to-chk} we have that $\chkseq{c}{\basety}{\cdot}{\emptyset}$

    with $\patsubst = \cdot$, as required.
\end{proofcase}

\begin{proofcase}{T-Ann-App}
    Assumption:

    \begin{mathpar}
    \inferrule
    {
        \prog(f) = \fndef{f}{\seq{x : \tya}}{\tyb}{\tma} \\
        (\tseq{\tyenv_i}{\vala_i}{\tya_i})_{i \in 1..n}
    }
    { \anntmseq
        {\tyenv_1 + \cdots + \tyenv_n}
        {\fnapp{f}{\seq{\vala}}}
        {\tyb}
        {\fnapp{f}{\seq{\vala}}}
    }
    \end{mathpar}

    By (repeated) use of Lemma~\ref{lem:val-completeness}, we have that there
    exist some $\tyenv'_i$ such that $\tyenv_i \subtype \tyenv'_i$
    and $\chkseq{\vala_i}{\tya_i}{\tyenv'_i}{\emptyset}$ for $i \in 1..n$.

    By repeated use of Lemma~\ref{lem:combine-completeness}, we have that there
    $\combineseqmany{\tyenv'_1 + \cdots + \tyenv'_n}{\penv}{\constrs}$
    for some $\penv, \constrs$ and that there
    exists some usable solution $\patsubst$ of
    $\tyenv \subtype \aps{\penv}$.

    Thus by \textsc{TS-App} we can show

    \begin{mathpar}
            \inferrule
            {
                \proglookup{f} = {\tyfun{\seq{\tya}}{\tyb}} \\
                (\chkseq{\vala_i}{\tya_i}{\tyenv'_i}{\emptyset})_{i \in 1..n} \\
                \combineseqmany{\tyenv'_1 + \cdots + \tyenv'_n}{\penv}{\constrs}
            }
            { \synthseq
                { \fnapp{f}{\seq{\vala}} }
                {\tyb}
                {\penv}
                {\constrs}
            }
    \end{mathpar}
    and by Lemma~\ref{lem:synth-to-chk} we have that
    $
        \chkseq
            { \fnapp{f}{\seq{\vala}} }
            {\tyb}
            {\penv}
            {\constrs}
    $
    as required.
\end{proofcase}

\begin{proofcase}{T-Ann-Let}

    Assumption:
    \begin{mathpar}
    \inferrule
    {
        \anntmseq{\tyenv_1}{\tma}{\tya}{\ptma} \\
        \returnablepred{\tya} \\
        \anntmseq{\tyenv_2, x : \tya}{\tmb}{\tyb}{\ptmb}
    }
    { \anntmseq
        {\scombtwo{\tyenv_1}{\tyenv_2}}
        {\letin{x}{\tma}{\tmb}}
        {\tyb}
        {\letin[\tya]{x}{\ptma}{\ptmb}}
    }
    \end{mathpar}

    By the IH we have that:
    \begin{itemize}
        \item There exist some $\penv_1, \constrs_1$ and usable solution $\patsubst_1$ of $\constrs_1$ such that
            $\chkseq{\ptma}{\tya}{\penv_1}{\constrs_1}$ where $\tyenv_1
            \subtype \aps[\patsubst_1]{\penv_1}$
        \item There exist some $\penv_2, \constrs_2$ and usable solution $\patsubst_2$ of $\constrs_2$ such that
            $\chkseq
                {\ptmb}
                {\tyb}
                {\penv_2}
                {\constrs_2}$
            where
            $\tyenv_2, x : \tya \subtype \aps[\patsubst_2]{\penv_2}$
    \end{itemize}

    By Lemma~\ref{lem:join-completeness}, we have that
    $\joinseq{\penv_1}{(\penv_2 - x)}{\penv}{\constrs_3}$ and a usable
    solution $\patsubst \supseteq \patsubst_1 \cup \patsubst_2$ of $\constrs_3$ such that
    $\scombtwo{\tyenv_1}{\tyenv_2} \subtype \aps{\penv}$.

    By Lemma~\ref{lem:checkfn-subtype}, we have that
    $\checkenvseq{\penv_2}{x}{\tya}{\constrs_4}$ and $\patsubst$
    is a usable solution of $\constrs_4$.

    Since $\patsubst \supseteq \patsubst_1 \cup \patsubst_2$ and pattern variables in these
    subderivations are only introduced fresh (Lemma~\ref{lem:patvar-fresh}),
    we have that $\patsubst$ is also a usable solution of $\constrs_1$ and
    $\constrs_2$.

    Therefore, we have that $\patsubst$ is a
    usable solution of $\constrs_1 \cup \constrs_2 \cup \constrs_3 \cup
    \constrs_4$.

    Recomposing using \textsc{TC-Let}:
    \begin{mathpar}
    \inferrule
    {
        \chkseq{\ptma}{\tya}{\penv_1}{\constrs_1} \\
        \chkseq{\ptmb}{\tyb}{\penv_2}{\constrs_2} \\
        \returnablepred{\tya} \\\\
        \checkenvseq{\penv_2}{x}{\tya}{\constrs_4} \\
        \joinseq{\penv_1}{(\penv_2 - x)}{\penv}{\constrs_3}
    }
    {
        \chkseq
            {\letin[\tya]{x}{\ptma}{\ptmb}}
            { \tyb }
            { \penv }
            { \constrs_1 \cup \cdots \cup \constrs_4 }
    }
    \end{mathpar}

    where $\aps{\penv} \subtype \tyenv_1 \scomb \tyenv_2$
    and
    $\patsubst$ is a usable solution of
    $\constrs_1 \cup \constrs_2 \cup \constrs_3 \cup \constrs_4$, as required.
\end{proofcase}
\begin{proofcase}{T-Ann-Spawn}
    \begin{mathpar}
        \inferrule
        {
            \anntmseq{\tyenv}{\tma}{\one}{\ptma}
        }
        {
            \anntmseq
                {\makeusable{\tyenv}}
                {\spawn{\tma}}
                {\one}
                {\spawn{\ptma}}
        }
    \end{mathpar}

    By the IH
    $\chkseq{\tma}{\one}{\penv}{\constrs}$ for some $\penv,\constrs$,
    and a usable solution $\patsubst$ such that
    $\tyenv \subtype \aps{\constrs}$.

    Thus by \textsc{TS-Spawn}:
    \begin{mathpar}
        \inferrule
            {
                \chkseq{\ptma}{\one}{\penv}{\constrs}
            }
            { \synthseq{\spawn{\ptma}}{\one}{\makeusable{\penv}}{\constrs} }
    \end{mathpar}
    where $\tyenv \subtype \aps{\penv}$ and therefore $\makeusable{\tyenv}
    \subtype \makeusable{\aps{\penv}}$.

    Finally, by Lemma~\ref{lem:synth-to-chk}, we have that
    $\chkseq{\spawn{\ptma}}{\one}{\makeusable{\penv}}{\constrs}$ with usable
    solution $\patsubst$ of $\constrs$ as required.
\end{proofcase}
\begin{proofcase}{T-Ann-New}
    Assumption:
    \begin{mathpar}
        \inferrule
        { }
        { \anntmseq{\cdot}{\mbnew}{\retty{\mbrecvone}}{\mbnew} }
    \end{mathpar}

    By \textsc{TS-New} we have that
    $\synthseq{\mbnew}{\retty{\mbrecvone}}{\cdot}{\emptyset}$
    and by Lemma~\ref{lem:synth-to-chk} it follows that
    $\chkseq{\mbnew}{\retty{\mbrecvone}}{\cdot}{\emptyset}$; we can set
    solution $\patsubst = \cdot$, as required.
\end{proofcase}

\begin{proofcase}{T-Ann-Send}

    Assumption:

    \begin{mathpar}
        \inferrule
        {
          \siglookup{m} = \seq{\pretya} \\
          \tseq{\tyenv_{\var{target}}}{\vala}{\usablety{\tysend{\algomsg{m}}}} \\
          (\tseq{\tyenv'_i}{\valb_i}{\makeusable{\pretya_i}})_{i \in 1..n}
        }
        { \anntmseq
            {\tyenv_{\var{target}} + \tyenv'_1 + \cdots +  \tyenv'_n }
            {\send{\vala}{m}{\seq{\valb}}}
            {\one}
            {\send{\vala}{m}{\seq{\valb}}}
        }
    \end{mathpar}

    By the IH we have that:
    \begin{itemize}
        \item
            $\chkseq
                {\vala}
                {\usablety{\tysend{\algomsg{m}}}}
                {\penv_{\var{target}}}
                {\constrs_{\var{target}}}$
            for some $\penv_{\var{target}}$, $\constrs_{\var{target}}$ and some usable solution $\patsubst_{\var{target}}$ of $\constrs_{\var{target}}$
            such that $\tyenv_{\var{target}} \subtype \aps{\penv_{\var{target}}}$.
        \item
            $
              (\chkseq
                {\valb_i}
                {\makeusable{\pretya_i}}
                {\penv_i}
                {\constrs_i})_{i \in 1..n}
            $
            for $\penv_i$, $\constrs_i$ and usable solutions $\patsubst_i$ of $\constrs_i$ such that
            $\tyenv'_i \subtype \aps[\patsubst_i]{\penv_i}$
    \end{itemize}

    By repeated use of Lemma~\ref{lem:combine-completeness} we have that
    $\combineseqmany{\penv_{\var{target}} + \penv_1 + \cdots + \penv_n}{\penv}{\constrs_{\var{env}}}$, with some usable solution $\patsubst_{\var{env}}$ of $\constrs_{\var{env}}$ such that $\tyenv + \tyenv'_1 + \cdots + \tyenv'_n \subtype \aps[\patsubst_{\var{env}}]{\penv}$.

    Since pattern variables are always chosen fresh (Lemma~\ref{lem:patvar-fresh}) we have that
    $\patsubst_{\var{target}} \cup \patsubst_{\var{env}} \cup \bigcup_{i \in 1..n} \patsubst'_i$ is a solution of $\constrs_{\var{target}} \cup \constrs_{\var{env}} \cup \bigcup_{i \in 1..n} \constrs'_i$.

    Thus we can show by \textsc{TS-Send} and Lemma~\ref{lem:synth-to-chk}:

    \begin{mathpar}
        \inferrule
        {
            \inferrule
            {
                \siglookup{m} = \seq{\pretya} \\
                \chkseq{\vala}{\usablety{\tysend{\algomsg{m}}}}{\penv_{\var{target}}}{\constrs_{\var{target}}} \\\\
                (\chkseq{\valb_i}{\makeusable{\pretya_i}}{\penv_i}{\constrs_i})_{i \in 1..n} \\
                \combineseqmany{\penv_{\var{target}} + \penv_1 + \cdots + \penv_n}{\penv}{\constrs_{\var{env}}} \\
            }
            { \synthseq
                {\send{\vala}{m}{\seq{\valb}}}
                {\one}
                {\penv}
                {\constrs_{\var{target}} \cup \constrs_1 \cup \cdots \cup \constrs_n \cup \constrs_{\var{env}}}
            }
        }
        {
            \chkseq
              {\send{\vala}{m}{\seq{\valb}}}
              {\one}
              {\penv}
              {\constrs_{\var{target}} \cup \constrs_1 \cup \cdots \cup \constrs_n \cup \constrs_{\var{env}}}
        }
    \end{mathpar}

    where $\patsubst$ is a usable solution of
    $\constrs_{\var{target}} \cup \constrs_1 \cup \cdots \cup \constrs_n \cup \constrs_{\var{env}}$
    and $\tyenv \subtype \aps{\penv}$, as required.
\end{proofcase}

\begin{proofcase}{T-Ann-Guard}

    Assumption:

    \begin{mathpar}
        \inferrule
        {
          \tseq{\tyenv_1}{\vala}{\retty{\tyrecv{\pata}}} \\
          \anngseq{\pata}{\tyenv_2}{\seq{\gd}}{\tya}{\seq{\pgd}}
          \\
          \pnf{\pata}
        }
        { \anntmseq
            {{\tyenv_1} + {\tyenv_2}}
            {\guard{\vala}{\seq{\gd}}}
            {\tya}
            {\guardann{\vala}{\pata}{\seq{\pgd}}}
        }
    \end{mathpar}

    By the IH (premise 2) we have that
    $\chkgseq{\pata}{\seq{\pgd}}{\tya}{\penv'}{\constrs}{\pata'}$
    where $\patsubst$ is a usable solution of $\constrs$, and
    $\tyenv_2 \subtype \penv'$, and $\pata \tyequiv \pata'$.

    By Lemma~\ref{lem:val-completeness} we have that
    $\chkseq{\vala}{\retty{\tyrecv{\pata'}}}{\tyenv'_1}{\emptyset}$ where
    $\tyenv_1 \subtype \tyenv'_1$.

    By Lemma~\ref{lem:combine-completeness} we have that
    $\combineseq{\penv'}{\tyenv'_1}{\penv}{\constrs_2}$
    with $\tyenv_1 + \tyenv_2 \subtype \aps{\penv}$ and where $\patsubst$ is a
    solution of $\constrs_2$.

    Recomposing:

    \begin{mathpar}
        \inferrule
        {
            \chkgseq
                {\pata}
                {\seq{\pgd}}
                {\tya}
                {\penv'}
                {\emptyset}
                {\pata'} \\\\
            \chkseq{\vala}{\tyrecv{\pata'}}{\tyenv'_1}{\emptyset} \\
            \combineseq{\penv'}{\tyenv'_1}{\penv}{\constrs}
        }
        { \chkseq
            {\guardann{\vala}{\pata}{\seq{\pgd}}}
            {\tya}
            {\penv}
            {\constrs \cup \set{\subpatconstr{\pata}{\pata'}}}
        }
    \end{mathpar}

    where $\patsubst$ is a usable solution of $\constrs_1 \cup \constrs_2$ and
    since $\pata' \tyequiv \pata$ it follows that $\pata \subpatone \pata'$, as required.
\end{proofcase}

\begin{proofcase}{T-Ann-Fail}
    Assumption:
    \begin{mathpar}
        \inferrule
        {
            \tseq{\tyenv}{\vala}{\retty{\mbrecvzero}}
        }
        {
            \anntmseq
                {\tyenv}
                {\sugarfail{\vala}}
                {\tya}
                {\sugarfail{\vala}}
        }
    \end{mathpar}

    By the IH (1) we have that there exists some $\penv$ and $\constrs$
    and usable solution $\patsubst$ of
    $\chkseq{\vala}{\retty{\mbrecvzero}}{\penv}{\constrs}$ such that
    $\tyenv \subtype \aps{\penv}$.

    Recomposing by \textsc{TC-Fail}:

    \begin{mathpar}
        \inferrule
        {
            \chkseq{\vala}{\retty{\mbrecvzero}}{\penv}{\constrs}
        }
        { \chkseq{\sugarfail{\vala}}{\tya}{\penv}{\constrs} }
    \end{mathpar}
    where $\patsubst$ is a usable solution of $\constrs$
    and $\tyenv \subtype \aps{\penv}$ as required.
\end{proofcase}

\begin{proofcase}{T-Ann-Sub}
    \begin{mathpar}
    \inferrule
    { \subtypetwo{\tyenv}{\tyenv'} \\
      \subtypetwo{\tya}{\tyb} \\
      \anntmseq{\tyenv'}{\tma}{\tya}{\ptma} }
      { \anntmseq{\tyenv}{\tma}{\tyb}{\ptma}  }
    \end{mathpar}

    By the IH, we have that there exist $\penv, \constrs$ and some usable solution $\patsubst$ of $\constrs$ such that $\tyenv' \subtype \aps{\penv}$ and $\chkseq{\tma}{\tya}{\penv}{\constrs}$.

    By Lemma~\ref{lem:supertype-checkability} we have that
    $\chkseq{\ptma}{\tyb}{\penv'}{\constrs'}$ where $\patsubst$ is a usable
    solution of $\constrs'$ and $\aps{\penv} \subtype \aps{\penv'}$.

    Recalling that $\tyenv\subtype \tyenv'$, and $\tyenv' \subtype \aps{\penv}$,
    and noting that $\aps{\penv} \subtype \aps{\penv'}$ and
    that $\aps{\penv} \subtype \aps{\penv'}$, by the transitivity of subtyping
    we have that $\tyenv \subtype \aps{\penv'}$.

    Therefore we have that:

    \begin{itemize}
        \item $\chkseq{\ptma}{\tyb}{\penv'}{\constrs'}$
        \item $\patsubst$ is a usable solution of $\constrs'$
        \item $\tyenv \subtype \aps{\penv'}$
    \end{itemize}

    as required.
\end{proofcase}

\paragraph*{Premise 2.}
\begin{proofcase}{TG-Ann-GuardSeq}
    \begin{mathpar}
      \inferrule
      {
          (\anngseq{\patb_i}{\tyenv}{\gd_i}{\tya}{\pgd_i})_{i \in I}
      }
      {
          \anngseq
              {\pata}
              {\tyenv}
              {\seq{\gd}}
              {\tya}
              {\seq{\pgd}}
      }
    \end{mathpar}
    where $\pata = \patb_1 \patplus \cdots \patplus \patb_n$.

    By repeated use of the IH (3) we have that
    $\chkgseq{\pata}{\pgd_i}{\tya}{\penv_i}{\constrs_i}{\patb'_i}$
    for some $\penv_i, \patsubst_i$ such that
    $\tyenv \subtype \aps[\patsubst_i]{\penv_i}$ and
    $\patb'_i \tyequiv \patb_i$ for each $i \in I$.

    By the definition of equivalence we can construct
    $\pata' = \patb'_1 \patplus \cdots \patplus \patb'_n \tyequiv \pata$.

    By Lemma~\ref{lem:merge-completeness} we have that
    $\manymergeseq{\penv_1}{\penv_n}{\penv_{\var{env}}}{\constrs_{\var{env}}}$
    for some $\penv_{\var{env}}$ and $\constrs_{\var{env}}$
    and some solution $\patsubst_{\var{env}}$ of $\constrs_{\var{env}}$ such that
    $\tyenv \subtype \aps[(\patsubst_1 \cup \cdots \cup \patsubst_n \cup \patsubst_{\var{env}})]{\penv_{\var{env}}}$.

    Recomposing by \textsc{TCG-Guards}:

    \begin{mathpar}
        \inferrule
        {
            (\chkgseq{\pata}{\pgd_i}{\tya}{\penv_i}{\constrs_i}{\patb_i})_{i \in 1..n} \\
            \manymergeseq
                {\penv_1}
                {\penv_n}
                {\penv_{\var{env}}}
                {\constrs_{\var{env}}}
        }
        { \chkgseq
            {\pata}
            {\seq{\pgd}}
            {\tya}
            {\penv_{\var{env}}}
            {\constrs_{\var{env}} \cup \constrs_1 \cup \cdots \cup \constrs_n}
            {\pata'}
        }
    \end{mathpar}

    Since pattern variables are generated fresh, we have that the pattern variables for each $\patsubst_i$ are disjoint.
    Therefore, we have that:

    \begin{itemize}
        \item $\patsubst = \bigcup_{i \in 1..n} \patsubst_i \cup \patsubst_{\var{env}}$ is a usable solution of $\constrs = (\bigcup_{i \in 1..n} \constrs_n) \cup \constrs_{\var{env}}$
        \item $\tyenv \subtype \aps{\penv_{\var{env}}}$
    \end{itemize}

    as required.

\end{proofcase}

    \paragraph*{Premise 3.}

    Recall that we have some $\pata = \patb_1 \patplus \cdots \patplus \patb_n$
    and we are considering some $\patb_i$.  In each of the following we consider
    without loss of generality that $i = 1$.

\begin{proofcase}{TG-Ann-Free}

    \begin{mathpar}
        \inferrule
        {
            \anntmseq{\tyenv}{\tma}{\tya}{\ptma}
        }
        { \anngseq{\mbone}{\tyenv}{\free{\tma}}{\tya}{\free{\ptma}} }
    \end{mathpar}

    By the IH (1) we have that there exist $\penv, \constrs$ and usable solution
    $\patsubst$ of $\constrs$ such that $\chkseq{\tma}{\tya}{\penv}{\constrs}$
    with $\tyenv \subtype \aps{\penv}$.

    Recomposing by \textsc{TCG-Free}:

    \begin{mathpar}
        \inferrule
        { \chkseq{\ptma}{\tya}{\penv}{\constrs} }
        { \chkgseq{\pata}{\free{\ptma}}{\tya}{\penv}{\constrs}{\mbone} }
    \end{mathpar}

    with $\tyenv \subtype \aps{\penv}$ as required.
\end{proofcase}
\begin{proofcase}{TG-Ann-Recv}

    Assumption:
    \begin{mathpar}
        \inferrule
        {
          \proglookup{\msgtag{m}} = \seq{\pretya} \\
          \mbbase{\seq{\pretya}} \vee \mbbase{\tyenv} \\
          \anntmseq
            {\tyenv,
                y : \usageann{\returnable}{\tyrecv{\patb}},
                \seq{x} : \seq{\makeusable{\pretya}}}
            {\tma}
            {\tyb}
            {\ptma}
        }
        {
          \anngseq
            {\algomsg{m} \patconcat \patb}
            {\tyenv}
            {\receive{m}{\seq{x}}{y}{\tma}}
            {\tyb}
            {\receive{m}{\seq{x}}{y}{\ptma}}
        }
    \end{mathpar}

    We also know that
    $\pata = (\algomsg{m} \patconcat \patb) \patplus \patb_2 \patplus \cdots \patplus \patb_n$
    and $\pnf{\pata}$.

    Since $\pnf{\pata}$ it follows from the definition of PNF that:

    \begin{mathpar}
        \inferrule
        { \patb \tyequiv \pata \without \algomsg{m} }
        { \pnflittwo{\pata}{(\algomsg{m} \patconcat \patb)} }
    \end{mathpar}

    Let $\tyenv' = \tyenv,
                y : \usageann{\returnable}{\tyrecv{\patb}},
                \seq{x} : \seq{\makeusable{\pretya}}$.

    By the IH (premise 1) we have that there exist $ \penv, \constrs$ and usable solution
    $\patsubst$ of $\constrs$ s.t.\ $\chkseq{\ptma}{\tya}{\penv, y :
    \retty{\tyrecv{\ppata}}}{\constrs}$ where
    $\dom{\patsubst} = \pv{\penv} \cup \pv{\constrs}$ and $\tyenv' \subtype
    \aps{\penv, y : \retty{\tyrecv{\ppata}}}$.

    We next need to show that $\mbbase{\seq{\pretya}} \vee
    \mbbase{\tyenv}$ implies that $\mbbase{\seq{\pretya}} \vee \mbbase{\penv}$. 
    It suffices to show that $\mbbase{\tyenv}$ implies
    $\mbbase{\penv}$, which follows directly from the fact that
    $\tyenv \subtype \aps{\penv}$: base types can only be subtypes of
    themselves, and by the definition of environment subtyping
    $\dom{\penv} \subseteq \dom{\tyenv}$ with any overlapping
    variables only being of base type.
    
    Next, since $\tyenv, y : \retty{\tyrecv{\patb}}, \seq{x} :
    \makeusable{\seq{\pretya}} \subtype \aps{\penv}, y :
    \retty{\tyrecv{\aps{\ppata}}}$
    it follows that $\tyenv, \seq{x} :
    \makeusable{\seq{\pretya}} \subtype \aps{\penv}$ and thus by
    Corollary~\ref{cor:checkfn-many-subtype} we have that
    $\checkenvseq{\penv}{\seq{x}}{\seq{\makeusable{\pretya}}}{\constrs_2}$ where
    $\patsubst$ is a usable solution of $\constrs_2$.

    Next, since $\tyenv, y : \retty{\tyrecv{\patb}} \subtype
    \aps{\penv}, y : \retty{\tyrecv{\aps{\ppata}}}$ it follows by the definition of
    subtyping that $\patb \subpatone \aps{\ppata}$.

    We have one final proof obligation: showing that $\patsubst$ solves
    $\subpatconstr{(\pata \without \msgtag{m})}{\ppata}$.

    Since $\pnflittwo{\pata}{\msgtag{m} \patconcat \patb}$ we have that $\patb
    \tyequiv \pata \without \msgtag{m}$ and therefore both $\patb \subpatone (\pata \without \msgtag{m})$
        and $(\pata \without \msgtag{m}) \subpatone \patb$.

    Since $\tyrecv{\patb} \subtype \tyrecv{\aps{\ppata}}$ we have that $\patb
    \subpatone \aps{\ppata}$. Thus by transitivity we have that $\pata \without
    \msgtag{m} \subpatone \patb \subpatone \aps{\ppata}$ and therefore that
    $\patsubst$ solves
    $\subpatconstr{(\pata \without \msgtag{m})}{\ppata}$ as necessary.

    Similarly since $\patb \tyequiv \pata \without \algomsg{m}$ it follows by
    Lemma~\ref{lem:subpat-precongruence} that
    $\algomsg{m} \patconcat (\pata \without \algomsg{m}) \tyequiv
    \algomsg{m} \patconcat \patb$.

    Thus, recomposing, we have:

    \begin{mathpar}
        \inferrule
        {
            \chkseq{\ptma}{\tya}{\penv, y : \retty{\tyrecv{\ppata}}}{\constrs_1} \\
            \siglookup{\msgtag{m}} = {\seq{\pretya}} \\
            \penv' = \penv \envwithout \seq{x} \\
            \mbbase{\seq{\pretya}} \vee \mbbase{\penv} \\
            \checkenvseq{\penv}{\seq{x}}{\seq{\makeusable{\pretya}}}{\constrs_2}
        }
        { \chkgseq
            {\pata}
            {\receive{m}{\seq{x}}{y}{\ptma}}
            {\tya}
            {\penv'}
            {\constrs_1 \cup \constrs_2 \cup
                \set{\subpatconstr{\pata \without \msgtag{m}}{\ppata}}
            }
            {\algomsg{m} \patconcat (\pata \without \msgtag{m})}
        }
    \end{mathpar}

    as required.

\end{proofcase}

\end{proof}
 
 \newpage
\ADDED{
\section{Supplement to \S\ref{sec:extensions}}\label{ap:extensions-proofs}

The key threat to soundness is a function that closes over variables that
initially satisfy the ordering invariant on quasi-linearity annotations, but
where these are violated after substitution, as shown in
\S\ref{sec:extensions}:

{\small
\[
    \bl
        \letintwo{\var{mb}}{\mbnew} \\
        \letintwo{f}{(\linlambda{}{\one}{\send{\var{mb}}{m}{}})} \\
        \guardanntwo{\var{mb}}{\msgtag{m}} \\
        \quad \receive{m}{}{\var{mb}}{\sugarfree{\var{mb}}} \\
        \}; \\
        \fnapp{f}{}
    \el
\]
}

As we have seen, this is avoided by ensuring that $\lambda$-abstractions close
only over values of returnable type.
This issue would manifest itself most clearly in
Lemma~\ref{lem:returnable-val-env} that shows that if a value has a returnable
type under a cruftless environment $\tyenv$, then $\tyenv$ must be returnable.
This lemma is used within the \textsc{E-Return} case of the preservation proof,
and would not hold without the restriction that $\lambda$-abstractions close
over only returnable values.

We can state an updated version of Lemma~\ref{lem:returnable-val-env}:

\begin{lemma}\label{lem:returnable-val-env-lambdas}
If $\vseq{\tyenv}{\vala}{\tya}$ where $\returnablepred{\tya}$ and $\tyenv$ is
cruftless for $\vala$, then $\returnablepred{\tyenv}$.
\end{lemma}
\begin{proof}
    By case analysis on the derivation of $\vseq{\tyenv}{\vala}{\tya}$.

    We additionally need to consider the case where
    $\vseq
        {\tyenv}
        {(\flaglambda{\seq{x : \tya}}{\tyb}{\tma})}
        {\tyannfun{\seq{\tya}}{\tyb}}$; in both of the cases where
        $\linann = \linear$ and $\linann = \unrestricted$, the typing rules
        \textsc{T-LinLambda} and \textsc{T-UnLambda} require that
        $\returnablepred{\tyenv}$ as required.
\end{proof}

We also need an updated version of the substitution lemma
(Lemma~\ref{lem:substitution}), which follows straightforwardly.

Finally, we need to consider the case for function
application:

\begin{theorem}[Preservation (\langname with first-class functions)]
    If $\progseq{\prog}$, and
    $\cseq[\prog]{\Gamma}{\config{C}}$ with $\tyenv$ reliable, and
    $\config{C} \cevalpnopad \config{D}$, then $\cseq[\prog]{\tyenv}{\config{D}}$.
\end{theorem}
\begin{proof}
    By induction on the derivation of $\config{C} \ceval \config{D}$.

    \[
        \thread{(\flaglambda{\seq{x : \tya}}{\tyb}{\tma})(\seq{\vala})}{\framestack}
        \ceval
        \thread{\subst{\tma}{\seq{\vala}}{\seq{x}}}{\framestack}
    \]

    The cases where $\linann = \unrestricted$ and $\linann = \linear$ are
    similar, so we consider the case where $\linann = \linear$.

    Assumption:

    {\scriptsize
    \begin{mathpar}
        \inferrule*
        {
            {
            \begin{array}{l}
                \plainenv = \strip{\scombtwo{\tyenvb_1}{\tyenvb_2}} \\
                \tyenvb_1 = \tyenv + \tyenv_1 + \cdots + \tyenv_n
            \end{array}
            }
            \\
            \inferrule
            {
                \inferrule*
                {
                    \inferrule*
                    {
                      \tseq{\tyenv', \seq{x : \tya'}}{\tma}{\tyb'} \\\\
                        \returnablepred{\tyenv'} 
                    }
                    {\tseq
                        {\tyenv'}
                        {\linlambda{\seq{x : \tya'}}{\tyb'}{\tma}}
                        {\tylinfun{\seq{\tya'}}{\tyb'}}
                    }
                }
                {
                    {\tseq
                        {\tyenv}
                        {\linlambda{\seq{x : \tya'}}{\tyb'}{\tma}}
                        {\tylinfun{\seq{\tya}}{\tyb}}
                    }
                }
                \\
(\tseq{\tyenv_i}{\vala_i}{\tya_i})_{i \in 1..n}
            }
            {
                \tseq
                    {\tyenv + \tyenv_1 + \cdots + \tyenv_n}
                    {\fnapp{(\linlambda{\seq{x : \tya'}}{\tyb'}{\tma})}{\seq{\vala}}}
                    {\tyb}
            }
            \\
            \stackseq{\tyenvb_2}{\tyb}{\framestack}
        }
        {
            \cseq
            { \plainenv }
            {
                \thread{(\linlambda{\seq{x : \tya'}}{\tyb'}{\tma})(\seq{\vala})}{\framestack}
            }
        }
    \end{mathpar}
}

where: 
\begin{itemize}
    \item $\tyenv \subtype \tyenv'$
    \item $\seq{\tya} \subtype \seq{\tya'}$
    \item $\tyb' \subtype \tyb$
\end{itemize}

By repeated applications of (a suitable generalisation of) Lemma~\ref{lem:substitution}, 
$
\tseq
                    {\tyenv' + \tyenv_1 + \cdots + \tyenv_n}
                    {\subst{\tma}{\seq{\vala}}{\seq{x}}}
                    {\tyb'}
$

Recomposing:

  {\footnotesize
    \begin{mathpar}
        \inferrule*
        {
            {
            \begin{array}{l}
                \plainenv = \strip{\scombtwo{\tyenvb_1}{\tyenvb_2}} \\
                \tyenvb_1 = \tyenv + \tyenv_1 + \cdots + \tyenv_n
            \end{array}
            }
            \\
            \inferrule*
            {
                \tseq
                    {\tyenv' + \tyenv_1 + \cdots + \tyenv_n}
                    {\subst{\tma}{\seq{\vala}}{\seq{x}}}
                    { \tyb' }
            }
            {
                \tseq
                    {\tyenv + \tyenv_1 + \cdots + \tyenv_n}
                    {\subst{\tma}{\seq{\vala}}{\seq{x}}}
                    {\tyb}
            }
            \\
            \stackseq{\tyenvb_2}{\tyb}{\framestack}
        }
        {
            \cseq
            { \plainenv }
            {
                \thread{\subst{\tma}{\seq{\vala}}{\seq{x}}}{\framestack}
            }
        }
    \end{mathpar}
}

as required.
\end{proof}
}
 
\end{document}